# The Nature of
# Gamma Ray Burst Supernovae

Zach Cano
Astrophysics Research Institute

A thesis submitted in partial fulfilment of the requirements of
Liverpool John Moores University
for the degree of
Doctor of Philosophy.

September 2011

Not only is the universe stranger than we imagine,
it is stranger than we can imagine.

- Arthur Eddington

# Declaration

The work presented in this thesis was carried out at the Astrophysics Research Institute, Liverpool John Moores University. Unless otherwise stated, it is the original work of the author.

While registered as a candidate for the degree of Doctor of Philosophy, for which submission is now made, the author has not been registered as a candidate for any other award. This thesis has not been submitted in whole, or in part, for any other degree.

Zach Cano
Astrophysics Research Institute
Liverpool John Moores University
Twelve Quays House
Egerton Wharf
Birkenhead
CH41 1LD
UK

JULY 2011

# Acknowledgements

There are many wonderful people that I would like to express my sincerest gratitude to, for without their support I would never have been able to start, let alone finish, my doctoral thesis.

First, I would like to thank all of the academics in the GRB collaboration, especially my supervisor David Bersier for his fierce and direct instruction that has given me a new perspective on how to approach the scientific process. I would also like to thank Shiho Kobayshi and Cristiano Guidorzi whom have offered me advice and help that is deeply appreciated.

Additionally, I would like to thank Dr. Robert Smith and Dr. Mike Hardiman at the University of Sussex, for whom I will be eternally grateful towards. Without their support and belief in me during my undergraduate degree I would not have chosen to undertake the PhD project in the first place.

The love, support and selflessness of my partner Miné has been a constant source of energy and inspiration to me. Her belief in my abilities gave me the confidence to complete my project, and her adoration towards me will never be forgotten.

I would especially like to thank my father Roy for his unselfish and generous support over the years, for without his financial assistance and endless round trips to the Isle of Thorn telescope in the Ashdown Forest, I would have not been able to even begin this doctoral project.

I would like to thank to my family for their love and support, and also the staff and fellow students here at the ARI. I would also like to acknowledge the financial support provided for my PhD by the Science and Technology Facilities Council.

The results in this thesis are based partially on observations made with the NASA/ESA Hubble Space Telescope, obtained at the Space Telescope Science Institute, which is operated by the Association of Universities for Research in Astronomy, Inc., under NASA contract NAS 5-26555. These observations are associated with programs # 10909 (GRB 060729) and # 11709 (XRF 100316D).

For Miné and her endless belief in me as a person.

For my father and his unselfish support.

For my family and their unwavering love.

# Abstract

Gamma Ray Bursts (GRBs) and Supernovae (SNe) are among the brightest and most energetic physical processes in the universe. It is known that core-collapse SNe arise from the gravitational collapse and subsequent explosion of massive stars (the progenitors of nearby core-collapse SNe have been imaged and unambiguously identified). It is also believed that the progenitors of long-duration GRBs (L-GRBs) are massive stars, mainly due to the occurrence and detection of very energetic core-collapse supernovae that happen both temporally and spatially coincident with most L-GRBs. However many outstanding questions regarding the nature of these events exist: How massive are the progenitors? What evolutionary stage are they at when they explode? Do they exist as single stars or in binary systems (or both, and to what fractions)?

The work presented in this thesis attempts to further our understanding at the types of progenitors that give rise to long-duration GRB supernovae (GRB-SNe). This work is based on optical photometry obtained for three GRB-SNe events: GRB 060729, GRB 090618 and XRF 100316D (an X-Ray Flash is similar to a L-GRB, but has a lower peak energy). For GRB 060729 and GRB 090618 we model the optical light curves and account for light coming from three sources: the host galaxy, the afterglow and the supernova. When we remove the host flux, and model the afterglow, the remaining flux resembles that of a SN, both in the shape of the light curve and the shape of the spectral energy distribution.

Our investigation of XRF 100316D and its spectroscopically-confirmed Ic-BL SN 2010bh is more detailed as we were able to obtain optical and infrared data in many filters, which we utilize to created a quasi-bolometric light curve that we model to determine physical parameters of the SN. We then apply our model to previous, spectroscopically-connected GRB-SNe and discuss our results in relation to those already published in the literature.

Finally, we collect from the literature almost all of the data available for the GRB-SNe, as well as a moderate sample of nearby Ibc SNe. We then perform two analyses: First, we compare the peak, absolute $V$-band magnitudes of the GRB-SNe with those of the local Ibc SNe. Next, we compare the widths of the GRB-SNe and Ibc SNe light curves in the $V$ and $R$ bands. When we make a few assumptions regarding the physical processes occurring in these events, we are able to make several conclusions regarding the nature of the progenitors of stripped-envelope, core-collapse SNe.

# Publications

In the course of completing the work presented in this thesis, the contents of Chapters 2, 3 & 4 have been submitted and accepted for publication in a refereed journal. Chapters 2 and 3, which investigates GRBs 060729 and 090618, are published in Paper (I), while Chapter 4, which is an in-depth analysis of XRF 100316D / SN 2010bh, has been published in Paper (II):

(I) Z. Cano, D. Bersier, et al. 2011, 'A tale of two GRB-SNe at a common redshift of $z = 0.54$', *Monthly Notices of the Royal Astronomical Society*, 413, 669C.

(II) Z. Cano, D. Bersier, et al. 2011, 'XRF 100316D / SN 2010bh and the Nature of Gamma-Ray Burst Supernovae', *The Astrophysical Journal* (in press, September 2011), arXiv:1104.5141

# Contents

# List of Figures

# List of Tables

Do not believe, just because wise men say so.
Do not believe, just because it has always been that way.
Do not believe, just because others may believe so.

Examine and experience yourself!

-The Buddha-

# 1

# Introduction

SECTION 1.1

## Supernovae

The study of supernovae (SNe) has opened a unique window for astronomers and astrophysicists to study the nature of the universe on a variety of scales: nucleosynthesis and particle acceleration provides a laboratory to study the very smallest scales while the use of type Ia SNe as standardisable candles has allowed cosmologists to ascertain the geometry and topology of the universe at the very largest scales.

Historically, SNe have been classified by their observed properties. Mostly, classifications are based upon their observed spectra but also to some extent on their optical light-curves (LCs). Many SNe are classified on their early spectra near maximum light, which consists of a thermal continuum and P-Cygni profiles of lines formed by resonant scattering. This means that the SN type are categorized on the basis of the chemical and physical properties of the outermost layers of the expanding ejecta.

The two classes of SNe are based around the detection (or non-detection) of hydrogen in the observed spectra (Minkowski 1941): type I SNe are lacking in hydrogen while type II SNe are not. Type I are further divided into subclasses Ia, Ib and Ic. Type Ia SNe are characterized by the presence of a strong absorption trough near 6150Å which is attributed to Si II. Type Ib SNe lack this absorption feature but display helium features while type Ic lack (or have extremely weak) Si II, H and He features. The latter two subtypes have only been distinguished relatively recently in history (Elias et al. 1985; Gaskell et al. 1986; Uomoto & Kirshner 1985; Wheeler & Levreault 1985) and are frequently referred to under the more general label of type Ibc. Type II SNe, which do have hydrogen in their spectra, are further classified according to additional observational features. A few sub-classes are usually mentioned in the literature where IIP (Plateau) and IIL (Linear) constitute the majority of type II SNe (see Figure 1.2). Additional subgroups include type IIb, which are thought to be transitional objects between type II and Ib due to the detection of $H_\alpha$ and $H_\beta$ P-Cygni features at early-times that then fade and are replaced by He lines akin to type Ib SNe (Filippenko et al. 1993). Type IIn SNe (narrow emission lines, usually H$\alpha$ at early-times; e.g. Schlegel et al. 1990) which were once thought to represent an additional sub-class of type II SNe, are now accepted to be "interacting SNe" (i.e. a SN *of any type* occuring within a dense hydrogen shell/envelope and are thus related to the amount of circumstellar material surrounding the progenitor rather than representing a specific type of explosion (e.g. Kotak et al. 2004).

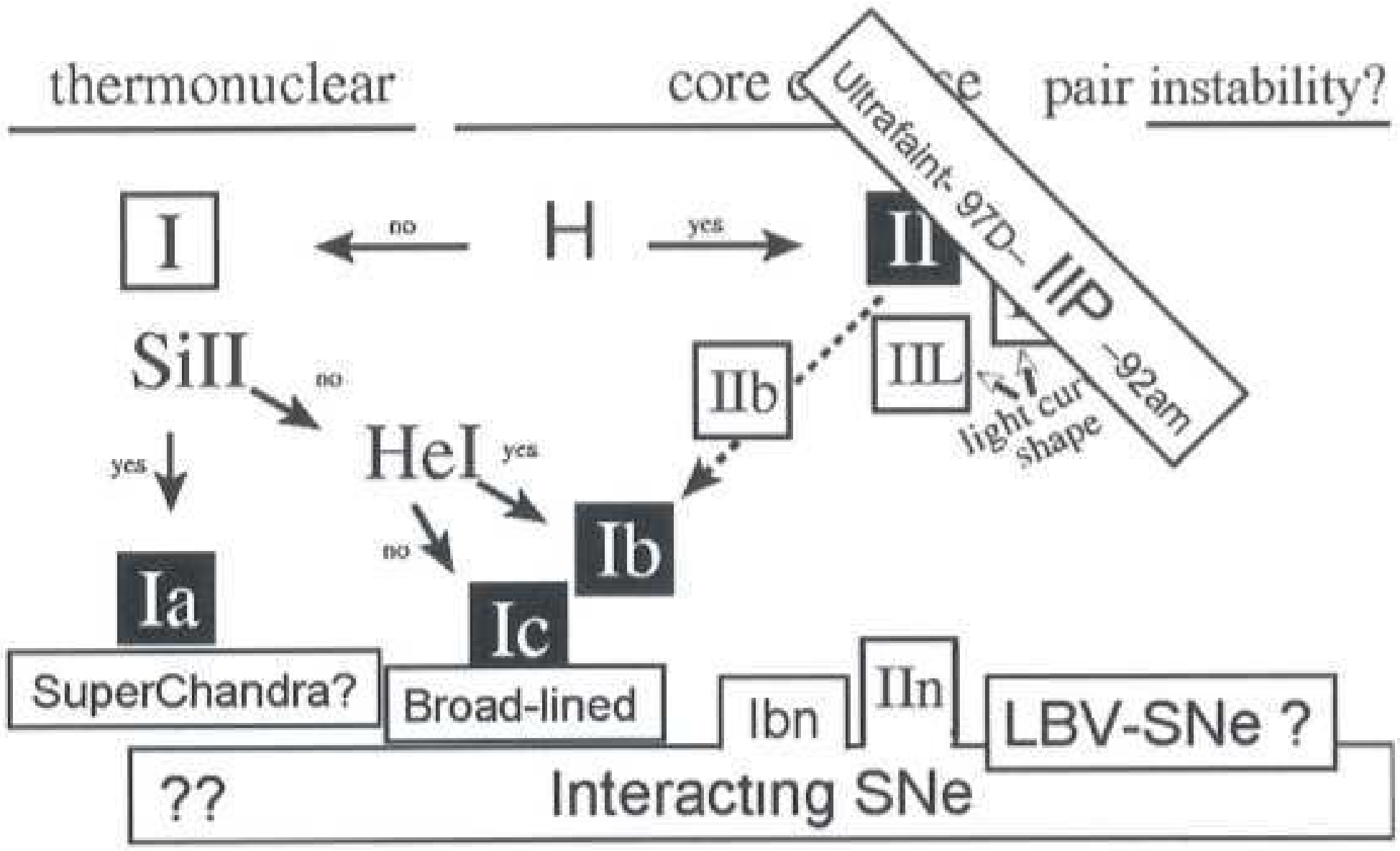

Figure 1.1 SNe classification. (Turatto 2007; revised version of original image from Turatto 2003).

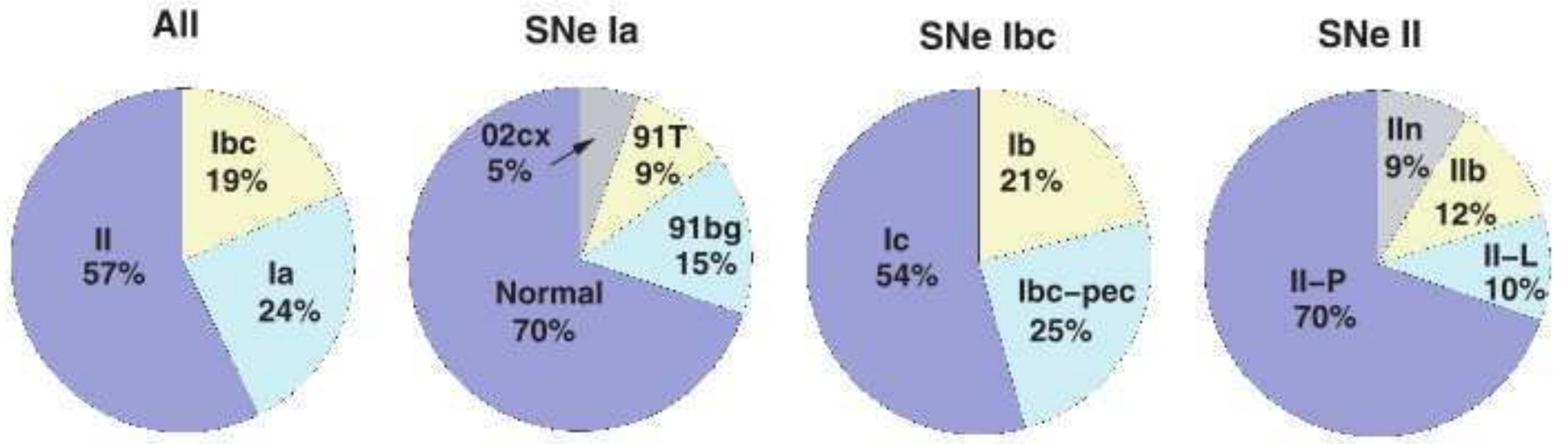

Figure 1.2 Observed fractions of the subclasses of SNe in a volume-limited sample as observed by Li et al. (2011) from the Lick Observatory Supernova Search.

Performing a nucleosynthesis study in 1960, Hoyle & Fowler first recognized two physically different groups of SN: Core-Collapse Supernovae (types Ibc & II), and Thermonuclear SNe (type Ia). Core-Collapse SNe (ccSNe) are the result of the gravitational collapse of the stellar nucleus of a massive star, where the conversion of gravitational potential energy into kinetic energy destroys the progenitor star. Thermonuclear-runaway SNe are the result of a thermonuclear explosion of a white dwarf (WD) where the accretion of material onto the WD takes it over the Chandrasekhar mass and an eventual SN is created.

However neat and structured the classification of SNe may at first appear, the reality is that there are many peculiar SNe that avoid any attempts to place them into tidy pigeon holes. Further still, the *actual* physical mechanisms that produce the SNe are not completely understood, and to date theorists still struggle to get their models to explode as the SNe we observe in the universe around us (e.g. Janka et al. 2007; Bruenn et al. 2009; Burrows et al. 2006). While many hours may have been consumed in a frustrated stupor, the current lack of clarity provides astronomers the exciting prospect of further understanding in the future.

SECTION 1.2

## Thermonuclear Supernovae

As mentioned above, type Ia SNe are thought to originate from the thermonuclear explosion of an accreting white dwarf (e.g. Branch et al. 1995). There are two sce-narios proposed in the literature to explain the explosion, both of which occur in a binary system. The *single-degenerate* scenario (Nomoto et al. 1982; Whelan & Iben 1973) is envisioned where a WD accretes hydrogen-rich material from either a Main-Sequence (MS) or Red-Giant companion until the accretion takes the WD over the Chandrasekhar mass and explodes. The *double-degenerate* scenario (Iben & Tutukov 1984; Webbink 1984) is thought to arise from the merging of two WDs through the loss of angular momentum from gravitational wave emission.

In both scenarios carbon ignition occurs in degenerate conditions at the centre/core of a carbon-oxygen WD. As the WD approaches the Chandrasekhar mass the stellar density and temperature increase rapidly, which eventually allows carbon fusion to occur. Convective burning of carbon continues in the stellar core until a region of the WD becomes unstable and is provoked into thermonuclear runaway, which marks the beginning of the SN explosion.

While the exact scenario of how the fuel is consumed and how the thermonuclear flame propagates through the degenerate stellar material is still yet to be fully appreciated theoretically, it is generally accepted that the propagation of the thermonuclear runaway can proceed in one of two ways: either by subsonic deflagration of the wave, where the flame is governed by electron thermal conduction (Nomoto et al. 1976), or by supersonic detonation where the flame is driven by shock waves (Arnett 1969).

The WD scenarios explains the majority of the main observational features of these objects and the fact they occur in all types of galaxies including ellipticals (i.e. it may take a Hubble time to accrete the critical mass (e.g. Woosley & Weaver 1986)), however the lack of hydrogen in their spectra is still difficult to explain within the single-degenerate model. The homogeneity of the explosion scenarios is seen in their observed properties (i.e. spectra and light-curve shape and brightness), which have led to their use as *standardizable candles*, and their application in cosmology is now an important part of scientific history. Two independent studies by Riess et al. (1998) and Perlmutter et al. (1999) used Type Ia SNe to show that the universe is not only expanding, but doing so at an accelerated rate. The use of the light-curves of type Ia SNe as standard candles has been developed during recent years, resulting in the derivation of empirical relationships such as the $\varnothing m_{15}$ relation (Phillips et al. 1993), the Multi-colour Light-Curve Shape method (MLCS) (Riess et al. 1996) and the stretch-method (Perlmutter et al. 1997).

SECTION 1.3

## Core-Collapse Supernovae

The ccSNe population consist of all of the type II SNe as well as types Ib and Ic. As mentioned previously, it is widely believed that ccSNe take place in massive stars with Zero Age Main Sequence (ZAMS) masses $\geq 8M_{\odot}$ at the end of a series of central nuclear burning episodes, the final burning process creating an iron core. The core eventually collapses and a compact object such as a neutron star or black hole is formed. The different configurations of the progenitors prior to explosion, as well as the different energies produced during the SNe and the interaction of the SN ejecta with circumstellar material produce the large variety of optical behaviour seen by astronomers.

Due to their association with massive stars, ccSNe are observed almost exclusively in star-forming galaxies, including spiral arms (e.g. Maza & van den Bergh 1976) and HII regions (e.g. van Dyk 1992) and have not been observed in elliptical galaxies.

1.3.1

### Type II Supernovae

Type II SNe are the most abundant type of ccSNe. Type II SNe display a wide variety of properties in their light curves and spectra (see Filippenko 1997 for a review). Four subclasses of type II SNe are usually mentioned in the literature: IIP (Plateau) and IIL (Linear) constitute the majority of all type II SNe (e.g. Barbon et al. 1979), and types IIn (narrow line) and IIb (an intermediate SN, with early features of type II SNe that are replaced by type Ib features at late times).

## 1. Introduction

The luminosity of type IIP SNe stops declining after maximum light and forms a plateau lasting roughly 100 days, during which it is believed that a recombination wave moves through the massive hydrogen envelope releasing its internal energy and powering the light curve (e.g. Chevalier 1976). In contrast, type IIL SNe show a linear decline in luminosity, likely due to the progenitor possessing a lower-mass envelope before exploding. These two classifications of SNe are not homogeneous and many intermediate cases exist with shorter plateau phases (e.g. SN 1992H; Clocchiatti et al. 1996B), indicating a continuous transition in shapes from type IIL to IIP (e.g. Young & Branch 1989). Since the duration of the plateau is likely related to the progenitor's stellar envelope (Litvinova & Nadyozhin 1983), it is generally accepted that envelopes of different masses and thicknesses constitute the continuous transition of light curves from IIL to IIP. Finally, it is worth noting that there are not any spectral differences between types IIP and IIL.

Type IIb SNe are thought to be transitional objects between type II ccSNe and Ib SNe as they possess hydrogen features in the early-time spectra (i.e. type II features) which are replaced by helium features in late-time spectra which are similar to type Ib SNe. Examples of type IIb SN are SN 1987K and SN 1993J in M81. Intriguingly, there are some events such as SN 2001ig and 2003bg that display these spectral features in the *opposite* order (i.e. they display helium features in the early-time spectra that are then replaced by hydrogen features in the late-time spectra).

Many peculiar type II SNe are grouped into the class of IIn. The spectra of these objects have a slow evolution and are dominated by strong Balmer emission lines but lack broad absorption features. It is commonly believed that the features we observe in these events are the result of interactions between the ejecta and a dense circumstellar medium (CSM), which transforms mechanical energy of the ejecta into radiation. The interaction of the fast-moving ejecta with the CSM generates a forward shock in the CSM and a reverse shock in the ejecta. An example of a type IIn SN is SN 2006gy (e.g. Smith et al. 2007).

### 1.3.2 Type Ibc Supernovae

Type Ib and Ic SNe have lost their hydrogen (Ib) and helium (Ic) envelopes prior to explosion, and are thus termed *stripped-envelope* SNe. The mass-loss is thought to arise from the increased mass of the progenitor relative to the progenitors of type II SNe, where higher stellar masses imply higher stellar winds (NB. increased metal content of the progenitor can also explain the higher mass-loss through line-driven winds; e.g. Puls et al. 1996; Kudritzki & Puls 2000; Mokiem 2007, thus suggesting that the progenitors of type Ibc SNe could also have higher metallicity than type II SNe), that blow away the outer layers of the progenitor star. However, it is also expected that a moderate to high percentage of Ibc SNe arise from progenitors that exist in binary systems (e.g. Smartt 2009). The role of binarity provides a plausible route for type Ibc SNe formation as interactions in the binary could strip the outer envelopes of the star prior to explosion (Podsiadlowski 1992).

The leading candidate progenitor of Ibc SNe are Wolf-Rayet stars (e.g. Gaskell et al. 1986). If some Ibc SNe arise from single, massive stars, then the large mass of Wolf-Rayet (WR) stars, as well as the large mass-loss observed (e.g. Barlow et al. 1981; Crowther et al. 2002), and their occurrence in binary systems (e.g. Pollock 1987) make them a natural candidate.

The environments of Ibc SNe also hold clues to their possible progenitors. Type Ibc SNe occur almost exclusively in late-type galaxies (e.g. van den Bergh et al. 2005), implying they arise from young, massive stellar progenitors that are likely to be more massive than type II progenitors.

Over the past decade or so, a particular subclass of type Ic SNe displaying very broad lines (Ic-BL) indicating high expansion velocities, is receiving attention due to their association with GRBs (e.g. GRB 980425 and its association with Ic-BL SN 1998bw; Galama et al. 1998). However, there are several Ic-BL events that, despite being as energetic as GRB-supernovae, though not as asymmetric, were not accompanied by a GRB-trigger (e.g. Ic-BL SN 2009bb; Pignata et al. 2011).

### 1.3.3 Faint & Peculiar Core-Collapse Supernovae

As previously mentioned, not all SNe observed fit into the aforementioned categories. Countless faint and peculiar SNe have been detected, that display either faint optical light-curves (e.g. peculiar SN 2008ha was only $M_V = -14.2$ at peak; Foley et al. 2009), or spectral features that are unlike the conventional ccSNe spectra. For instance, SN 1993R (Filippenko 1997) showed hybrid spectral features of type Ia *and* type Ibc at late epochs. Other examples include SN 2000er which is suspected to be a ccSN that lost its helium envelope shortly before explosion. SN 2000er which was classified as peculiar (Clocchiatti & Wheeler 2000; Maury et al. 2000) as it showed a broad maximum at $M_V \approx -19$, followed by a rapid decline of 5 mag in 50 days and an unusual spectrum that included iron and silicon species.

The observed properties of faint ccSNe are consistent with very small ejected nickel masses ($< 10^{-3}\ M_{\odot}$) and low explosion energies ($\ll 10^{51}$ erg) (Zampieri et al. 2003). This might suggest high-mass progenitors ($M_{\rm ZAMS} \geq 20 - 25 M_{\odot}$) for which material, which was initially expelled, cannot overcome the gravitational potential and falls back upon the newly-formed compact object (e.g. Colgate 1971). Fall-back is also thought to occur if the explosion energy is small. The inner part of the star falls back upon the central remnant and only the outer part of the star overcomes the gravitational potential. This outer layer may be ejected and observed as a SN (e.g. Fryer et al. 2007; Fryer et al. 2009). As this ejecta has a kinetic energy just above that needed to escape the gravitational potential, it is expected to have very low energy.

The existence of these faint SNe has been suggested in the framework of GRBs (see Section 1.11.3), for which there are several examples of nearby long-duration (e.g. GRBs 060505 and 060614; Fynbo et al. 2006) for which no SNe has been found to very deep optical limits (no brighter than $M_V = -13.5$). Tominaga et al. (2007) have shown that faint SNe with low energies and little nickel production are compatible with relativistic jet-induced black-hole forming explosions of massive stars.

SECTION 1.4

## The Core-Collapse Explosion Scenario

It is believed that the general explosion scenario applies to all ccSNe with stellar masses $M_{ZAMS} \geq 8M_{\odot}$. At the end of the helium-burning episode in the stellar core, the central temperature and density are high enough for the onset of carbon fusion. This scenario differs in stars that are less massive as carbon fusion cannot proceed and instead the fusion stops and the star eventually cools into a carbon-oxygen white dwarf.

In the more massive stars, carbon burning is followed by the burning of heavier elements up to iron, where a maximum is reached in the nuclear binding energy per nucleon, meaning no further energy can be released via nuclear fusion. At this stage the star can be envisioned as being comprised of concentric shells of ashes left over from the numerous stages of nuclear burning.

Once the iron core is formed, it continues to grow through silicon shell burning until it reaches its Chandrasekhar mass, and electron degeneracy cannot support it against collapse. The collapse is triggered at temperatures $T \sim 10^{10}$ K and densities $\rho \sim 10^{10}$ g cm$^{-3}$, where photo-disintegration of iron into free nucleons and alpha particles, as well as electron capture by protons, removes energy and electrons from the core (that helped support against collapse), as well as creating neutrinos, all of which accelerate the core-collapse. An inner core with mass of $0.6 - 0.8\ M_{\odot}$ begins to contract almost into a free-fall regime, and in a few hundredths of a second the central density reaches $\rho \sim 10^{14}$ g cm$^{-3}$ and neutron degeneracy halts the collapse. The outermost layers, still in a state close to free-fall, collides with the decelerated inner core, creating a shock-wave at the boundary of the inner and outer envelope.

Here is where theory debates the exact mechanism driving the expulsion of material from the star. Originally it was thought that the shock-wave created from the in-falling material was energetic enough to expel the outer layers of the star, but exten-sive modelling showed that there was not enough energy available to explain existing observations (e.g. Bethe 1990; Woosley & Weaver 1986, 1995). It was also theorized that the enormous neutrino flux (e.g. Colgate & White 1966) could power the supernova, but the cross-section of the interaction with the surrounding material is perhaps too small. Nevertheless, by $\sim 0.1$ s after the onset of the collapse, a "proto-neutron star", with a radius of $\sim 30$ km and mass $1.4 M_{\odot}$ is formed, and soon after, a BH may be formed if the proto-neutron star accretes enough material.

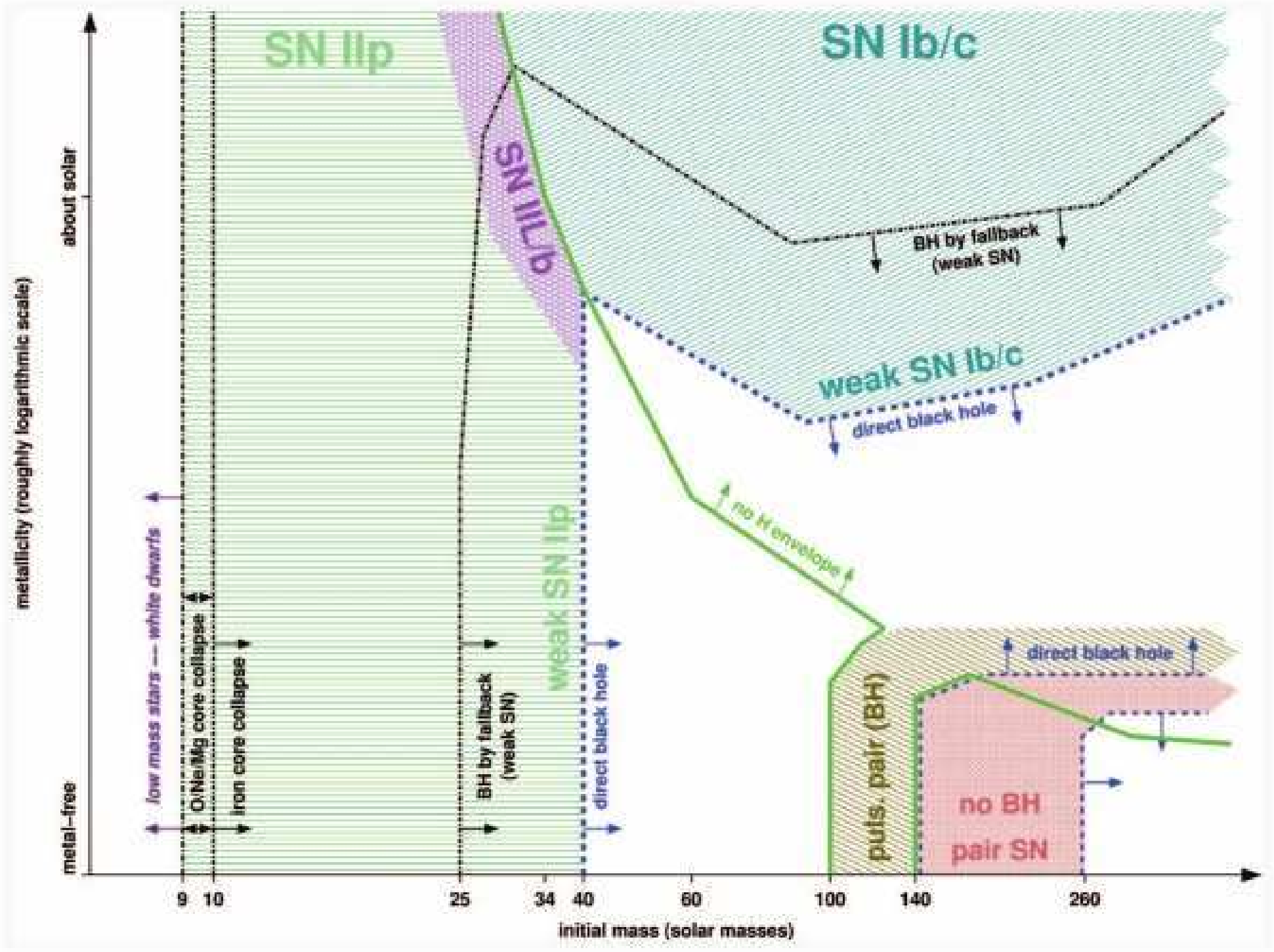

Figure 1.3 Supernova types that result from the collapse of non-rotating massive stars (Heger et al. 2003) as a function of main-sequence mass and progenitor metallicity. One key assumption of these models is that zero mass-loss is assumed for low-metallicity stars.

SECTION 1.5

## Progenitors of Core-Collapse Supernovae

As previously mentioned, the progenitors of all ccSNe are thought to arise from stars $\mathrm{M_{ZAMS}} \geq 8 M_{\odot}$. Hydrogen-rich type IIP SNe have been associated with red super-giant progenitors (Smartt 2009 and references therein) by direct observations of the progenitor stars in pre-explosion images (Smartt et al. 2004; Maund et al. 2005; Hendry et al. 2006; Li et al. 2006; Gal-Yam et al. 2007; Li et al. 2007; Crockett et al. 2008).

It is expected that the progenitors of type IIb and Ibc SNe arise from more massive progenitors than those of type II SNe. The larger mass (as well as increased metallicity) implies that more mass is lost by the progenitor before exploding, probably via strong, line-driven stellar winds (see Section 1.3.2). Figure 1.3 is from Heger et al. (2003) that determined the type of remnant and supernova that would result from the collapse of non-rotating, massive stars with varying metallicity and main-sequence mass. In these models the progenitors of type II SNe arise from progenitors with lower mass (regardless of progenitor metallicity) than those of Ibc SNe. Indeed in their models, Ibc SNe only arise from massive stars ($M_{ZAMS} \geq 25-30 M_{\odot}$) with a metal content roughly $\sim 0.5-1.0$ solar.

However, only the progenitors of type IIP SNe have been directly detected despite numerous attempts to find those of types IIL, Ibc and IIb (e.g. Gal-Yam et al. 2005; Maund et al. 2005; Crockett et al. 2008; Smartt et al. 2009). The exact identity of the progenitor stars of IIb and Ibc SNe remains a mystery, though indirect measurements as well as theoretical expectations suggest that high-mass ($M_{ZAMS} \geq 30-35 M_{\odot}$) Wolf-Rayet (WR) stars, as well as massive stars (that are less massive than those that are usually associated with WR stars) in binary-systems are likely candidates.

Indirect attempts to characterise the progenitors of these SNe are performed via statistical analyses of massive-star populations. Two methods are exploited in the literature, the first being the measurement of the number ratio of type Ibc SNe to type II SNe as a function of host-galaxy metallicity, using host luminosity as a metallicity proxy (e.g. Prantzos & Boissier 2003; Boissier & Prantzos 2009), and the second by direct metallicity measurements (e.g. Prieto et al. 2008). Both methods show that the ratio of Ibc to II SNe increases with increasing host metallicity, suggesting wind-driven mass loss plays an important role in the formation of stripped envelope WR stars exploding as type Ibc SNe and that at lower metallicities massive stars that might have ended as type Ibc SNe explode instead as type II SNe.

Attempts to extend these studies to quantify metallicity differences in the progenitors of Ib vs. Ic SNe have proved inconclusive, with two independent studies by Anderson et al. (2010) and Leloudas et al. (2011) showing no statistically significant difference in the metallicities at the sites of Ib and Ic SNe. This is at odds to Modjaz et al. (2011) who find that the sites of Ic SNe are systematically more metal-rich than those of Ib SNe, and Ic-BL (broad-lined Ic SNe) occuring at metallicities between Ib and Ic. Thus with no clear understanding whether the progenitors of Ic SNe are more metal-rich than those of Ib SNe, it is still uncertain whether the progenitor stars of Ic SNe have an increased rate of mass-loss via line-driven stellar winds than those of Ib SNe.

The location of a SN in its host galaxy can also be an indicator of the mass and metallicity of the stellar progenitor. Studies have shown that type Ibc SN are preferentially found in more central regions of their host than type II SNe (van den Bergh 1997; Tsvetkov et al. 2004; Kelly et al. 2008; Hakobyan 2008; Anderson & James 2009) with type Ic the most centrally concentrated. This implies a progenitor-metallicity sequence from type II-Ib-Ic, and supports the results of the number ratio analyses that type Ibc progenitors would explode as type II SNe in low metallicity regions.

However, the progenitors of GRB-SNe arise from a very special population of progenitors, which among other properties, are metal poor. It has been observed that all of the spectroscopically-linked GRB-SNe have been of type Ic-BL (i.e. Ic SNe with relativistically-moving ejecta), but arise from environments that are very metal-poor (see Section 1.10). Observations have showed that long-duration GRBs occur in faint, metal-poor galaxies, and Fruchter et al. (2006) showed that they occur in the regions of highest surface brightness of their host galaxy. This result was extended by Kelly et al. (2008), showing that Ic SNe also occur in the regions of highest surface brightness of their host galaxy, providing an indirect link between the two events. These results would suggest that some type Ic progenitor analogues in low-metallicity environments tend to explode as broad-lined Ic SNe, that are accompanied by a long-GRB, rather than a type II SNe.

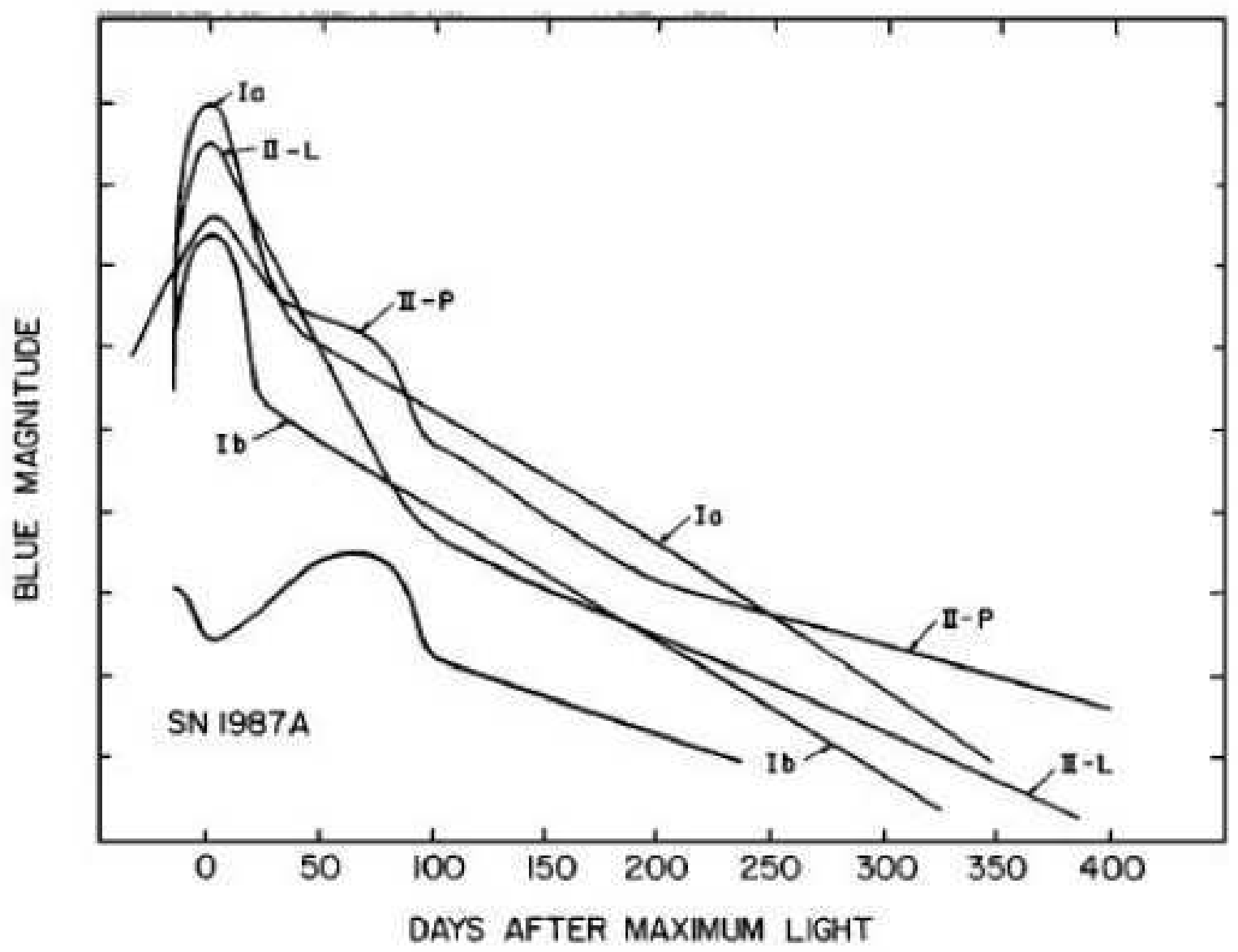

Figure 1.4 Schematic SN LCs (Filippenko 1997).

SECTION 1.6

## Supernovae Light Curves

The light curve (LC) of a supernova acts a signature of the energy released during the event (see Figure 1.4 for schematic light curves), and correct modelling of photometric and spectroscopic data can provide information of the progenitor mass and radius, the amount of mass ejected during the supernova and the amount of nickel produced during the explosion.

As the shock-wave erupts from the surface of the star, electromagnetic radiation is created, initially as a UV flash (or soft X-rays). The matter is highly ionised and electron scattering causes the stellar material to become opaque. As the material expands the surface area increases, causing a rise in the LC. The peak in the LC occurs as the temperature in the outer layers starts to decrease as the material expands and the diffusing radiation is cooled. Numerous early peaks that have been attributed to the shock-heated, expanding envelope including: type IIP SN 1987A (Early peaks in optical LCs: Hamuy et al. 1988, and P Cygni-like feature around 1500Å in UV light echo off of a dust cloud $\sim$ 300 pc from the SN: Gilmozzi & Panagia 1999); type

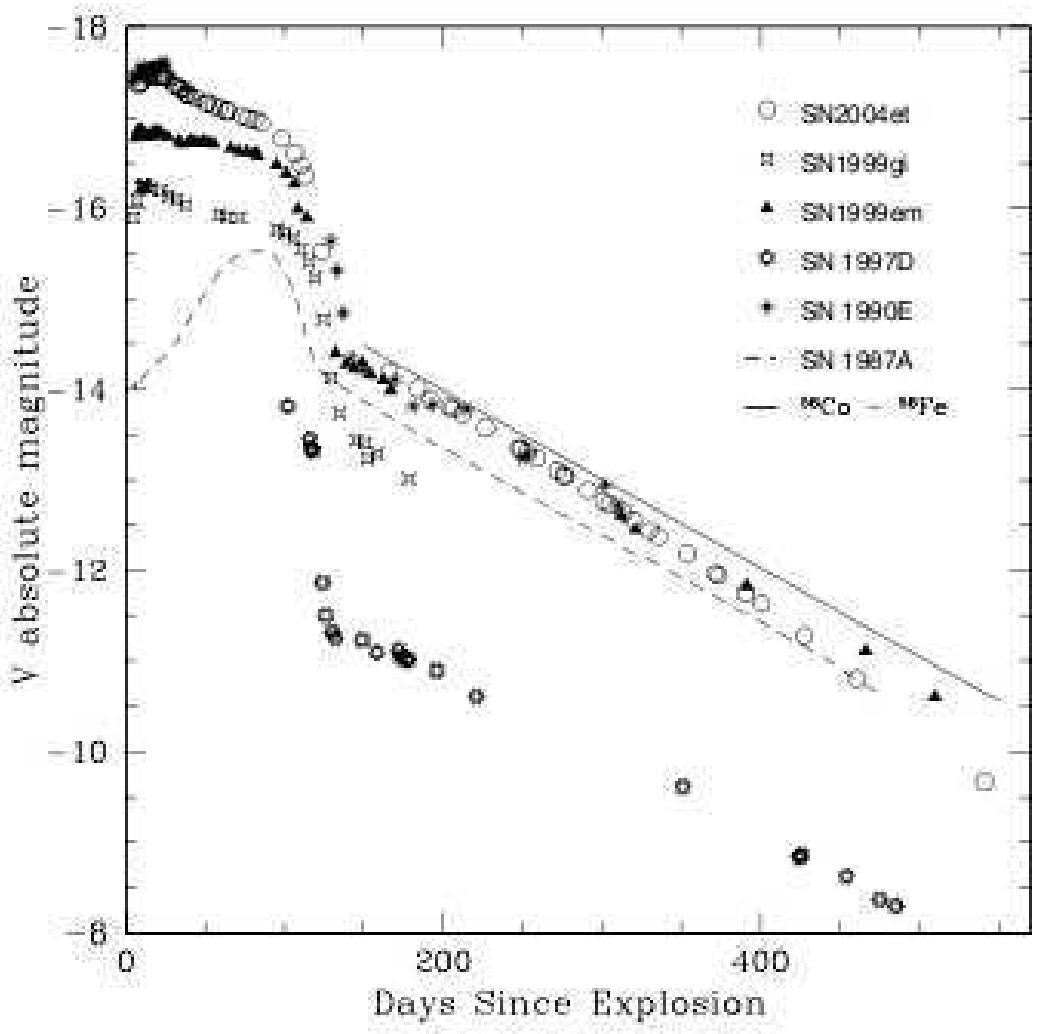

Figure 1.5 Type IIP SN 2004et along with several other type IIP SNe (Sahu et al. 2006). The length of the plateau phase corresponds to the time it takes for the hydrogen recombination wave the propagate through the dense stellar envelope. Afterwards the LC is powered purely by the radioactive decay of cobalt.

IIb SN 1993J (Early peak in optical light curves: Wheeler et al. 1993, Schmidt et al. 1993, Richmond et al. 1994, Lewis et al. 1994); type Ibc SN 1999ex (early optical peaks: Stritzinger et al. 2002); type IIP SNe SNLS-04D2dc & SNLS-06D1jd (UV-flash: Gezari et al. 2008); type IIP GALEX supernova SNLS-04D2dc (UV-flash: Schawinski et al. 2008); type IIn PTF 09UJ (UV-flash: Ofek et al. 2010); and type IIP SN 2010aq (early UV & optical peaks: Gezari et al. 2010).

For type IIP SNe (or any supernovae with an extended envelope of hydrogen), the plateau commences as hydrogen-rich zones expand and cool at temperatures $T \leq$ 5500 K. For typical densities, hydrogen recombines and releases trapped radiation. This *recombination wave* propagates inward in radius maintaining an approximately constant effective temperature. All the while the star is continuously expanding, the photosphere recedes deeper into the star and successive regions cool to the temperature of recombination. Since this temperature remains almost constant as the photosphere propagates through the hydrogen envelope, a plateau is created in the LC. The length of the plateau depends on the depth of the hydrogen envelope, and stars with smaller radii will have shorter and fainter plateaus.

After the hydrogen has recombined, the electromagnetic display from the energy deposited from the shock-wave quickly declines. The observed LCs are now powered solely by the radioactive decay of Ni-Co-Fe. Decay of Ni into Co deposits roughly $6 \times 10^{48}$ erg/(0.1 $M_{\odot}$) (Woosley, Heger & Weaver 2002) with a half-life of 6.1 days. The further decay of Co into Fe produces roughly $10^{49}$ erg/(0.1 $M_{\odot}$) with a half-life of 77.3 days. Most of the energy from nickel decay goes into accelerating the expansion of the interior of the supernova, with very little escaping. However, the decay of Co into Fe is much more significant. In Red Super-Giants cobalt decay is seen as the *radioactive tail* while in bluer stars (e.g. 1987A) cobalt decay dominates the LC from an early time ($>$ 20 days) and is responsible for the peak. For massive stars that still retain a small hydrogen envelope, a brief plateau merges into the cobalt-powered tail, producing the type IIL SNe LCs.

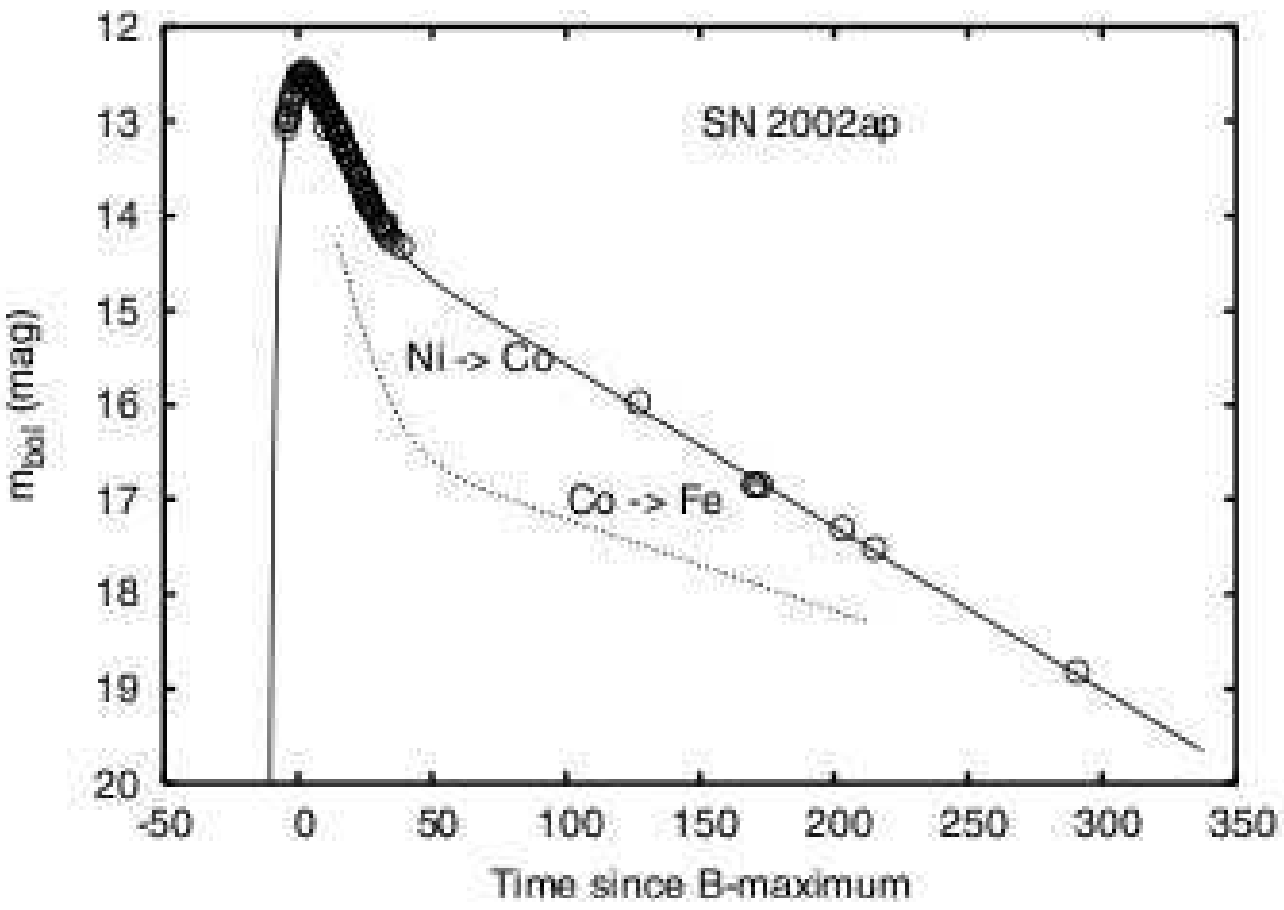

Figure 1.6 Bolometric LC of SN 2002ap (Vinkó et al. 2004) which is an energetic, broad-lined Ic SNe. The solid line is the analytical model of Vinkó et al. (2004) while the dotted line shows the rate of energy input by the decay of nickel into cobalt and cobalt into iron.

In stripped-envelope SNe (i.e. types Ib and Ic) nickel and cobalt decay powers the entire LC. Since the stellar progenitor has lost its hydrogen (Ib) and helium (Ic) envelopes, there is no plateau in the LC. Owing to the small progenitor radius ($\sim 1\ R_{\odot}$) the breakout transient is brief, faint and hard. The subsequent LC is then powered entirely by radioactivity, similar in this respect to type Ia SNe. The light curves of type Ib SNe are consistent with the production of $\sim 0.15\ M_{\odot}$ of nickel (Woosley, Heger & Weaver 2002), which is about a quarter of that produced in type Ia events. Because of their larger mass and lower average velocities, type Ib SNe trap the gamma-rays produced from cobalt decay very effectively, and in some cases their radioactive tails may track its half-life (e.g. Clocchiatti & Wheeler 1997). The LCs of type Ic SNe are very similar to type Ib SNe, with their main observational differences due to different spectral features.

Figure 1.7 Mosaic of GRB pulses, which are available from the public BATSE archive (http://gammaray.msfc.nasa.gov/batse/grb/catalog/).

SECTION 1.7

## Gamma Ray Bursts

More than forty years have passed since the serendipitous discovery of gamma ray bursts (GRBs) by USA military spy satellites (which were designed to detect gamma radiation emitted by the detonation of nuclear weapons in space, and thus contravening the Limited Nuclear Test Ban Treaty). Though GRBs were originally detected in the late 1960s, their existence was not made public until 1973 (Klebesadel et al. 1973; Strong et al. 1974). During the following decades many theories were developed to explain the nature of the progenitors of GRBs, sparking passionate debates regarding whether they occur at cosmological distances or within our galaxy/galactic halo. For many years the latter explanation was preferred due to the almost impossible energies implied if they did originate from other side of the universe.

No two gamma ray bursts are the same, and the high-energy, gamma-ray emission can last from a few milliseconds to thousands of seconds. Some pulses are smoother and exhibit little variability, while others consists of many sharp spikes. GRBs occur at random locations in the sky, and it was shown in the 1990s, with the Burst and Transient Source Experiment (BATSE) aboard the NASA satellite the Compton Gamma Ray Observatory, that GRBs are isotropically distributed in the sky, and thus probably located at cosmological distances (Meegan et al. 1992). Using the same data collected by BATSE, it was shown that GRBs have a bimodal temporal and spectral distribution (Kouveliotou et al. 1993) and can be roughly grouped/classified as either: (1) Short, Hard Bursts (S-GRBs) having a burst duration $< 2$ s and a harder spectrum, or (2) Long, Soft Bursts (L-GRBs) with a burst duration $> 2$ s and a softer spectrum. Although the exact nature and number of sub-classes is still debatable (for example, X-Ray Flashes, or XRFs, are thought to be low-energy "cousins" of long, soft GRBs, with even softer spectra but durations similar to long, soft GRBs), there is general agreement that short, hard GRBs are a distinctly different sub-class to long, soft GRBs. Recently, additional evidence for a distinction between these sub-classes has come from an investigation of an energy-dependent pulse delay/lag in the time histories of short GRBs (e.g. Norris et al. 2010), for which none was found. This is in contrast to long, soft GRBs for which a delay is clearly observed (NB: by definition, a positive spectral delay/lag occurs when high-energy photons arrive before the low-energy ones).

Great strides in understanding the GRB phenomena were made in 1997, which proved to be a landmark year thanks to the launch of BeppoSAX, an Italian-Dutch satellite. BeppoSAX was constructed with an X-ray telescope that, once a GRB was detected, could precisely localize an afterglow at X-ray wavelengths (to within a few arc-seconds) and relay the afterglow position to ground-based telescopes. This was first done successfully for GRB 970228, which lead to the first detected optical afterglow (Costa et al. 1997), and heralded the start of the "Afterglow Era".

A few months later the redshift of GRB 970508 was measured ($z = 0.835$, Metzger et al. 1997), finally proving beyond doubt that GRBs lie at cosmological distances. The same burst also proved to be important for understanding the physical processes that occur during and after a GRB. Radio emission for GRB 970508 was detected, and initially showed large interstellar scintillation that was later suppressed. This presented clear evidence of superluminal expansion of the ejecta that produced the GRB (Frail et al. 1997).

Substantial progress regarding the nature of GRB progenitors was made the next year when GRB 980425 was spectroscopically & photometrically linked with type Ic-

Figure 1.8 Anatomy of a GRB. A GRB can be loosely broken down to five main components: (1) the progenitor, (2) the central engine, (3) the outflow properties, (4) the prompt emission, and (5) the long-lasting afterglow. Original cartoon was first published in Scientific American, Gehrels et al., Dec. 2002.

BL SN 1998bw (Galama et al. 1998). This was the first direct evidence for the Gamma Ray Burst-Supernova (GRB-SN) connection, which was predicted by theory contemporaneously with the first GRB detection in the 1960s (Colgate 1968). This discovery was further advocated in 2003 with the discovery and spectroscopic & photometric links between GRB 030329 / SN 2003dh (Stanek et al. 2003, Hjorth et al. 2003) and GRB 031203 / SN 2003lw (Malesani et al. 2004). These three bursts provided clear, irrefutable evidence that the progenitors of some, if not all long, soft GRBs, are due to the explosion of a massive star resulting in a very energetic, type Ic supernova.

SECTION 1.8

## Anatomy of a GRB

The production of a burst of gamma rays is largely independent of the progenitor. In Section 1.9 the most widely-accepted progenitors models for short and long GRBs will be presented, while in this section we will look at the physical scenario that leads to the production of a GRB.

The physical processes that ultimately lead to the production of a GRB are displayed in Figure 1.8, and can be loosely broken down into: (1) the progenitor, (2) the central engine, (3) the outflow properties, (4) the prompt emission, and (5) the long-lasting afterglow.

A general picture has been developed over the past of couple decades, with the most widely-accepted physical model for GRB-production being the "Fireball Internal/External Shock Model" (see Piran 2004 for an extensive review), where the initial GRB and subsequent afterglow radiation is created by the dissipation of collisionless internal (prompt emission) and external (long-lasting X-ray, optical & radio afterglow) shocks. This model is largely independent of the true nature of the exact progenitor, and only requires that an extremely large amount of energy is deposited into a very small volume of space. This then inevitably leads to the production of a highly energetic and relativistic outflow that eventually interacts with the surrounding medium.

In the model, fireball shells are ejected by the central engine and propagate away from the central engine. Multiple shells are ejected with different energies and moving at different (relativistic) velocities. The interaction of overtaking shells causes them to heat up via "internal" shocks (Rees & Mesźaros 1994; Paczyński & Xu 1994), ultimately leading to the production of gamma rays, which is termed the "prompt emission". The emitted radiation is not thermal, and is usually attributed to being synchrotron (e.g. Rees & Mesźaros 1992; Mesźaros & Rees 1993), though additional processes such as inverse Compton radiation might also contribute to the observed gamma-ray radiation. Electrons in the shock-heated outflow are accelerated and then cool, radiating their energy in the form of synchrotron radiation. The transient, and long-lived afterglow that is observed at all wavelengths from X-ray to radio is attributed to the interaction of the relativistic outflow with the surrounding medium. "External" shocks are created and the electrons in the shocked material radiate their gained energy as synchrotron radiation.

Although the launch of the *Swift* satellite has revealed many complexities of GRB behaviour that are not yet adequately explained, a large fraction of GRB behaviour is explained by attributing the radiation to being synchrotron in origin. This assumption explains the temporal and spectral power-law behaviour of the GRB afterglows, where the observed flux is related to the frequency and time by $f_{\nu,t} \propto \nu^{-\beta} t^{-\alpha}$.

The central engine that is produced either by the collapse of a massive star (long, soft GRBs) or from the merger of binary compact objects (short, hard GRBs) is likely to be a compact object such as a rapidly rotating neutron star (NS) or a black hole (BH) that accretes material from a debris disk.

Many current models propose the presence of a stellar black hole of several solar masses that accretes material from a debris disk/torus at a rate of $0.01 - 10 M_\odot$ s$^{-1}$ (e.g. Popham et al. 1999). One advantage of these models is that BH-accretion is known to produce relativistic jets in other systems such as Active Galactic Nuclei (AGN) and micro-quasars. The millisecond magnetar model (see below) is another popular central engine candidate. In this scenario, the reservoir of rotational energy of the rapidly-spinning magnetar is in the range required for GRB-production, and the winds produced by the newly-formed magnetar are highly relativistic. The advantage of these types of models is due to the physics of these scenarios being well understood, though the total energy left to power a GRB and a supernova is less than that for BH central engines.

Figure 1.9 GRB 060729 (Grupe et al. 2007). A plateau phase is seen in the X-ray and optical LCs up to $\sim 10^5$ s, and is explained by Xu et al. (2009) as the consequence of prolonged energy injection by a millisecond magnetar.

Regardless of the model invoked, one outcome of many years of observing GRBs has been the realization for the need of *prolonged central engine activity* to explain many observations. That is to say, the need for energy injection into the fireball that continues for time-scales much longer than the prompt emission. Observationally, this manifests itself as highly-variable, early-time LCs, such as X-Ray Flares, as well as LC "plateaus" (e.g. GRB 060729, Grupe et al. 2007; Xu et al. 2009; see Figure 1.9). The millisecond magnetar model provides a natural explanation for prolonged central engine activity via the strongly-magnetized winds, while extended episodes of fall-back of material onto a central BH can also provide prolonged energy injection into the expanding fireball.

A key ingredient of all models is that the material/outflow responsible for GRB production must be moving relativistically. This is a consensus of *all* models. The argument requiring relativistic motion is for a low optical depth for two-photon pair production ($\gamma\gamma \rightarrow e^- e^+$). If one considers the size of the emitting region, one might naively expect (without relativistic motion) the emission area to be $c\delta t$ (where $\delta t$ is the observed time-scale of the prompt emission). If a fraction of the emitted photons pos-sess energies higher than the two-photon pair production threshold, the optical depth is very large, implying that all of the gamma rays should be attenuated before reaching the earth (see Zhang & Mesźaros 2004 for a detailed calculation). This apparent paradox is referred to as the “compactness problem” and can be overcome in two ways by requiring that the emitting material be moving relativistically. First, if the emitting region is moving relativistically towards the observer, the photon energy is blue shifted so that the observed gamma-rays are emitted as X-rays in the co-moving frame, and thus drastically reducing the amount of photons above the pair production threshold. A second effect is that the real physical size of the emitting region is larger by a factor of $\Gamma^2$, where $\Gamma$ is the bulk Lorentz factor of the outflow.

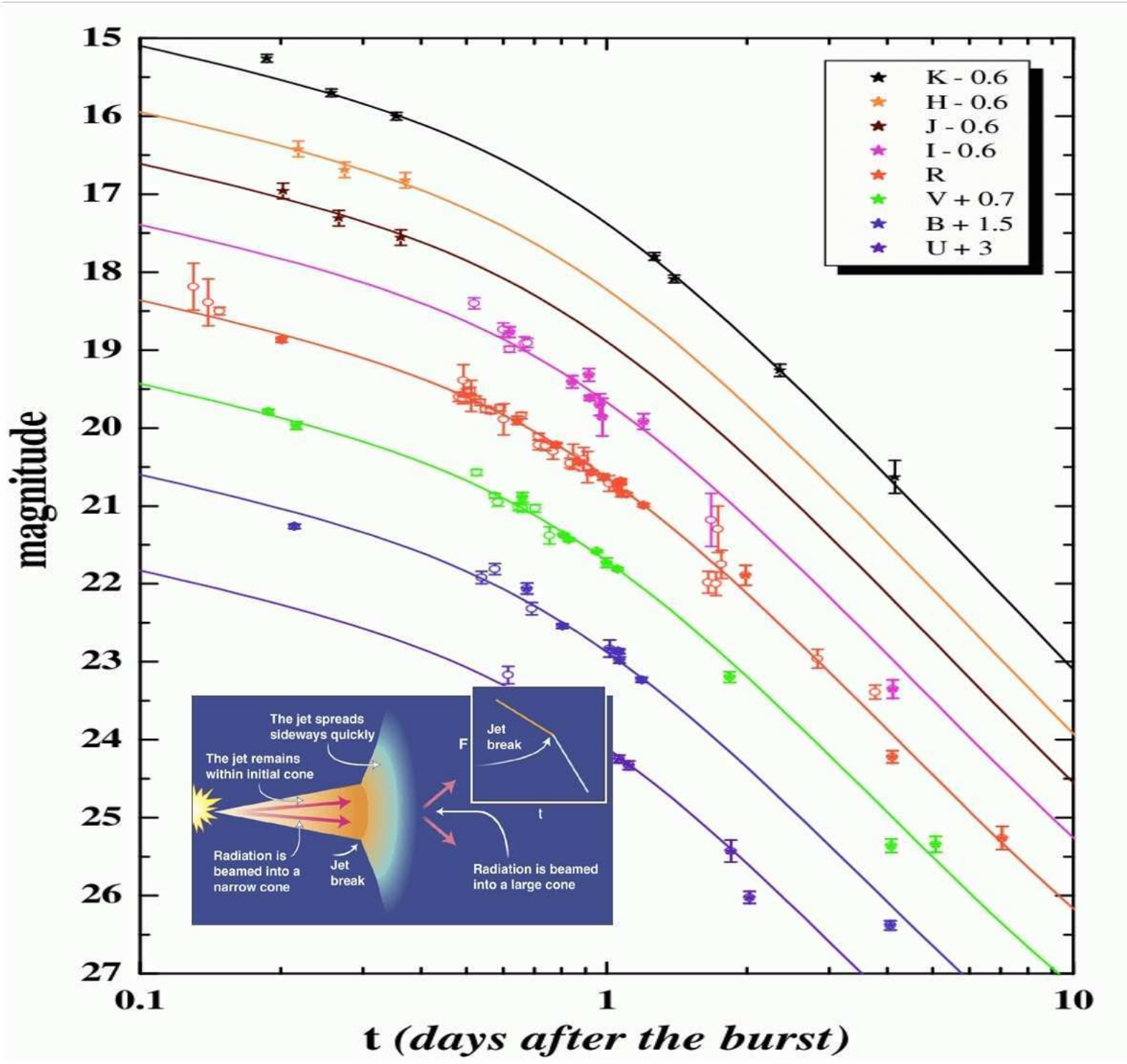

Figure 1.10 GRB 030226 (Klose et al. 2004). An achromatic break is seen in the optical and infrared LCs that is attributed to the sideways expansion of the jet. *Inset*: cartoon of the effect that the sideways expansion of the jet has on the optical LC. Original cartoon was first published in Science, February 2002.

Like other high-energy phenomena such as AGNs, it is widely accepted that the gamma ray and afterglow emission is collimated (i.e. “beamed” into a jet). The observed evolution of X-ray and optical LCs often displays achromatic breaks that are attributed to the jet-like geometry of the outflow (e.g. Rhoads 1999; Sari et al. 1999; see Figure 1.10). The break is explained by the deceleration of the collimated outflow, which slows down to a point where the jet expands sideways. The expansion of the jet causes more radiation to be emitted away from the line of sight, causing a break in the observed LC due to the detection of less photons. In addition, the implication that GRBs are beamed relaxes the total energy emitted by a factor of $\frac{\theta^2}{2}$, where $\theta$ is the opening angle of the jet, and also implies that the true rate of GRBs throughout the universe is substantially larger than once thought (Frail et al. 2001).

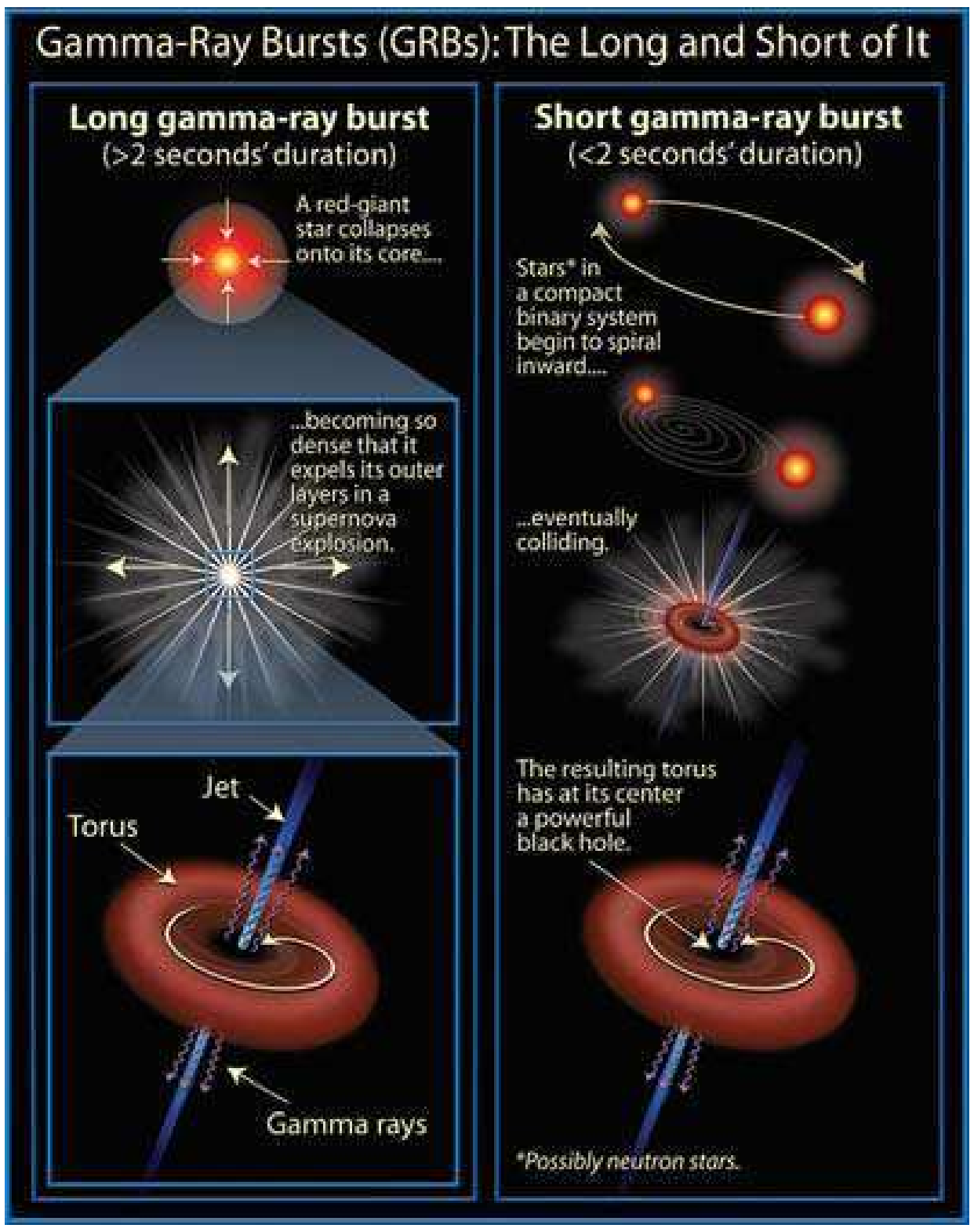

Figure 1.11 Cartoon of the likely progenitors of long and short duration GRBs. Short GRBs are thought to arise from the merger of binary compact objects such as NSs and/or BHs. Long GRBs are thought to be produced during the collapse of a massive star. Image courtesy of http://hubblesite.org.

SECTION 1.9

# GRB Progenitor Models

1.9.1

## Short GRBs

Due to the sparseness of observational data of short, hard GRBs, progress towards understanding their progenitors has been limited. In 2005 a breakthrough occurred following the detection of the first short GRB afterglows (e.g. Gehrels et al. 2005; Castro-Tirado et al. 2005) following their detection by *Swift* and *HETE-2*. Detections of short GRB afterglows led to observations of the host galaxies (and measurements of the host redshifts) in which this sub-class of GRBs occur. Observations of short GRBs have established that they are cosmological relativistic sources that do not arise from the collapse of massive stars, and most likely arise from a different physical phenomena (Nakar 2007).

One of the big open questions regarding GRB phenomenology is the nature of the progenitors of short GRBs and the processes that lead to the formation of the GRB and the central engine (Nakar 2007). Observations of the environments of short GRBs (see below) have led to recent developments in the progenitor models. The leading progenitor model describes the coalescence of compact binary objects, such as a binary neutron star or a neutron star-BH binary system. The merger of compact objects was first proposed in the 1980s as a possible progenitor of GRBs (e.g. Blinnikov et al. 1984; Paczýnski 1986; Goodman 1986; Eichler et al. 1989) and developed further by other groups (Narayan et al. 1992; Paczýnski 1991; Mochkovitch et al. 1993).

Binary mergers are natural candidates as they occur at reasonable rates and the amount of energy that is liberated is large enough to power a GRB. Additionally, in current models the central engine that is formed is similar to that in long GRBs (i.e. an accreting BH). The similarity in the central engines partially explains the similarities in the observational properties between the two classes of GRBs. It is interesting to note that the duration of the GRB is governed by the lifetime of the accretion disk, with the disk formed via merging binaries being consumed in a fraction of a second, while in the collapsar model (see below) in-falling material from the collapsing star feeds the disk for a longer time.

### 1.9.2 Long GRBs

Due to the vast amounts of observational data, much more progress has been made towards understanding the progenitors that give rise to long, soft GRBs. The most popular models (see Woosley & Bloom 2006 for an extensive review) fall under the general banner of "Massive Star Models", with the leading candidates being: (1) the Collapsar Model (Woosley 1993; MacFadyen & Woosley 1999) and (2) the Millisecond Magnetar Model (e.g.Usov 1992; Thompson 1994; Meszáros & Rees 1997; Wheeler et al. 2000).

#### Collapsars

To form a "collapsar", it is necessary to form a BH in the middle of a massive star with sufficient angular momentum to form an accretion disk. The required angular momentum is at least that of the last stable orbit around a BH of several solar masses. Providing a disk and BH form, the greatest uncertainty in this model is the mechanism for turning the binding energy of the disk (or BH rotational energy) into beamed relativistic outflows (Woosley & Bloom 2006). The most likely mechanisms are neutrinos (e.g. Woosley 1993; Popham et al. 1999), magnetic instabilities in the disk (e.g. Blandford & Payne 1982; Proga et al. 2003) and magneto-hydrodynamic extraction of the rotational energy of the BH (e.g. Blandford & Znajek 1977).

In the collapsar model the GRB and SNe derive their energies from different sources. The nickel that powers the supernova is produced by a disk "wind" (e.g.

MacFayden & Woosley 1999; MacFadyen 2003) with a velocity of $\sim 0.1\ c$. A GRB is produced when roughly 1 $M_{\odot}$ of material accretes onto the BH and gets ejected. The material is ejected into a jet, with the nickel created in the outflow being ejected in a larger jet surrounding the ejecta that leads to a GRB (though nickel can also be created at other angles during the explosion).

A prediction of the collapsar model is that the central engine remains active for a time much longer than the duration of the GRB, and may contribute to the GRB afterglow at later times (e.g. Burrows et al. 2005). This is because the jet and disk wind are inefficient at ejecting all of the material, and some falls back and accretes onto the BH at later times (MacFadyen et al. 2001).

## Millisecond Magnetars

The millisecond magnetar model has been developed by many groups (e.g. Usov 1992; Thompson 1994; Mesźaros & Rees 1997; Wheeler et al. 2000; Drenkhahn & Spruit 2002; Lyutikov & Blackman 2001; Lyutikov & Blandford 2003). The general idea is that the energy source for GRB production comes via the rotational energy of a highly-magnetised neutron star with an initial period of $\sim$ one millisecond (i.e. rotating near breakup). For the expected rotational velocity and magnetic field strengths, it is expected that there is enough energy available to power a GRB as well as an accompanying supernova. The strengths of such models arise from the relation between GRBs and known objects such as neutron stars, as well as energy scales that are about right for a neutron star rotating near breakup. Further developments are still needed however to explain how accretion can be reversed so as to not form a BH (e.g. Fryer & Warren 2004).

SECTION 1.10

# Environments of GRBs

1.10.1

## Short GRBs

The environments/host galaxies of short GRBs provide clues to their progenitors. Short GRBs occur in both early- and late-type galaxies as well as field and cluster galaxies (Nakar 2007). This is in contrast to host galaxies of long, soft GRBs that occur mostly in small, blue, irregular, star-forming galaxies. It is also seen that the specific star formation rate of the host galaxies of short GRBs are much lower than that seen for long GRBs, with the latter having $\sim 10 M_{\odot}$ yr$^{-1}$ $(L/L_{*})^{-1}$ (e.g. Christensen, Hjorth & Gorosabel 2004), and the former having $\leq 1 M_{\odot}$ yr$^{-1}$ $(L/L_{*})^{-1}$ (e.g. Nakar 2007). (NB: $L_{*}$ is a characteristic galaxy luminosity. An $L_{*}$ galaxy is a bright galaxy, roughly similar to the luminosity of the Milky Way. A galaxy with $< 0.1 L_{*}$ is a dwarf). The differences in the environments of short and long GRBs are a main reason for believing that they arise from different physical phenomena.

Short GRBs that occur in early-type galaxies are highly likely to be associated with old stellar populations, and even the presence of short GRBs in late-type galaxies does not necessarily imply that they arise from younger progenitors. For example, the host galaxies of short GRBs 050709 & 051221 (Covino et al. 2006; Soderberg et al. 2006, respectively) both showed evidence for a considerable population of stars older than $\sim 1$ Gyr, while for short GRB 050709 (Fox et al. 2005), though it occurred in a late-type galaxy, it was shown that it was not associated with any of the star-forming regions within the galaxy.

The offset of the occurrence of short GRBs from the host centres also supports the notion that short GRBs arise from different progenitors than long GRBs. In a recent paper by Fong et al. (2010), the authors exploited *HST* observations of 10 short GRB host galaxies, finding that on average, short GRBs occur *five* times further away from the apparent centre of the host galaxy than long GRBs. The increased offset is in agreement with expectations that short GRBs arise from the merger of compact objects such as NS-NS or NS-BH that merge very far from where they originated. By the time a binary neutron star merges it has travelled many kpc from its origin, having acquired a large velocity as a consequence from two SN explosions (Paczyński 1998). This is in contrast to the massive progenitors of long GRBs, which given their much shorter lifespans ($\sim 5 - 10$ Myr), do not travel very far from their birth place. However, since host galaxies have only been found for roughly 50% of short GRBs detected by *Swift*, the sample of short GRBs likely suffers from selection effects.

### 1.10.2 Long GRBs

Long, soft GRBs are found almost entirely in blue, dwarf, irregular galaxies that are undergoing star-formation. There is also spectroscopic evidence that typical GRB host galaxies are perhaps forming stars at a higher rate per unit mass than field galaxies (e.g. Djorgovski et al. 2001; Christensen Hjorth & Gorosabel 2004). While most GRB hosts are small and blue, a few nearby XRFs and GRBs have been found in spiral galaxies (e.g. GRB 980425), and they have been found to occur near or in the star-forming regions of the host galaxy. It has also been seen that long GRBs occur preferentially in regions of highest surface brightness of the host galaxy (Fruchter et al. 2006), which is similar to that observed for type Ic SNe (Kelly et al. 2008), thus providing another indirect connection between the two events.

A theoretical expectation is that the progenitors of GRBs have a low metal content. Mass-loss in massive stars comes via line-driven winds, and as mass-loss also removes angular momentum (which is needed to power a GRB), one early prediction was that GRBs arise from progenitors with low metallicity. Indeed low metallicity has been observed for several local long GRB sites (e.g. Sollerman et al. 2005; Stanek et al. 2006; Modjaz et al. 2008), as well as more distant host galaxies (e.g. Fynbo et al. 2003; Gorosabel et al. 2005; Fruchter et al. 2006). A recent paper by Savaglio et al. (2009) found evidence for sub-solar metallicity of 17 host galaxies of long GRBs, with an average metallicity of $\frac{1}{6}$ solar. Additionally, the SNe that accompany long GRBs also have a lower metal content than broad-lined Ic (Ic-BL) SNe, which is an interesting result as Ic-BL SNe have most of the requirements for becoming a GRB such as large stellar mass, loss of hydrogen envelope, and asymmetry.

It has also been observed that a large fraction of the host galaxy systems have either recently undergone, or are currently undergoing some interaction with a neighbouring galaxy (e.g. GRB 970828, Djorgovski et al. 2001; GRB 060923A, Tanvir et al. 2008; GRB 051022, Graham et al. 2009; see also Wainwright et al. 2007 for a review). It is known that galaxy-galaxy interactions are known to trigger intense periods of star-formation (e.g. Joseph et al. 1984) as well as producing a wide-range in galactic morphology, as observed by Conselice et al. (2005).

### 1.10.3 Dark GRBs

Only a few months after the detection of the first optical afterglow in 1997, the discovery of GRB 970828 (Groot et al. 1998) showed that not all GRBs have an optical afterglow. Since then it has been noted that as many as 50% of all GRBs do not have an optical or infrared afterglow, and are termed "dark GRBs". It seems that there is not a single reason for the occurrence of dark bursts, but several, including: (1) heavy dust extinction in the region or along the line of sight of a GRB (2) instrument detection limits (either slow response or too faint to be detected) (3) intrinsically faint GRBs, and (4) high-redshift bursts, where photons redshifted into the optical passbands are absorbed by neutral hydrogen in the host galaxy and inter-galactic medium.

Countless papers have explored these issues, and have come up with ways to classify a burst as dark (e.g. Jakobsson et al. 2004, Van Der Horst et al. 2009). More often then not, dust obscuration is indicated as a main cause for the non-detection of a GRB at optical wavelengths (e.g. Piro et al. 2002, Perley et al. 2009, Pellizza et al. 2006, Nakagawa et al. 2006, Djorgovski et al. 2001). In cases where the host galaxy was identified and observed it was generally found that dust and molecular clouds are unevenly distributed around the progenitor (Perley et al. 2009, Reichart & Price 2002, Piro et al. 2002). Such clouds can obscure light coming from an optical afterglow, leading to their non-detection on earth. Dark GRB host galaxies have also been detected as very red in colour, (e.g. GRB 020127, Berger et al. 2007; GRB 030115, Levan et al. 2006) giving further indications of the dusty nature of GRB host environments.

The effect of galaxy-galaxy interactions also adds to the amount of obscuring material between the GRB and earth, as it is seen that interactions between galaxies strip large amounts of gas & dust from the dust layers in the galactic centers and can create long streams of material between galaxies, as seen in the host galaxy system of GRB 970828 (Djorgovski et al. 2001).

However, extinction by dust is not the only cause of dark GRBs. The dark nature of GRB 051022 (Castro-Tirado et al. 2007) and GRB 000210 (Gorosabel et al. 2003) could not be explained by dust, where low values for rest-frame and foreground extinction were derived. Neither did these events occur at a high redshift (GRB 051022: z=0.809; GRB 000210: z=0.842), where the optical flux would be absorbed by intergalactic neutral hydrogen. Thus, these bursts could imply a separate class of intrinsically-faint GRBs, as supported by the conclusions of the analysis of dark GRB 051028 (Urata et al. 2007) and dark GRB 060108 (Oates et al. 2006).

SECTION 1.11

## The GRB-SN Connection

To date a large and convincing amount of direct and indirect data have been obtained that provide a connection between long, soft GRBs with a unique sub-class of stripped-envelope, core-collapse Ibc SNe. To date, all of the spectroscopically-connected GRB-SNe have been of type Ic-BL. However the converse is not true as there are many Ic-BL SNe that have not had an accompanying GRB-trigger.

1.11.1

## Direct Evidence

The first direct connection between a long GRB and a Ic-BL SNe came in 1998 when SN 1998bw was seen to occur both spatially and temporally coincident with GRB 980425 (Galama et al. 1998). Spectroscopy of SN 1998bw reveals broad emission lines indicative of material moving at velocities as high as $60,000$ km s$^{-1}$ a couple of days after the burst, and at velocities $\sim 20,000$ km s$^{-1}$ near peak.

However, there was a lot of controversy surrounding the nature of GRB 980425 in relation to other GRBs observed to date. The high energy emission of GRB 980425 was significantly less than was observed for "cosmological" GRBs. For GRB 980425, which occurred in a nearby, late-type galaxy ($z = 0.0085$, Tinney et al. 1998) it is estimated that the isotropic emission in gamma-rays is $\sim 1 \times 10^{48}$ erg (Amati 2006), whereas for a "typical" GRB, the isotropic emission is usually in the region of $\sim 10^{51} - 10^{53}$ erg. Thus GRB 980425 was under-luminous by a factor of $100 - 1000$! It was suggested at the time that perhaps GRB 980425 was not representative of the general cosmological GRB population, but rather was a sub-class of GRB that occur only in the local universe.

A long, five-year wait ensued until 2003, when SN 2003dh was seen to coincide with low-redshift (z=0.1685; Greiner et al. 2003) GRB 030329 (Hjorth et al, 2003; Stanek et al. 2003; Matheson et al. 2003). Spectroscopy clearly showed that SN 2003dh was a Ic-BL SN, and it was observed that the isotropic gamma-ray emission was typical of other long GRBs ($\sim 1.7 \times 10^{52}$ erg; Amati 2006). Thus the "smoking-gun" observation that connected cosmological, long GRBs with the collapse of massive stars has been made.

To date five GRBs and XRFs have been spectroscopically connected to Ic-BL SNe: GRB 980425 & SN 1998bw (Galama et al. 1998; Patat et al. 2001), GRB 030329 & SN 2003dh (Hjorth et al. 2003; Stanek et al. 2003; Matheson et al. 2003), GRB 031203 & SN 2003lw (Malesani et al. 2004), XRF 060218 & SN 2006aj (Pian

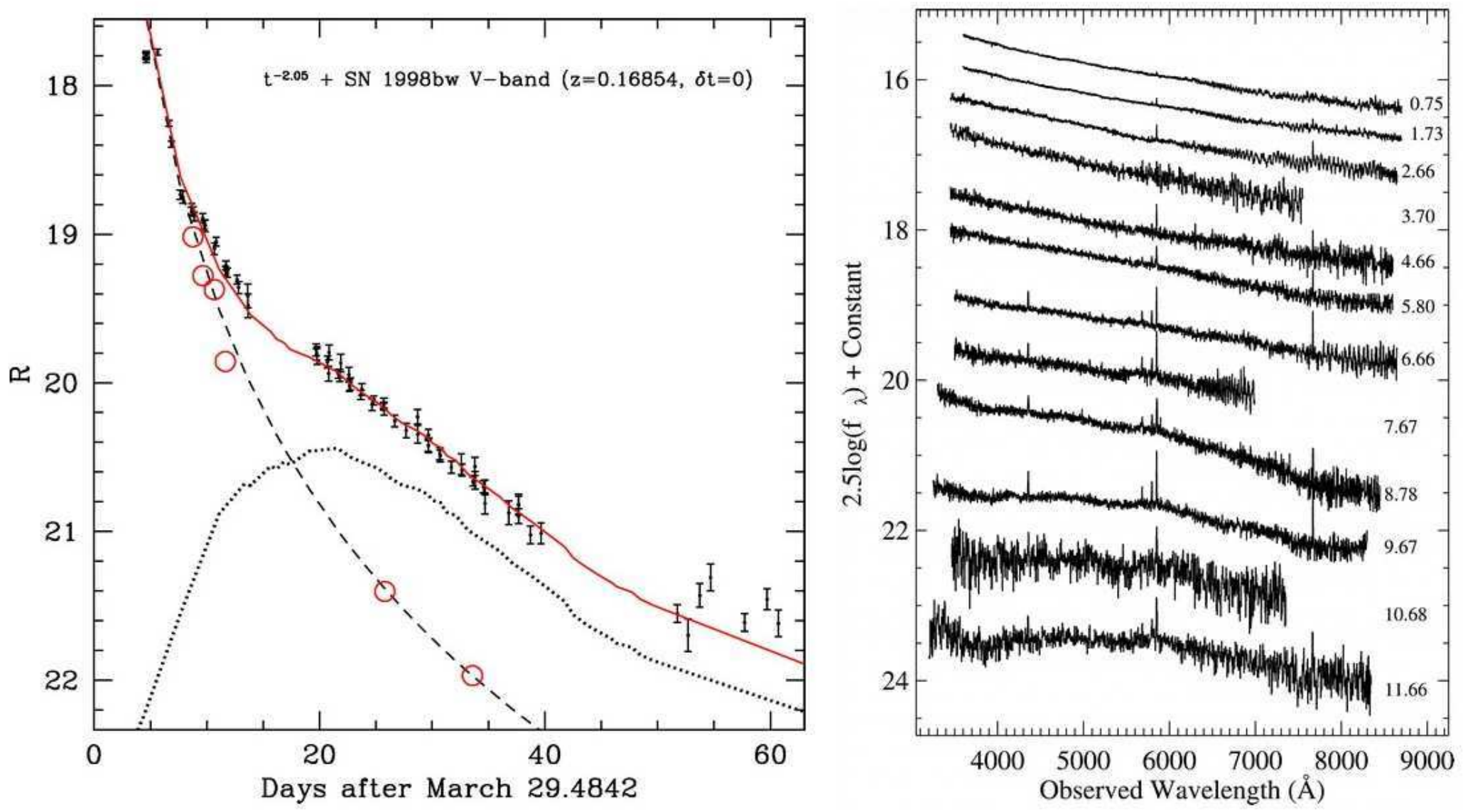

Figure 1.12 GRB 030329: *Left: R*-band LC. The dashed line displays the rate at which the afterglow fades, as determined from the optical spectra of Matheson et al. (2003). The dotted line displays light coming from a 1998bw-like SN. The solid, red line is the sum of flux coming from the afterglow and the 1998bw-like SN. *Right:* Sequence of optical spectra. Early epochs reveal flat spectra, indicative of light coming from an afterglow, however the emergence of SN features after several days proves the existence of the accompanying SN (Matheson et al. 2003).

et al. 2006; Mazzali et al. 2006b) and XRF 100316D & SN 2010bh (Starling et al. 2011; Chornock et al. 2010; Cano et al. 2011). For each of these events spectroscopic observations provide irrefutable evidence of an accompanying SNe with the GRB.

Additional evidence for the presence of an accompanying SN to a long GRB are the photometric observations of “red bumps” in optical and infrared LCs. A red bump was seen in the LC of GRB 980326 (Bloom et al. 1999) and was interpreted as due to a coincident SN at a redshift of $\sim 1$. Many other SN bumps were observed for other GRB and XRF events such as GRB 041006 (Stanek et al. 2005) and XRF 020903 (Bersier et al. 2006). In the latter case, in addition to the red, SN bump, a curved broadband spectrum was presented of the SN, which was similar in shape (though with less curvature) to SN 1998bw at a similar epoch, thus giving two lines of photometric evidence for an accompanying SNe with the XRF.

The observations made to date have shown that the SN that accompany long GRBs are quite unique. Spectroscopy has shown through the absence of hydrogen lines that all are type Ic, and the broad spectral features that indicate ejecta moving at velocities near $0.1c$ at peak light indicate that the SNe are of sub-type Ic-BL. It has also been shown through spectroscopy and photometric modeling that large amounts of nickel are created during the explosion, e.g. in SN 1998bw it is expected that $\sim 0.4 - 0.5 M_{\odot}$ of nickel was created. This is of order the amount of nickel created in Ia SNe, and more than is typically measured for other ccSNe.

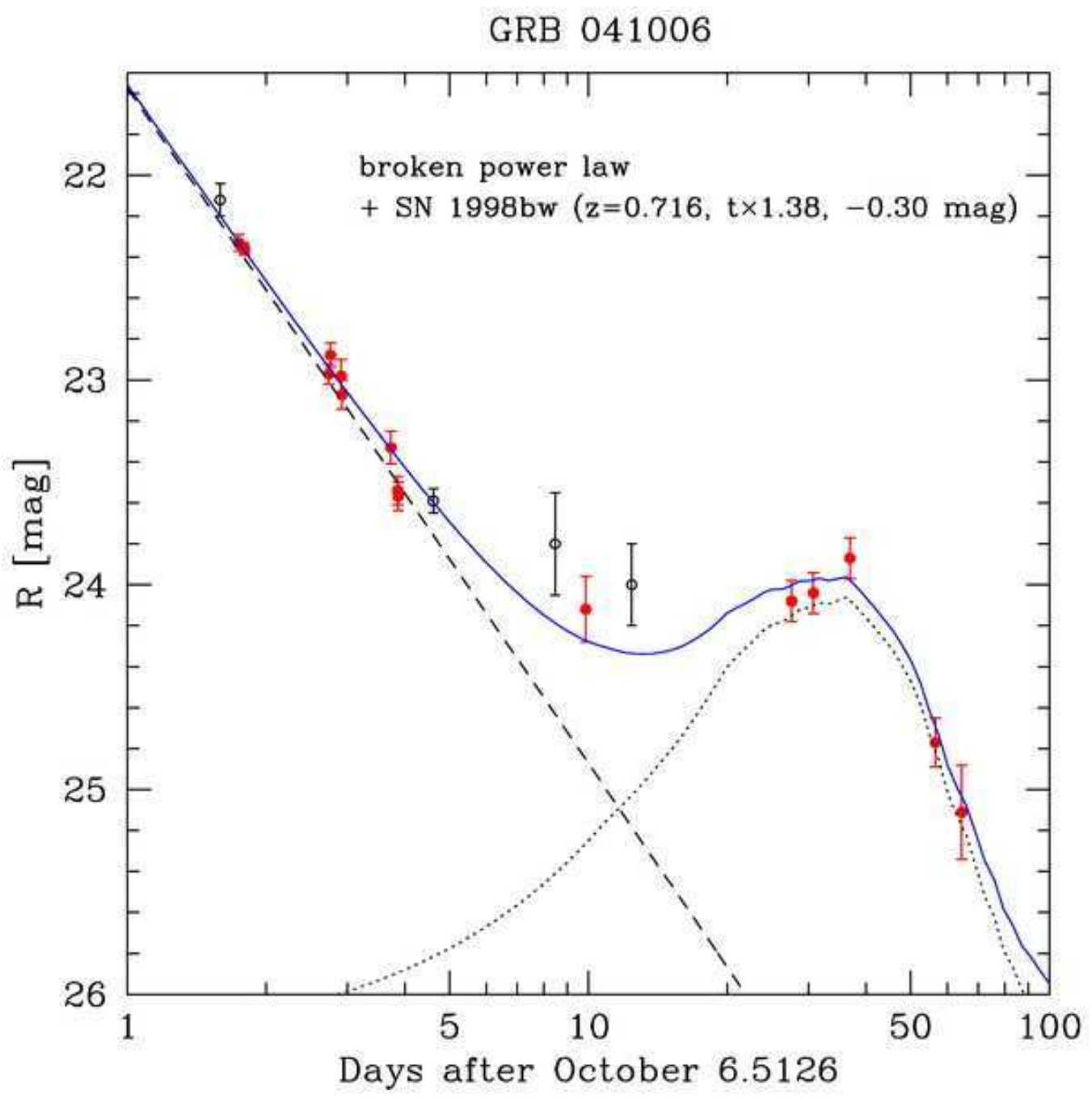

Figure 1.13 GRB 041006: "Bumps" are seen in the $R$-band LC, which is attributed to light coming from an accompanying supernova (Stanek et al. 2005).

Another distinguishing feature is that GRB-SNe have a large amount of the explosion energy concentrated into relativistic ejecta. Quite interestingly and crucially, this does not necessarily require that the accompanying SNe need to be optically bright or even exceptionally energetic (e.g. SN 2006aj; Mazzali et al. 2006), though some quite clearly are (e.g. SN 1998bw & SN 2003dh). It also allows for the existence of SNe powered by a similar energy source (i.e. a BH with an accretion disk) but without a GRB trigger. But to produce the GRB as much energy that is released in the supernova ejecta must also be available for the gamma-rays and the afterglow.

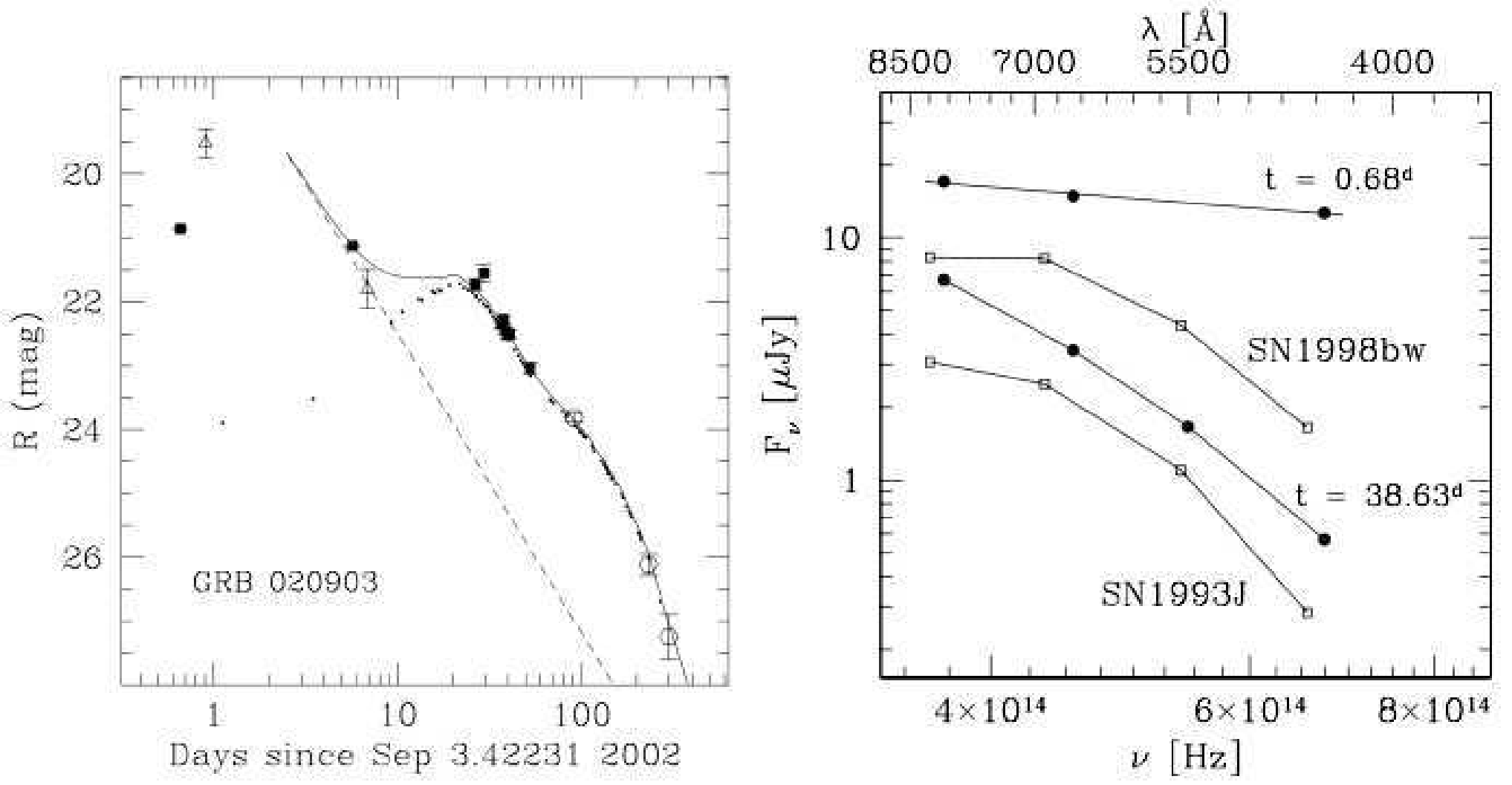

Figure 1.14 XRF 020903: *Left:* "Bumps" are seen in the *R*-band LC. *Right:* The curved SED of XRF 020903 resembles that of GRB-SN 1998bw at a common epoch, albeit with less curvature. Both of these observations provide strong evidence for light coming not only from a GRB afterglow, but also from a supernova (Bersier et al. 2006).

### 1.11.2 Indirect Evidence

Before the connection between GRB 980425 & SN 1998bw, observations of the locations of GRBs 970228, 970508 and 970828 in star-forming regions of their host galaxies led Pacyzński (1998) to suggest that the progenitors of long GRBs are massive stars and are not formed via the merger of compact objects. As binary neutron stars will be ejected to distances far from their birth places, the association of long GRBs with star forming regions is highly suggestive that the progenitors are massive stars that shine very brightly and use their fuel very rapidly. Their short lifespans do not allow the progenitors to travel very far during their lifetime and thus they explode very close to the site where they were formed.

Bloom et al. (2002) showed, with a sample of 20 long GRBs, that the median offset from the centre of their host galaxies was only 1.7 kpc. The authors also showed that there was a strong connection between the position of the GRB location within the UV light of the host galaxies, thus providing another clue between long GRBs and massive stars.

Fruchter et al. (2006) showed via high-resolution *HST* observations that long GRBs preferentially occur in regions of highest surface brightness of their host galaxies. Kelly et al. (2008) furthered this result to show that Ic SNe also occur in the brightest locations of the host galaxies, thus providing an interesting, indirect link between the two events, as well as implying that GRB-SNe and Ic SNe arise from massive progenitors.

### 1.11.3 Non-detections

The growing wealth of GRB-SN associations have led many astronomers to wonder if *all* long GRBs have an accompanying optically-bright SN. However in 2006, despite many observing campaigns to find accompanying SNe, none were found for nearby long GRBs 060505 and 060614 (e.g. Fynbo et al. 2006). While both were located cosmologically nearby (GRB 060505: z=0.089, Ofek et al. 2007; GRB 060614: z=0.125, Gal-Yam et al. 2006, Della Valle et al. 2006a), the dedicated observing campaigns did not detect any light from an accompanying SN. While it now generally accepted that GRB 060614 is likely to be a short GRB (Norris et al. 2010), the cause of the non detection of an optically-bright SN with GRB 060505 is still debated. Some authors debate that GRB 060505 may be a short GRB (e.g. Ofek et al. 2007), while other authors surmise that the accompanying SN was simply optically faint (e.g. Della Valle et al 2006a).

Despite these non-detections, the general consensus is still that all long GRBs are formed during the collapse of a massive star. Whether or not an optically bright SN is seen after the initial GRB is perhaps not a problem as there are several reasons why a SN may not be seen. We have already discussed in Section 1.3.3 the existence of very faint ccSNe such as SN 2008ha that have very faint peak magnitudes ($M_V = -14.2$). Valenti et al. (2008) claim that SN 2008ha, which is a faint and hydrogen-poor SN, is the result of a weak core-collapse explosion (though see Foley et al. 2009, 2010 who claim that SN 2008ha is a thermonuclear explosion). Similar faint peak magnitudes were seen for type II SNe 1997D and 1999br (Zampieri et al. 2003), which is indicative of very small amounts of nickel being created during the explosion. The authors conclude that these events were under-energetic with respect with other type II SNe. Despite these events arising from massive progenitors that ejected a large amount of material, the faint nature of the SN is due to significant fallback of material onto the BH remnant.

Thus observations show that it is possible to have faint ccSN, with corresponding low amounts of nickel and low explosion energies. Theory also suggests that optically-faint SN should occur (e.g. Tominaga et al. 2007). The available energy in these events would be very small, and perhaps either just enough or not even enough to overcome the gravitational potential, thus material that would normally be expelled to power the optically-bright SN instead falls back onto the central object (i.e. NS or BH).

Additionally, other factors may also contribute to the non-detection of an accompanying SN such as dust extinction (either foreground or rest-frame) and high redshift. Some events such as GRB 090417B (Holland et al. 2010) at $z = 0.345$ have shown that very large amounts of rest-frame dust exists for some events ($A_V \geq 12$ mag). Such dust can obviously extinguish light coming from a SN (this effect is more profound at shorter/bluer wavelengths and less so at longer/redder wavelengths). Large redshift also limits the distance to which an accompanying SN can be detected. It has been possible to infer the existence of an accompanying SN to a GRB up to redshift unity (e.g. GRB 080319B: z=0.937; Vreeswijk et al. 2008). Tanvir et al. (2010) showed that SN bumps are present in the optical and infrared LCs of GRB 080319B. However beyond these redshifts it is very hard to detect light from a GRB-SN because the peak light is too faint to be detected by modern instrumentation. Host brightness can also contaminate SN light at late times, though image subtraction methods can usually overcome this problem.

SECTION 1.12

# Thesis Outline

1.12.1

## The GRB-SNe Connection

The main aim of this doctoral project is to further understanding of the connection between long, soft GRBs and XRFs and type Ic-BL SNe by determining physical properties of the progenitors and SNe of these enigmatic events.

### Spectroscopically-connected events

Up to the start of 2010 there were only four GRB- and XRF-SNe that had been spectroscopically connected (i.e. events where SN features are unambiguously detected in the spectra of SN that are seen to occur spatially and temporally coincident with a long, soft GRB):

1. GRB 980425 / SN 1998bw (Galama et al. 1998; Patat et al. 2001)
2. GRB 030329 / SN 2003dh (Stanek et al. 2003; Hjorth et al. 2003; Matheson et al. 2003)
3. GRB 031203 / SN 2003lw (Malesani et al. 2004)
4. XRF 060218 / SN 2006aj (Pian et al. 2006; Mazzali et al. 2006)

There was also the curious case of X-Ray Transient (XRT) 080109 and SN 2008D (Soderberg et al. 2008; Mazzali et al. 2008). On the $9^{th}$ of January, 2008, while *Swift* was observing type Ib SN 2007uy (Nakano et al. 2008; Blondin et al. 2008) in nearby spiral galaxy NGC2770 ($z = 0.007$), a bright X-Ray transient was detected serendipitously. The subsequent power-law spectrum and LC shape of SN 2008D was reminiscent of GRBs and XRFs, but with a high-energy release more than two orders of magnitude less than that detected for cosmological GRBs and XRFs (e.g. Amati 2002, Ghirlanda et al. 2004). Spectroscopy obtained by many groups (e.g. Malesani et al. 2008) showed that the accompanying SN 2008D was of type Ib.

There was much controversy surrounding the nature of this event. The SN type (Ib) hinted that this event may not have originated from the same type of progenitors as GRBs and XRFs, where to date all of the accompanying SNe are of type Ic-BL. However the nature and origin of the high-energy properties of this event are what split opinion the most. There are two explanations in the literature for the origin of the high-energy emission: Soderberg et al. (2008) interpret the hot, blackbody X-Ray spectrum and early peaks in the optical LCs as being due to the shock breakout. Conversely Mazzali et al. (2008) attributes the emission to a "choked" jet. Interestingly, a recent paper by Van der Horst et al. (2011) has perhaps put to rest the heated debate. Using observations of the radio emission of SN 2008D for the first year after explosion, Van der Horst et al. (2011) showed that there was no evidence of a relativistic jet contributing to the observed radio flux, thus suggesting that the high-energy emission was from a shock breakout and not due to a GRB- or XRF-like event. Thus, it was only by obtaining vast amounts of observations in many decades of frequency (i.e. X-ray to radio) that the nature of this enigmatic event was able to be ascertained.

Similarly, all of the GRB- and XRF-SNe have undergone intense scrutiny. Detailed optical and infrared LCs have been published for all events, as well as extensive amounts of spectra. The observations have been used by many authors to determine properties of the progenitors that give rise to GRB/XRF-SNe. Table 1.1 summarizes some of the results of the detailed modeling. It is seen that the progenitors of GRB-SNe arise from massive stars, with Zero Age Main Sequence (ZAMS) masses in the range $35-45M_{\odot}$. However, the sole XRF-SN is seen to arise from a less massive progenitor ($\sim 20M_{\odot}$). The GRB-SNe are also seen to ejecta much more mass than the XRF-SN.

Additionally, of all of the spectroscopically-connected GRB-SNe, only GRB 030329 has an energy release in gamma-rays that is similar to "cosmological" GRBs (i.e. energies $\sim 10^{51}-10^{53}$ erg, and that obey the Amati Relation; Amati 2002). GRBs

Table 1.1 Ejected Masses and ZAMS Masses of type Ibc & GRB-SNe

| SN | Type | $M_{ej}$ ($M_{\odot}$) | error ($M_{ej}$) | $M_{ZAMS}$ ($M_{\odot}$) | error ($M_{ZAMS}$) | Ref. |
|---|---|---|---|---|---|---|
| 1998bw | GRB-SN | 10.4 | 1.0 | 40 | 5 | (1), (2), (3), (4) |
| 2003dh | GRB-SN | 8.0 | 2.0 | 35 | 5 | (5) |
| 2003lw | GRB-SN | 13.0 | 2.0 | 45 | 5 | (6) |
| 2006aj | XRF-SN | 1.8 | 0.8 | 20 | 2 | (7), (8) |

(1) Galama et al. (1998), (2) Iwamoto et al. (1998), (3) Nakamura et al. (2001), (4) Maeda et al. (2006), (5) Deng et al. (2005), (6) Mazzali et al. (2006a), (7) Mazzali et al. (2006b), (8) Modjaz et al. (2006).

980425, 031203 and XRF 060218 are all under-luminous in gamma-rays relative to other GRBs, prompting many authors to suggest that they may represent a sub-class of GRBs and XRFs in the local universe.

## Supernova Bumps

While the amount of spectroscopically-connected events number only a few, many photometric inferences of GRB/XRF-SNe have been reported in the literature. Up to 2009, the (arguably) best evidence for the detection of light coming from an accompanying SNe was shown for GRB 041006 (Stanek et al. 2005) and XRF 020903 (Bersier et al. 2006) (see Figures 1.13 and 1.14 respectively), though there are as many as 20 events that claim to detect light from a SN via late-time bumps in optical and IR LCs. However, in many of these events the claim of the presence of a SN was limited to only a few datapoints (e.g. Zeh et al. 2004; Ferrero et al. 2006).

While the lack of broadband photometry and spectroscopy limits what can be determined regarding the physical properties of the progenitors in each of the events where a SN bump has been detected, the inclusion of these events with the spectroscopically-connected events does allow the possibility for some statistical analysis such as comparing LC shapes and peak brightnesses.

## Progenitor Models

As already described in some detail in Section 1.9, the leading progenitor models of long, soft GRBs fall under the general title of “massive star models”, with the collapsar model (Woosley 1992) the leading candidate. Within the collapsar model two types of models that describe the collapse of a single, massive star exist, and differ only with their inclusion or exclusion of rapid rotation. The former models have been developed by Yoon & Langer (2005) and Woosley & Heger (2006), while the latter have been considered by th models of Heger et al. (2003).

Rapid rotation is included in the models as a way to impart enough angular momentum to the stellar remnant and accretion disk so as to then power a GRB. A consequence of the rapid rotation seen in the models of Yoon & Langer (2005) and Woosley & Heger (2006) is that the progenitor star undergoes extensive mixing of the stellar interior, leading to a composition that is almost completely homogeneous. This mixing also means that instead of losing their outer envelope of hydrogen to strong stellar winds (as in the case of the non-rotating models), the bulk of the hydrogen envelope is retained and burned into helium.

## 1.12.2 Thesis Research

Thus to date the field of GRB-SNe stands on secure footing, though more data are needed to determine the type of progenitors that these events arise from. Additional information is needed about individual nearby events for which detailed photometry and spectroscopy can be obtained, and then modeled to determine properties of the supernova, such as ejecta mass, composition, brightness and energetics. Determining properties of the SNe then allows astronomers to infer properties of the progenitor star.

Additional information regarding the nature of GRB/XRF-SNe can also be obtained by making statistical comparisons of the existing data published in the literature, for which roughly 22 GRB-SNe have been detected up to 2011.

First, investigations into GRBs 060729 and 090618 are presented (Chapters 2 and 3 respectively), where we use the optical data obtained on the Faulkes Telescope South, the Hubble Space Telescope (*HST*) and other telescopes to show the presence of SN bumps in the optical LCs, with the latter providing the most detailed SN bump to date in the literature. For these events we have determined both the shape and brightness of the accompanying SNe.

In Chapter 4 we present our considerable optical and IR photometry of the spectroscopically-confirmed (Chornock et al. 2010; Bufano et al. 2011) Ic-BL SN 2010bh. We use our optical and IR photometry to construct a quasi-bolometric LC of SN 2010bh and then, using an analytical model developed by Arnett (1982) for type I SNe, determine properties of the SN explosion including the explosion energy and ejected mass. We then analyse the existing GRB/XRF-SN dataset to look for the presence of a possible correlation between the shape and peak brightness of the LCs.

In Chapter 5 we use all of the existing photometry for the GRB/XRF-SNe, as well as a large dataset of non-GRB Ibc SNe and compare: (1) the peak brightness of the $V$-band SN LCs, and (2) the widths of the $V$- and $R$-band LCs. We use the former to test whether peak brightness can be used as a differentiating factor between GRB/XRF and non-GRB/XRF Ibc SNe. The latter analysis is used to obtain mass ratios between the different sub-classes of Ibc SNe. We then compare our results with those obtained via modeling of photometric and spectroscopic data of GRB/XRF-SNe and non-GRB/XRF Ibc SNe. The results of our statistical comparison are then related to the theoretical progenitor models, where we make several tentative conclusions surrounding the nature of the progenitors of GRB/XRF-SNe.

It is well worth mentioning that this thesis is heavily based in obtaining observations of GRBs *if* and *when* they occur. The author is fortunate to have joined an established group at the Astrophysics Research Institute in Liverpool that has access and time granted on many telescopes, including three, 2-metre robotic telescopes that respond automatically and very rapidly to GRB alerts: the Liverpool Telescope (LT) in La Palma, and the two Faulkes Telescopes: South (FTS), in Australia, and North (FTN) in Hawaii. Thus the amount of data, especially early-time data (i.e. only a few minutes after the GRB) that has been collected has mostly been due to the fortunate occurrence of GRB 090618 and XRF 100316D during the doctoral project, as well as their visibility from one of these three telescopes.

Additionally, key members of the group also have time granted on larger telescopes such as the 2.5m Isaac Newton Telescope (INT), the 4.2m William-Herschel Telescope (WHT), Gemini, and *HST*, all of which, and especially the latter, has been of vital importance to the impact of the results published in this thesis and in leading astrophysical journals (MNRAS and ApJ). The author is all the more grateful and in-debted to the existence of proposals having been accepted and large amounts of observing time available on these telescopes, which have led to the impressive LCs shown here. The importance of being part of this GRB group is fully appreciated by the author and should not be understated.

Finally, the photometric methods employed to obtain the data presented in this thesis are described in detail the Appendix.

Throughout this thesis observer-frame times are used unless specified otherwise. The respective decay and energy spectral indices $\alpha$ and $\beta$ are defined by $f_\nu \propto (t - t_0)^{-\alpha} \nu^{-\beta}$, where $t_0$ is the time of burst and $\nu$ is the frequency. We adopt a flat $\Lambda$CDM cosmology with $H_0 = 71$ km/s/Mpc, $\Omega_M = 0.27$, and $\Omega_\Lambda = 1 - \Omega_M = 0.73$. For this cosmology a redshift of $z = 0.54$ (i.e. GRB 060729 and GRB 090618) corresponds to a luminosity distance of $d_L = 3099$ Mpc and a distance modulus of 42.45 mag. For XRF 100316D / SN 2010bh, which is located at a redshift of $z = 0.0591$, this corresponds to a luminosity distance of $\mathrm{d_L} = 261$ Mpc and a distance modulus of 37.08 mag.

# 2

# GRB 060729

The *Swift* Burst Alert Telescope (BAT) detected the long-duration ($T_{90} \approx 115$ s) GRB 060729 on July 29, 2006 (Grupe et al. 2006) with a remarkably bright and long-lasting X-ray afterglow (Grupe et al. 2007; Grupe et al. 2010). The redshift was measured by Thöne et al. (2006) to be $z = 0.54$, and a later spectroscopic analysis by Fynbo et al. (2009) measured the redshift to be $z = 0.5428$.

Foreground reddening has been corrected for using the dust maps of Schlegel et al. (1998), from which we find $E(B-V) = 0.055$ mag for GRB 060729.

The results in this chapter are based partially on observations made with the NASA/ESA Hubble Space Telescope, obtained at the Space Telescope Science Institute, which is operated by the Association of Universities for Research in Astronomy, Inc., under NASA contract NAS 5-26555. These observations are associated with program # 10909.

Table 2.1 Ground-based Photometry of GRB 060729

| $T - T_o$ (days) | Telescope[a] | Filter | Mag | $\sigma$ (mag) | Calibrated to |
|---|---|---|---|---|---|
| 0.452 | Prompt | $B$ | 18.22 | 0.09 | Vega |
| 0.452 | Prompt | $z$ | 17.36 | 0.08 | AB |
| 0.481 | Prompt | $z$ | 17.29 | 0.07 | AB |
| 0.515 | Prompt | $z$ | 17.22 | 0.06 | AB |
| 0.551 | Prompt | $z$ | 17.49 | 0.07 | AB |
| 0.571 | Prompt | $B$ | 18.40 | 0.04 | Vega |
| 0.594 | Prompt | $z$ | 17.49 | 0.07 | AB |
| 1.574 | Prompt | $z$ | 18.43 | 0.17 | AB |
| 4.601 | Gemini-S | $R_c$ | 20.24 | 0.04 | Vega |
| 4.613 | Gemini-S | $g$ | 20.91 | 0.04 | AB |
| 4.626 | Gemini-S | $I_c$ | 19.92 | 0.03 | Vega |
| 4.638 | Gemini-S | $z$ | 20.51 | 0.04 | AB |
| 17.598 | CTIO | $R_c$ | 21.90 | 0.06 | Vega |
| 46.583 | Gemini-S | $R_c$ | 23.29 | 0.06 | Vega |
| 49.554 | Gemini-S | $R_c$ | 23.32 | 0.05 | Vega |

[a]Telescope key: CTIO: 4m Cerro Tololo Blanco Telescope; Prompt: 0.41m Prompt Telescope; Gemini-S: 8.1m Gemini-South Telescope. All magnitudes have been corrected for foreground extinction.

SECTION 2.1

# Observations & Photometry

2.1.1

## Ground-based Data

We obtained data with four ground-based telescopes: Panchromatic Robotic Optical Monitoring and Polarimetry Telescopes (PROMPT), Cerro Tololo Inter-American Observatory (CTIO), Faulkes Telescope South (FTS) & Gemini-South. Two epochs of data were taken with PROMPT $\sim 0.5$ & 1.5 days after the burst. The first epoch yielded images in $B$ and $z$, while the second epoch was only in $z$. One epoch of data was obtained on the Cerro Tololo (CTIO) Blanco 4m telescope 17.6 days after the burst, yielding an $R_c$ image. Three epochs of data were obtained on the Gemini-South (Gemini-S) 8.1m telescope, the first one was 4.6 days post-burst in *griz*, and two additional epochs in $r$ at 46.5 and 49.5 days post-burst. Finally, one further epoch of imaging was obtained on $5^{th}$ April, 2007 ($\sim$ 250 days after the burst) on the 2m Faulkes Telescope South (FTS). Observations were made of the GRB field (well after the GRB and SN had faded below the instrument detection limit) in $BVR_ci$, as well as images taken of Landolt photometric standard regions (Landolt 1992) in the same filters. The observations taken by FTS that are used in our calibration were made under photometric conditions.

Aperture photometry was performed on all images using standard routines in IRAF[1] . A small aperture was used, and an aperture correction was computed and applied. The aperture-corrected, instrumental magnitudes were then calibrated via standard star photometry into magnitudes in $BVR_cI_c$. Using the images of Landolt standards, IRAF routines were used to solve transformation equations of the form:

$$m_{inst} = z_p + M + a_1X + a_2(B-V) \tag{2.1}$$

where $m_{inst}$ is the aperture-corrected, instrumental magnitude, $M$ is the standard magnitude, $z_p$ the zero-point, $a_1$ the extinction, $X$ the airmass, $a_2$ the colour-term and $(B-V)$ (and variations thereof) the colour.

The validity of the transformation equations was checked by using the solutions on the Landolt standard stars in the CCD images, which revealed computed magnitudes that were consistent with those in the Landolt Catalogue (1992) within the magnitude errorbars (typical $\sigma \sim 0.04$ mag).

The solutions from the standard star photometry were then applied to a sequence of secondary standards in the field of GRB 060729. The Gemini-S observations are calibrated against these stars using a zero-point and colour term in filters $gR_cI_cz$. The $g$ and $z$ magnitudes of secondary standards in the GRB field were calculated using transformation equations from Jordi et al. (2006), and $r$ and $i$ were calibrated to $R_c$ and $I_c$. The single CTIO epoch was calibrated directly against the stars in the GRB field using a zero-point (no colour term). Images taken by PROMPT are calibrated against standard stars observed by PROMPT in $B$ and $z$, using a zero-point and colour term. All of the ground-based photometry, which has been corrected for foreground extinction, is presented in Table 2.1.

[1] IRAF is distributed by the National Optical Astronomy Observatory, which is operated by the Association of Universities for Research in Astronomy, Inc., under cooperative agreement with the National Science Foundation.

Table 2.2 *HST* Photometry of GRB 060729

| $T - T_o$ (days) | Filter[a] | Mag[b] | $\sigma$ (mag) | Calibrated to |
|---|---|---|---|---|
| 8.983 | $F330W$ | 22.00 | 0.03 | AB |
| 17.051 | $F330W$ | 23.40 | 0.05 | AB |
| 26.125 | $F330W$ | 24.08 | 0.10 | AB |
| 9.042 | $F625W$ | 21.09 | 0.01 | $R_c$ (Vega) |
| 16.909 | $F625W$ | 21.94 | 0.02 | $R_c$ (Vega) |
| 26.250 | $F625W$ | 22.57 | 0.03 | $R_c$ (Vega) |
| 48.689 | $F625W$ | 24.07 | 0.07 | $R_c$ (Vega) |
| 139.888 | $F625W$ | 26.20 | 0.62 | $R_c$ (Vega) |
| 9.059 | $F850LP$ | 20.90 | 0.01 | $I_c$ (Vega) |
| 16.926 | $F850LP$ | 21.74 | 0.02 | $I_c$ (Vega) |
| 26.266 | $F850LP$ | 22.24 | 0.03 | $I_c$ (Vega) |
| 48.757 | $F850LP$ | 23.28 | 0.05 | $I_c$ (Vega) |
| 139.905 | $F850LP$ | 25.65 | 0.57 | $I_c$ (Vega) |
| 9.139 | $F160W$ | 21.14 | 0.02 | AB |
| 16.500 | $F160W$ | 22.16 | 0.04 | AB |
| 27.198 | $F160W$ | 22.85 | 0.05 | AB |
| 9.124 | $F222M$ | 20.38 | 0.30 | AB |
| 16.460 | $F222M$ | $> 22.1$ | - | AB |
| 27.127 | $F222M$ | $> 21.7$ | - | AB |

[a]Filter key: $F330W$, $F625W$, $F850LP$: ACS; $F160W$, $F222M$: NICMOS.
[b]Apparent magnitudes are of the OT from the subtracted images *except for* the ACS $F330W$ and NICMOS $F222M$ filters (for which no image subtraction occurred). All magnitudes have been corrected for foreground extinction.

### 2.1.2 Hubble Space Telescope Data

We obtained a total of six epochs of data with *HST*, using the Advanced Camera for Surveys (ACS) and the High-Resolution Channel (HRC) in $F330W$; and the Wide Field Channel (WFC) in $F625W$ and $F850LP$; the Near Infrared Camera and Multi-Object Spectrograph (NICMOS) in $F160W$ and $F222M$; and the Wide-Field Planetary

Camera 2 (WFPC2) in $F300W$, $F622W$ and $F850LP$ for the last epoch. The images were reduced and drizzled in a standard manner using *Multidrizzle* (Koekemoer et al. 2002). The use of WFPC2 images in the last epoch was a consequence of the ACS being broken at this time. While we were wary that images from two different cameras and two different filters might disrupt our intention of using the last epoch as a template image for use in image subtraction, we were able to work around the problem of the differing cameras and filters by carefully choosing the stars used in the image-subtraction procedure (see section 2.1.3). The ACS images that were used for image subtraction were drizzled to the WFPC2 scale ($0.10''$), while the remaining images were drizzled to the native scale of the individual cameras ($0.2''$ and $0.10''$ for NICMOS and WFPC2 respectively).

Image subtraction was not carried out on the $F330W$ images because the host is not visible in the final WFPC2 ($F300W$) image, thus we used the procedure described in Appendix B of Sirianni et al. (2005) for the ACS data, i.e.: (a) perform aperture photometry using aperture of radius $0.15''$, (b) apply an aperture correction for apertures of radius from $0.15''$ to $0.50''$, (c) apply an aperture correction for $F330W$ for aperture of radius $0.50''$ to infinity (Table 5 of Sirianni et al. 2005), (d) compute and correct for a CCD charge transfer efficiency (CTE) uncertainty, (e) used AB zeropoint for each filter.

### 2.1.3 Image Subtraction

Using ISIS (Alard 2000) we performed image subtraction on the ACS $F625W$ and $F850LP$ images, using the WFPC2 images in $F622W$ and $F850LP$ as the respective templates. The ACS images were drizzled to the WFPC2 scale ($0.10''$) prior to subtraction. Image subtraction was also performed on the NICMOS $F160W$ images, using the last epoch as the template. Subtracting the WFPC2 image from the first five ACS images gave a clear detection of the optical transient (OT). As the reference image was taken with a different camera and different filters (with the corresponding differences in the filters' transmission curves and different CCD responses) we checked for the presence of a colour-term in the residuals in the subtracted images. For very red objects ($F622W - F850LP > 1.5$) a small colour dependent effect was seen, however when we restricted the colour range to include only stars with $0 \leq F622W - F850LP \leq 1.0$ we found no statistically-significant colour-term in the residuals. The colour of the OT was never beyond this colour range, thus by using the stars in this colour range to match the ACS frames to the WFPC2 template in our image-subtraction procedure we are confident of the validity of our results.

Aperture photometry was then performed on the subtracted images using an aperture of radius $0.15''$, and correcting for steps (a) - (c) as described previously but for the WFPC2 images (Holtzman et al. 1995). The instrumental magnitudes were then calibrated to $R_c$ and $I_c$ using a zero-point and colour-term. Aperture photometry was performed on the NICMOS $F160W$ images using the procedures prescribed in the NICMOS data handbook (Thatte et al. 2009). We used an aperture of radius $0.40''$ and included in our procedure: (a) an aperture correction for apertures of radius from $0.40''$ to $1.00''$, (b) corrected from $1.00''$ aperture radius to an infinite aperture by multiplying the flux (counts) by 1.075, (c) used AB zeropoint for each filter. All of the *HST* magnitudes, which have been corrected for foreground extinction, are listed in Table 2.2.

### 2.1.4 Host Photometry

The host was clearly visible and extended in all of the final epoch images taken by WFPC2, apart from the $F300W$ image. Aperture photometry was performed on the images: WFPC2 ($F622W$ and $F850LP$) and NICMOS ($F160W$), using an aperture of $0.7''$, $0.4''$ and $0.6''$ respectively. As the host is clearly extended, CTE effects have been neglected when determining the host magnitudes as the effects due to CTE on WFPC2 images are only well understood for point sources. The corrected instrumental magnitudes in $F622W$ and $F850LP$ were then calibrated to $R_c$ and $I_c$ using a zero-point and colour-term, while the $F160W$ and $F222M$ magnitudes are given as AB magnitudes. All of the host magnitudes, which have been corrected for foreground extinction, are listed in Table 2.3.

Table 2.3 *HST* Photometry of the Host Galaxy of GRB 060729

| $T - T_o$ (days) | Filter[a] | Mag | $\sigma$ (mag) | Calibrated to |
|---|---|---|---|---|
| 425.298 | $F300W$ | $> 24.80$ | - | AB |
| 428.984 | $F622W$ | 24.42 | 0.20 | $R_c$ (Vega) |
| 429.181 | $F850LP$ | 24.26 | 0.20 | $I_c$ (Vega) |
| 425.363 | $F160W$ | 22.82 | 0.06 | AB |
| 424.806 | $F222M$ | $> 22.10$ | - | AB |

[a]Filter key: $F622W$ & $F850LP$: WFPC2; $F160W$: NICMOS.
All magnitudes have been corrected for foreground extinction.

SECTION 2.2

## Results & Discussion

2.2.1

### The Afterglow

Grupe et al. (2007) found that the optical and X-ray light curves of GRB 060729 displayed similar temporal behaviour (i.e., the decay constants and time of break in the X-ray, UV and optical filters were consistent up to late times; see Figure 8 in Grupe et al. 2007). Thus, assuming at late times that the decay constant will be approximately the same in all optical passbands, we used our *HST*(ACS) $F330W$ magnitudes to characterise the temporal behaviour of the GRB afterglow at longer wavelengths (i.e., $R_c$ and $I_c$). The reason for using the $F330W$ magnitudes is that any SN contribution at these wavelengths is expected to be negligible with respect to the afterglow (e.g., see Figure 2.1).

Using the Ultraviolet and Optical Telescope (UVOT) aboard *Swift*, Grupe et al. (2007) found that the optical afterglow was visible in the UVOT-$U$ filter up to 7.66 days after the burst. We clearly detect the GRB afterglow up to our third *HST* epoch (26.12 days after the burst) in the *HST*(ACS) *F*330*W* filter. Thus, by combining our *HST* data with that obtained by *Swift*, we were able to extend the time-domain of the analysis and more accurately determine the temporal behaviour of the afterglow.

To enable us to compare the *F*330*W* magnitudes obtained with *HST*(ACS) to those determined by *Swift*, we used IRAF/Synphot/Calcphot to transform our *F*330*W* magnitudes into $U$ and corrected for foreground extinction. Next, using transformation equations determined by *Swift* (Poole et al. 2008), we transformed the UVOT-$U$ magnitudes into $U$. All of the $U$-band data (*HST* and *Swift*) were thus calibrated to $U$.

A broken power-law was fitted to the data (Figure 2.1), using the form:

$$m_v(t) = -2.5 \times \log\left(\left(\left(\frac{t}{T_{break}}\right)^{\alpha_1} + \left(\frac{t}{T_{break}}\right)^{\alpha_2}\right)^{-1}\right) + B \qquad (2.2)$$

where $t$ is the time since the burst and $T_{break}$ is the time when the power-law changes from temporal index $\alpha_1$ to $\alpha_2$.

The values of the fit ($\chi^2/dof = 1.31$) are: $\alpha_1 = 0.01 \pm 0.03$; $\alpha_2 = 1.65 \pm 0.05$ and $T_{break} = 0.75 \pm 0.08$ days. When we restrict the upper limit to the time range to 8 days (i.e., up to the last UVOT-$U$ detection) we find ($\chi^2/dof = 1.33$): $\alpha_2 = 1.47 \pm 0.11$ and $T_{break} = 0.58 \pm 0.10$ days, which are consistent with the decay constant and time of break found by Grupe et al. (2007) for the UVOT-$U$ data (see Table 5 of Grupe et al. 2007). We note that we originally fit the data with the complete Beuermann Function (Beuermann et al. 1999) and let $n$ vary (where $n$ is a measure of the "smoothness" of the transition from $\alpha_1$ to $\alpha_2$). The results of our fit gave $n \approx 1$ (i.e., a smooth transition), so we fixed $n = 1$ when performing the final fit.

In Figure 2.1, variability is seen in the early-time data. To check that the early-time data does not affect our empirical fit at late-times, we fit a single power-law to the data at $t - t_o > 2.0$ days. It was seen that by limiting the fit to data after this time, we found $\alpha = 1.64 \pm 0.07$, fully consistent with the value of $\alpha_2$ found previously.

The value of $\alpha_2 = 1.65 \pm 0.05$ is consistent with that found by Grupe et al. (2010)

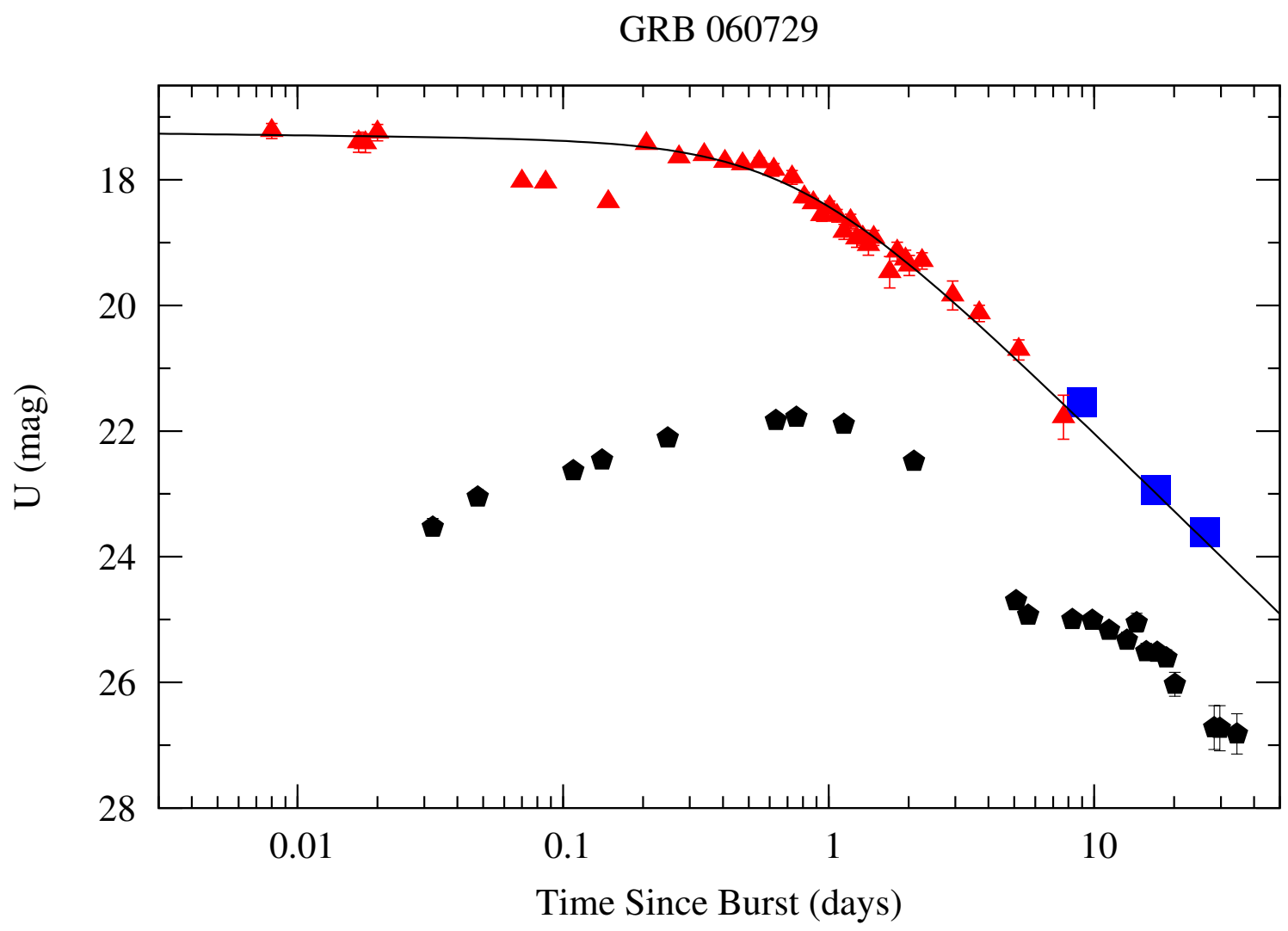

Figure 2.1 GRB 060729: Light-curve of *UVOT (U)* (triangles) and *HST (F330W)* (squares) magnitudes transformed into $U$. The afterglow model is shown as a broken power-law (solid line), where $\alpha_1 = 0.01 \pm 0.03$; $\alpha_2 = 1.65 \pm 0.05$ and $T_{break} = 0.75 \pm 0.08$ days. Plotted for comparison (pentagons) is the *UVOT* ($UVW1$) LC of XRF 060218/ SN 2006aj (Brown et al. 2009) as it would appear at $z = 0.54$. All magnitudes have been corrected for foreground extinction and the LC of SN 2006aj has also been corrected for host extinction. At these wavelengths the contribution of flux from an accompanying SN is expected to be negligible in comparison to the afterglow. This assumption is confirmed by the *HST* observations.

for the X-ray light curve ($\alpha = 1.61^{+0.10}_{-0.06}$) from $1.2 \times 10^6$ s (13.8 days) to $38 \times 10^6$ s (439.8 days), though we note that the optical and X-ray decay constants do not necessarily *have* to be the same, as seen for many bursts (e.g., Melandri et al. 2008).

Plotted for comparison in Figure 2.1 is the *UVOT (UVW1)* LC of XRF 060218 / SN 2006aj. At a redshift of $z = 0.54$, the light emitted in the restframe at *UVOT (UVW1)* wavelengths will be redshifted into observer-frame $U$. We can see that the broken power-law fits the $U$ data well and we conclude that the contribution of flux at this wavelength from an accompanying SN is negligible in comparison to the flux from the afterglow.

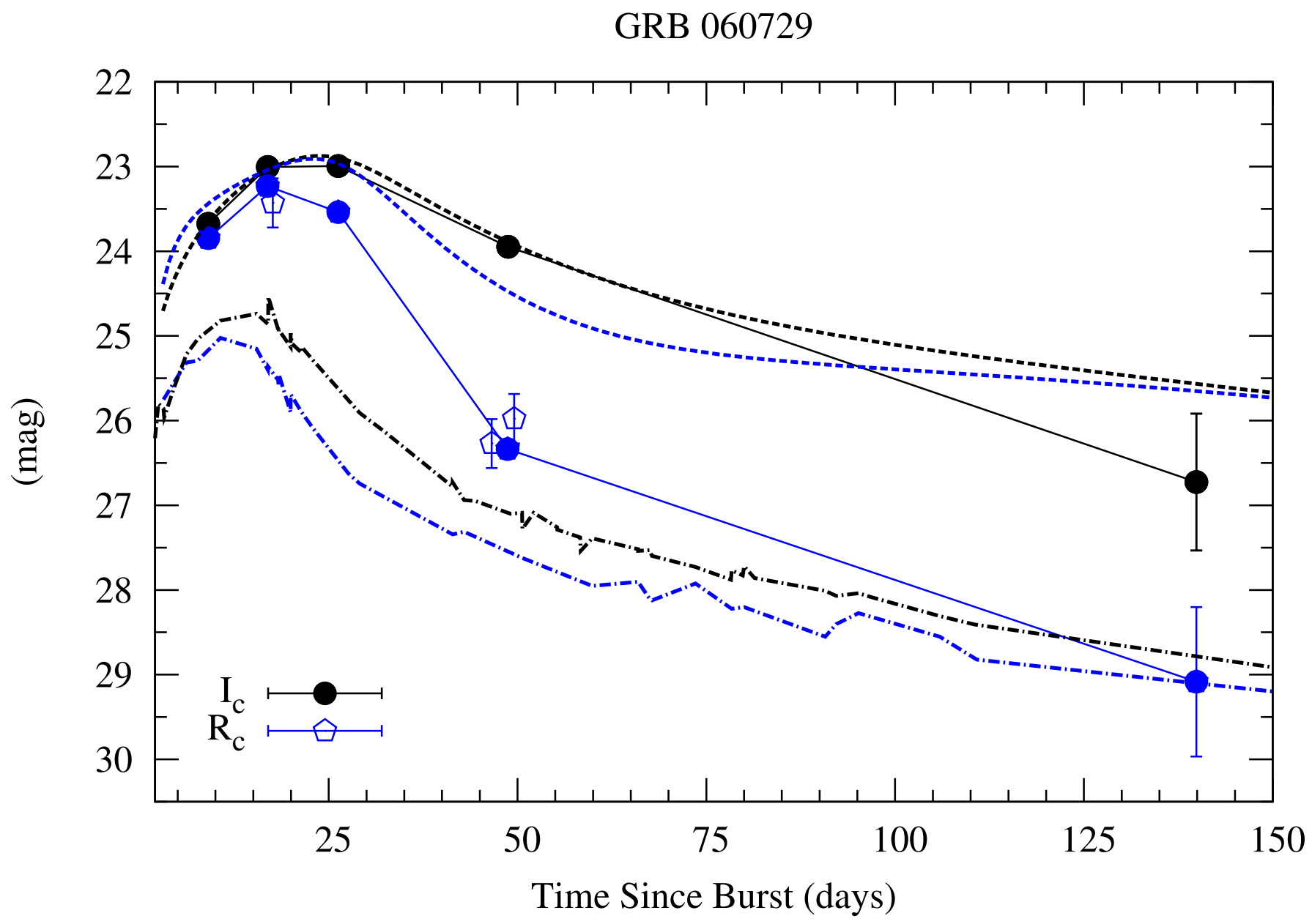

Figure 2.2 GRB 060729: Afterglow-subtracted, supernova $R_c$ (blue) and $I_c$ (black) light curves. The solid points are derived from the image-subtracted magnitudes from the *HST* data while the open points are derived from the magnitudes obtained by mathematically subtracting the host flux from the Gemini-S and CTIO epochs (see text). Plotted for comparison are the LCs of SN 1998bw (dashed) and SN 1994I (dot-dashed) in the same filters, as they would appear at $z = 0.54$. The brightness and time-evolution of the SN resembles that of SN 1998bw up to $\sim$ 30 days, rather than SN 1994I. All magnitudes have been corrected for foreground and host extinction.

### 2.2.2 The Supernova

For the complete GRB event, we assume that flux is coming from up to three sources: the afterglow, the host galaxy and a (possible) SN. We have already photometrically removed the flux due to the host in the *HST* images via image subtraction, however we were not able to perform image subtraction on the Gemini-S and CTIO images. Instead we mathematically subtracted the contribution of host flux (which was determined from the WFPC2 images) from the three Gemini-S and CTIO images, after all of the magnitudes were converted into fluxes using the zero-points from Fukugita et al. (1995). Thus, after removing the flux due to the host, and then subtracting the flux due to the GRB afterglow (which was characterized by Equation 2), any remaining flux could be due to an accompanying SN.

Our host- and afterglow-subtracted light curves are shown in Figure 2.2, where the quoted errors have been calculated in quadrature. Plotted for comparison are the distance and foreground-corrected light curves of SN 1998bw and SN 1994I. The SN 1998bw LCs are *model* light curves (i.e., interpolated/extrapolated through frequency)[2] however we have used the restframe $B$ and $V$ LCs of SN 1994I to represent the observer-frame $R_c$ and $I_c$ LCs respectively. It is seen, even despite the paucity of data-points, that the SN associated with GRB 060729 resembles more closely that of SN 1998bw up to $\sim$ 30 days, than SN 1994I in regards to both time evolution and brightness.

We find peak apparent magnitudes of the associated SN of $R_c = 23.80 \pm 0.08$ and $I_c = 23.20 \pm 0.06$. To make an accurate estimate of the peak absolute magnitude of the SN, we have determined the amount of host extinction (Section 2.2.4) to be $E(B-V) = 0.61$ mag. In comparison, Grupe et al. (2007) determined from their optical to X-ray spectral energy distribution (SED) analysis of the afterglow a restframe extinction local to the GRB of $E(B-V) = 0.34$ mag. However, using the same $UVOT$ data-set as Grupe et al. (2007) as well as additional $R_c$ data, Schady et al. (2010) found a restframe extinction of $E(B-V) \leq 0.06$. We note that our value of the host reddening is poorly constrained by our SED template fitting (though it is worth noting that the best two fits to the host SED both imply large amounts of reddening). However, it has been seen that the host extinction for many bursts is not always the same as the extinction local to the event (e.g., GRB 000210 (Gorosabel et al. 2003), where the local reddening was very high, but the host extinction was negligible. Further examples are seen in Modjaz et al. 2008 and Levesque et al. 2010). Thus it is more reasonable to use the extinction derived by Schady et al. (2010) as it is local to the event and is a more complete analysis than that performed by Grupe et al. (2007).

As the majority of long-duration GRB host galaxies contain dust that is Small Magellanic Cloud (SMC)-like (e.g., Starling et al. 2007; Schady et al. 2007; Kann et al. 2010), we applied the SMC extinction law ($R_V = 2.93$; Pei 1992) to the value of $E(B-V)$ found by Schady et al. (2010) to estimate the restframe extinction. We then approximated the restframe $V$-band magnitudes from the observer-frame $I_c$-band magnitudes, and find an absolute magnitude of the SN associated with GRB 060729 of $M_V = -19.43 \pm 0.06$. This implies that the SN associated with GRB 060729 has approximately the same peak brightness as 1998bw in the $V$-band (see Table 5.1).

[2] Note that the synthetic SN 1998bw LCs used throughout this thesis have been created by Cristiano Guidorzi. In Chapters 4 and 5, the synthetic LCs have been modified by the author when determining the stretch factors of various GRB/XRF-SNe.

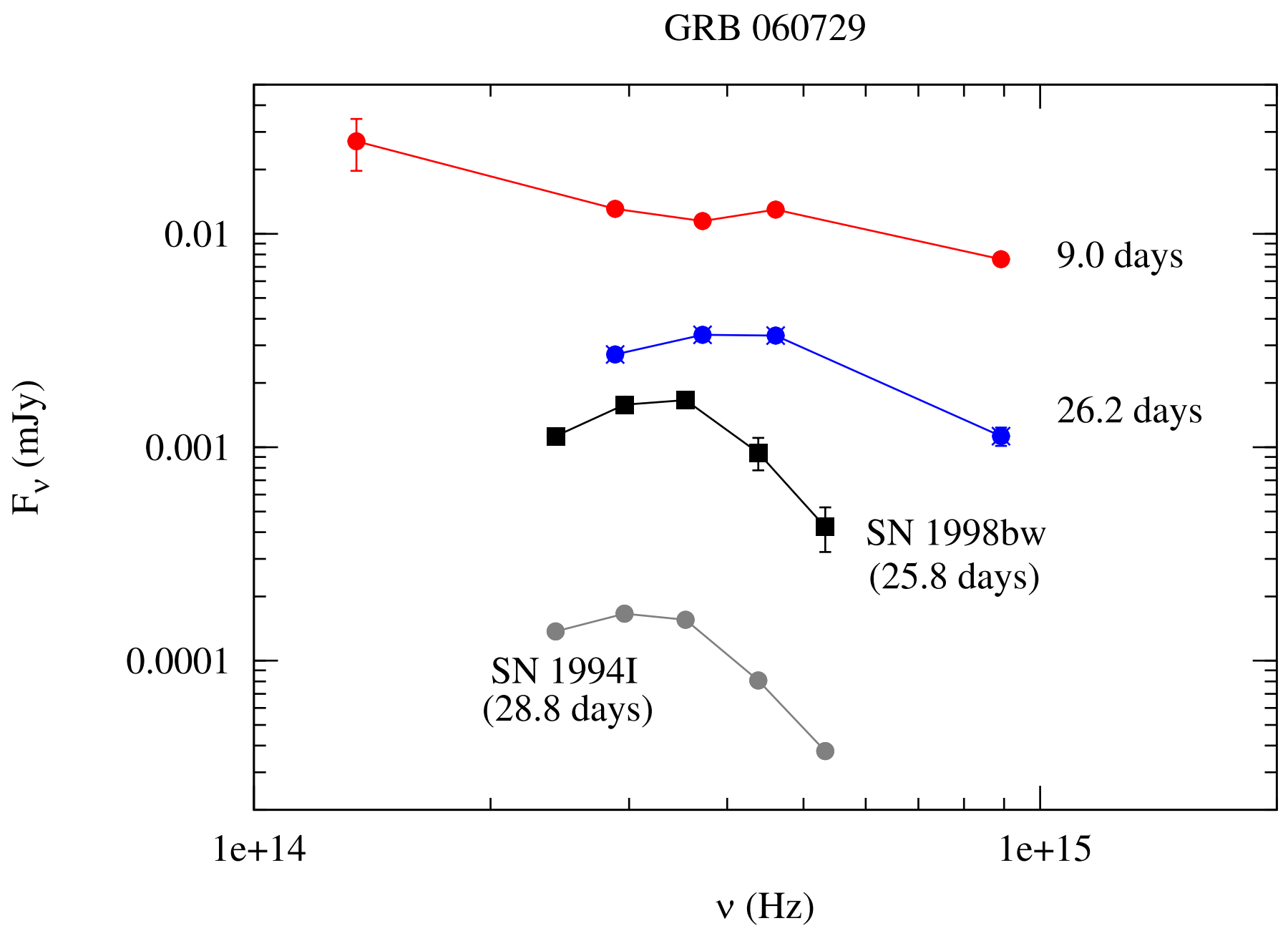

Figure 2.3 GRB 060729: Observer-Frame SED. The host-subtracted *HST* data (filled circles) of the OT (afterglow and supernova) has been corrected for foreground and host extinction. Plotted for comparison are the (see text) *UBVRI* data of SN 1998bw at 25.8 days and the *UBVRI* data for SN 1994I at 28.8 days, which have been corrected for foreground and host extinction. At 9.0 days the SED is fitted by a power-law with spectral index $\beta = 0.59 \pm 0.02$, however at 26.2 days the SED resembles that of the local Ic SNe, although with less curvature.

### 2.2.3 The Spectral Energy Distribution

We determined the observer-frame spectral energy distribution of the OT (afterglow and SN) for two epochs: 9.0 and 26.2 days after the burst (Figure 2.3). We used the zero-points from Fukugita et al. (1995) to convert the magnitudes to fluxes, and corrected for foreground and host extinction. Plotted for comparison are the observer-frame SEDs for SN 1998bw (25.8 days) and SN 1994I (28.8 days), which have been corrected for distance as well as foreground and host extinction. At 9.0 days the SED is fitted by a power-law with spectral index $\beta = 0.59 \pm 0.02$. However, the most striking feature of Figure 2.3 is at 26.2 days, where the SED resembles that of 1998bw and 1994I at a similar epoch, although with less curvature.

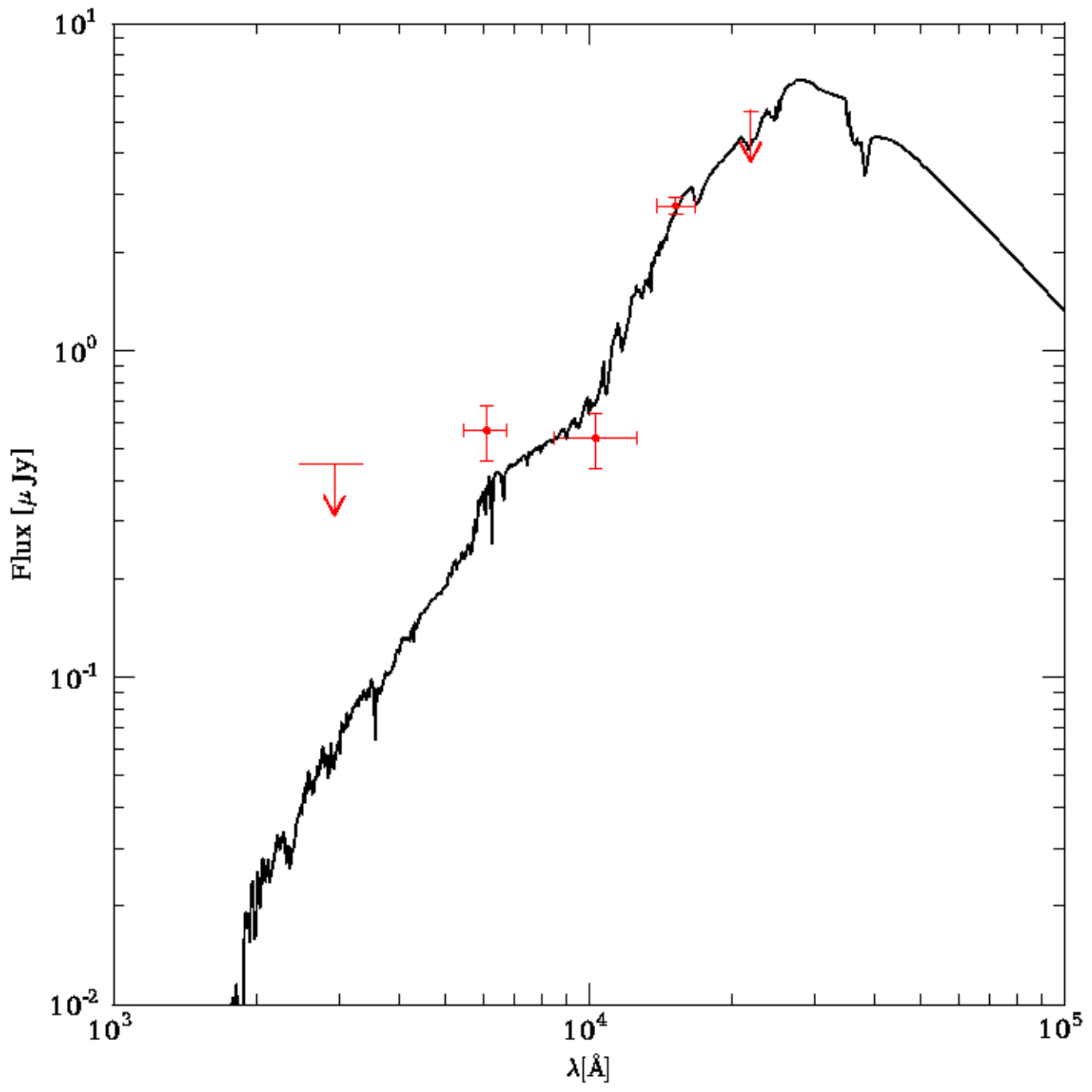

Figure 2.4 GRB 060729: SED (red points) of the host galaxy. The best-fit model is for a galaxy with a young stellar population and low metallicity.

### 2.2.4 The Host Galaxy

The detection of the host galaxy with WFPC2 allowed us to construct its optical SED (Figure 2.4). Magnitudes of the host, which are corrected for foreground reddening, were transformed to AB magnitudes, giving: $F622W = 24.54 \pm 0.20$, $F850LP = 24.60 \pm 0.20$, $F160W = 24.12 \pm 0.06$[3] , and upper limits: $F330 > 24.80$, $F222M > 22.10$ mag.

[3] A recent image of the host galaxy has been taken with the Wide Field Camera 3 aboard *HST* in filter $F160W$ by A. Levan, who confirms that the host brightness in this filter has not changed between this epoch and our epoch taken at $\approx 425$ days (A. Levan, private communication).

We used the procedure described in Svensson et al. (2010) to calculate the stellar mass, star-formation rate and the specific star-formation rate of the host galaxy (NB. The procedure was completed by K. Svensson). Using a redshift of $z = 0.54$, we fitted our *HST* magnitudes to synthetic galaxy evolution models (Bruzual & Charlot 2003). None of the synthetic models were very well constrained, however the best fit template ($\chi^2/dof = 6.58$) is for a galaxy with a young stellar population and low metallicity. We note that the next best-fit model is for a M82-type galaxy with large amounts of reddening.

The magnitudes of the best-fitting template are: $M_U = -16.72$, $M_B = -17.13$, $M_V = -17.49$, $M_K = -19.94$ and $E(B-V) \approx 0.61$ mag. The magnitudes are not host-extinction corrected. We also find a star-formation rate of $\log(SFR[M_\odot/yr]) = -0.89^{+0.04}_{-0.07}$; a stellar mass of $\log(Mass[M_\odot]) = 9.13^{+0.04}_{-0.08}$ and a specific star-formation rate of $\log(SSFR[Gyr^{-1}]) = -1.03^{+0.18}_{-0.04}$. The latter values of the star-formation rates and mass are estimated from an extinction-corrected SED. The quoted errors are $1\sigma$ and are systematic only and are calculated from the distributions of SFR, mass, etc., given by all templates (i.e., the errors reflect how good the best-fit parameters are compared with all of the considered models).

We were able to determine the position of GRB 060729 in the host galaxy using the *HST* images taken by WFPC2. We find for the position of the OT in our subtracted images (error $= 0.02''$): $06^h21^m31.77^s$, $-62^d22^m12.21^s$; which is offset from the apparent centre of the host by $0.33'' \pm 0.02''$ ($2.1 \pm 0.1$ kpc).

Secondly, we used the pixel statistic created by Fruchter et al. (2006), which calculates the percentage of light ($F_{light}$) contained in regions of lower surface brightness than the region containing a GRB or SN (where a value of $F_{light} = 100.0$ corresponds to the event occurring in the brightest region of the host galaxy). For GRB 060729, $F_{light} = 39.0$ in the $F622W$ filter, which is somewhat smaller than that seen in other GRBs by Fruchter et al. (2006).

Thus we conclude that the host galaxy of GRB 060729 has a young stellar population with a modest stellar mass ($\sim 10^9\, M_\odot$) and star-formation rate ($SFR \sim 0.13\, M_\odot yr^{-1}$), suggesting that the host is typical of other GRB host galaxies (e.g., Wainwright et al. 2005; Fruchter et al. 2006; Savaglio et al. 2009; Svensson et al. 2010; Christensen et al. 2004).

SECTION 2.3

## Conclusions

We have presented photometric evidence for a supernova associated with GRB 060729. For this event (and any long, soft GRB event) we attribute light coming from three sources: (1) the afterglow, (2) the supernova, and (3) the host galaxy. First, we performed image subtraction on our *HST* images to remove the constant source of flux from the host galaxy. Next, we determined the behaviour of the optical afterglow using the $U$-band data (Figure 2.1), and subtracted it from the light of the OT. The "left-over"/residual light resembles that of a SN, both in shape (Figure 2.2) and in its SED (Figure 2.3).

For the SN associated with GRB 060729, we found peak apparent magnitudes of $R_c = 23.80 \pm 0.08$ and $I_c = 23.20 \pm 0.06$. When restframe extinction was considered using the analysis performed by Schady et al. (2010) (who found $E(B-V)_{restframe} \leq 0.06$ mag), the peak, restframe absolute $V$-band magnitude was shown to be $M_V = -19.43 \pm 0.06$, which is approximately the same peak brightness as SN 1998bw (in the $V$-band).

Finally, we used our *HST* observations at $\mathrm{t} - \mathrm{t_o} \sim 425$ days in filters $F222W$, $F850LP$ and $F160W$ to constrain the SED of the host galaxy of GRB 060729. The paucity of optical points didn't allow us to constrain the SED very well, however the best-fitting template is for a galaxy with a young stellar population, low global metallicity, a modest stellar mass ($\sim 10^9 M_\odot$) and a star formation rate (SFR $\sim 0.13 M_\odot$ yr$^{-1}$) that is typical of many other GRB host galaxies.

# 3

# GRB 090618

The *Swift* Burst Alert Telescope (BAT) detected the long-duration ($T_{90} = 113.2 \pm 6$ *s*) GRB 090618 on June 18, 2009 at 08:28:29 UT (Schady et al. 2009) which was followed up by many teams with ground-based telescopes at optical wavelengths, and published in the GCN Circulars.[1]

The redshift was measured to be $z = 0.54$ by Cenko et al. (2009) and Fatkhullin et al. (2009).

Foreground reddening has been corrected for using the dust maps of Schlegel et al. (1998), from which we find $E(B-V) = 0.085$ mag for GRB 090618.

[1] The data presented in this chapter supersedes those in the GCN Circulars: FTN: Melandri et al. 2009 and Cano et al. 2009; LOAO: Im et al. 2009a; OAUV-0.4m telescope: Fernandez-Soto et al. 2009; Shajn: Rumyantsev & Pozanenko 2009; BOAO: Im et al. 2009b; SAO-RAS: Fatkhullin et al. 2009 (photometry & spectra); HCT: Anupama, Gurugubelli & Sahu 2009; Mondy: Klunko, Volnova & Pozanenko 2009. GRB 090618 was also detected at radio wavelengths and published in GCN Circulars: AMI: Pooley 2009a, 2009b; VLA: Chandra & Frail 2009; WSRT: Kamble, van der Horst & Wijiers 2009.

SECTION 3.1

## Observations & Photometry

3.1.1

### Optical Data

We obtained optical observations with many ground-based telescopes: the 2m FTN; the 2m LT; the 8.1m Gemini-North; the 4.2m WHT; the 2.5m INT; the 2.6m Shajn telescope of CrAO; the 1.25m AZT-11 telescope of CrAO; the 1.5m Maidanak AZT-22 telescope with the SNUCAM CCD camera (Im et al. 2010); the 1.5m AZT-33IK telescope of Sayan observatory, Mondy; the 6m BTA-6 optical telescope and the 1m Zeiss-1000 telescope of the Special Astrophysical Observatory of the Russian Academy of Sciences (SAO-RAS); the 1m LOAO telescope (Lee, Im & Urata 2010) at Mt. Lemmon (Arizona, USA) which is operated by the Korea Astronomy Space Science Institute; 15.5cm telescope at Bohyunsan Optical Astronomy Observatory (BOAO) in Korea which is operated by the Korea Astronomy Space Science Institute; the 2m Himalayan Chandra Telescope (HCT) and the T40 telescope at the Observatorio de Aras de los Olmos, operated by the Observatori Astronómic de la Universitat de València.

Aperture photometry was performed on all images using standard routines in IRAF. A small aperture was used, and an aperture correction was computed and applied. The aperture-corrected, instrumental magnitudes of stars in the GRB field were calibrated to the *SDSS* catalogue (in filters $BVR_ci$, where non-SDSS magnitudes were calculated using transformation equations from Jordi et al. (2006) using Equation 1, initially using a zero-point and a colour term. However, it was found that the contribution of a colour term to the calibration was negligible, hence only a zero-point was used in all cases. All optical magnitudes have been corrected for foreground extinction and are listed in Table 3.1.

Table 3.1. Ground-based Optical Photometry of GRB 090618

| $T-T_o$ (days) | Filter | $m_v$[a] | $\sigma$ (mag) | Telescope[b] | $T-T_o$ (days) | Filter | $m_v$[a] | $\sigma$ (mag) | Telescope[b] |
|---|---|---|---|---|---|---|---|---|---|
| 0.06 | $B$ | 17.35 | 0.06 | FTN | 0.11 | $R_c$ | 17.22 | 0.02 | FTN |
| 0.06 | $B$ | 17.38 | 0.05 | FTN | 0.11 | $R_c$ | 17.30 | 0.02 | FTN |
| 0.06 | $B$ | 17.43 | 0.05 | FTN | 0.12 | $R_c$ | 17.26 | 0.01 | FTN |
| 0.07 | $B$ | 17.51 | 0.07 | FTN | 0.12 | $R_c$ | 17.22 | 0.06 | LOAO |
| 0.07 | $B$ | 17.61 | 0.06 | FTN | 0.12 | $R_c$ | 17.40 | 0.01 | FTN |
| 0.08 | $B$ | 17.69 | 0.06 | FTN | 0.13 | $R_c$ | 17.38 | 0.01 | FTN |
| 0.09 | $B$ | 17.81 | 0.17 | LOAO | 0.14 | $R_c$ | 17.38 | 0.02 | FTN |
| 0.09 | $B$ | 17.76 | 0.04 | FTN | 0.15 | $R_c$ | 17.45 | 0.05 | FTN |
| 0.10 | $B$ | 17.83 | 0.02 | FTN | 0.15 | $R_c$ | 17.56 | 0.03 | FTN |
| 0.10 | $B$ | 17.92 | 0.06 | FTN | 0.16 | $R_c$ | 17.62 | 0.02 | FTN |
| 0.11 | $B$ | 17.83 | 0.05 | FTN | 0.16 | $R_c$ | 17.67 | 0.01 | FTN |
| 0.11 | $B$ | 17.88 | 0.14 | LOAO | 0.17 | $R_c$ | 17.68 | 0.01 | FTN |
| 0.11 | $B$ | 17.91 | 0.04 | FTN | 0.18 | $R_c$ | 17.76 | 0.02 | FTN |
| 0.11 | $B$ | 17.97 | 0.04 | FTN | 0.18 | $R_c$ | 17.87 | 0.02 | FTN |
| 0.12 | $B$ | 18.05 | 0.05 | FTN | 0.20 | $R_c$ | 17.86 | 0.04 | FTN |
| 0.13 | $B$ | 18.11 | 0.04 | FTN | 0.20 | $R_c$ | 17.88 | 0.03 | FTN |
| 0.14 | $B$ | 18.20 | 0.03 | FTN | 0.21 | $R_c$ | 17.88 | 0.02 | FTN |
| 0.15 | $B$ | 18.31 | 0.08 | FTN | 0.21 | $R_c$ | 17.94 | 0.02 | FTN |
| 0.15 | $B$ | 18.32 | 0.08 | FTN | 0.22 | $R_c$ | 17.96 | 0.02 | FTN |
| 0.15 | $B$ | 18.24 | 0.03 | FTN | 0.23 | $R_c$ | 17.94 | 0.02 | FTN |
| 0.16 | $B$ | 18.32 | 0.03 | FTN | 0.24 | $R_c$ | 17.97 | 0.08 | FTN |
| 0.16 | $B$ | 18.39 | 0.05 | FTN | 0.24 | $R_c$ | 18.13 | 0.05 | FTN |
| 0.17 | $B$ | 18.42 | 0.03 | FTN | 0.24 | $R_c$ | 18.13 | 0.03 | FTN |
| 0.18 | $B$ | 18.43 | 0.05 | FTN | 0.25 | $R_c$ | 18.04 | 0.02 | FTN |
| 0.19 | $B$ | 18.49 | 0.04 | FTN | 0.26 | $R_c$ | 18.09 | 0.02 | FTN |
| 0.19 | $B$ | 18.47 | 0.09 | FTN | 0.41 | $R_c$ | 18.77 | 0.13 | Mondy |
| 0.20 | $B$ | 18.62 | 0.06 | FTN | 0.42 | $R_c$ | 18.81 | 0.13 | Mondy |
| 0.20 | $B$ | 18.60 | 0.05 | FTN | 0.46 | $R_c$ | 18.85 | 0.04 | HCT |
| 0.20 | $B$ | 18.56 | 0.04 | FTN | 0.46 | $R_c$ | 18.88 | 0.04 | HCT |
| 0.21 | $B$ | 18.61 | 0.03 | FTN | 0.48 | $R_c$ | 18.84 | 0.07 | Shajn |
| 0.22 | $B$ | 18.69 | 0.04 | FTN | 0.48 | $R_c$ | 18.84 | 0.06 | Shajn |
| 0.22 | $B$ | 18.66 | 0.02 | FTN | 0.48 | $R_c$ | 18.75 | 0.08 | Shajn |
| 0.23 | $B$ | 18.75 | 0.04 | FTN | 0.48 | $R_c$ | 18.89 | 0.02 | Maidanak |
| 0.24 | $B$ | 18.84 | 0.12 | FTN | 0.49 | $R_c$ | 18.89 | 0.02 | Maidanak |
| 0.24 | $B$ | 18.73 | 0.07 | FTN | 0.49 | $R_c$ | 18.81 | 0.09 | Shajn |
| 0.24 | $B$ | 18.76 | 0.05 | FTN | 0.49 | $R_c$ | 18.77 | 0.10 | Shajn |
| 0.25 | $B$ | 18.79 | 0.03 | FTN | 0.50 | $R_c$ | 18.96 | 0.02 | Maidanak |
| 0.25 | $B$ | 18.88 | 0.05 | FTN | 0.50 | $R_c$ | 19.01 | 0.11 | Shajn |
| 0.48 | $B$ | 19.84 | 0.17 | Shajn | 0.50 | $R_c$ | 18.94 | 0.02 | Maidanak |

Table 3.1—Continued

| $T - T_o$ (days) | Filter | $m_v$[a] | $\sigma$ (mag) | Telescope[b] | $T - T_o$ (days) | Filter | $m_v$[a] | $\sigma$ (mag) | Telescope[b] |
|---|---|---|---|---|---|---|---|---|---|
| 0.48 | $B$ | 19.79 | 0.16 | Shajn | 0.51 | $R_c$ | 18.81 | 0.10 | Shajn |
| 0.48 | $B$ | 19.38 | 0.13 | Shajn | 0.51 | $R_c$ | 18.90 | 0.10 | Shajn |
| 0.49 | $B$ | 19.61 | 0.23 | Shajn | 0.51 | $R_c$ | 18.77 | 0.10 | Shajn |
| 0.49 | $B$ | 19.52 | 0.02 | Maidanak | 0.51 | $R_c$ | 18.92 | 0.10 | Shajn |
| 0.49 | $B$ | 19.53 | 0.23 | Shajn | 0.52 | $R_c$ | 18.99 | 0.02 | Maidanak |
| 0.49 | $B$ | 19.61 | 0.29 | Shajn | 0.52 | $R_c$ | 19.01 | 0.04 | Shajn |
| 0.50 | $B$ | 19.54 | 0.02 | Maidanak | 0.52 | $R_c$ | 18.99 | 0.05 | Shajn |
| 0.50 | $B$ | 19.51 | 0.30 | Shajn | 0.52 | $R_c$ | 18.97 | 0.02 | Maidanak |
| 0.51 | $B$ | 19.61 | 0.20 | Shajn | 0.52 | $R_c$ | 18.94 | 0.05 | Shajn |
| 0.51 | $B$ | 19.57 | 0.03 | Maidanak | 0.53 | $R_c$ | 18.97 | 0.03 | Shajn |
| 0.51 | $B$ | 19.59 | 0.16 | Shajn | 0.53 | $R_c$ | 19.00 | 0.02 | Maidanak |
| 0.51 | $B$ | 19.57 | 0.02 | Maidanak | 0.53 | $R_c$ | 19.02 | 0.04 | Shajn |
| 0.51 | $B$ | 19.73 | 0.30 | Shajn | 0.54 | $R_c$ | 19.05 | 0.03 | Shajn |
| 0.52 | $B$ | 19.88 | 0.17 | Shajn | 0.54 | $R_c$ | 19.02 | 0.01 | Maidanak |
| 0.52 | $B$ | 19.60 | 0.12 | Shajn | 0.54 | $R_c$ | 19.08 | 0.04 | Shajn |
| 0.52 | $B$ | 19.69 | 0.08 | Shajn | 0.54 | $R_c$ | 19.05 | 0.03 | Shajn |
| 0.53 | $B$ | 19.62 | 0.03 | Maidanak | 0.55 | $R_c$ | 19.03 | 0.03 | Shajn |
| 0.53 | $B$ | 19.61 | 0.03 | Maidanak | 0.55 | $R_c$ | 19.05 | 0.02 | Maidanak |
| 0.53 | $B$ | 19.90 | 0.08 | Shajn | 0.55 | $R_c$ | 19.03 | 0.04 | Shajn |
| 0.53 | $B$ | 19.74 | 0.09 | Shajn | 0.56 | $R_c$ | 19.16 | 0.04 | Shajn |
| 0.53 | $B$ | 19.77 | 0.08 | Shajn | 0.56 | $R_c$ | 19.10 | 0.04 | Shajn |
| 0.54 | $B$ | 19.76 | 0.08 | Shajn | 0.56 | $R_c$ | 19.08 | 0.02 | LT |
| 0.54 | $B$ | 19.63 | 0.02 | Maidanak | 0.56 | $R_c$ | 19.13 | 0.03 | Shajn |
| 0.54 | $B$ | 19.84 | 0.10 | Shajn | 0.56 | $R_c$ | 19.10 | 0.02 | LT |
| 0.54 | $B$ | 19.97 | 0.10 | Shajn | 0.57 | $R_c$ | 19.07 | 0.03 | Shajn |
| 0.54 | $B$ | 19.65 | 0.03 | Maidanak | 0.57 | $R_c$ | 19.07 | 0.02 | LT |
| 0.55 | $B$ | 19.97 | 0.09 | Shajn | 0.57 | $R_c$ | 19.08 | 0.04 | Shajn |
| 0.55 | $B$ | 19.86 | 0.09 | Shajn | 0.57 | $R_c$ | 19.13 | 0.04 | Shajn |
| 0.55 | $B$ | 19.68 | 0.03 | Maidanak | 0.57 | $R_c$ | 19.16 | 0.05 | Shajn |
| 0.55 | $B$ | 19.85 | 0.09 | Shajn | 0.58 | $R_c$ | 19.14 | 0.04 | Shajn |
| 0.56 | $B$ | 19.87 | 0.09 | Shajn | 0.58 | $R_c$ | 19.14 | 0.05 | Shajn |
| 0.56 | $B$ | 19.83 | 0.07 | Shajn | 0.58 | $R_c$ | 19.16 | 0.04 | Shajn |
| 0.56 | $B$ | 19.86 | 0.08 | Shajn | 0.59 | $R_c$ | 19.14 | 0.04 | Shajn |
| 0.57 | $B$ | 19.96 | 0.09 | Shajn | 0.59 | $R_c$ | 19.16 | 0.05 | Shajn |
| 0.57 | $B$ | 19.95 | 0.11 | Shajn | 0.59 | $R_c$ | 19.13 | 0.04 | Shajn |
| 0.57 | $B$ | 19.92 | 0.09 | Shajn | 0.60 | $R_c$ | 19.20 | 0.05 | Shajn |
| 0.58 | $B$ | 19.87 | 0.08 | Shajn | 0.60 | $R_c$ | 19.17 | 0.05 | Shajn |
| 0.58 | $B$ | 19.84 | 0.09 | Shajn | 0.60 | $R_c$ | 19.16 | 0.05 | Shajn |
| 0.58 | $B$ | 19.91 | 0.10 | Shajn | 0.60 | $R_c$ | 19.17 | 0.02 | LT |

Table 3.1—Continued

| $T - T_o$ (days) | Filter | $m_v$[a] | $\sigma$ (mag) | Telescope[b] | $T - T_o$ (days) | Filter | $m_v$[a] | $\sigma$ (mag) | Telescope[b] |
|---|---|---|---|---|---|---|---|---|---|
| 0.58 | $B$ | 20.07 | 0.09 | Shajn | 0.60 | $R_c$ | 19.21 | 0.06 | Shajn |
| 0.59 | $B$ | 20.12 | 0.10 | Shajn | 0.61 | $R_c$ | 19.17 | 0.02 | LT |
| 0.59 | $B$ | 20.06 | 0.11 | Shajn | 0.61 | $R_c$ | 19.15 | 0.06 | Shajn |
| 0.59 | $B$ | 19.93 | 0.08 | Shajn | 0.61 | $R_c$ | 19.18 | 0.02 | LT |
| 0.60 | $B$ | 19.99 | 0.11 | Shajn | 0.61 | $R_c$ | 19.15 | 0.06 | Shajn |
| 0.60 | $B$ | 19.99 | 0.14 | Shajn | 0.61 | $R_c$ | 19.18 | 0.06 | Shajn |
| 0.60 | $B$ | 20.26 | 0.19 | Shajn | 0.62 | $R_c$ | 19.14 | 0.06 | Shajn |
| 0.61 | $B$ | 19.99 | 0.21 | Shajn | 0.62 | $R_c$ | 19.24 | 0.07 | Shajn |
| 0.61 | $B$ | 20.03 | 0.18 | Shajn | 0.62 | $R_c$ | 19.32 | 0.07 | Shajn |
| 0.61 | $B$ | 20.03 | 0.14 | Shajn | 0.63 | $R_c$ | 19.23 | 0.06 | Shajn |
| 0.62 | $B$ | 20.08 | 0.15 | Shajn | 0.63 | $R_c$ | 19.26 | 0.05 | Shajn |
| 0.62 | $B$ | 20.06 | 0.17 | Shajn | 0.63 | $R_c$ | 19.37 | 0.07 | Shajn |
| 0.62 | $B$ | 20.06 | 0.19 | Shajn | 0.64 | $R_c$ | 19.28 | 0.06 | Shajn |
| 0.62 | $B$ | 19.89 | 0.14 | Shajn | 0.64 | $R_c$ | 19.34 | 0.06 | Shajn |
| 0.63 | $B$ | 19.96 | 0.13 | Shajn | 0.64 | $R_c$ | 19.28 | 0.07 | Shajn |
| 0.63 | $B$ | 20.13 | 0.14 | Shajn | 0.64 | $R_c$ | 19.31 | 0.05 | Shajn |
| 0.63 | $B$ | 20.14 | 0.12 | Shajn | 0.65 | $R_c$ | 19.36 | 0.08 | Shajn |
| 0.64 | $B$ | 20.17 | 0.10 | Shajn | 0.65 | $R_c$ | 19.39 | 0.05 | Shajn |
| 0.64 | $B$ | 20.02 | 0.11 | Shajn | 0.65 | $R_c$ | 19.25 | 0.02 | LT |
| 0.64 | $B$ | 20.02 | 0.09 | Shajn | 0.65 | $R_c$ | 19.26 | 0.04 | Shajn |
| 0.65 | $B$ | 20.15 | 0.14 | Shajn | 0.65 | $R_c$ | 19.25 | 0.02 | LT |
| 0.65 | $B$ | 20.09 | 0.12 | Shajn | 0.66 | $R_c$ | 19.28 | 0.04 | Shajn |
| 0.65 | $B$ | 20.24 | 0.10 | Shajn | 0.66 | $R_c$ | 19.29 | 0.02 | LT |
| 0.65 | $B$ | 20.04 | 0.10 | Shajn | 0.66 | $R_c$ | 19.38 | 0.05 | Shajn |
| 0.66 | $B$ | 20.17 | 0.11 | Shajn | 0.66 | $R_c$ | 19.29 | 0.05 | Shajn |
| 0.66 | $B$ | 20.23 | 0.16 | Shajn | 0.67 | $R_c$ | 19.39 | 0.02 | LT |
| 0.66 | $B$ | 20.05 | 0.13 | Shajn | 0.67 | $R_c$ | 19.38 | 0.06 | Shajn |
| 0.67 | $B$ | 20.19 | 0.18 | Shajn | 0.67 | $R_c$ | 19.40 | 0.02 | LT |
| 0.67 | $B$ | 20.05 | 0.24 | Shajn | 0.67 | $R_c$ | 19.42 | 0.07 | Shajn |
| 0.67 | $B$ | 20.17 | 0.22 | Shajn | 0.67 | $R_c$ | 19.33 | 0.07 | Shajn |
| 0.68 | $B$ | 20.08 | 0.28 | Shajn | 0.67 | $R_c$ | 19.41 | 0.02 | LT |
| 0.68 | $B$ | 19.97 | 0.33 | Shajn | 0.67 | $R_c$ | 19.20 | 0.07 | Shajn |
| 1.46 | $B$ | 21.12 | 0.05 | Maidanak | 0.68 | $R_c$ | 19.25 | 0.10 | Shajn |
| 1.46 | $B$ | 21.05 | 0.05 | Maidanak | 0.68 | $R_c$ | 19.38 | 0.11 | Shajn |
| 1.47 | $B$ | 21.11 | 0.05 | Maidanak | 0.68 | $R_c$ | 19.47 | 0.19 | Shajn |
| 1.48 | $B$ | 21.10 | 0.05 | Maidanak | 0.74 | $R_c$ | 19.42 | 0.03 | LT |
| 1.49 | $B$ | 21.02 | 0.07 | Maidanak | 0.74 | $R_c$ | 19.39 | 0.03 | LT |
| 1.49 | $B$ | 21.02 | 0.05 | Maidanak | 0.75 | $R_c$ | 19.41 | 0.03 | LT |
| 1.50 | $B$ | 21.14 | 0.07 | Maidanak | 1.46 | $R_c$ | 20.38 | 0.04 | Maidanak |

Table 3.1—Continued

| $T - T_o$ (days) | Filter | $m_v$[a] | $\sigma$ (mag) | Telescope[b] | $T - T_o$ (days) | Filter | $m_v$[a] | $\sigma$ (mag) | Telescope[b] |
|---|---|---|---|---|---|---|---|---|---|
| 1.65 | $B$ | 21.20 | 0.09 | BTA-6 | 1.46 | $R_c$ | 20.44 | 0.12 | AZT-11 |
| 3.47 | $B$ | 22.52 | 0.06 | Maidanak | 1.47 | $R_c$ | 20.42 | 0.04 | Maidanak |
| 0.06 | $V$ | 16.92 | 0.06 | FTN | 1.48 | $R_c$ | 20.41 | 0.04 | Maidanak |
| 0.10 | $V$ | 17.46 | 0.01 | FTN | 1.49 | $R_c$ | 20.50 | 0.06 | Maidanak |
| 0.15 | $V$ | 17.72 | 0.32 | BOAO | 1.57 | $R_c$ | 20.43 | 0.07 | HCT |
| 0.15 | $V$ | 17.80 | 0.02 | FTN | 1.58 | $R_c$ | 20.42 | 0.06 | BTA-6 |
| 0.15 | $V$ | 17.69 | 0.25 | BOAO | 1.59 | $R_c$ | 20.51 | 0.02 | LT |
| 0.16 | $V$ | 17.84 | 0.25 | BOAO | 1.80 | $R_c$ | 20.66 | 0.02 | LT |
| 0.16 | $V$ | 17.95 | 0.28 | BOAO | 2.55 | $R_c$ | 21.26 | 0.06 | LT |
| 0.19 | $V$ | 18.14 | 0.24 | BOAO | 2.56 | $R_c$ | 21.33 | 0.23 | AZT-11 |
| 0.19 | $V$ | 18.02 | 0.02 | FTN | 3.74 | $R_c$ | 21.73 | 0.10 | Zeiss-1000 |
| 0.24 | $V$ | 18.35 | 0.09 | FTN | 4.47 | $R_c$ | 21.94 | 0.07 | Maidanak |
| 0.48 | $V$ | 19.15 | 0.09 | Shajn | 4.75 | $R_c$ | 22.18 | 0.09 | Zeiss-1000 |
| 0.48 | $V$ | 19.03 | 0.08 | Shajn | 4.98 | $R_c$ | 22.15 | 0.14 | FTN |
| 0.48 | $V$ | 19.05 | 0.10 | Shajn | 5.23 | $R_c$ | 22.13 | 0.04 | LT |
| 0.49 | $V$ | 19.20 | 0.15 | Shajn | 5.44 | $R_c$ | 22.27 | 0.05 | Maidanak |
| 0.49 | $V$ | 19.26 | 0.15 | Shajn | 5.78 | $R_c$ | 22.23 | 0.16 | Zeiss-1000 |
| 0.50 | $V$ | 19.22 | 0.21 | Shajn | 8.80 | $R_c$ | 22.66 | 0.17 | Zeiss-1000 |
| 0.50 | $V$ | 19.43 | 0.24 | Shajn | 10.97 | $R_c$ | 22.73 | 0.14 | FTN |
| 0.50 | $V$ | 19.34 | 0.24 | Shajn | 18.53 | $R_c$ | 22.75 | 0.18 | FTN |
| 0.50 | $V$ | 19.23 | 0.16 | Shajn | 20.79 | $R_c$ | 22.67 | 0.13 | LT |
| 0.51 | $V$ | 19.22 | 0.14 | Shajn | 25.68 | $R_c$ | 22.71 | 0.05 | WHT |
| 0.51 | $V$ | 19.15 | 0.12 | Shajn | 23.47 | $R_c$ | 22.75 | 0.18 | FTN |
| 0.51 | $V$ | 19.14 | 0.13 | Shajn | 28.70 | $R_c$ | 22.76 | 0.08 | LT |
| 0.52 | $V$ | 19.23 | 0.07 | Shajn | 29.06 | $R_c$ | 22.77 | 0.02 | Gemini (N) |
| 0.52 | $V$ | 19.21 | 0.06 | Shajn | 34.56 | $R_c$ | 22.92 | 0.08 | Shajn |
| 0.52 | $V$ | 19.22 | 0.06 | Shajn | 41.02 | $R_c$ | 23.09 | 0.02 | Gemini (N) |
| 0.53 | $V$ | 19.31 | 0.04 | Shajn | 52.58 | $R_c$ | 23.24 | 0.08 | Shajn |
| 0.53 | $V$ | 19.37 | 0.04 | Shajn | 53.97 | $R_c$ | 23.21 | 0.03 | Gemini (N) |
| 0.54 | $V$ | 19.33 | 0.04 | Shajn | 63.92 | $R_c$ | 23.34 | 0.08 | FTN |
| 0.54 | $V$ | 19.28 | 0.05 | Shajn | 98.03 | $R_c$ | 23.34 | 0.04 | INT |
| 0.54 | $V$ | 19.33 | 0.04 | Shajn | 354.76 | $R_c$ | 23.45 | 0.06 | WHT |
| 0.54 | $V$ | 19.32 | 0.14 | OAUV | 0.06 | $i$ | 16.72 | 0.02 | FTN |
| 0.54 | $V$ | 19.32 | 0.04 | Shajn | 0.06 | $i$ | 16.78 | 0.01 | FTN |
| 0.55 | $V$ | 19.38 | 0.05 | Shajn | 0.07 | $i$ | 16.82 | 0.01 | FTN |
| 0.55 | $V$ | 19.34 | 0.05 | Shajn | 0.07 | $i$ | 16.90 | 0.01 | FTN |
| 0.55 | $V$ | 19.33 | 0.04 | Shajn | 0.08 | $i$ | 16.97 | 0.00 | FTN |
| 0.56 | $V$ | 19.22 | 0.09 | Shajn | 0.09 | $i$ | 17.05 | 0.01 | FTN |
| 0.56 | $V$ | 19.30 | 0.03 | OAUV | 0.09 | $i$ | 17.11 | 0.01 | FTN |

Table 3.1—Continued

| $T - T_o$ (days) | Filter | $m_v$[a] | $\sigma$ (mag) | Telescope[b] | $T - T_o$ (days) | Filter | $m_v$[a] | $\sigma$ (mag) | Telescope[b] |
|---|---|---|---|---|---|---|---|---|---|
| 0.56 | *V* | 19.34 | 0.04 | Shajn | 0.10 | *i* | 17.18 | 0.02 | FTN |
| 0.56 | *V* | 19.40 | 0.06 | Shajn | 0.11 | *i* | 17.23 | 0.01 | FTN |
| 0.57 | *V* | 19.44 | 0.06 | Shajn | 0.11 | *i* | 17.29 | 0.01 | FTN |
| 0.57 | *V* | 19.40 | 0.06 | Shajn | 0.12 | *i* | 17.32 | 0.01 | FTN |
| 0.57 | *V* | 19.43 | 0.07 | Shajn | 0.12 | *i* | 17.37 | 0.01 | FTN |
| 0.58 | *V* | 19.41 | 0.05 | Shajn | 0.13 | *i* | 17.40 | 0.01 | FTN |
| 0.58 | *V* | 19.38 | 0.05 | Shajn | 0.14 | *i* | 17.47 | 0.01 | FTN |
| 0.58 | *V* | 19.40 | 0.04 | Shajn | 0.15 | *i* | 17.53 | 0.03 | FTN |
| 0.59 | *V* | 19.53 | 0.05 | Shajn | 0.15 | *i* | 17.59 | 0.02 | FTN |
| 0.59 | *V* | 19.51 | 0.08 | Shajn | 0.16 | *i* | 17.60 | 0.01 | FTN |
| 0.59 | *V* | 19.45 | 0.07 | Shajn | 0.16 | *i* | 17.64 | 0.01 | FTN |
| 0.59 | *V* | 19.27 | 0.09 | OAUV | 0.17 | *i* | 17.67 | 0.01 | FTN |
| 0.59 | *V* | 19.52 | 0.08 | Shajn | 0.18 | *i* | 17.73 | 0.01 | FTN |
| 0.60 | *V* | 19.52 | 0.08 | Shajn | 0.18 | *i* | 17.76 | 0.01 | FTN |
| 0.60 | *V* | 19.48 | 0.08 | Shajn | 0.20 | *i* | 17.72 | 0.02 | FTN |
| 0.60 | *V* | 19.70 | 0.12 | Shajn | 0.20 | *i* | 17.85 | 0.02 | FTN |
| 0.61 | *V* | 19.65 | 0.12 | Shajn | 0.21 | *i* | 17.90 | 0.01 | FTN |
| 0.61 | *V* | 19.48 | 0.10 | Shajn | 0.21 | *i* | 17.93 | 0.01 | FTN |
| 0.61 | *V* | 19.47 | 0.09 | Shajn | 0.22 | *i* | 17.94 | 0.01 | FTN |
| 0.62 | *V* | 19.46 | 0.09 | Shajn | 0.23 | *i* | 17.98 | 0.01 | FTN |
| 0.62 | *V* | 19.51 | 0.09 | Shajn | 0.24 | *i* | 17.91 | 0.04 | FTN |
| 0.62 | *V* | 19.58 | 0.11 | OAUV | 0.24 | *i* | 17.95 | 0.02 | FTN |
| 0.62 | *V* | 19.45 | 0.10 | Shajn | 0.24 | *i* | 18.02 | 0.02 | FTN |
| 0.63 | *V* | 19.58 | 0.10 | Shajn | 0.25 | *i* | 18.09 | 0.01 | FTN |
| 0.63 | *V* | 19.58 | 0.09 | Shajn | 0.26 | *i* | 18.11 | 0.01 | FTN |
| 0.63 | *V* | 19.69 | 0.09 | Shajn | 0.57 | *i* | 19.06 | 0.02 | LT |
| 0.63 | *V* | 19.55 | 0.06 | Shajn | 0.58 | *i* | 19.04 | 0.02 | LT |
| 0.64 | *V* | 19.53 | 0.06 | OAUV | 0.58 | *i* | 19.08 | 0.01 | LT |
| 0.64 | *V* | 19.61 | 0.13 | Shajn | 0.61 | *i* | 19.09 | 0.02 | LT |
| 0.64 | *V* | 19.48 | 0.07 | Shajn | 0.62 | *i* | 19.06 | 0.02 | LT |
| 0.64 | *V* | 19.65 | 0.09 | Shajn | 0.62 | *i* | 19.07 | 0.01 | LT |
| 0.65 | *V* | 19.52 | 0.08 | Shajn | 0.68 | *i* | 19.23 | 0.01 | LT |
| 0.65 | *V* | 19.53 | 0.06 | Shajn | 0.68 | *i* | 19.19 | 0.01 | LT |
| 0.65 | *V* | 19.56 | 0.06 | Shajn | 0.68 | *i* | 19.20 | 0.01 | LT |
| 0.66 | *V* | 19.60 | 0.08 | Shajn | 0.75 | *i* | 19.36 | 0.01 | LT |
| 0.66 | *V* | 19.68 | 0.08 | Shajn | 0.75 | *i* | 19.36 | 0.01 | LT |
| 0.66 | *V* | 19.69 | 0.08 | Shajn | 0.76 | *i* | 19.41 | 0.02 | LT |
| 0.66 | *V* | 19.72 | 0.09 | Shajn | 1.62 | *i* | 20.41 | 0.02 | LT |
| 0.67 | *V* | 19.58 | 0.07 | Shajn | 1.65 | *i* | 20.50 | 0.07 | BTA-6 |

Table 3.1—Continued

| $T-T_o$ (days) | Filter | $m_v$[a] | $\sigma$ (mag) | Telescope[b] | $T-T_o$ (days) | Filter | $m_v$[a] | $\sigma$ (mag) | Telescope[b] |
|---|---|---|---|---|---|---|---|---|---|
| 0.67 | $V$ | 19.52 | 0.11 | Shajn | 1.82 | $i$ | 20.61 | 0.02 | LT |
| 0.67 | $V$ | 19.73 | 0.13 | Shajn | 2.59 | $i$ | 21.07 | 0.04 | LT |
| 0.68 | $V$ | 19.63 | 0.14 | Shajn | 4.73 | $i$ | 21.92 | 0.09 | FTN |
| 0.68 | $V$ | 19.62 | 0.16 | Shajn | 5.26 | $i$ | 22.09 | 0.03 | LT |
| 0.68 | $V$ | 19.73 | 0.27 | Shajn | 11.01 | $i$ | 22.41 | 0.07 | FTN |
| 3.74 | $V$ | 22.22 | 0.14 | Zeiss-1000 | 18.97 | $i$ | 22.34 | 0.08 | FTN |
| 0.04 | $R_c$ | 16.46 | 0.08 | LOAO | 22.78 | $i$ | 22.53 | 0.11 | LT |
| 0.04 | $R_c$ | 16.82 | 0.19 | LOAO | 25.71 | $i$ | 22.33 | 0.05 | WHT |
| 0.06 | $R_c$ | 16.70 | 0.03 | FTN | 27.11 | $i$ | 22.38 | 0.07 | FTN |
| 0.06 | $R_c$ | 16.81 | 0.02 | FTN | 27.75 | $i$ | 22.43 | 0.06 | LT |
| 0.06 | $R_c$ | 16.82 | 0.01 | FTN | 29.09 | $i$ | 22.45 | 0.01 | Gemini (N) |
| 0.07 | $R_c$ | 16.89 | 0.01 | FTN | 41.05 | $i$ | 22.63 | 0.02 | Gemini (N) |
| 0.07 | $R_c$ | 16.90 | 0.08 | LOAO | 54.00 | $i$ | 22.79 | 0.02 | Gemini (N) |
| 0.08 | $R_c$ | 16.99 | 0.01 | FTN | 60.66 | $i$ | 22.88 | 0.07 | WHT |
| 0.08 | $R_c$ | 17.04 | 0.01 | FTN | 63.50 | $i$ | 22.96 | 0.05 | FTN |
| 0.09 | $R_c$ | 17.09 | 0.01 | FTN | 99.55 | $i$ | 23.11 | 0.04 | INT |
| 0.10 | $R_c$ | 17.10 | 0.11 | LOAO | 354.79 | $i$ | 23.22 | 0.06 | WHT |

[a]The apparent magnitude of the GRB+SN+HOST.

[b]Telescope key: AZT-11: 1.25m AZT-11 Telescope of CrAO; BOAO: 15.5cm BOAO Telescope; BTA-6: 6m SAO-RAS Telescope FTN: 2m Faulkes Telescope North; Gemini (N): 8.1m Gemini North; HCT: 2m Himalayan Telescope; INT: 2.5m Isaac Newton Telescope; LOAO: 1m LOAO Telescope; LT: 2m Liverpool Telescope; Maidanak: 1.5m Maidanak Telescope; Mondy: 1.5m AZT-33IK Telescope; OAUV: 0.4m OAUV Telescope; Shajn: 2.6m Shajn Telescope of CrAO; WHT: 4.2m William Herschel Telescope.

## 3.1.2 X-ray Data

The X-ray afterglow was detected and monitored by *Swift*-XRT (Schady et al. 2009; Evans et al. 2009; Beardmore & Schady 2009). The XRT data have been processed[2] with the HEASOFT package v.6.9 and the corresponding calibration files (standard filtering and screening criteria have been applied). For $t \leq 240$ s the data are strongly affected by pile-up: piled-up Window Timing (WT) mode data have been corrected by eliminating a strip of data from the original rectangular region of event extraction: the strip size has been estimated from the grade 0 distribution (according to the *Swift* nomenclature: see Burrows et al. 2005; Romano et al. 2006). Piled-up Photon Counting (PC) data have been extracted from an annular region whose inner radius has been derived comparing the observed to the nominal point spread function (PSF, Moretti et al. 2005; Vaughan et al. 2006). The background is estimated from a source-free por-tion of the sky and then subtracted. The 0.3-10 keV background-subtracted, PSF- and vignetting-corrected light curve has been re-binned so as to ensure a minimum number of 25 re-constructed counts. The count-rate light curves are calibrated into flux and luminosity light curves using a time dependent count-to-flux conversion factor as detailed in Margutti et al. (2010). In this way the strong spectral evolution detected at $t \leq 240$ s is properly accounted for.

[2] The X-ray data was obtained and reduced by C. Guidorzi and R. Margutti. Fitting of the X-ray LC was performed by the author.

Table 3.2 Radio Observations[a] of GRB 090618

| $T - T_o$ (days) | $F_{2.3\ GHz}$ | $\sigma$ | $F_{4.8\ GHz}$ | $\sigma$ | $F_{8.46\ GHz}$ | $\sigma$ | Telescope[b] |
|---|---|---|---|---|---|---|---|
| 0.17 | - | - | - | - | - | - | RT-22 |
| 0.82 | - | - | - | - | - | - | AMI |
| 0.98 | - | - | - | - | 0.38 | 0.05 | VLA[c] |
| 1.68 | - | - | 0.39 | 0.03 | - | - | WSRT |
| 3.20 | <3 | - | - | - | <3 | - | RT-22 |
| 4.67 | - | - | 0.09 | 0.05 | - | - | WSRT |
| 4.70 | 0.04 | 0.05 | - | - | - | - | WSRT |
| 8.68 | - | - | 0.08 | 0.03 | - | - | WSRT |
| 10.26 | <3 | - | - | - | <8.900 | - | RT-22 |

| $T - T_o$ (days) | $F_{15\ GHz}$ | $\sigma$ | $F_{22\ GHz}$ | $\sigma$ | Telescope[b] |
|---|---|---|---|---|---|
| 0.17 | - | - | <5 | - | RT-22 |
| 0.82 | 0.62 | 0.12 | - | - | AMI |
| 0.98 | - | - | - | - | VLA[c] |
| 1.68 | - | - | - | - | WSRT |
| 3.20 | - | - | - | - | RT-22 |
| 4.67 | - | - | - | - | WSRT |
| 4.70 | - | - | - | - | WSRT |
| 8.68 | - | - | - | - | WSRT |
| 10.26 | - | - | - | - | RT-22 |

[a]all measurements in mJy.
[b]Telescope key: AMI: AMI Large Array; RT-22: 22-m RT-22 dish; WSRT: Westerbork Synthesis Radio Telescope; VLA: Very Large Array.
[c]Observations from GCN 9533 (Chandra & Frail 2009).

### 3.1.3 Radio Data

Radio observations were performed with three radio telescopes[3] : the Arcminute Microkelvin Imager (AMI) Large Array, part of the Cavendish Astrophysics Group, Mullard

[3] Radio data were collected and reduced by G. Pooley, A. Pozenenko and A. Van der Horst. Light curve fitting was performed by the author.

Radio Astronomy Observatory, Cambridge UK; the 22m RT-22 radio telescope, in Ukraine; and the Westerbork Synthesis Radio Telescope (WSRT), in the Netherlands.

One epoch was obtained on the AMI on 19 June 2009 in the 14.6 to 17.5 GHz band at 0.823 days post burst which led to a positive detection, reported in the GCN Circulars (Pooley 2009a). The typical noise level of the maps was 0.115 to 0.190 mJy, and resolution was typically 40 x 20 arcseconds.

Three epochs were obtained on the RT-22 radio telescope: 18 June 2009 (22 GHz), 21 June 2009 (2 GHz and 8 GHz) and 28 June 2009 (2 GHz and 8 GHz). The GRB was not detected in any observation, leading to upper limits only.

Additional radio observations were made with WSRT at 2.3 GHz and 4.8 GHz. We used the Multi Frequency Front Ends (Tan 1991) in combination with the IVC+DZB back end[4] in continuum mode, with a bandwidth of 8x20 MHz. Gain and phase calibrations were performed with the calibrators 3C 48 and 3C 286, at 2.3 and 4.8 GHz, respectively. The observations have been analyzed using the Multichannel Image Reconstruction Image Analysis and Display (MIRIAD; Sault et al. 1995) software package.

The first observation at 4.8 GHz, 1.7 days after the burst, was reported in Kamble et al. (2009), but a careful reanalysis of the data resulted in a lower flux value and lower uncertainty in that value. All of the radio observations are listed in Table 3.2.

SECTION 3.2

## Results & Discussion

3.2.1

### The Optical, X-ray & Radio Afterglow

Over 5 hours of optical data were obtained on FTN in $BVR_ci$ starting 1.3 hours after the initial GRB trigger. This was supplemented by data collected on the other telescopes, resulting in well-sampled light curves. Figure 3.1 shows the first six days of optical data: this includes all of the detections in $B$ and $V$, and the first six days of data in $R_c$ and $i$. Several days after the burst the onset of light from an accompanying SN starts to become significant, thus the data after this time cannot be reliably used to characterise *just* the afterglow as it would incorrectly imply a slower afterglow decay.

[4] See Sect. 5.2 at http://www.astron.nl/radio-observatory/astronomers/wsrt-guide-observations

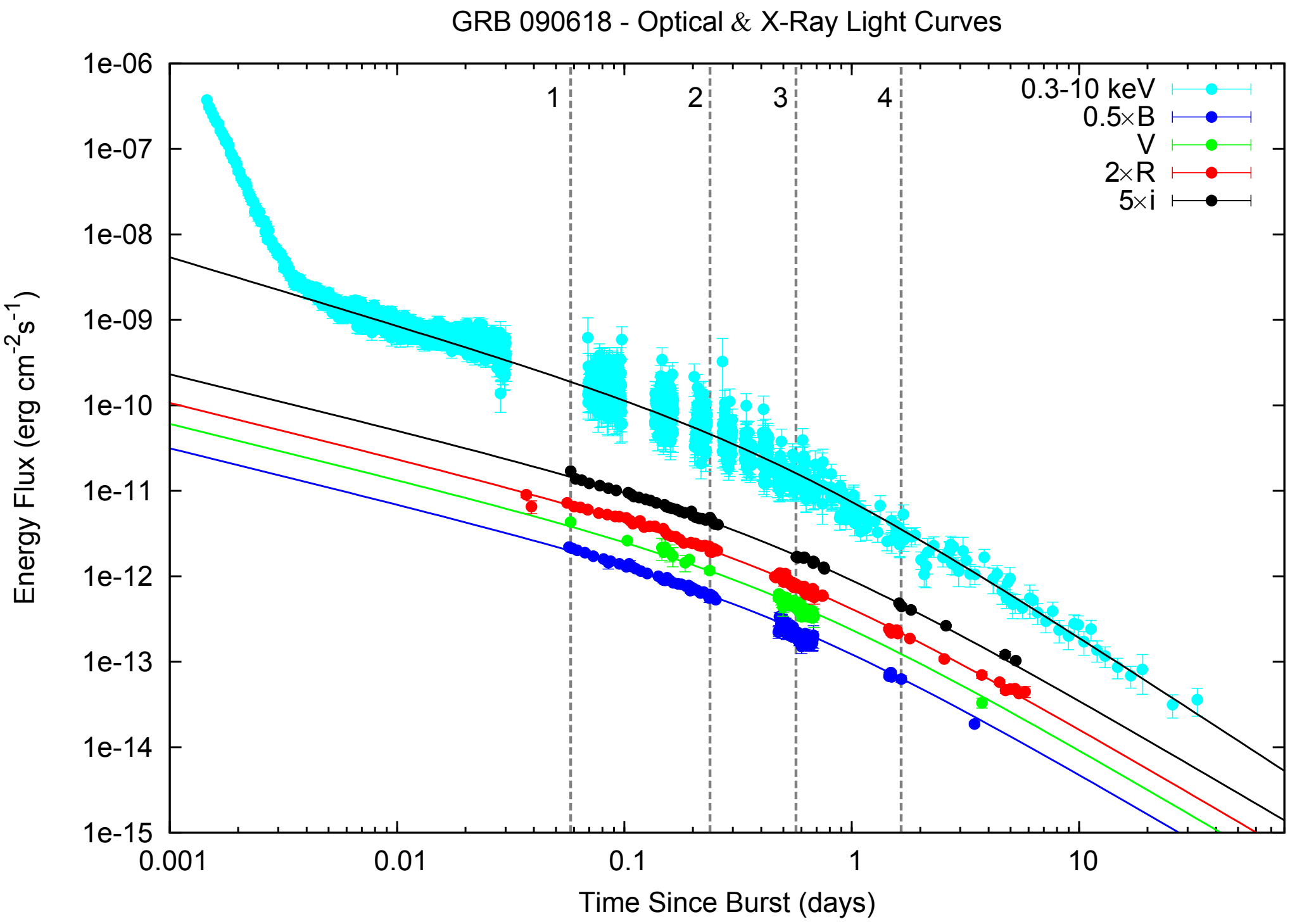

Figure 3.1 GRB 090618: The X-ray and optical light curve. A broken power-law has been fitted separately to the X-ray ($\chi^2/dof = 1.17$: $\alpha_1 = 0.79 \pm 0.01$; $\alpha_2 = 1.74 \pm 0.04$ and $T_{break} = 0.48 \pm 0.08$ days) and optical data ($\chi^2/dof = 1.34$: $\alpha_1 = 0.65 \pm 0.07$; $\alpha_2 = 1.57 \pm 0.07$ and $T_{break} = 0.50 \pm 0.11$ days). The dashed gray lines represent the times of the SEDs.

Inspection of the panchromatic light curve clearly shows an achromatic break. Broken power-laws have been simultaneously fit to the $BVR_ci$ data from 0.03 days to 6 days using Equation (2)[5] , resulting in the following values ($\chi^2/dof = 1.38$): $\alpha_1 = 0.65 \pm 0.08$; $\alpha_2 = 1.57 \pm 0.07$ and $T_{break} = 0.50 \pm 0.11$ days.

The 0.3-10 keV light curve, also shown on Figure 3.1, has been corrected for foreground and host $N_H$ absorption ($N_H = 4.0 \pm 0.9 \times 10^{21}$ cm$^{-2}$, see below) and dis-plays the common "steep-shallow-steep" temporal behaviour (e.g., Nousek et al. 2006; Zhang et al. 2006; Evans et al. 2009) seen for many other GRBs. The LC is quite featureless (i.e., no flares, etc), which is typical of other classical GRBs with confirmed SNe (e.g., Figure 7 in Starling et al. 2010). The data at times greater than $\mathrm{t} - \mathrm{t_o} = 0.006$ days were fit with a broken power-law (Equation 2) resulting in the following values ($\chi^2/dof = 1.17$): $\alpha_1 = 0.79 \pm 0.01$; $\alpha_2 = 1.74 \pm 0.04$ and $T_{break} = 0.48 \pm 0.08$ days. While the values of $\alpha_2$ derived from the optical and X-ray data are relatively similar, they differ by $\approx 2\sigma$, which could be explained by the presence of a frequency break (i.e., cooling break) between the two wavelength regimes (see section 3.2.3 and Figure 3.7).

[5] Similar to our analysis of the afterglow of GRB 060729, we originally fit the optical and X-ray data with the complete Beuermann Function (Beuermann et al. 1999), letting $n$ vary. Again, the results of our fit found $n \approx 1$, so we fixed $n = 1$ when performing the final fit.

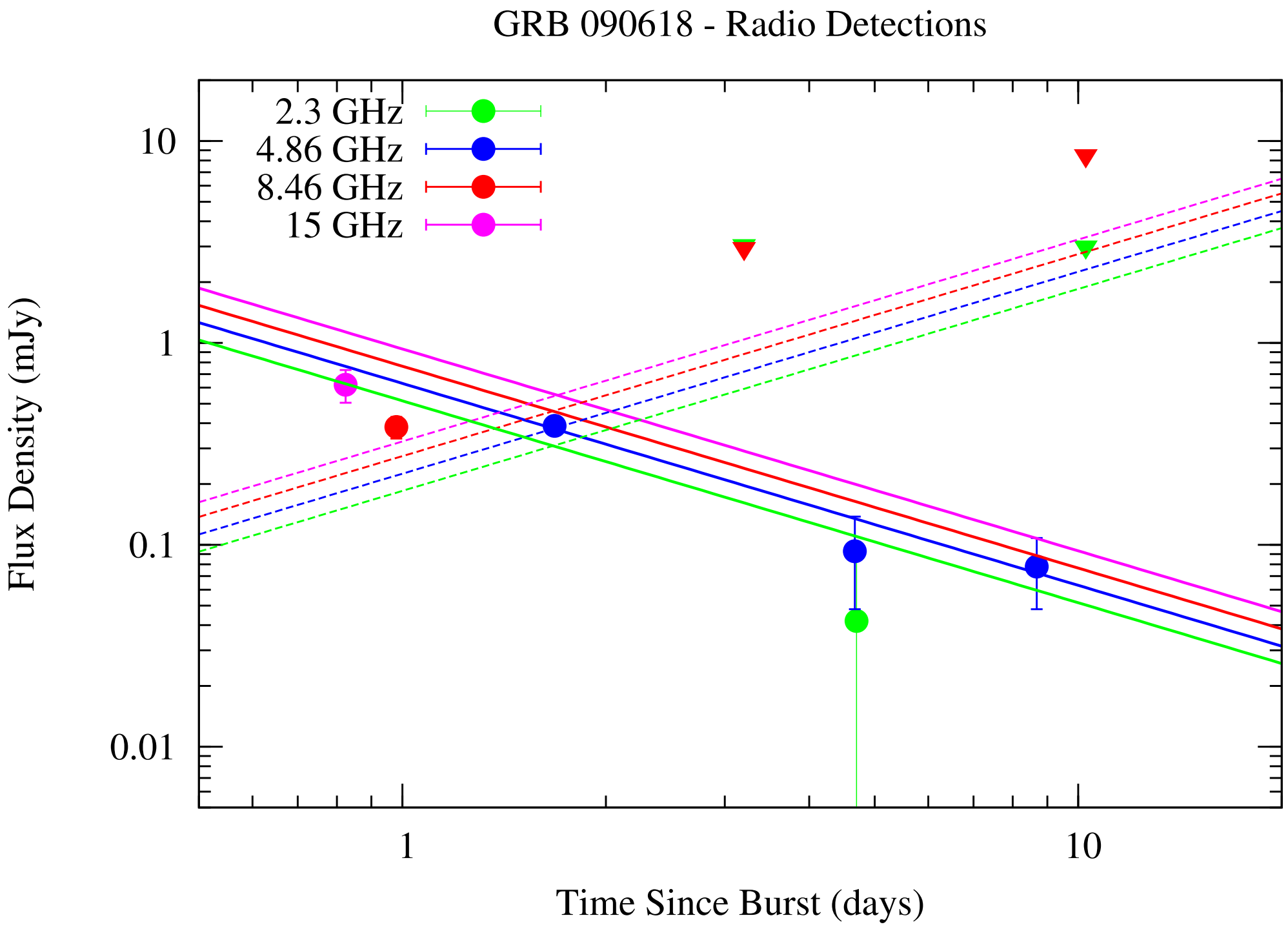

Figure 3.2 GRB 090618: Radio light curves at 2.3 GHz, 4.8 GHz, 8.46 GHz and 16 GHz, where upper limits are denoted by filled triangles. Also plotted are the derived radio light curves for a spherical ($f_\nu(t) \propto t^{1/2}$) (dotted) and jet-like ($f_\nu(t) \propto t^{-1/3}$) (solid) evolution of the afterglow.

We note that the post-break decay index of $\alpha < 2$ is not "permitted" in the most simple theories. However, shallow post-break decays are not a very rare phenomenon, and several examples are seen in the literature (e.g. Zeh et al. 2006).

The column density along the line of sight to the GRB within the host galaxy was estimated from the absorption in the late time X-ray spectrum: the Galactic contribution, $N_H = 5.8 \times 10^{20}$ cm$^{-2}$, was fixed to the value measured from 21cm line radio surveys (Kalberla et al. 2005), while the intrinsic (GRB restframe) $N_H$ was found to be $N_H = 2.7 \pm 0.3 \times 10^{21}$ cm$^{-2}$. The Galactic and intrinsic $N_H$ were obtained with the tbabs (which allows the molecular hydrogen column to be varied) and ztbabs (similar to tbabs but allows the redshift to be fixed) models under xspec respectively. Using $N_H = 2.7 \pm 0.3 \times 10^{21}$ cm$^{-2}$ as the restframe column density we estimate the amount of extinction expected at optical wavelengths using Equation (4) and Table 2 from Pei (1992) for SMC interstellar dust parameters, finding a modest value of $A_V \approx 0.29$ mag. Similarly low values are also derived for Milky Way (MW) and Large Magellanic Cloud (LMC)-type templates.

The radio detections and upper limits are plotted in Figure 3.2. Also plotted are our predicted radio LCs (see section 3.2.3). The radio detections are consistent with a jet-like evolution of the afterglow into an ISM environment. The modelling of the radio detections will be discussed further in section 3.2.4.

### 3.2.2 The Supernova

We measured the host magnitudes from our late-time WHT images (354.7 days) to be $R_c = 23.44 \pm 0.06$ and $i = 23.22 \pm 0.06$. The flux contribution from the host was mathematically subtracted from all epochs, after all of the magnitudes were converted into fluxes using the zero-points from Fukugita et al. (1995). The resulting light curves, which have been corrected for foreground extinction, are displayed in Figure 3.3, and the quoted errors have been calculated in quadrature. Bumps are clearly detected in both light curves.

We then subtracted the flux due to the GRB from the "host-subtracted" light curves using Equation (2), producing the "SN" light curves shown in Figure 3.4. Also plotted are the distance and extinction (foreground) corrected light curves of 1998bw and 1994I. As was seen for the SN associated with 060729, the time evolution and brightness of the SN associated with 090618 resembles more closely that of SN 1998bw[6] than SN 1994I.

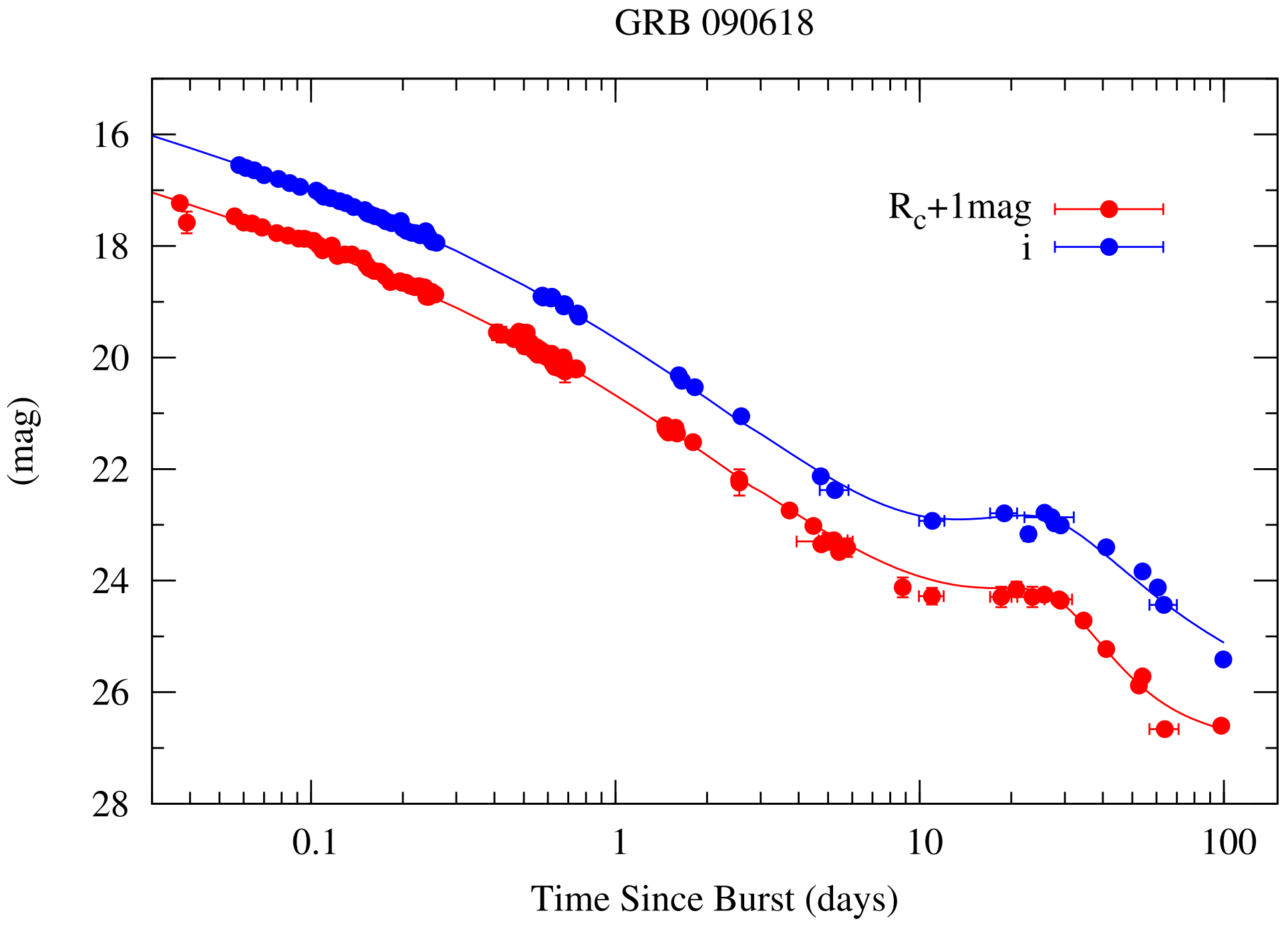

Figure 3.3 GRB 090618: $R_c$ (red) and $i$ (blue) light curves of the host-subtracted magnitudes (see text), which have been corrected for foreground extinction. The solid line is our model (afterglow and SN 1998bw-type SN). The afterglow is modelled with a broken power-law with parameters: $\alpha_1 = 0.65 \pm 0.07$; $\alpha_2 = 1.57 \pm 0.07$ and $T_{break} = 0.50 \pm 0.11$ days, and the template SN (SN 1998bw) is dimmed by 0.75 and 0.5 mag in $R_c$ and $i$ respectively.

We find for the SN associated with GRB 090618 peak apparent magnitudes of $R_c = 23.45 \pm 0.14$ and $i = 23.00 \pm 0.09$. These values, as well as the peak time, are in good agreement with the values predicted by Dado & Dar (2010). Using the observer-frame $i$-band magnitudes to approximate the restframe $V$-band magnitudes, as well as our estimation of the restframe extinction ($A_V \sim 0.3 \pm 0.1$; see section 3.2.3), we find $M_V = -19.75 \pm 0.13$, which is $\sim 0.3$ mag brighter than SN 1998bw in the $V$-band when host extinction is considered (see Table 5.1).

[6] NB: We also considered comparing the accompanying SN with that of SN 2003dh, however it was adjudged that this would not be a useful comparison as the LC of SN 2003dh is not smooth like SN 1998bw as the afterglow of GRB 030329 was very bright and makes the actual evolution of SN 2003dh difficult to accerately determine.

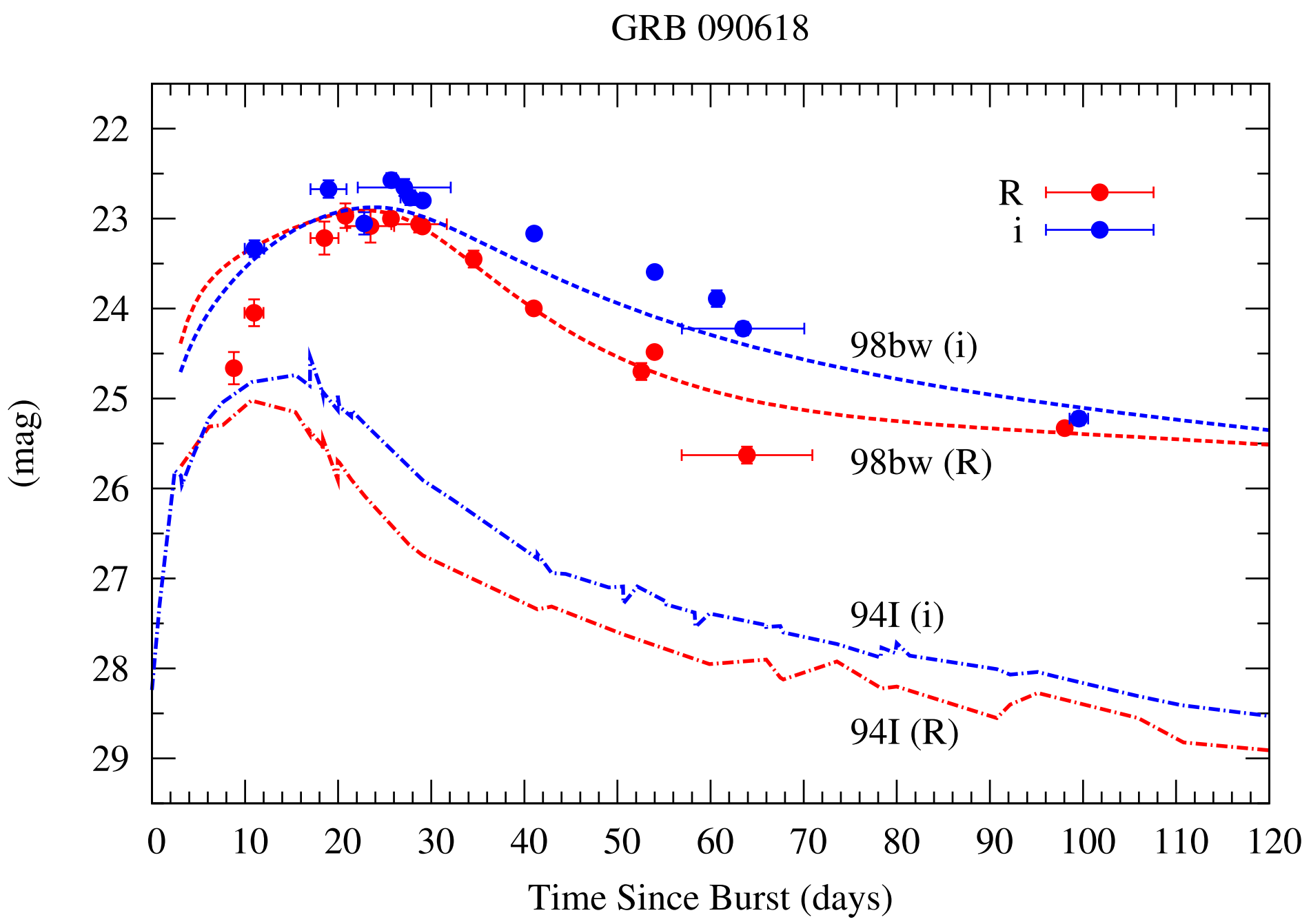

Figure 3.4 GRB 090618: $R_c$ (red) and $i$ (blue) supernova light curves of the host and GRB-subtracted (see text) magnitudes. Plotted for comparison are the light curves of SN 1998bw (dashed) and SN 1994I (dot-dashed) plotted in the same colours, as they would appear at $z = 0.54$. As seen for the 060729-SN, the brightness and time-evolution of the 090618-SN resembles that of SN 1998bw rather than SN 1994I. All magnitudes have been corrected for foreground and host extinction

.

Additional evidence for an association of a SN with GRB 090618 is the change in $R_c - i$ (of the afterglow, SN and host galaxy) over time (Figure 3.5). At early times a blue colour is seen ($R_c - i \approx 0.1$), which we attribute to the afterglow. However, over time the colour index increases, which is not expected for an afterglow (i.e., solely synchrotron radiation), but is indicative of a component of light coming from a core-collapse supernova. At late times as light from the supernova fades away, the colour index decreases and approaches that of the host galaxy ($R_c - i \approx 0.22$, which is typical of GRB host galaxies, e.g., Savaglio et al. 2009, their Table 8).

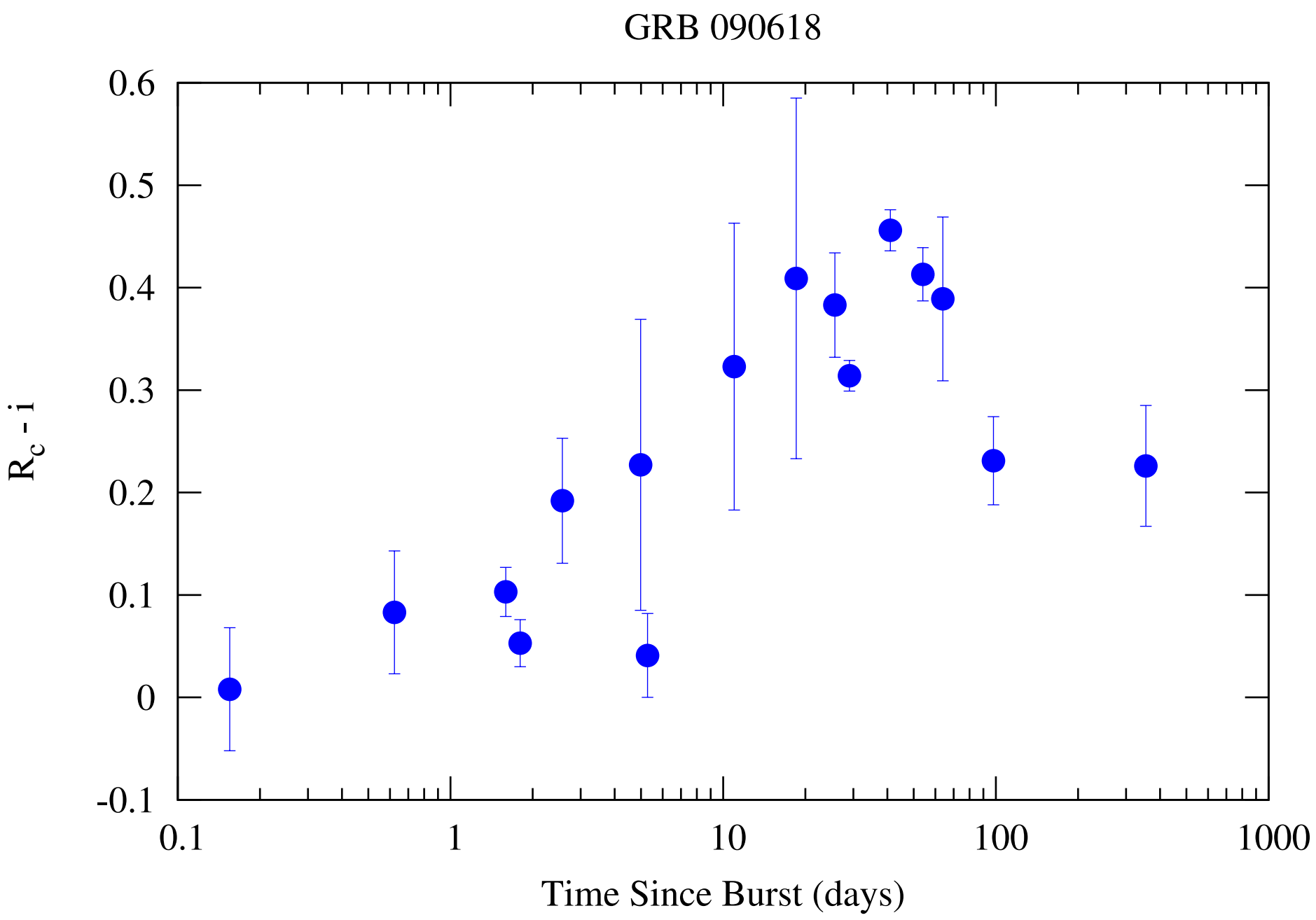

Figure 3.5 GRB 090618: $R_c - i$ for the afterglow, SN and host. The observed colour is not constant (as is expected for light from only a GRB) but increases with time. Such behaviour would be expected if a component of the flux was coming from a core-collapse SN. At late times the colour is that of the host galaxy ($R_c - i \approx 0.22$).

### 3.2.3 The Spectral Energy Distribution

The restframe X-ray to optical SEDs were determined for four epochs: $\mathrm{t} - \mathrm{t_o} = 0.059, 0.238, 0.568$ and 1.68 days[7] . The times of their occurrence are highlighted on Figure 3.1 and are referred to in Figure 3.6 as SED1 to SED4. For SED1, where there is no contemporaneous X-ray and optical data, the X-ray data have been interpolated to the time of the optical observations. For all of the SEDs, the X-ray data have been derived from a broader interval so as to collect enough photons, where a normalization coefficient was determined and the interpolation performed through a power-law, which proved to be a good approximation to the data.

The paucity of optical points prevents us from discriminating between the different optical extinction profiles (i.e., MW, LMC & SMC), thus we considered only the most successful template, i.e., an SMC-type template. The SEDs were fit using the SMC dust extinction profile from Pei (1992) and combined with a parametrization of the photometric cross-section to account for the soft X-ray absorption. They were then applied to a range of models, in particular a simple power-law and a broken power-law. The simple power-law case was immediately rejected. Taking the first epoch as an example (similar behaviour was also seen in the other three epochs), the slope of the optical points is $\beta_{opt} \approx 0.3$ while that of the X-rays is $\beta_X \approx 1.0$. These values are not consistent with a common value for $\beta$. If the optical points had a larger value of $\beta$ than the X-ray, in principle the two values could be reconciled when considering dust effects. However the opposite case is seen here, indicating the only way to reconcile the two frequency regimes is by invoking a frequency break (i.e., cooling break, $\nu_c$) between the two.

[7] The SED fitting was performed by C. Guidorzi. Interpretation of the results of the SED fitting was performed by the author.

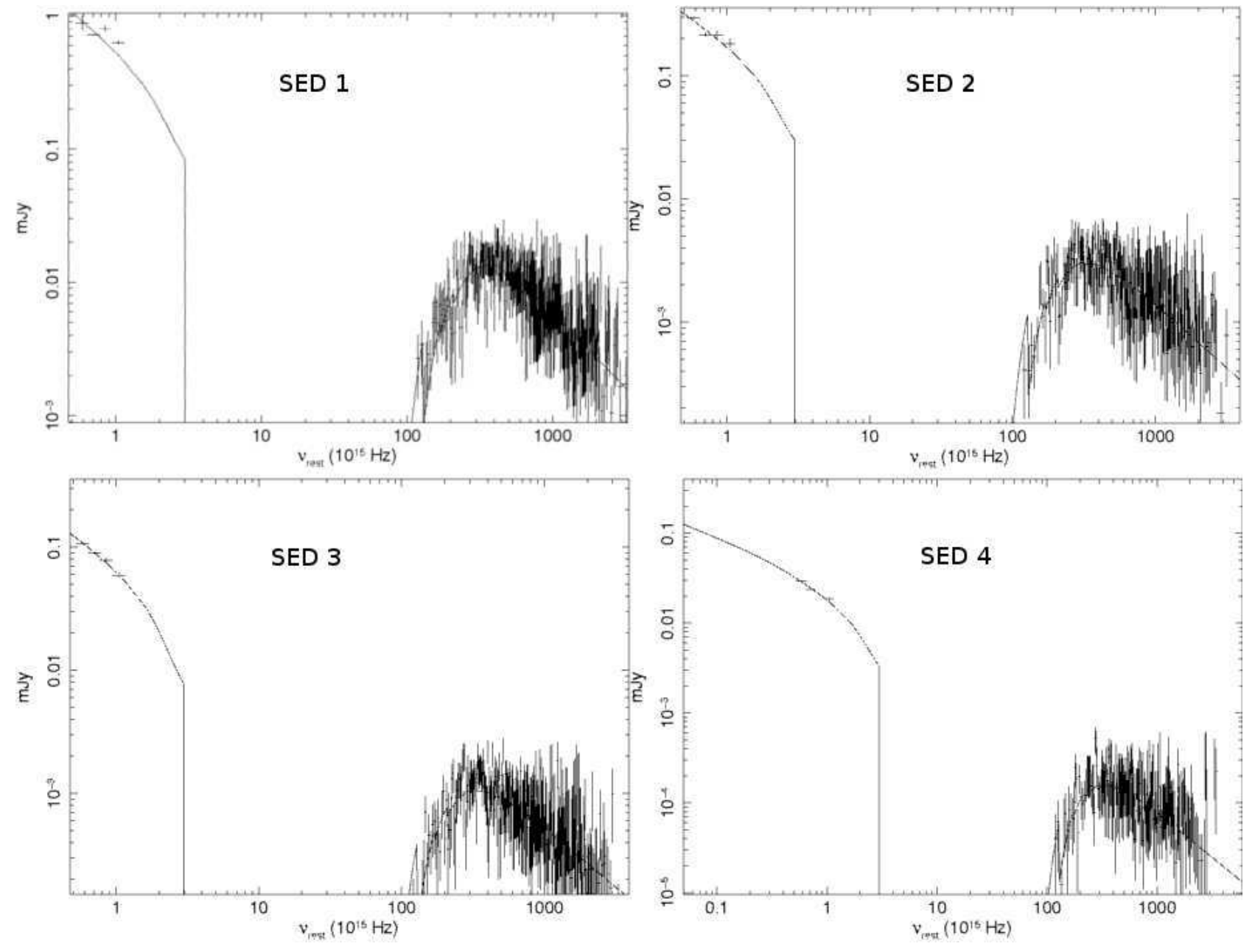

Figure 3.6 GRB 090618: Rest-frame X-ray to optical SEDs: SED1 ($\mathrm{t}-\mathrm{t_o} = 0.059$ days); SED2 ($\mathrm{t}-\mathrm{t_o} = 0.238$ days); SED3: ($\mathrm{t}-\mathrm{t_o} = 0.568$ days); SED4: ($\mathrm{t}-\mathrm{t_o} = 1.68$ days). The results of the SED modelling are: (1) small restframe dust extinction, with $A_V \sim 0.3 \pm 0.1$; (2) Each epoch is well fit by a broken power-law (i.e., with a cooling break between the optical and X-ray energy bands) with $\beta_{optical} \sim 0.5$ and $\beta_{xray} = \beta_{optical} + 0.5$; (3) the break frequency is clearly decreasing with time, which indicates an ISM environment rather than a wind environment.

Thus the optical and X-ray spectral indices, as well as the local extinction were allowed to vary during the fit. After comparison of the initial results it was seen that the results of the fit for each epoch indicated a low value of the extinction, each with a slightly different value. We thus constrained the extinction within the same range for each epoch (as it is physically unlikely that the extinction would change from epoch to epoch), taking the best determined value from SED3, and allowing it to vary within its range ($0.15 < A_V < 0.42$ mag) and re-fitted the other epochs. The best-fitting results are displayed in Table 3.3.

Table 3.3 Best-fitting results to SED modelling for GRB 090618

| SED | $T-T_o$ (days) | $\beta_{optical}$ | $\beta_{xray}$ | $\nu_c$ (Hz) | $A_{V,rest}$ (mag) | $\chi^2/dof$ |
|---|---|---|---|---|---|---|
| SED1 | 0.06 | $0.64^{+0.02}_{-0.01}$ | $1.14^{+0.02}_{-0.01}$ | $3.9^{+0.8}_{-1.2}\times10^{17}$ | $0.24^{+0.09}_{-0.09}$ | 1.0 |
| SED2 | 0.24 | $0.55^{+0.06}_{-0.07}$ | $1.05^{+0.06}_{-0.07}$ | $5.4^{+4.6}_{-3.5}\times10^{16}$ | $0.24^{+0.09}_{-0.09}$ | 0.96 |
| SED3 | 0.57 | $0.48^{+0.06}_{-0.06}$ | $0.98^{+0.06}_{-0.06}$ | $1.6^{+2.4}_{-1.0}\times10^{16}$ | $0.33^{+0.09}_{-0.09}$ | 0.92 |
| SED4 | 1.68 | $0.50^{+0.10}_{-0.11}$ | $1.00^{+0.10}_{-0.11}$ | $7.6^{+6.3}_{-6.2}\times10^{15}$ | $0.24^{+0.09}_{-0.09}$ | 0.96 |

The temporal evolution of the break frequency was also investigated (Table 3.3, Column 5). Sari, Piran & Halpern (1999) have shown that the cooling frequency evolves differently depending on the geometry of the afterglow: $\nu_c \propto t^{-1/2}$ for a spherical evolution, and $\nu_c \approx$ *constant* in a jet-like evolution. It was seen from the X-ray and optical light curves that a break occurred around $\mathrm{t}-\mathrm{t_o} \approx 0.5$ days, thus SED1 and SED2 occur before the break (i.e., spherical evolution), while SED3 and SED4 occur after the break (i.e., a jet-like evolution). For the first two epochs, a power-law was fit to the data (i.e., $\nu_c = A \times t^{\lambda}$). The minimum value of the power-law index that can be reasonably fit to the data is $\nu_c \propto t^{-0.72\pm0.18}$ ($\chi^2/dof = 2.08$), which is consistent with that expected from theory. For the latter two epochs, the large uncertainties derived from the SED fitting (Table 3.3) do not allow us to investigate whether the frequency of the cooling break is still evolving or not, thus it can be constant as expected within the derived errors.

The main conclusions we have drawn from the SED modelling are: (1) small restframe dust extinction, with $A_V \approx 0.3 \pm 0.1$; (2) Each epoch is well fit by a broken power-law (i.e., with a cooling break between the optical and X-ray energy bands) with $\beta_{optical} \sim 0.5$ and $\beta_{xray} = \beta_{optical} + 0.5$; (3) the break frequency is clearly decreasing with time, indicating an ISM environment rather than a magnetised wind environment (as one expects the break frequency to increase in a magnetised wind environment. This is because if the magnetic energy density decreases with time, like in a wind, the synchrotron power also decreases, and thus $\nu_c$ increases. In an ISM environment, which has a constant magnetic energy density, $\nu_c$ decreases as the electrons cool).

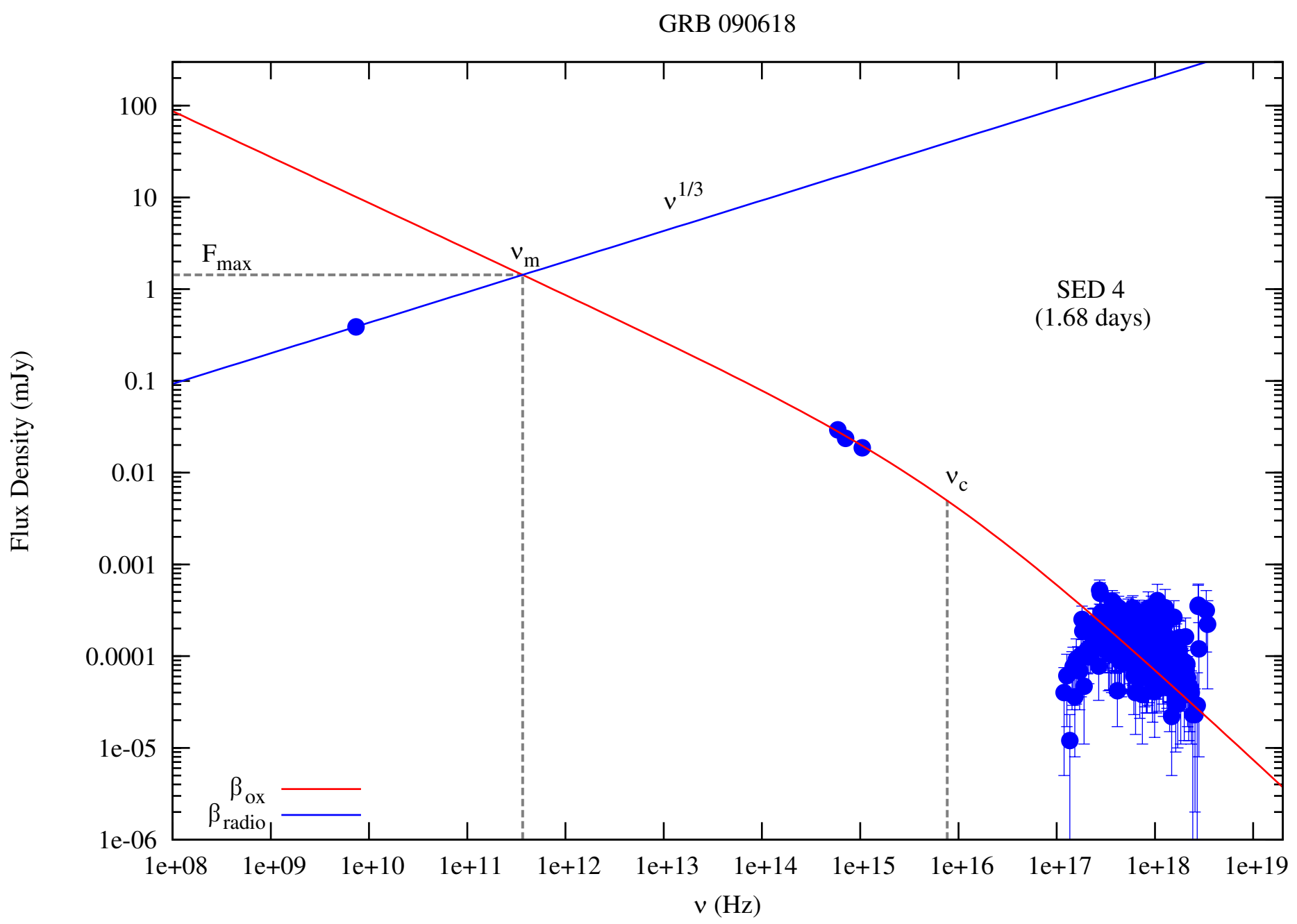

Figure 3.7 GRB 090618: Rest-frame X-ray to radio SED at $t - t_o = 1.68$ days. The typical synchrotron frequency has been estimated from our data, where the optical and X-ray spectral indices have been found from the SED fitting to be: $\beta_{optical} = 0.5$; $\beta_x = \beta_{optical} + 0.5$. We have used the theoretically-expected value of $\beta_{radio} = 1/3$ and found $\nu_m = 3.66 \times 10^{11}$ Hz. The maximum flux was found to be $F_{max} = 1.43$ mJy.

The results of the SED fitting appear quite typical of general afterglow broadband spectra, as does its observed time evolution.

### 3.2.4 The Jet-break, GRB energetics & Predicted Radio Light-Curves

The restframe $1 - 10^4$ keV isotropic energy release in gamma-rays was determined by Ghirlanda, Nava & Ghisellini (2010) in their Table 1: $E_{\gamma,iso} \approx 2.57 \times 10^{53}$ ergs.

A break was seen in the panchromatic LC (Figure 3.1), occurring at both optical and X-ray wavelengths at $\mathrm{t - t_o} \sim 0.5$ days. If the break is interpreted in the context of a single jet (double sided with sharp edges), using a simple model based upon the standard theoretical framework (e.g., Rhoads 1997; Sari, Piran & Halpern 1999) and assuming: (1) forward-shock emission; (2) synchrotron radiation; (3) adiabatic evolution of the afterglow; (4) no self-absorption; (5) canonical medium density of $n \sim 1$ cm$^{-3}$; and (6) an isotropic equivalent kinetic energy comparable to that observed in gamma-rays, as well as the definition for $\theta_{jet}$ from Sari, Piran & Halpern (1999), we estimate an opening angle of $\theta_{jet} \sim 1.5^o$. This in turn implies a corrected, observed gamma-ray emission of $E_{\gamma,\theta} \approx 8.16 \times 10^{49}$ ergs. If the break is indeed due to a jet a homogeneous ISM is favoured, as a clear break is not expected for a wind environment (e.g., Kumar & Panaitescu 2000).

We have been fortunate with our contemporaneous radio detection at 4.86 GHz at $\mathrm{t - t_o} = 1.68$ days, and have modelled our radio, optical and X-ray data as the SED shown in Figure 3.7. We have assumed a jet-like evolution of the afterglow for $\mathrm{t - t_o} >$ 0.5 days (i.e., $f(t) \propto t^{-1.57}$), and neglected effects due to self-absorption, as well as assuming $f_\nu \propto \nu^{1/3}$ for frequencies below the typical frequency (i.e., $\nu < \nu_m$). From our SED modelling at $\mathrm{t - t_o} = 1.68$ days we find $\nu_{sa} \leq \nu_{radio} < \nu_m < \nu_{optical} < \nu_c < \nu_{xray}$ and have estimated the typical synchrotron frequency from our data to be $\nu_m = 3.66 \times 10^{11}$ Hz and the peak flux to be $F_{max} = 1.43$ mJy (Figure 3.7). Using our above assumptions, as well as the derived energetics of the GRB, we have calculated the expected typical frequency of the synchrotron photons and compared it with the observationally-derived value. In the case where the magnetic field energy density is of order the electron energy density (i.e., $\varepsilon_B \approx \varepsilon_e$) we estimate $\varepsilon_B \approx \varepsilon_e \approx 0.045$. Alternatively, if $\varepsilon_B \approx 0.1\varepsilon_e$, we estimate $\varepsilon_B \approx 0.007$ and $\varepsilon_e \approx 0.07$. Additionally, if we use the value of the cooling/break frequency at $t - t_o = 1.68$ days ($\nu_c = 7.6^{+6.3}_{-6.2} \times 10^{15}$ Hz) that was derived from the SED modelling to estimate $\varepsilon_B$ and the medium density $n$, we derive $\varepsilon_B \approx 0.001$ if we assume $n \approx 1$. Alternatively, if we assume $\varepsilon_B \approx 0.01$, we derive a medium density $n \approx 0.05$ proton $cm^{-3}$. These micro-physical values are similar to the assumed values used in the literature (e.g., Zhang et al. 2006).

Using our simple model we have neglected self-absorption effects. However, this assumption may not be completely reasonable, and the effect of self-absorption would result in a lower typical frequency and a higher peak flux. We calculate, using typical values of the micro-physical parameters as well as the definition for $\nu_{sa}$ from Granot, Piran & Sari (1999), a restframe self-absorption frequency of $\nu \approx 2.25 \times 10^{10}$ Hz, which is located in the radio frequency range of our detections.

Assuming that the afterglow has a jet-like evolution after $\approx 0.5$ days, we can estimate the electron energy distribution index $p$ from the optical and X-ray LCs by assuming $t^{-p}$ (i.e., $\alpha = p$) after the break (Sari, Piran & Halpern 1999). The optical data imply $p \approx 1.6$ and the X-ray data imply $p \approx 1.75$. These values are consistent with derived values of $p$ before the break (i.e., for a spherical evolution of the fireball). Using the X-ray data before the break (i.e., $t^{-3p/4+1/2}$ above the cooling frequency), the electron index is found to be $p \approx 1.72$, while from the optical data before the break (i.e., $t^{-3(p-1)/4}$ below the cooling frequency), the value of $\alpha$ implies $p \approx 1.87$. While these values are lower than the predicted universal value of $p \sim 2.2 - 2.3$ (Kirk et al 2000; Achterberg et al. 2001), our derived value of $p$ can be accommodated, when it is considered that $p$ cannot have a single value for all GRBs (e.g., Kann et al. 2006; Shen et al. 2006; Starling et al. 2008; Curran et al. 2010), but has a wide range of possible values (e.g., Curran et al. 2010 used X-ray spectral indices of GRB afterglows to parametrize the underlying distribution of $p$, finding the range consistent with a Gaussian distribution centered on $p = 2.36$ and having a width of 0.59. Our derived value of $p \approx 1.7 - 1.8$ falls at the lower end of this distribution).

The initial Lorentz factor of the ejecta can be estimated by considering the deceleration time. Assuming that the onset of the afterglow is masked by the steep decay component (i.e., the tail of the prompt emission), using the definition from Sari & Pi-ran (1999), and correcting for observer time, we estimate an initial Lorentz factor of $\gamma_o > 100$.

The derived radio light curves are shown in Figure 3.2, where the flux scalings have been derived from the X-ray to radio SED at $\mathrm{t - t_o} = 1.68$ days (Figure 3.7). Both a spherical ($f_\nu(t) \propto t^{1/2}$) and jet-like ($f_\nu(t) \propto t^{-1/3}$) evolution of the afterglow are considered, and we conclude that all of the radio observations can be suitably explained by a jet-like evolution of the fireball, though we note that the scarcity of radio detections limits our ability to confidently determine the actual evolution of the afterglow beyond doubt.

SECTION 3.3

## Conclusions

We have presented optical, X-ray and radio data for GRB 090618. The large optical dataset has allowed us to clearly detect supernova bumps in the $R_c$- and $i$-band LCs. As for GRB 060729, we attributed light coming from three sources: (1) the afterglow, (2) the supernova, and (3) the host galaxy. This time, we have not been able to carry out image-subtraction on our optical images, however we have measured the host brightness at late times (i.e. well after the GRB and SN have faded away), converted the host magnitudes into fluxes, and subtracted them from the earlier observations, producing LCs of just the OT (Figure 3.3). We then modeled the early-time optical data to determine the rate at which the afterglow faded, and then subtracted this model from the OT LC to obtain "afterglow-subtracted", supernova light curves (Figure 3.4).

We used the value of the restframe extinction determined from the optical-to-X-ray SEDs (Figure 3.6) to determine the peak, restframe, absolute $V$-band magnitude of the associated SN. We found peak apparent magnitudes of the SN to be $R_c = 23.45 \pm 0.08$ and $i = 23.00 \pm 0.06$, which at a redshift of $z = 0.54$, corresponds to $M_V = -19.75 \pm 0.13$, which is $\sim 0.3$ mag brighter than SN 1998bw in the $V$-band.

In addition to constraining the restframe extinction using our optical-to-X-ray

SEDs, we also concluded that: (1) each epoch was well described by a broken power-law, with a cooling break between the optical and X-ray energy bands, and (2) the cooling frequency clearly decreased in time, indicating an ISM environment (rather than a wind environment). These conclusions, coupled with the results of modeling the radio data, where we found that the observations were consistent with a jet-like evolution of the fireball, support the notion that GRB 090618 is typical of other long, soft GRBs.

# 4

# XRF 100316D / SN 2010bh

XRF 100316D was detected by *Swift* on the $16^{th}$ March, 2010 at 12:44:50 UT. The redshift was measured to be $z = 0.0591 \pm 0.0001$ (Vergani et al. 2010a; Starling et al. 2011; Chornock et al. 2010). Spectroscopy presented by Chornock et al. (2010) revealed similar behaviour to previous GRB-SNe, where the spectra are featureless during the first few days, and with the emergence of broad spectral features after several days. As for other GRB-SNe, Chornock et al. (2010) reported that there was no evidence for helium in their spectroscopy, indicating that the progenitor is a highly-stripped star. However, spectroscopy taken by Bufano et al. (2011) revealed a broad He absorption feature at 10830Å, a spectral line that has only been observed one other time in a GRB-SNe (SN 1998bw, Patat et al. 2001). Bufano et al. (2011) also detected He in their early spectra at 5876Å. These detections of He indicate that the progenitor star still retained at least a small fraction of its He envelope prior to explosion, indicating that SN 2010bh is a type Ibc SNe.

Starling et al. (2011) found that the high-energy properties of XRF 100316D / SN 2010bh are quite different from those of other GRB-SNe, but are similar to the high energy properties of XRF 060218 / SN 2006aj. Starling et al. (2011) showed in the X-ray spectrum at 144 s $\leq \mathrm{t} - \mathrm{t_o} \leq$ 737 s the presence of a soft, hot, black-body component that contributes $\approx 3\%$ to the total 0.3 – 10.0 keV flux (though see also Fan et al. 2011 who claim that both a Cutoff Power-law model or a Cutoff Power-law model + black-body component provide equally acceptable fits). The presence of the black-body component ($kT = 0.14$ keV; $T \approx 1.6 \times 10^6$ K) is similar in temperature to the soft, hot, black-body component seen in the X-ray spectrum of XRF 060218 / SN 2006aj ($kT \approx 0.17$ keV; $T \approx 2.0 \times 10^6$ K; Campana et al. 2006).

A detailed discussion of the host galaxy metallicity at the explosion site of SN 2010bh was performed by Levesque et al. (2011). The authors presented evidence, which was also reported by Chornock et al. (2010) and Starling et al. (2011), that SN 2010bh occurred in a star-forming region of the host galaxy that has a star-formation rate of $\sim 1.7 M_{\odot}$ yr$^{-1}$ (at the site of the supernova) and low metallicity ($Z \leq 0.4 Z_{\odot}$; Chornock et al. (2010), Levesque et al. (2011)).

In this chapter we present in Section 4.1 optical and infrared photometry of XRF 100316D / SN 2010bh obtained on FTS, Gemini-S and *HST*. In Section 4.2 we use our optical light curves (LCs) to derive the peak optical properties of SN 2010bh, showing that SN 2010bh evolves much quicker than SN 1998bw, and at a rate similar to SN 2006aj. We also discuss in Section 4.2.4 the origin of the flux at $\mathrm{t} - \mathrm{t_o} = 0.598$ days, concluding that it is not synchrotron but is likely coming from the shock breakout. In Section 4.3 we make an estimate of the rest frame extinction, and derive peak absolute magnitudes of SN 2010bh. In Section 4.4 we use our optical and infrared (IR) photometry, along with our estimation of the host extinction, to construct the bolometric light curve, and use an analytical model to extract physical parameters of the SN. In Section 4.5 we discuss SN 2010bh in relation to SN 2006aj, as well as other GRB-SNe and local type Ic SNe that were not accompanied by a GRB. In the discussion we investigate the light curve properties of most of the spectroscopically and photometrically-linked GRB-SNe. By making simple assumptions, we obtain host- and afterglow-subtracted "supernova" light curves, and compare them with SN 1998bw, deriving stretch ($s$) and luminosity ($k$) factors of the GRB-SNe in relation to SN 1998bw. We then check for a correlation between the stretch and luminosity factors, concluding that no statistically-significant correlation exists.

Throughout this chapter foreground reddening of $E(B-V) = 0.117$ mag has been corrected for using the dust maps of Schlegel, Finkbeiner & Davis (1998).

SECTION 4.1

## Observations & Photometry

4.1.1

### Data Acquisition & Calibration

***Faulkes Telescope South & Gemini-South***

We obtained forty-one epochs of photometric data between March and May 2010 using the 2m FTS. Observations were made using the $4.6'$ x $4.6'$ field of view Merope Camera in filters $BVR_ci$. A $42^{nd}$ and final epoch of observations was made in December 2010 in all filters, which were used as template/reference images for image subtraction (see Section 2.2). Images were also taken of Landolt photometric standard regions (Landolt 1992) in $BVR_ci$ on the $2^{nd}$ of May, 2010 using the Spectral Camera, which with its wider field of view ($10.5'$ x $10.5'$), allowed us to calibrate over 140 secondary standards in the field of XRF 100316D. Subsequent observations made on Gemini-S and *HST* are calibrated to these secondary standards.

Five epochs of photometric data were obtained on Gemini-S[1] in filters *griz*, the first starting only $\approx 0.5$d after the initial burst. A final epoch was obtained on the $28^{th}$ of January, 2011 in all filters, which were used as template images for image subtraction (see Section 2.2).

***Hubble Space Telescope***

We obtained five epochs of photometric data with *HST*, using the Wide Field

[1] The data presented here supersedes those published in the GCN Circulars: GCNs 10513 (Vergani et al. 2010b); 10523 (Levan et al. 2010); 10525 (Wiersema et al. 2010).

Camera 3 (WFC3). The first three epochs of data were obtained in filters $F336W$, $F555W$, $F814W$, $F125W$ and $F160W$, while the fourth epoch yielded images in filters $F336W$, $F555W$, $F814W$, and the fifth epoch in filters $F555W$ and $F814W$. We find for the optical transient (OT) associated with XRF 100316D a position of: $07^h10^m30.54^s(\pm0.02)^s$, $-56^d15'19.80''(\pm0.10)''$.

Aperture photometry was performed on the *HST* images using an aperture of radius 2 pixels and standard routines in IRAF[2] . A small aperture was used, and an aperture correction for an isolated star was computed and applied. The aperture-corrected, instrumental magnitudes were then calibrated via standard star photometry to secondary standards in the field of XRF 100316D. Images taken in filters $F555W$ and $F814W$ are calibrated respectively to filters $V$ and $i$ using a zero-point and a colour term. The remaining filters are calibrated to AB magnitudes using the appropriate AB zero-point for each filter that are listed in the WFC3 instrument and data-handbook.[3] The *HST* magnitudes, which have been corrected for foreground extinction, are listed in Table 4.1. The quoted errors, which are added in quadrature, are derived from the uncertainties associated with the photometry and calibration.

### 4.1.2 Image Subtraction

Using ISIS (Alard 2000), we performed image subtraction on the FTS and Gemini-S images, using the last epoch of images taken in each filter and on each telescope as the respective templates. Subtracting the template from the early FTS images gave a clear detection of the optical transient (OT) and an accurate position, and the same was also observed for the Gemini-S images.

Aperture photometry was then performed on the subtracted images using a small aperture and an aperture correction was computed from the pre-subtracted images and

[2] IRAF is distributed by the National Optical Astronomy Observatory, which is operated by the Association of Universities for Research in Astronomy, Inc., under cooperative agreement with the National Science Foundation.

[3] http://www.stsci.edu/hst/wfc3

Table 4.1 *HST* Photometry of XRF 100316D / SN 2010bh

| Filter | $t - t_o$ (days) | Mag | $\sigma$ (mag) | Calibrated to |
|---|---|---|---|---|
| $F336W$ | 8.650 | 21.86 | 0.06 | AB |
| $F336W$ | 15.810 | 22.80 | 0.08 | AB |
| $F336W$ | 24.860 | 23.56 | 0.11 | AB |
| $F336W$ | 47.340 | 24.05 | 0.14 | AB |
| $V$ | 8.663 | 19.50 | 0.01 | Vega |
| $V$ | 15.822 | 19.81 | 0.01 | Vega |
| $V$ | 24.869 | 20.52 | 0.02 | Vega |
| $V$ | 47.352 | 21.63 | 0.03 | Vega |
| $V$ | 137.490 | 23.07 | 0.08 | Vega |
| $i$ | 8.675 | 19.00 | 0.01 | AB |
| $i$ | 15.844 | 19.12 | 0.01 | AB |
| $i$ | 24.881 | 19.59 | 0.01 | AB |
| $i$ | 47.370 | 20.87 | 0.02 | AB |
| $i$ | 137.518 | 22.70 | 0.06 | AB |
| $F125W$ | 8.720 | 19.89 | 0.07 | AB |
| $F125W$ | 15.870 | 20.09 | 0.07 | AB |
| $F125W$ | 25.480 | 20.10 | 0.07 | AB |
| $F160W$ | 8.730 | 19.93 | 0.08 | AB |
| $F160W$ | 15.880 | 20.18 | 0.08 | AB |
| $F160W$ | 25.520 | 20.32 | 0.07 | AB |

All magnitudes have been corrected for foreground extinction.

applied. The instrumental magnitudes obtained from the FTS subtracted images are calibrated to $BVR_ci$. The Gemini-S subtracted images taken in filters *griz* are calibrated to $BR_ciz$ respectively. Images in filters *gr* are calibrated using a zero-point and a colour term, while images taken in *i* are calibrated using only a zero-point. Magnitudes of the secondary standards in *z* are computed using transformation equations from Jester et al. (2005), and the Gemini-S data are calibrated to these using a zero-point. All of the image-subtracted, foreground-corrected magnitudes obtained from the FTS and Gemini-S data are displayed in Table 4.2, where the quoted errors (added in quadrature) are determined from the photometry and calibration.

Table 4.2 Ground-based Photometry of XRF 100316D / SN 2010bh

| Filter | $t - t_o$ (days) | Mag | $\sigma$ (mag) | Calibrated to | Telescope[a] |
|---|---|---|---|---|---|
| $B$ | 0.60 | 20.45 | 0.03 | Vega | GS |
| $B$ | 3.49 | 20.53 | 0.01 | Vega | GS |
| $B$ | 5.47 | 20.22 | 0.01 | Vega | GS |
| $B$ | 7.06 | 20.25 | 0.05 | Vega | FTS |
| $B$ | 7.94 | 20.14 | 0.03 | Vega | FTS |
| $B$ | 9.96 | 20.13 | 0.03 | Vega | FTS |
| $B$ | 10.97 | 20.19 | 0.06 | Vega | FTS |
| $B$ | 16.99 | 21.18 | 0.07 | Vega | FTS |
| $B$ | 17.86 | 21.30 | 0.05 | Vega | FTS |
| $B$ | 19.96 | 21.44 | 0.13 | Vega | FTS |
| $B$ | 22.99 | 21.83 | 0.11 | Vega | FTS |
| $B$ | 25.96 | 21.88 | 0.16 | Vega | FTS |
| $B$ | 30.97 | 22.27 | 0.10 | Vega | FTS |
| $B$ | 33.91 | 22.09 | 0.14 | Vega | FTS |
| $V$ | 5.01 | 19.64 | 0.02 | Vega | FTS |
| $V$ | 7.09 | 19.49 | 0.03 | Vega | FTS |
| $V$ | 7.96 | 19.48 | 0.02 | Vega | FTS |
| $V$ | 8.94 | 19.47 | 0.02 | Vega | FTS |
| $V$ | 9.97 | 19.47 | 0.03 | Vega | FTS |
| $V$ | 17.00 | 19.91 | 0.04 | Vega | FTS |
| $V$ | 17.90 | 19.96 | 0.02 | Vega | FTS |
| $V$ | 19.01 | 20.03 | 0.07 | Vega | FTS |
| $V$ | 23.03 | 20.29 | 0.04 | Vega | FTS |
| $V$ | 25.94 | 20.61 | 0.04 | Vega | FTS |
| $V$ | 35.42 | 21.16 | 0.06 | Vega | FTS |
| $R_c$ | 0.47 | 20.21 | 0.01 | Vega | GS |
| $R_c$ | 1.47 | 20.09 | 0.01 | Vega | GS |
| $R_c$ | 2.03 | 20.03 | 0.03 | Vega | FTS |
| $R_c$ | 2.46 | 19.96 | 0.01 | Vega | GS |
| $R_c$ | 2.89 | 19.87 | 0.02 | Vega | FTS |
| $R_c$ | 3.49 | 19.77 | 0.01 | Vega | GS |
| $R_c$ | 5.00 | 19.47 | 0.02 | Vega | FTS |
| $R_c$ | 5.47 | 19.46 | 0.01 | Vega | GS |
| $R_c$ | 7.89 | 19.33 | 0.02 | Vega | FTS |
| $R_c$ | 9.07 | 19.29 | 0.02 | Vega | FTS |
| $R_c$ | 11.02 | 19.33 | 0.02 | Vega | FTS |
| $R_c$ | 11.99 | 19.32 | 0.04 | Vega | FTS |
| $R_c$ | 16.05 | 19.49 | 0.04 | Vega | FTS |
| $R_c$ | 17.02 | 19.52 | 0.02 | Vega | FTS |
| $R_c$ | 19.92 | 19.71 | 0.02 | Vega | FTS |
| $R_c$ | 22.98 | 19.83 | 0.02 | Vega | FTS |
| $R_c$ | 25.96 | 20.07 | 0.03 | Vega | FTS |
| $R_c$ | 29.94 | 20.54 | 0.04 | Vega | FTS |
| $R_c$ | 33.90 | 20.83 | 0.04 | Vega | FTS |
| $R_c$ | 36.90 | 21.08 | 0.07 | Vega | FTS |
| $R_c$ | 37.99 | 21.20 | 0.13 | Vega | FTS |
| $R_c$ | 45.92 | 21.36 | 0.08 | Vega | FTS |
| $R_c$ | 48.87 | 21.45 | 0.07 | Vega | FTS |
| $R_c$ | 54.85 | 21.57 | 0.07 | Vega | FTS |
| $R_c$ | 57.36 | 21.64 | 0.07 | Vega | FTS |
| $i$ | 0.48 | 20.04 | 0.01 | AB | GS |
| $i$ | 3.50 | 19.44 | 0.01 | AB | GS |
| $i$ | 5.00 | 19.14 | 0.02 | AB | FTS |
| $i$ | 5.48 | 19.12 | 0.01 | AB | GS |
| $i$ | 8.97 | 19.04 | 0.01 | AB | FTS |
| $i$ | 16.03 | 19.12 | 0.03 | AB | FTS |
| $i$ | 17.01 | 19.18 | 0.02 | AB | FTS |
| $i$ | 18.96 | 19.27 | 0.05 | AB | FTS |
| $i$ | 19.94 | 19.34 | 0.02 | AB | FTS |
| $i$ | 23.01 | 19.55 | 0.03 | AB | FTS |
| $i$ | 25.98 | 19.59 | 0.05 | AB | FTS |
| $i$ | 33.89 | 20.26 | 0.08 | AB | FTS |
| $i$ | 37.01 | 20.63 | 0.08 | AB | FTS |
| $i$ | 46.46 | 20.85 | 0.07 | AB | FTS |
| $i$ | 49.39 | 20.99 | 0.06 | AB | FTS |
| $i$ | 56.89 | 21.20 | 0.09 | AB | FTS |
| $z$ | 0.47 | 19.91 | 0.11 | AB | GS |
| $z$ | 3.49 | 19.40 | 0.01 | AB | GS |
| $z$ | 5.47 | 19.15 | 0.01 | AB | GS |

[a]Telescope key: FTS: 2m Faulkes Telescope South; GS: 8.1m Gemini-South Telescope.
All magnitudes have been corrected for foreground extinction.

Table 4.3 Gemini-S Host Photometry of XRF 100316D / SN 2010bh

| Filter | Mag | $\sigma$ (mag) | Telescope |
|---|---|---|---|
| $g$ | 17.40 | 0.08 | Gemini-S |
| $r$ | 17.14 | 0.07 | Gemini-S |
| $i$ | 17.07 | 0.06 | Gemini-S |
| $z$ | 17.03 | 0.06 | Gemini-S |

All magnitudes have been corrected for foreground extinction.

### 4.1.3 Host Photometry

As already noted by Starling et al. (2011), the host galaxy system of XRF 100316D / SN 2010bh has a highly disturbed morphology, with a bright central region, a possible spiral arm and a number of bright knots (see Figure 8 in Starling et al. 2011). SN 2010bh is located on top of a bright knot close to the bright central region, which is perhaps the galactic nucleus. In our final epoch of Gemini-S images, the host-complex appears as a blended, elliptical object due to the lower resolution of Gemini-S relative to *HST*.

We have performed photometry on the final Gemini-S images using SExtractor (Bertin & Arnouts 1996). We obtained (1) corrected isophotal magnitudes (MAG_ISOCOR), and (2) Kron-like elliptical aperture magnitudes (MAG_AUTO), with the best of these values being used (MAG_BEST). The host magnitudes in filters *griz* have been calibrated to secondary standards in the field of XRF 100316D, where the magnitudes of the secondary standards in filters *grz* are computed using transformation equations from Jester et al. (2005), and the host photometry is calibrated to these using a zeropoint. The Gemini-S photometry, which has been corrected for foreground extinction, is listed in Table 4.3, where the quoted errors of the Gemini-S magnitudes are added in quadrature and are derived from the photometry and calibration.

SECTION 4.2

## Optical & Infrared Light Curves and colour curves

4.2.1

### Light Curves

The template-subtracted light curves of XRF 100316D / SN2010bh, which have been corrected for foreground and host extinction (see Section 4.3), are displayed in Figure 4.1. Plotted for comparison are the multi-wavelength light curves of two known GRB-SNe: SN 1998bw and SN 2006aj, as well as SN 1994I which is a local type Ic SN that was not associated with a GRB. All of the comparison light curves have been shifted in magnitude to match SN 2010bh around maximum light. No other changes have been made to the light curves apart from the light curves of SN 1998bw, which have been stretched in each filter by the values listed in Table 4.5 (see Section 3.5), and the early-time data ($\mathrm{t}-\mathrm{t_o} < 3$ days) for SN 2006aj, for which the light was dominated by flux coming from the very bright shock break-out, was not included to avoid confusion between different filters. The stretch-factors have been applied to SN 1998bw for clarity.

The peak times and magnitudes in each filter are estimated from low-order polynomial fits. We have determined that SN 2010bh peaked at $\mathrm{t}-\mathrm{t_o} = 8.02 \pm 0.12$ days in the $B$ filter, and at subsequently later times in the redder filters. The quoted errors are statistical and are determined from the distribution of peak times determined from the different order polynomials. The behaviour of the LC peaking at later times in redder filters is typical behaviour for a core-collapse SNe, and SN 2010bh appears no different in this respect. The main photometric parameters of XRF 100316D / SN 2010bh in filters $BVR_ci$ are presented in Table 4.4.

Comparing the optical light curves of SN 2010bh with those of SN 2006aj, it appears they evolve at roughly the same rate in the $B$ filter. However it appears that the $U$-band light curve of SN 2006aj evolves faster than SN 2010bh, while in the

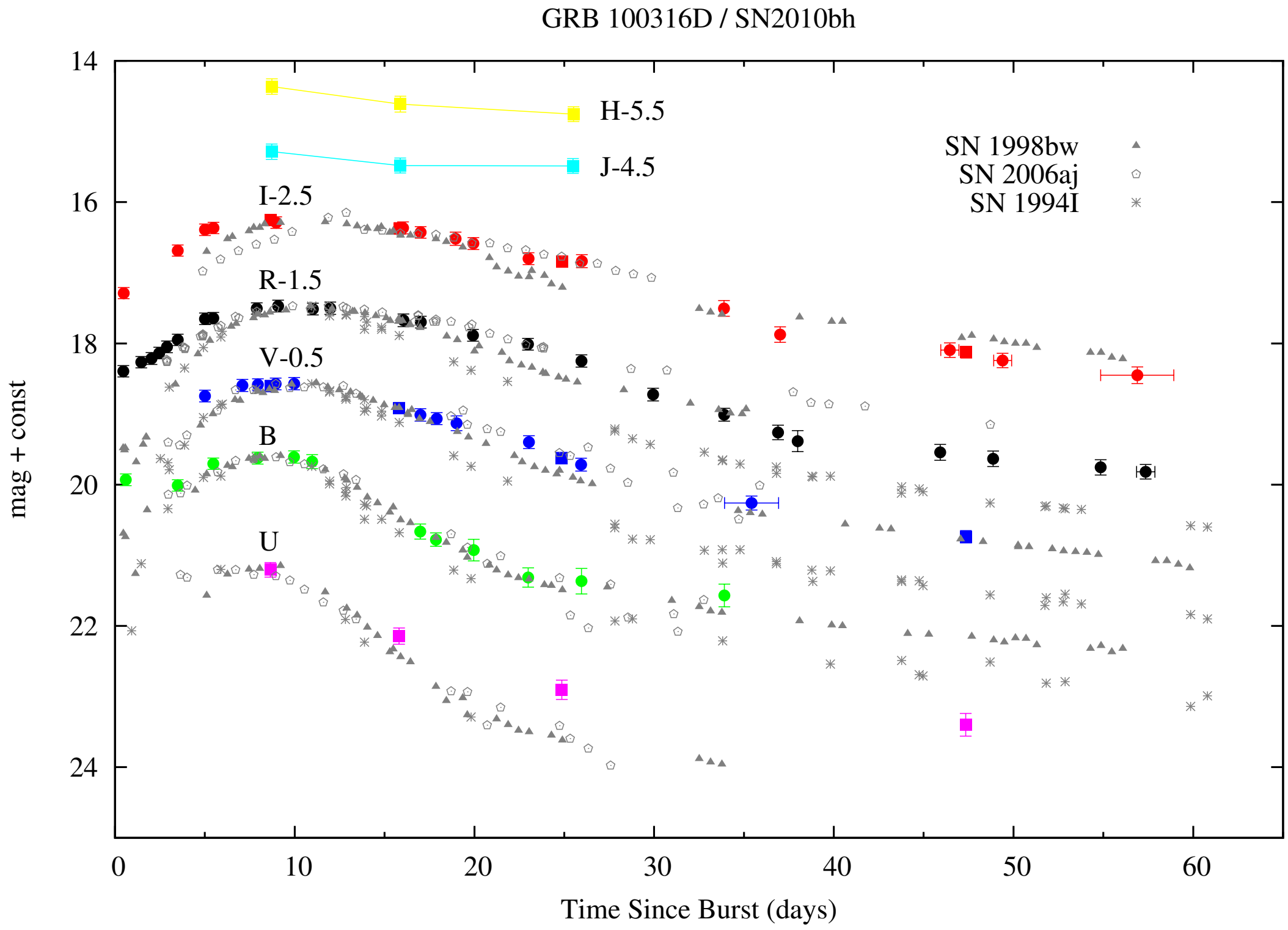

Figure 4.1 *UBVR$_c$iJH* light curves of XRF 100316D /SN 2010bh (solid points (ground-based data) and filled squares (*HST*)). Plotted for comparison are the light curves of two known GRB-SNe: SN 1998bw (Galama et al. 1998a; Patat et al. 2001; Sollerman et al. 2000) and SN 2006aj (Campana et al. 2006; Sollerman et al. 2006; Ferrero et al. 2006), as well as a local SN that is not associated with a GRB: SN 1994I (Richmond et al. 1996). All of the SN 2010bh light curves have been corrected for foreground and host extinction. The light curves of the comparison SNe have been shifted by different amounts in each filter to match in peak brightness. No other changes have been made to the light curves apart from the light curves of SN 1998bw which have been stretched in each filter by a factor of $\sim 0.6$ (see Table 4.5) and the early-time data ($t - t_0 < 3$ days) for SN 2006aj, for which the light was dominated by flux coming from the shock break-out, was not included to avoid confusion between different filters.

Table 4.4 Main Photometric parameters of XRF 100316D / SN 2010bh

| | $B$ | $V$ | $R_c$ | $i$ |
|---|---|---|---|---|
| peak apparent magnitude | $20.13 \pm 0.08$ | $19.47 \pm 0.08$ | $19.29 \pm 0.08$ | $19.05 \pm 0.08$ |
| peak absolute magnitude | $-18.27 \pm 0.08$ | $-18.62 \pm 0.08$ | $-18.60 \pm 0.08$ | $-18.62 \pm 0.08$ |
| $T_{\rm peak}$ (days) | $8.02 \pm 0.12$ | $8.86 \pm 0.07$ | $11.18 \pm 0.05$ | $11.21 \pm 0.22$ |
| $A_{v,\rm fore+host}$ (mag) | $1.32 \pm 0.08$ | $1.01 \pm 0.08$ | $0.81 \pm 0.08$ | $0.59 \pm 0.08$ |
| $\varnothing m_{15}$ (mag) | $1.80 \pm 0.14$ | $0.95 \pm 0.10$ | $0.90 \pm 0.05$ | $0.77 \pm 0.20$ |

Apparent magnitudes are not corrected for foreground or host extinction.
Absolute magnitudes *are* extinction corrected (foreground & host).

redder $V$, $R_c$ and $i$ filters it evolves more slowly. It is also worth noting that the $U$-band measurement of SN 2010bh at $\rm t - t_o = 24.8$ days is $\approx 0.5$ mag brighter than the shifted LC of SN 2006aj, and perhaps even more at $\rm t - t_o = 47.3$ days. As the fluxes are measured in a fixed, and albeit small, aperture, the flux we are detecting is from both the SN and the underlying host galaxy. When performing photometry on the third and fourth epoch $U$-band images, it appears that the background/host flux is becoming dominant and affecting the overall shape of the LC, (i.e. making it appear that it is fading *slower* than is expected).

## 4.2.2 Colour Curves

The colour curves of SN 2010bh are plotted in Figure 4.2, and have been corrected for foreground and host extinction (see Section 4). Plotted for comparison are the colour curves of GRB-SNe 1998bw and 2006aj, as well as local SN 1994I, all of which have been corrected for foreground and host extinction using the values quoted in the literature: SN 1994I: $E(B-V)_{\rm total} = 0.45$ mag (Richmond et al. 1996; Iwamoto et al. 1996; though see Sauer et al. 2006 who find $E(B-V) = 0.30$ mag from spectral modelling); SN 1998bw: $E(B-V)_{\rm total} = 0.07$ mag (Patat et al. 2001); SN 2006aj: $E(B-V)_{\rm total} = 0.14$ mag (Sollerman et al. 2006; though see Campana et al. 2006 who find $E(B-V)_{\rm host} = 0.20$ mag from their *UVOT* obsverations.). No other changes have been made to the colour curves apart from those of SN 1998bw, which have been stretched by $s = 0.63$ in $B-V$ and $V-R_c$ and by $s = 0.61$ in $R_c - i$ (see Table 4.5). The stretch factors, which are the average values of $s$ determined in the individual filters (i.e. $s = 0.62, 0.64$ for respective filters $B$ and $V$; thus the average value is $s = 0.63$), have been applied so as to present a consistent analysis of the properties between SN 2010bh and SN 1998bw (i.e. as the stretch factors were also applied to the LCs of SN 1998bw in Figure 4.1).

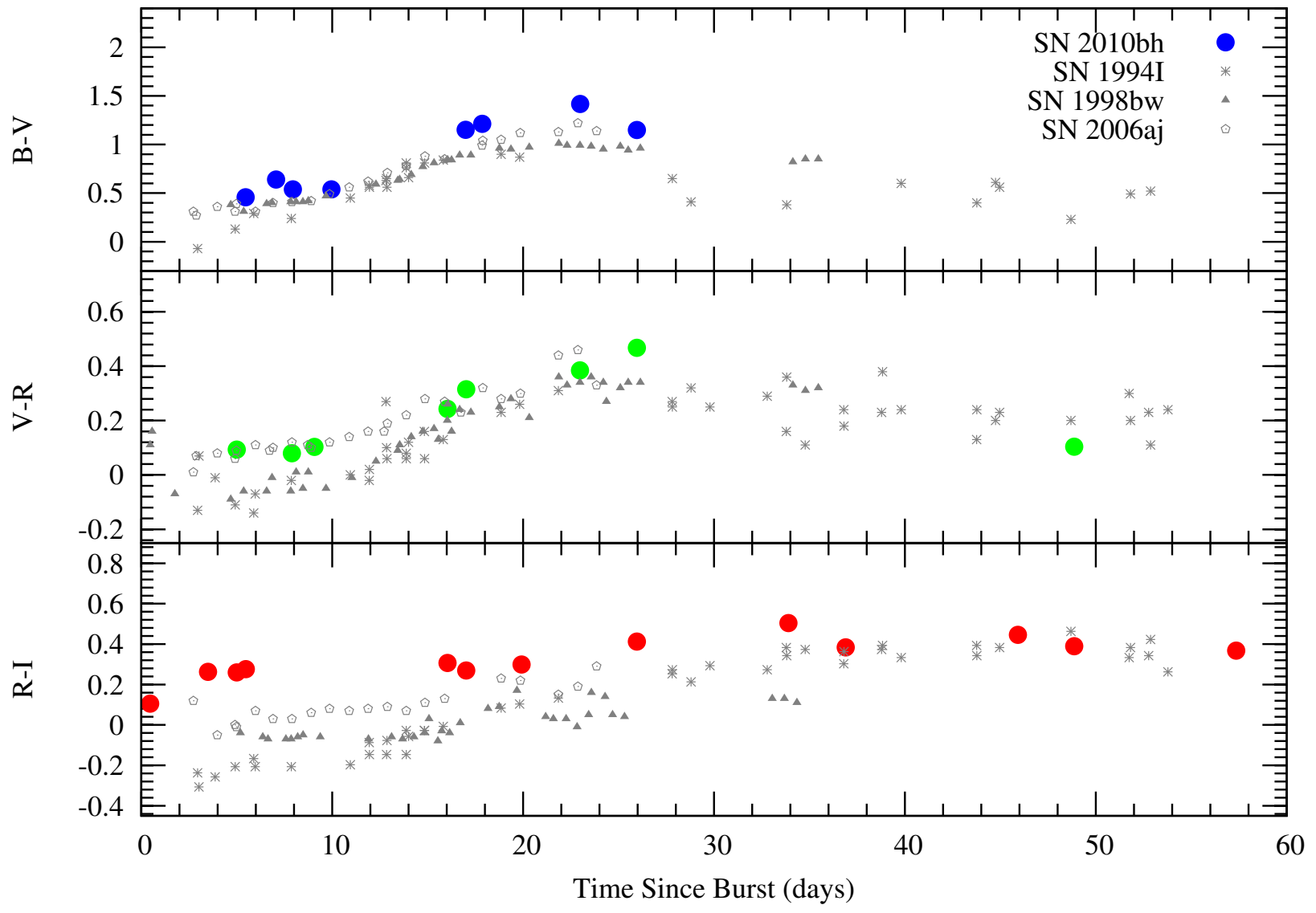

Figure 4.2 Colour curves of XRF 100316D/SN 2010bh, which are corrected for foreground and host extinction. Plotted for comparison are the colour curves of known GRB-SNe: 1998bw and 2006aj as well as local type Ic SN 1994I that was not accompanied by a GRB trigger. No other changes have been made to the colour curves apart from those of SN 1998bw, which have been stretched by $s = 0.63$ in $B-V$ and $V-R$ and by $s = 0.61$ in $R_c - i$ (see Section 3.5). The stretch factors, which are the average values of $s$ determined in the individual filters, have been applied so as to present a consistent analysis of the properties between SN 2010bh and SN 1998bw.

Starting before maximum light, all of the colours of SN 2010bh become redder over time, with similar behaviour seen in the comparison SNe. Additionally, SN 2010bh appears to be redder than SN 1998bw and SN 1994I in the pre-maximum phases, while in the post-maximum phase SN 2010bh appears to have colours similar to those seen in SN 1994I.

### 4.2.3 $\varnothing m_{15}$

We have determined the $\varnothing m_{15}$ parameter (i.e. the amount that the light curve fades by from peak to fifteen days later) of the $BVR_ci$ light curves of XRF 100316D / SN 2010bh. We have used low-order polynomials to determine the peak times of each light curve, and in turn the $\varnothing m_{15}$ parameter. It is seen that the light curves decay more slowly in the redder filters, with the $\varnothing m_{15}$ parameter in filters $BVR_ci$, respectively, being $1.80 \pm 0.14$, $0.95 \pm 0.10$, $0.90 \pm 0.05$ and $0.77 \pm 0.20$ (which are also listed in Table 4.5). The uncertainties in the $\varnothing m_{15}$ parameters are derived from the uncertainties in determining the peak times from the photometry, which, for example, in the $i$ band are less certain due to the sparseness of photometry around and just after the peak, as well as from the distribution of the peak times determined from fitting different order polynomials.

The $V$- and $R$-band values of $\varnothing m_{15}$ measured for SN 2010bh are somewhat larger (i.e. fade faster) than the average values measured for a sample of local Ibc SNe observed in the $V$ band and $R$ band by Drout et al. (2011) (their Table 4). It is also seen that SN 2010bh fades faster, on average, than Ia SNe in the $B$-band (e.g. Phillips 1993; Riess et al. 1998).

### 4.2.4 Shock Break-out or optical afterglow?

The break-out of the blast/shock-wave and/or emission coming from the shock-heated stellar envelope, has been detected in both type Ibc and type II SNe, including: type IIp SN 1987A (Early peaks in optical LCs: Hamuy et al. 1988, and P Cygni-like feature around 1500Å in UV light echo off of a dust cloud $\sim$ 300 pc from the SN: Gilmozzi & Panagia 1999); type IIb SN 1993J (Early peak in optical light curves: Wheeler et al. 1993, Schmidt et al. 1993, Richmond et al. 1994, Lewis et al. 1994); type Ibc SN 1999ex (early optical peaks: Stritzinger et al. 2002); type IIP SNe SNLS-04D2dc

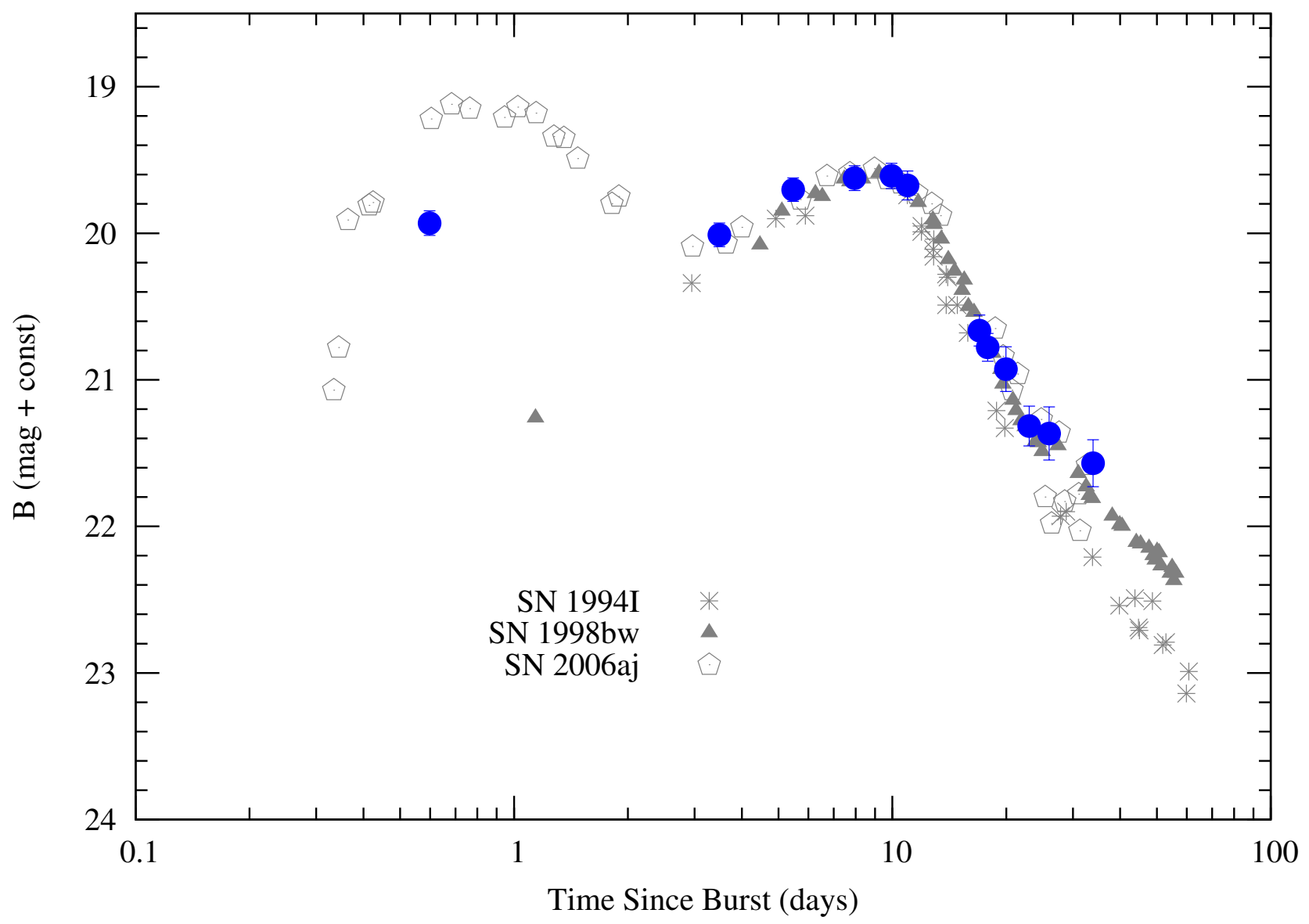

Figure 4.3 *B*-band light curve of XRF 100316D / SN 2010bh (blue points), which has been corrected for foreground and host extinction. Plotted for comparison are the *B*-band light curves of two known GRB-SNe: 1998bw and 2006aj. The light curves of the comparison SNe have been shifted by different amounts in each filter to match in peak brightness. No other changes have been made to the light curves apart for the light curves of SN 1998bw which have been stretched by a factor $s = 0.62$ (see Table 4.5). It is seen that SN 2010bh did not increase linearly in brightness from first detection like SN 1998bw, and at $\mathrm{t - t_o} = 0.59$ days is of comparable brightness as the shifted light curve of SN 2006aj.

& SNLS-06D1jd (UV-flash: Gezari et al. 2008); type IIP GALEX supernova SNLS-04D2dc (UV-flash: Schawinski et al. 2008); type IIn PTF 09UJ (UV-flash: Ofek et al. 2010); and type IIP SN 2010aq (early UV & optical peaks: Gezari et al. 2010).

One SN in particular has been the subject of much interest and debate when intrepreting the high energy properties of the event. SN 2008D (Modjaz et al. 2008; Soderberg et al. 2008; Mazzali et al. 2008; Malesani et al. 2009) occurred in nearby galaxy NGC 2770 while *Swift* was observing SN 2007uy within the same galaxy. The resultant high-energy emission detected by *Swift* was interpreted differently in the literature, with the hot, black-body X-ray spectrum & early peak in optical LC being intrepreted by Soderberg et al. 2008 as being due to the shock break-out, while Mazzali et al. (2008) suggests that the emission is due to a “choked” jet. Interestingly, a recent paper by Van der Horst et al. (2011) has perhaps put the debate to rest. Using observations of the radio emission of SN 2008D during the first year after the explo-sion, the authors showed that there was no evidence of a relativistic jet contributing to the observed radio flux, thus strongly suggesting that the high-energy emission was due to the shock break-out and not to a GRB- or XRF-like event.

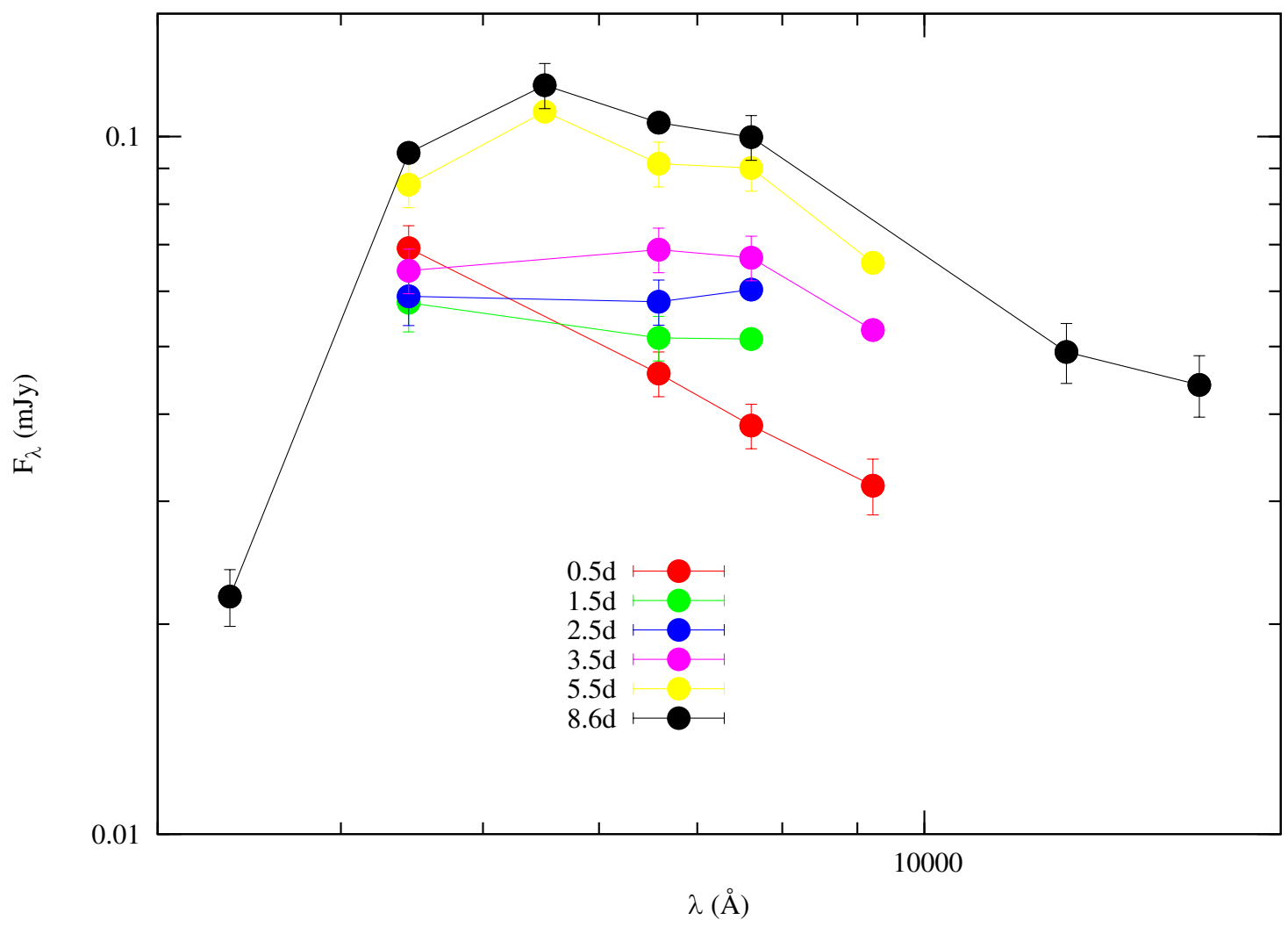

Figure 4.4 Spectral Energy Distributions (SEDs) of XRF 100316D / SN 2010bh over the first 8.6 days. All photometry has been corrected for foreground and host extinction. The first five epochs ($\mathrm{t} - \mathrm{t_o} \approx 0.5$, 1.5, 2.5, 3.5 and 5.5 days) are from the Gemini-S detections, while the SED at $t - t_o = 8.5$ days is from the first *HST* data. The power-law index (i.e. $F_\nu \propto \nu^\beta$) of the first SED is $\beta = +0.94 \pm 0.05$; $\chi^2/\mathrm{d.o.f} = 2.17/2$ (best-fitting line not plotted). The slope of the spectrum is steeper than that expected for light emitted as synchrotron radiation, where, when neglecting effects due to self-absorption, a maximum value of $\beta = +1/3$ is allowed in synchrotron theory (e.g. Sari et al. 1998). Thus, the origin of light is not expected to be synchrotron but rather is coming from the shock break-out from the stellar surface and dense stellar wind.

Emission coming from the shock-heated, expanding stellar evelope as also been detected in GRB/XRF-SNe, though the topic is still contentiously debated in the literature. For XRF 060218, the hot, black-body X-ray spectrum & early peak in UV and optical LCs was interpreted by Campana et al. 2006 (C06 here-on), and later Waxman et al. (2007), as being due to the shock break-out. However, the emission was intrepreted by Ghisellini et al. (2006) as being due to a relativistic jet.

Figure 4.3 shows the *B*-band LC of SN 2010bh, which has been corrected for foreground and host extinction. Plotted for comparison are the light curves of two GRB-SNe: 1998bw and 2006aj, which have been shifted to match SN 2010bh in peak brightness, and the LC of SN 1998bw has also been stretched by a factor $s = 0.62$ (see

Table 4.5).

Upon inspection, similarities between SN 2010bh and SN 2006aj are seen in the early-time LC, where a bright component is detected at $\mathrm{t}-\mathrm{t_o} = 0.598$ days. The magnitude at this epoch is similar to that at $\mathrm{t}-\mathrm{t_o} = 3.489$ days, indicating that SN 2010bh did not increase linearly in brightness from detection, as was seen for SN 1998bw. In the case of SN 2006aj, early peaks in the UV and optical LCs ($\mathrm{t}-\mathrm{t_o} \leq 10^4$ s) were explained by C06 as light coming from the low energy tail of the thermal X-ray emission (see below) produced by the radiation shock driven into the stellar wind. At later times, the optical and UV emission is attributed to the expanding envelope of the progenitor star that is heated as the shock-wave passes through it. Initially the envelope is opaque due to the dense stellar wind, but as the star and wind expands, the photosphere propagates inward exposing the shocked stellar material. For SN 2010bh, though we are not able to ascertain the exact shape of the early *B*-band light curve due to the paucity of our early-time observations, it is not unreasonable to envision a scenario similar to that observed for SN 2006aj in which the brightness of the flux at this early epoch is due to flux coming from the shock-heated, expanding stellar envelope.

In addition to the bright *B*-band detection, inspection of the spectral energy distribution (SED) at this epoch, which has been corrected for foreground and host extinction, gives additional clues to the origin of the flux. At the first epoch the SED is very blue and does not resemble the shape of the SEDs at later times (see Fig 4.4).

To determine whether the flux at this time is synchrotron in origin, the optical SED was fit with a power-law (i.e. $F_\nu \propto \nu^\beta$), for which we find $\beta = +0.94 \pm 0.05$ ($\chi^2$/d.o.f $= 2.17/2$). The slope of the spectrum is harder than that expected for light emitted as synchrotron radiation, where, when neglecting effects due to self-absorption (see below), a maximum value of $\beta = +1/3$ is allowed (e.g. Sari et al. 1998). Thus, the shape of the SED at this early epoch is consistent with emission coming from the shock-heated stellar envelope.

Additionally, we have constructed SEDs for SN 2006aj at $\mathrm{t}-\mathrm{t_o} = 0.50$ days and 0.75 days and fit them with a single power-law. Quite interestingly, at $\mathrm{t}-\mathrm{t_o} = 0.50$ days we find $\beta = +0.97 \pm 0.05$ ($\chi^2$/d.o.f $= 2.13/2$), while at $\mathrm{t - t_o} = 0.75$ days we find $\beta = +0.94 \pm 0.03$ ($\chi^2$/d.o.f $= 2.03/2$). These values for the spectral index are akin to that found for SN 2010bh at a similar epoch, and are both harder than that expected for synchrotron radiation.

The assumption that the optical flux is above the self-absorption frequency ($\nu_a$) cannot be proved with our current set of observations, however it has been seen in other GRB events that the synchrotron self-absorption frequency at typical observing epochs of about 1 day after the GRB is expected to be of order $\sim 10^9 - 10^{10}$ Hz (i.e. at radio frequencies). In the literature there are several events where there has been enough multi-wavelength observations to determine the self-absorption frequency, including: GRB 970508: $\nu_a \sim 3 \times 10^9$ Hz (Granot et al. 1999); $\nu_a \sim 2.5 \times 10^9$ Hz (Galama et al. 1998); GRB 980329: $\nu_a \sim 13 \times 10^9$ Hz (Taylor et al. 1998); GRB 991208 $\nu_a \sim 4 - 11 \times 10^9$ Hz (Galama et al. 2000). As the frequency range of the optical observations lie between $\sim 3 - 10 \times 10^{14}$ Hz (i.e. $\sim 10^4 - 10^5$ times higher frequency), it is unlikely that the optical observations suffer from self-absorption effects.

Further evidence that the light at $\mathrm{t - t_o} = 0.598$ days is coming from the shock-heated stellar envelope is the analysis performed by Starling et al. (2011) on the X-ray spectrum at 144 s $\leq \mathrm{t - t_o} \leq$ 737 s. The authors found evidence for a soft, hot, black-body component that contributes $\approx 3\%$ to the total 0.3 – 10.0 keV flux (though see also Fan et al. 2011 for a different interpretation). With this in mind, the presence of the black-body component, with $kT = 0.14$ keV ($T \approx 1.6 \times 10^6$ K) is similar in temperature to the soft, hot, black-body component seen in the X-ray spectrum of XRF 060218 / SN 2006aj ($kT \approx 0.17$ keV; $T \approx 2.0 \times 10^6$ K; C06).

According to C06, the thermal component is key to understanding XRF 060218. It is known that the signature of shock break-out is a hot black-body X-ray spectrum immediately after the explosion, and C06 argued that the hot temperature of the thermal X-ray component is indicative of radiation emitted by a shock-heated plasma. Furthermore, the characteristic radius of the emitting region was $\sim 5 \times 10^{12}$ cm (Waxman el al. 2007 also find the emitting region to be at a radius of $7.8 \times 10^{12}$ cm), which though quite large, (i.e. of order of the radius of a blue supergiant), was explained by C06 and Waxman et al. (2007) by the presence of a massive, dense stellar wind surrounding the progenitor, which is common for Wolf-Rayet stars (the supposed progenitors of type Ibc SNe and possibly long-duration GRBs). Both authors argue that the thermal radiation is observed once the shock, that is driven into the wind, reaches a radius where the wind becomes optically thin.

For XRF 100316D / SN 2010bh, the temperature of the black-body component in the X-ray spectrum implies an emitting radius of $\sim 8 \times 10^{11}$ cm, which is almost an order of magnitude larger than the radius of a Wolf-Rayet star ($\sim 10^{11}$ cm). As for XRF 060218, the large radius of the X-ray emitting region could be the result of the shock being driven into the dense stellar wind, and the thermal radiation is observed once the wind density decreases and becomes optically thin.

Thus, the brightness of the *B*-band detection at $\mathrm{t}-\mathrm{t_o} = 0.598$ days, when taken in tandem with the hard value of the power-law index ($\beta = +0.94 \pm 0.05$) of the SED at this epoch, as well as the presence of the extremely hot thermal component ($T \approx 1.6 \times 10^6$ K) in the X-ray spectrum at 144 s $\leq \mathrm{t}-\mathrm{t_o} \leq$ 737 s, suggests that we have detected optical emission coming from the cooling, expanding envelope that was heated by the passage of the shock-wave through it.

### 4.2.5 Stretch Factor relative to SN 1998bw

We have determined the stretch factor ($s$) and luminosity factor ($k$) of SN 2010bh in relation to the archetype GRB-SN, SN 1998bw. To do this we have created synthetic flux SN 1998bw light curves in filters $BVR_ci$ as they would appear if they occurred at $z = 0.0591$. At each epoch, an SED of the observed SN 1998bw light curves is built and interpolated using polynomials. This allows us to cover the temporal-frequency space for SN 1998bw and build synthetic light curves for other specified observed frequencies. When constructing the synthetic LCs we have accounted for the inverse-squared distance (i.e. luminosity distance) suppression of flux, as well as restframe extinction and time dilation effects. We then fit the flux SN 1998bw light curves with

Table 4.5 Luminosity ($k$) and Stretch ($s$) Factors of SN 2010bh

| Filter | $s$ | $k$ |
|---|---|---|
| $B$ | $0.62 \pm 0.01$ | $0.41 \pm 0.01$ |
| $V$ | $0.64 \pm 0.01$ | $0.43 \pm 0.01$ |
| $R_c$ | $0.62 \pm 0.02$ | $0.40 \pm 0.01$ |
| $i$ | $0.60 \pm 0.01$ | $0.48 \pm 0.01$ |

an empirical relation of the form:

$$U(t) = A + \rho t \left( \frac{e^{\left(\frac{-t^{\alpha_1}}{F}\right)}}{1 + e^{\left(\frac{p-t}{R}\right)}} \right) + t^{\alpha_2} \log(t^{-\alpha_3}) \tag{4.1}$$

where $A$ is the intercept of the line and $\rho$ and $F$ are related to the respective amplitude and width of the function. The exponential cut-off function for the rise has a characteristic time $R$ and a phase zero-point $p$, while $\alpha_1$, $\alpha_2$ and $\alpha_3$ are free parameters. All of the parameters are allowed to vary during the fit.

Once the flux light curve of SN 1998bw was fit by Equation (4.1), the stretch and luminosity factors of SN 2010bh were determined by fitting the following equation:

$$W(t) = k \times U(t/s) \tag{4.2}$$

to the flux light curve of SN 2010bh in each filter.

The stretch factors of SN 2010bh relative to SN 1998bw were determined using data taken several days after the initial burst ($\mathrm{t} - \mathrm{t_o} \geq 7.0$ days) as the early light curves, whose shapes are altered by light coming from an additional shock break-out component, are very different to that of SN 1998bw. The results of the fit are listed in Table 4.5, where the quoted errors are statistical only.

For GRB and XRF events where an optically bright SN has been detected (i.e. not including events such as SN-less GRBs 060505 and 060614; e.g. Fynbo et al. 2006), it is seen that SN 2010bh is the faintest SN to date that has been linked spectroscopically with a GRB or XRF (see Table 4.8 to compare the stretch factors between the GRB/XRF-SNe; see also Tables 4.4 and 5.1 to compare the peak SN magnitudes), and is of similar peak brightness as the SN associated with GRB 970228 ($k = 0.40 \pm 0.29$, though the host extinction is unknown), but not as faint as the supernova that was photometrically-linked to GRB 101225A ($k = 0.08 \pm 0.03$, Thöne et al. 2011). The slower evolution of SN 2006aj relative to SN 2010bh seen upon inspection of the optical light curves is echoed in the computed stretch factors for the two SNe. In the $B$ filters they are approximately the same value (SN 2010bh: $s = 0.62 \pm 0.01$; SN 2006aj: $s = 0.60 \pm 0.01$), however in the redder filters SN 2010bh has smaller stretch factors relative to SN 2006aj.

It should be mentioned that SN 1998bw is not the ideal template for SN 2010bh. At early times light from the shock break-out changes the shape of the light curve, which affects how well SN 1998bw, for which no shock break-out was observed, can be used as a template. Even at late times it is seen in Figure 4.1 that SN 2010bh decays more slowly than the stretched LC of SN 1998bw, thus showing that the two SNe evolve quite differently. The limitations of using SN 1998bw as template was also noted by Ferrero et al. (2006), where they used an additional power-law component when fitting their SN 1998bw template to their multi-filter observations of SN 2006aj.

SECTION 4.3

## Reddening & Absolute Magnitudes

Determining the absolute magnitude of SN 2010bh requires an estimate of the amount of extinction local to the site of the SN explosion. Drout et al. (2010) found for a sample of local type Ibc SNe that the mean values of the $(V - R)$ colour at ten days after maximum are tightly distributed. The authors found at ten days after $V$ maximum $\langle (V-R)_{V10} \rangle = 0.26 \pm 0.06$ mag, and ten days after $R$ maximum $\langle (V-R)_{R10} \rangle = 0.29 \pm$ 0.08 mag. For SN 2010bh we found the $(V-R)$ colour at ten days after $V$ and $R$ maximum to be, respectively, $\langle (V-R)_{V10} \rangle = 0.38 \pm 0.05$ mag, and $\langle (V-R)_{R10} \rangle = 0.43 \pm 0.05$ mag. The difference in the colours of the sample found by Drout et al. (2010) in comparison with SN 201bh implies a colour excess of $E(V-R) \simeq 0.12$ mag, which in turn implies $E(B-V) \simeq 0.18$ mag (using the relative extinction values in Table 6 of Schlegel et al. 1998).

For comparison, using the Balmer H$\alpha$ and H$\beta$ lines fluxes to measure the Balmer decrement, Starling et al. (2011) found a combined reddening (host and foreground) of $E(B-V) = 0.178$ mag for a nearby HII region ("Source A"; see Figure 11 in Starling et al. 2011). The foreground extinction found from the dust maps of Schlegel et al. (1998) is $E(B-V) = 0.117$ mag, which implies a restframe, host extinction of $E(B-V) = 0.061$ mag. This value is consistent with the value found via comparing the colour curves, albeit somewhat smaller, but provides additional credence to the method described by Drout et al. (2010). We note that the value for the colour excess found by Starling et al. (2011) was not for the specific explosion site of SN 2010bh as the SN was still bright during their observations, thus the authors made the colour excess measurement of a nearby bright knot/blob instead. Nevertheless, these independent analyses imply a small colour excess for the location near and around SN 2010bh. Using the value $E(B-V) = 0.18 \pm 0.08$ mag, we have calculated the absolute magnitudes for SN 2010bh (Table 4.4).

In comparison with the peak magnitudes of almost all of the previously detected GRB-SN tabulated by Cano et al. (2011), it is seen that the peak absolute $V$-band magnitude of SN 2010bh ($M_V = -18.62 \pm 0.08$) is the faintest ever detected for a spectroscopically-detected GRB-SNe and is $\sim 0.8$ mag fainter than SN 1998bw at peak brightness. Cano et al. (2011) found for a sample of 22 GRB-SNe (the authors excluded events where only an upper limit to the SN brightness was obtained; e.g. GRBs 060505 & 060614, Fynbo et al. 2006), where measurements of the host extinction have been made and included in the absolute magnitude calculation, an average, absolute, $V$-band magnitude of $\langle M_V \rangle = -19.00$, with a standard deviation of $\sigma = 0.77$ mag. In a separate analysis of the peak GRB-SNe magnitudes, Richardson (2009) de-termined for a sample of 14 GRB-SNe with known values for the host extinction, an average absolute $V$-band magnitude of $\langle M_V \rangle = -19.20$, with a standard deviation of $\sigma = 0.70$ mag. In comparison with these analyses it appears that SN 2010bh lies at the lower end of the peak brightness distribution.

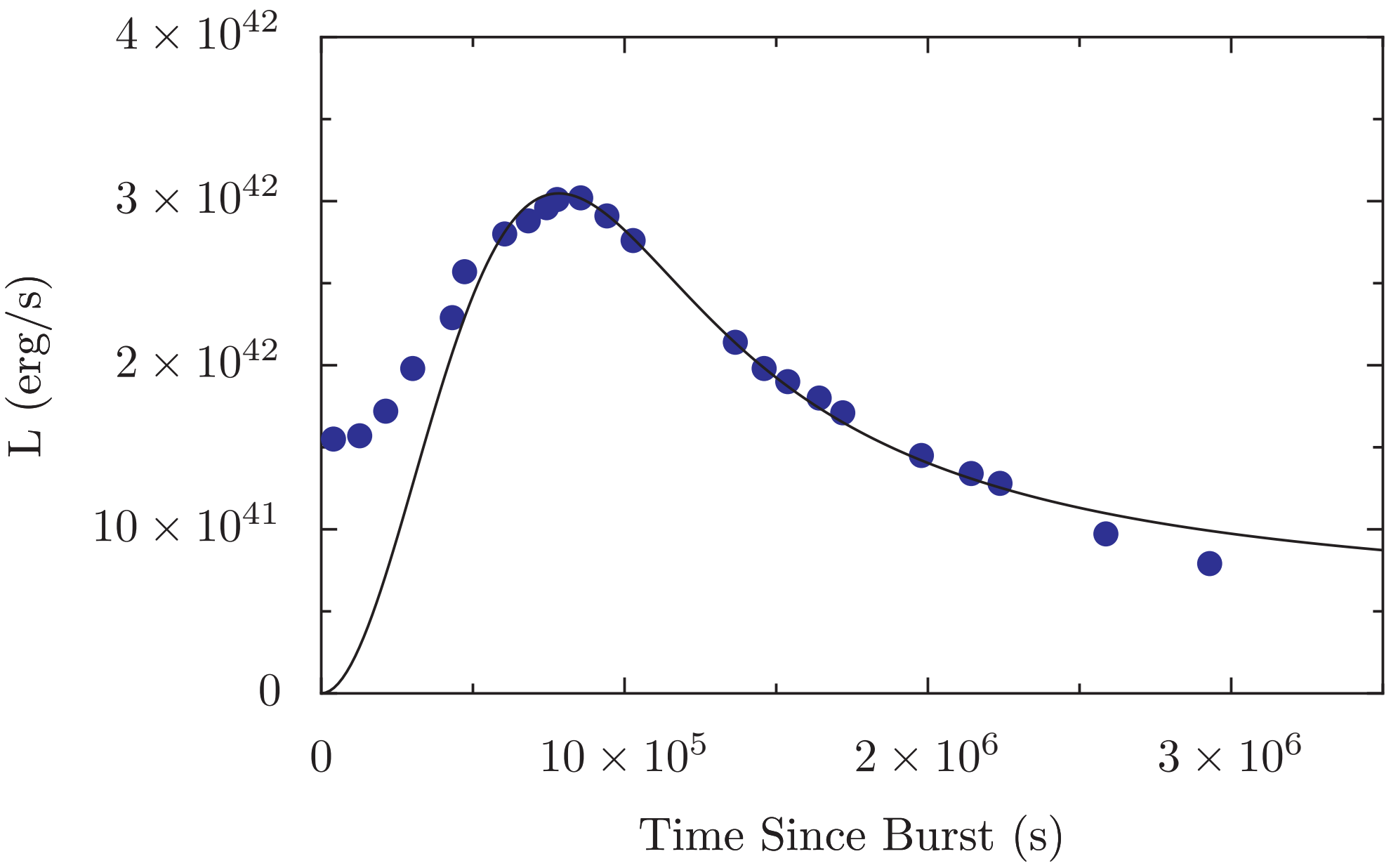

Figure 4.5 The quasi-bolometric light curve for XRF 100316D / SN 2010bh in the 3,000Å – 16,600Å wavelength range, which was obtained by integrating the flux in the optical and infrared filters $UBVRiJH$. The model (solid line) that is used to determine the nickel mass and ejecta mass is also plotted. Peak bolometric light is found to be at $t - t_o = 8.57 \pm 0.04$ days.

SECTION 4.4

## Explosion parameters

We have constructed a quasi-bolometric light curve of XRF 100316D / SN 2010bh (Figure 4.5) in the 3,000Å – 16,600Å wavelength range (i.e. $UBVR_ciJH$) by taking the following steps: (1) converting all magnitudes to monochromatic fluxes after correcting for foreground and host extinction; (2) the SED at each epoch was integrated over frequency. At epochs where there were no contemporaneous data, the individual light curves in each filter were extrapolated after fitting Equation (4.1) to the individual light curves in each filter. For epochs where there were no IR data, we estimated the amount of IR flux as a fraction of the total bolometric flux using the existing data and extrapolating between epochs.

For epochs before the first IR detection (i.e. $t - t_o = 8.6$ days), we have assumed a constant fraction of flux at IR wavelengths of 20%. We note that this assumption of a constant amount of flux at IR wavelengths is valid only for the first dozen or so days, as it is seen that the fraction of light at IR wavelengths actually increases with time and can be as high as 45% at late times (see Figure 4.8 in this work, as well as Figure 9 of Modjaz et al. 2009). The increase in the amount of flux at redder wavelengths is expected however, as it is seen that core-collapse SNe become redder over time. We estimate the uncertainty in the bolometric magnitude, which is dominated by systematic errors described in the method above, to be of order 0.2 mag.

We have calculated the time of peak bolometric light by fitting a series of different order polynomials to the bolometric light curve. We find peak bolometric light to be at $t - t_o = 8.57 \pm 0.04$ days, where the error is statistical and determined from the distribution of peak times determined from the different order polynomials.

Here we have determined a quasi-bolometric light curve of XRF 100316D / SN 2010bh up to $\approx 35$ days. This time interval corresponds to the *photospheric phase*. To attempt to extract physical parameters of the progenitor star we have used a simple analytical model that was developed by Arnett (1982) for type I SNe, which was later expanded by Valenti et al. (2008).

Arnett-like models have been used by many authors to intrepret the light curves of numerous Ibc SNe and GRB-SNe events, including: Ib SN 1999dn, Benetti et al. (2011); Ic-BL SN 2003jd, Valenti et al. (2008); Ic SN 2004aw, Taubenberger et al. (2006); Ic-BL SN 2009bb, Pignata et al. (2011); GRB-SNe 1998bw, Iwamoto et al. (1998); Richardson et al. (2006) for a sample of Ibc SNe; Drout et al. (2011) for a sample of Ibc SNe. The model assumes: (1) homologous expansion of the ejecta, (2) spherical symmetry, (3) all of the radioactive nickel ($^{56}$Ni) is located at the centre of the explosion and does not mix, (4) radiation-pressure dominance, (5) a small initial radius before explosion ($R_o \rightarrow 0$), and (6) the applicability of the diffusion approximation for photons (i.e. the *photospheric phase*). We also assume a constant opacity $\kappa = 0.07$ cm$^2$g$^{-1}$ (e.g. Chugai 2000), which is justified if electron scattering is the dominant opacity source[4] (e.g. Chevalier 1992). Finally, with respect to Arnett (1982) who only considered the energy produced by the decay of nickel into cobalt ($^{56}$Ni $\rightarrow$ $^{56}$Co), we also include, using the methodology described in Valenti et al. (2008), the energy produced by the decay of cobalt into iron ($^{56}$Co $\rightarrow$ $^{56}$Fe):

$$\mathrm{L(t) = M_{Ni}e^{-x^2} \times \left( (\varepsilon_{Ni} - \varepsilon_{Co}) \int_0^x A(z)dz + \varepsilon_{Co} \int_0^x B(z)dz \right)} \qquad (4.3)$$

where

$$A(z) = 2ze^{-2zy+z^2}, B(z) = 2ze^{-2zy+2zs+z^2} \qquad (4.4)$$

and $x \equiv t/\tau_m$, $y \equiv \tau_m/(2\tau_{Ni})$, and $s \equiv (\tau_m(\tau_{Co} - \tau_{Ni})/(2\tau_{Co}\tau_{Ni}))$. The energy release in one second by one gram of $^{56}$Ni and $^{56}$Co are, respectively, $\varepsilon_{Ni} = 3.90 \times 10^{10}$ erg s$^{-1}$ g$^{-1}$ and $\varepsilon_{Co} = 6.78 \times 10^9$ erg s$^{-1}$ g$^{-1}$ (Sutherland & Wheeler 1984; Cappellaro et al. 1997). The decay times of $^{56}$Ni and $^{56}$Co, respectively, are $\tau_{Ni} = 8.77$ days (Taubenbeger et al. 2006 and references therein) and $\tau_{Co} = 111.3$ days (Martin 1987).

$\tau_m$ is the effective diffusion time and determines the width of the bolometric light curve. $\tau_m$ is expressed as a function of the opacity $\kappa$ and the ejecta mass $\mathrm{M_{ej}}$, as well as the photospheric velocity $\mathrm{v_{ph}}$ at the time of bolometric maximum and the kinetic energy of the ejecta $\mathrm{E_k}$:

[4] Metal lines also contribute to the total opacity, and strictly speaking the electron scattering opacity is not actually constant in time as the ionization state changes. For triply ionized ejecta, $\kappa \approx 0.07$.

$$\tau_m \approx \left(\frac{\kappa}{\beta c}\right)^{1/2} \left(\frac{M_{ej}^3}{E_k}\right)^{1/4} \tag{4.5}$$

where $\beta \approx 13.8$ is a constant of integration (Arnett 1982). Thus, for SNe with similar ejecta velocities, a wider bolometric light curve implies more ejecta mass and higher kinetic energies, while for SNe with similar amount of ejected material, a wider light curve implies slower ejecta velocities.

We have fit the data at times $> 5$ days (as to not include flux that may also be coming from the shock-heated, expanding stellar envelope). We have taken the photospheric velocity around peak light for XRF 100316D / SN 2010bh to be $v_{ph} \approx 25,000$ km s$^{-1}$ (determined from the Si II absorption lines measured by Chornock et al. 2010; their Figure 3), the analytical model yields $M_{Ni} = 0.10 \pm 0.01 M_{\odot}$, $M_{ej} = 2.24 \pm 0.08 M_{\odot}$, $E_k = 1.39 \pm 0.06 \times 10^{52}$ erg. Our results are displayed in Table 4.6, where they are compared with other Ibc and GRB-SNe.

The quoted errors are statistical and include the uncertainties in the photometry, extinction, and calibration. However, the dominant sources of error are systematic. For example, our assumption of a constant value of the opacity is perhaps valid during the photospheric phase, but the actual value may differ from our assumed value of $\kappa = 0.07$ cm$^2$g$^{-1}$. Using Equation 4.5 and the values of the explosion energy and ejecta mass from the literature for SN 1998bw, we calculate an opacity of $\kappa \approx 0.05$ cm$^2$g$^{-1}$. Since the total opacity depends on the metal content of the ejecta as well as that from electron scattering, it is quite possible that the value of the opacity may have a slightly higher or lower value than that used here. A small change in the value of the opacity leads to an uncertainity in the ejected mass of ($M_{ej}$) of $25\% - 30\%$ and an uncertainty in the kinetic energy ($E_k$) of $60\% - 75\%$. The uncertainty in the nickel mass is less, as it is mainly affected by the uncertainty in the restframe extinction, and to a lesser extent in the photometric errors.

As an alternative method to check how much mass was ejected during SN 2010bh,

we have compared the width of the bolometric LC of XRF 100316D / SN 2010bh with those of three other spectroscopically-confirmed GRB-SNe: SN 1998bw, SN 2003dh and SN 2006aj. If we assume that the opacity is the same in each event, we can solve Equation (4.5) to find the ejected mass for SN 2010bh:

$$\mathrm{M_{ej,10bh}} \approx \mathrm{M_{ej,GRB}} \left(\frac{\mathrm{v_{10bh}}}{\mathrm{v_{GRB}}}\right) \left(\frac{\tau_{\mathrm{m,10bh}}}{\tau_{\mathrm{m,GRB}}}\right)^2 \tag{4.6}$$

First, using SN 1998bw, from the literature (e.g. Iwamoto et al. 1998; Patat et al. 2001) it is expected that $\approx 8-10M_{\odot}$ of material was ejected. We have measured $\tau_{m,98bw}$ from our bolometric LC of SN 1998bw (see Section 7.2, where we have also measured $\tau_m$ for SN 2006aj and SN 2003dh on the way to deriving physical parameters using our analytical model) and thus derive $\mathrm{M_{ej,10bh}} \approx 3.4M_{\odot}$. For SN 2003dh, it is expected that $\approx 7 \pm 3M_{\odot}$ of material was ejected (e.g. Deng et al. 2005), which in turn implies for SN 2010bh an ejected mass of $\mathrm{M_{ej,10bh}} \approx 3.3M_{\odot}$. For SN 2006aj, it was determined that $\approx 2M_{\odot}$ of material was ejected during the explosion (e.g. Mazzali et al. 2006). This implies an ejected mass for SN 2010bh of $\mathrm{M_{ej,10bh}} \approx 2.7M_{\odot}$. In turn, these values for the ejected mass imply, respectively, explosion energies for SN 2010bh of $\mathrm{E_{k,10bh}} \approx 2.1 \times 10^{52}$ erg, $\mathrm{E_{k,10bh}} \approx 2.0 \times 10^{52}$ erg, and $\mathrm{E_{k,10bh}} \approx 1.7 \times 10^{52}$ erg. This method gives larger values for the ejected mass of SN 2010bh which imply larger explosion energies. On average, these values are larger than those determined directly from our fit.

It is important to consider the limitations of our model and the implied caveats from our assumptions. While it appears that our model is able to accurately reproduce the nickel masses in each of these events, the energy estimates have probably the largest uncertainities. As mentioned, Arnett (1982) considers spherical geometry, while GRB-SNe are thought to be asymmetric and may well have a non-homologous distribution of nickel mass in the ejecta mass. Furthermore, this model calculates an explosion energy by assuming that all of the ejecta are moving at the photospheric velocity. In the case of XRF 100316D / SN 2010bh, while Chornock et al. (2010) observed absorptions in the spectra at 25,000 km $s^{-1}$, it does not mean that *all* of the ejecta are moving at that velocity. It is quite possible that only a small or modest fraction of the mass is expelled at high velocities (either isotropically or collimated into a jet) creating the absorption lines in the spectra, while most of the mass moves at slower velocities. A more accurate way of determining the actual explosion energy would be to understand the way the mass is distributed in the ejecta (i.e. via detailed modeling), and integrate over the density-velocity profile.

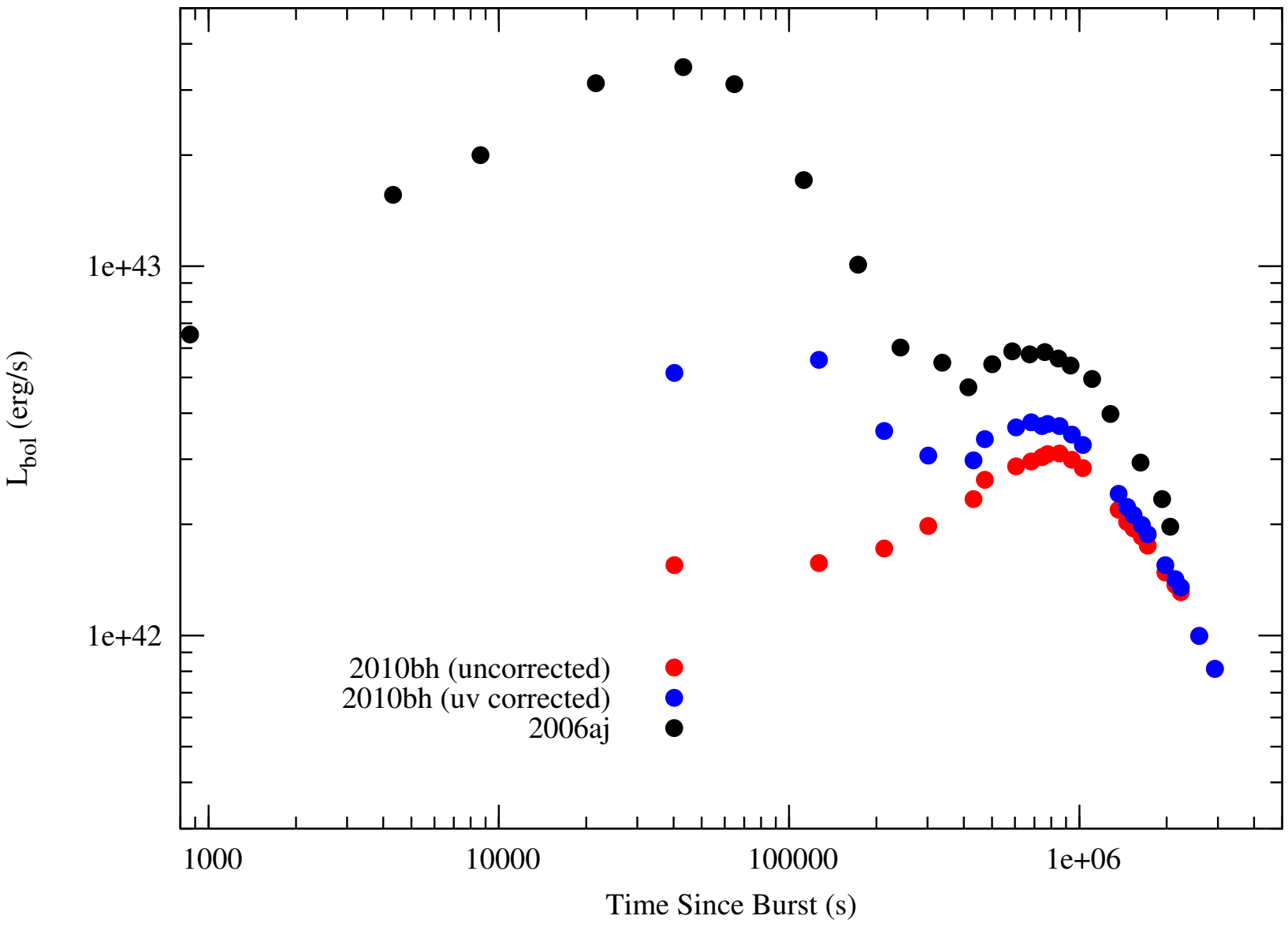

Figure 4.6 The UVOIR bolometric light curves, in the 1,900Å – 16,660Å wavelength range, of XRF 100316D / SN 2010bh (red and blue filled circles) and XRF 060218 / SN 2006aj (black filled circles). When constructing the bolometric light curve of SN 2010bh, we have assumed that the same fraction of the bolometric flux emitted at UV wavelengths is the same for SN 2010bh and for SN 2006aj. Plotted for comparison is the $UBVR_ciJH$ bolometric light curve of SN 2010bh (open circles). All photometry has been corrected for foreground and host extinction.

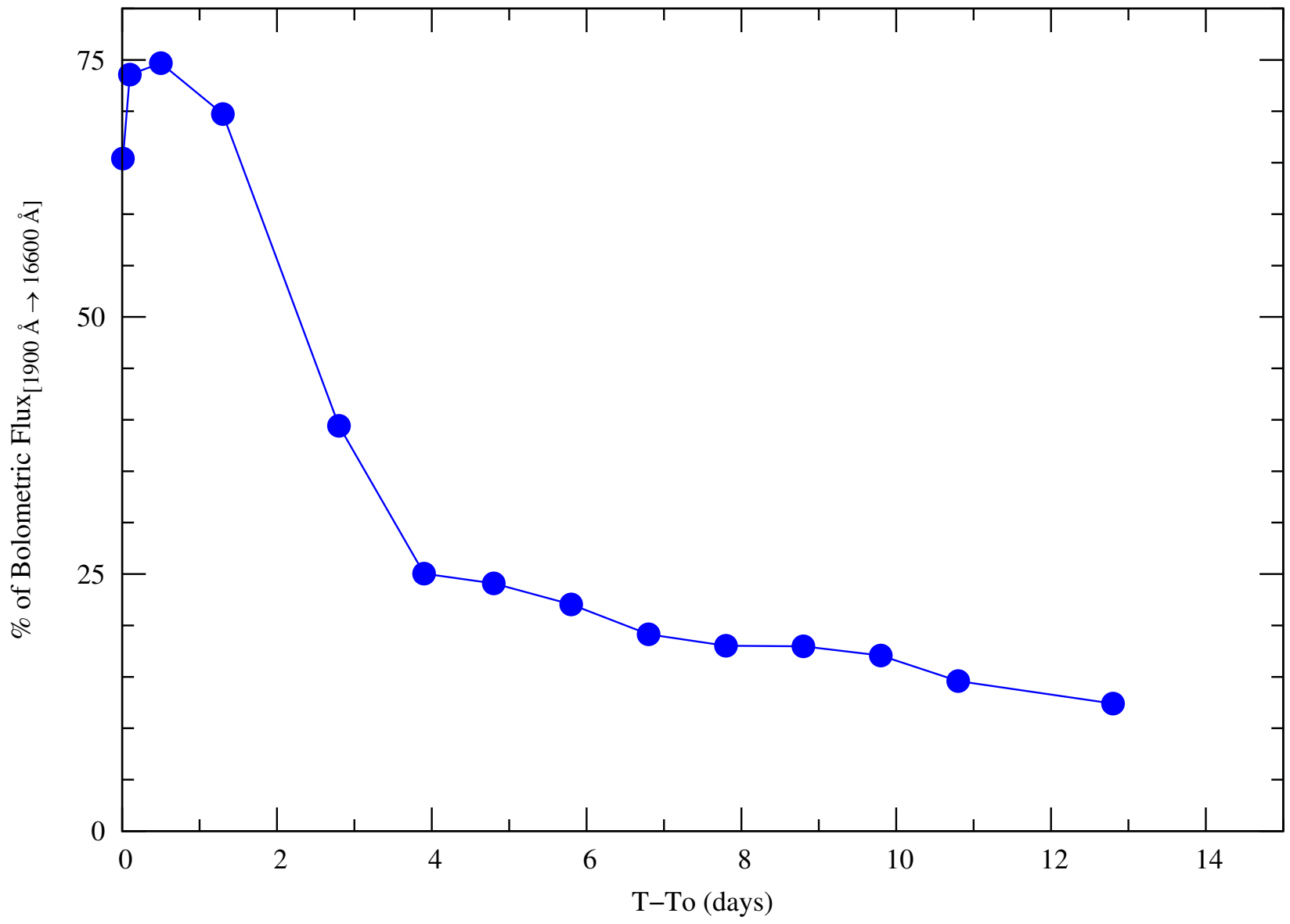

Figure 4.7 XRF 060218 / SN 2006aj: The amount of UV flux as a fraction of the total bolometric flux in the 1,900Å – 16,660Å wavelength regime, plotted over time. It is seen that up to a few days after the initial explosion, the amount of UV flux is considerable. More importantly, even at late times ($\geq$ 10 days) the amount of flux at UV wavelengths is a non-negligible fraction of the total bolometric flux.

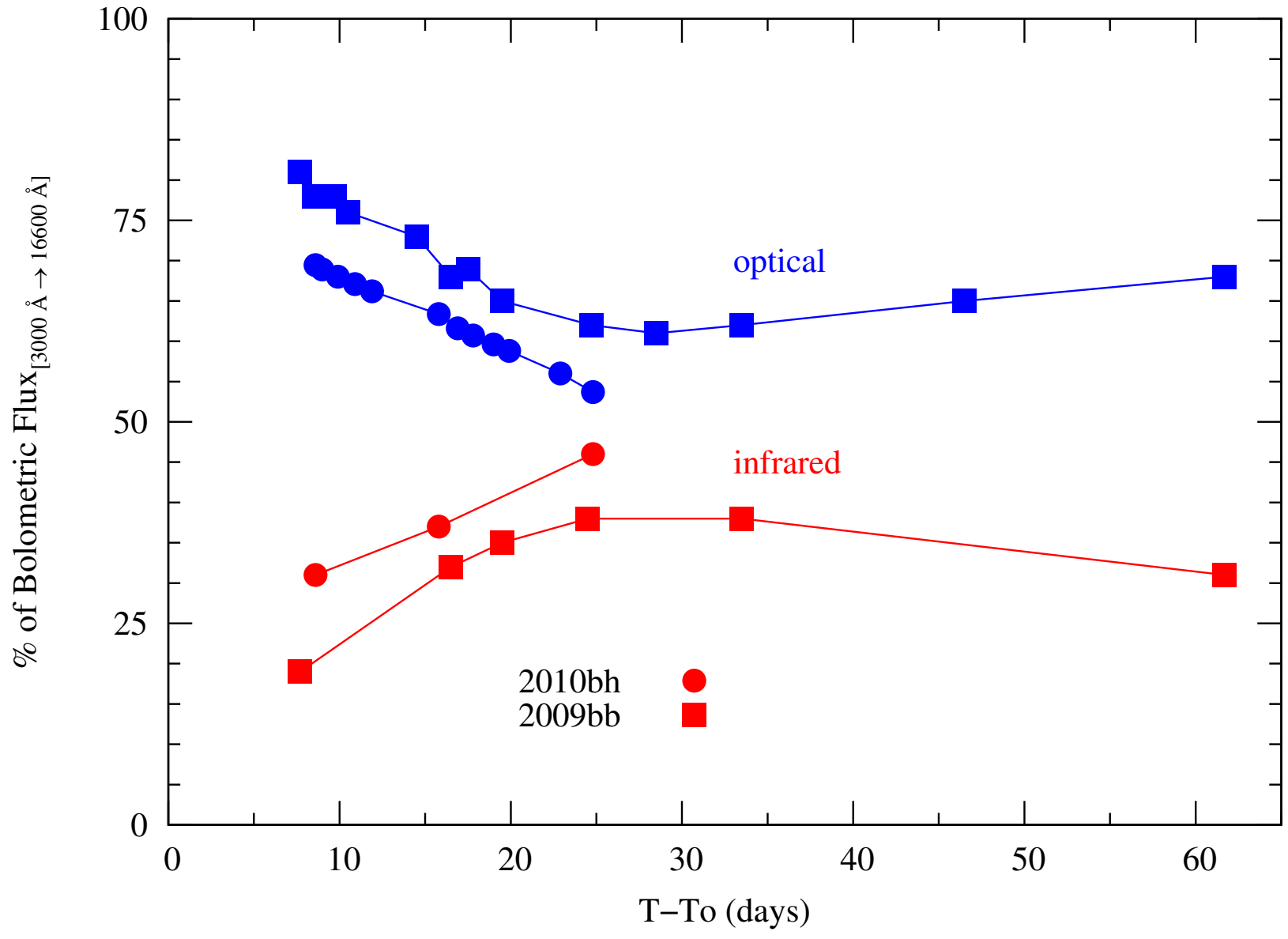

Figure 4.8 The amount of flux as a fraction of the total bolometric flux (in filters $UBVR_ciJH$) that is obtained by integrating the optical (blue) and infrared (red) flux for SN 2010bh and SN 2009bb. It is seen that the behaviour is similar in each SN event.

SECTION 4.5

# Discussion and Further Analysis

4.5.1

## SN 2010bh in relation to SN 2006aj

It was shown in the preceding section that, despite ejecting only a moderate amount of material, SN 2010bh was an energetic explosion, mainly due to the large photospheric velocities. To put the results of our bolometric light curve modelling into context, we first compare SN 2010bh with SN 2006aj and then with the other GRB-SNe.

The ultra-violet, optical and infrared (UVOIR) bolometric light curve (in the $1,900Å - 16,660Å$ wavelength regime) of XRF 060218 / SN 2006aj, which has been constructed from the data published in the literature (Campana et al. 2006; Sollerman et al. 2006; Cobb et al. 2006; Ferrero et al. 2006), is displayed in Figure 4.6. It is seen that during the first few days the fractional contribution of UV flux (i.e. light detected in the $1,900Å - 3,000Å$ wavelength regime) to the total bolometric LC (Figure 4.7) is considerable due to the shock break-out of the progenitor star (Campana et al. 2006), which decays at a linear rate after $\approx 4$ days. It is important to note that the contribution of flux at UV wavelengths is **not negligible**, and at roughly 10 days still contributes 15% of the total bolometric flux.

By making the assumption that the fraction of UV flux of the total bolometric flux at a given epoch is the same in SN 2010bh as it is in SN 2006aj, we have used SN 2006aj as a template/proxy for creating a "UV-corrected" bolometric light curve for SN 2010bh, which is also displayed in Figure 4.6. The bolometric light curve of SN 2010bh is shown twice: (1) first, by considering flux only from the optical ($UBVR_ci$) and infrared ($JH$) filters and, (2) by including a contribution of flux at UV wavelengths.

Our assumption that the fraction of UV flux is similar in each event is perhaps not unreasonable when we also consider the percentage of the bolometric flux detected in the optical and IR regimes for SN 2010bh and SN 2009bb. Plotted in Figure 4.8 is a plot of the amount of light detected in the optical and infrared as a fraction of the total bolometric flux in each epoch for SN 2009bb and SN 2010bh. The data included in this plot do not include all of the data obtained for these two events, but only those where there are contemporaneous optical and IR data. The first IR detection of SN 2010bh was during the first *HST* epoch at $t - t_o = 8.6$ days, and the first IR detection of SN 2009bb was at $t - t_o = 7.7$ days (Pignata et al. 2011). Inspection of the optical and infrared filter pairs reveal similar behaviour in both events, though it is seen that more flux is emitted at IR wavelengths in SN 2010bh relative to SN 2009bb.

The fractional contribution of the optical and IR flux to the bolometric light curves for these two events, as well as the large UV contribution seen in SN 2006aj, is similar to that seen in XRF 080109 / SN 2008D. Modjaz et al. (2009) also compared the relative contribution of the UV, optical, and IR flux in the bolometric LC of SN 2008D for the first $\approx 30$ days (Figure 9; Modjaz et al. 2009), and noted a large contribution of flux from 2008D at UV wavelengths ($\approx 40\%$ at $t - t_o \approx 0.8$ days). This is not quite as high as for SN 2006aj which is $\approx 70\%$ at $t - t_o \approx 0.8$ days, but their UV contribution does not include far-UV flux, whereas SN 2006aj does. Indeed at $t - t_o = 1.8$ days, when they do include the far-UV flux, the total UV flux is seen to increase. Modjaz et al. (2009) also showed that the amount of IR flux increased over time, while the optical flux increased linearly for the first ten days or so, then decreased. While there is scant data plotted in Figure 4.8 before $t - t_o = 10$ days, our plot does show that the optical contribution decreases during 8.6 days $\leq t - t_o \leq 24.8$ days for SN 2010bh, whereas for SN 2009bb, it decreases until $\approx 25 - 30$ days, then appears to contribute a nearly constant, if slightly increasing, amount to the total bolometric flux. Indeed, when comparing Figure 4.8 with Figure 9 of Modjaz et al. (2009), the general behaviour of SNe 2008D, 2009bb and 2010bh appear to follow similar trends in their relative bolometric fluxes during the first $\approx 30$ days.

Next, we applied our analytical model to our constructed bolometric light curve of XRF 060218 / SN 2006aj. Taking the peak photospheric velocity to be $v_{ph} \approx 20,000$ km s$^{-1}$ (Pian et al. 2006), we find: $M_{Ni} = 0.16 \pm 0.01 M_{\odot}$, $M_{ej} = 1.63 \pm 0.05 M_{\odot}$, $E_k =$ $6.47 \pm 0.20 \times 10^{51}$ erg. In comparison, Mazzali et al. (2006) found when modeling the spectra and photometry of SN 2006aj parameters of $M_{Ni} \approx 0.2 M_{\odot}$, $M_{ej} \approx 2 M_{\odot}$, $E_k \approx 2 \times 10^{51}$ erg. It appears that our simple model is consistent with the values of the physical parameters of the SN 2006aj explosion determined via detailed modeling, but over-estimates the explosion energy by a factor of $\sim 3$. As mentioned in Section (5), our estimate of the energy assumes that all of the mass moves at the photospheric velocity. Here is a clear example where this assumption *over*-estimates the explosion energy, implying that a bulk of the ejecta are moving at slower velocities.

Alternatively, if we exclude the contribution of flux at UV wavelengths, similar to the $UBVR_cI_cJH$ bolometric LCs presented by Pian et al. (2006), we find for SN 2006aj: $M_{Ni} = 0.14 \pm 0.01 M_{\odot}$, $M_{ej} = 1.92 \pm 0.12 M_{\odot}$, $E_k = 7.64 \pm 0.49 \times 10^{51}$ erg, which is again consistent with those found by Mazzali et al. (2006), but the explosion energy is again over-estimated, this time by a factor of $\sim 4$. In both cases the errors are estimated as per Section (5). We note that by including the contribution of flux at UV wavelengths, which is considerable at early times (i.e. $\leq 5$ days), this increases the total luminosity of the SN, but makes the bolometric light curve peak earlier, thus decreasing the overall width of the LC (i.e. less mass is ejected). The results of our modelling are displayed in Table 4.6.

When we apply our model to the bolometric light curve of 2010bh that includes a UV correction, we find $M_{Ni} = 0.12 \pm 0.01 M_{\odot}$, $M_{ej} = 1.93 \pm 0.07 M_{\odot}$, $E_k = 1.20 \pm 0.05 \times 10^{52}$ erg. The inclusion of the UV flux makes the overall bolometric light curve brighter, implying more $^{56}$Ni was created during the explosion. However, as for SN 2006aj, the inclusion of the UV flux, which is dominant at early times, causes the light curve to peak earlier, thus making the overall shape of the bolometric light curve narrower, and implying that less mass was ejected during the explosion (and thus a less energetic explosion).

Never the less, whether an estimated contribution of the UV flux is included or not, the results of bolometric LC modeling imply that SN 2010bh ejected a mass of $M_{ej} = 1.9 - 2.2 M_{\odot}$, has a nickel mass of $M_{Ni} = 0.10 - 0.12 M_{\odot}$, and a kinetic energy of roughly $E_k = 1.2 - 1.4 \times 10^{52}$ erg.

Table 4.6 Physical parameters of Ibc SNe and GRB-SNe

| SN | type | $v_{ph}$ (km $s^{-1}$) | $M_{Ni}$ ($M_{\odot}$) | $M_{ej}$ ($M_{\odot}$) | $E_k$ ($10^{52}$ erg) | Wavelength Range (Å) | Ref. | Note[a] |
|---|---|---|---|---|---|---|---|---|
| 1994I | Ic | 10,000 | $0.07 \pm 0.01$ | $0.88 \pm 0.05$ | $0.09 \pm 0.01$ | $4,400 - 8,000$ | (1) | **This Work** |
| 1994I | Ic | 10,000 | $\approx 0.07$ | $\approx 0.9$ | $0.10 \pm 0.01$ | $4,400 - 8,000$ | (12) | Literature |
| GRB 980425 / 1998bw | Ic-BL | 18,000 | $0.43 \pm 0.02$ | $7.04 \pm 0.57$ | $2.27 \pm 0.17$ | $3,000 - 16,600$ | (2), (3), (4), (5) | **This Work** |
| GRB 980425 / 1998bw | Ic-BL | 18,000 | $0.4 - 0.7$ | $8 \pm 2$ | $2 - 5$ | $3,000 - 16,600$ | (13), (14) (5) | Literature |
| GRB 980425 / 1998bw | Ic-BL | 18,000 | $0.52 \pm 0.02$ | $5.36 \pm 0.53$ | $1.73 \pm 0.17$ | $1,900 - 16,600$ | (2), (3), (4), (5) | **This Work** |
| 1999dn | Ib | - | $\approx 0.11$ | $4 - 6$ | 0.5 | $3,000 - 16,600$ | (18) | Literature |
| 1999ex | Ib | - | 0.25 | 0.9 | 0.03 | - | (16) | Literature |
| 2002ap | Ic-BL | - | $\approx 0.07$ | $2.5 - 5.0$ | $0.4 - 1.0$ | $3,000 - 16,600$ | (17) | Literature |
| GRB 030329 / 2003dh | Ic-BL | 20,000 | $0.39 \pm 0.03$ | $4.82 \pm 0.54$ | $1.92 \pm 0.21$ | $3,000 - 8,000$ | (6) | **This Work** |
| GRB 030329 / 2003dh | Ic-BL | 20,000 | $\approx 0.4$ | $7 \pm 3$ | $3.5 \pm 1.5$ | $3,000 - 8,000$ | (6) | Literature |
| 2003jd | Ic-BL | - | $0.36 \pm 0.04$ | $3.0 \pm 0.5$ | $\approx 0.7$ | $4,400 - 8,000$ | (19) | Literature |
| 2004aw | Ic | - | $0.2 - 0.3$ | $3 - 8$ | $\approx 0.8$ | $3,000 - 16,600$ | (19), (20) | Literature |
| XRF 060218 / 2006aj | Ic-BL | 20,000 | $0.14 \pm 0.01$ | $1.92 \pm 0.12$ | $0.76 \pm 0.05$ | $3,000 - 16,600$ | (7), (8), (9), (10) | **This Work** |
| XRF 060218 / 2006aj | Ic-BL | 20,000 | $\approx 0.2$ | $\approx 2$ | $\approx 0.2$ | $3,000 - 16,600$ | (15) | Literature |
| XRF 060218 / 2006aj | Ic-BL | 20,000 | $0.16 \pm 0.01$ | $1.63 \pm 0.05$ | $0.65 \pm 0.02$ | $1,900 - 16,600$ | (7), (8), (9), (10) | **This Work** |
| 2009bb | Ic-BL | 15,000 | $0.23 \pm 0.01$ | $3.86 \pm 0.13$ | $1.73 \pm 0.06$ | $3,000 - 16,600$ | (11) | **This Work** |
| 2009bb | Ic-BL | 15,000 | $0.22 \pm 0.06$ | $4.1 \pm 1.9$ | $1.8 \pm 0.7$ | $3,000 - 16,600$ | (11) | Literature |
| 2009jf | Ib | - | $0.17 \pm 0.03$ | $4 - 9$ | $0.3 - 0.8$ | $3,000 - 8,000$ | (21) | Literature |
| XRF 100316D / 2010bh | Ic-BL | 25,000 | $0.10 \pm 0.01$ | $2.24 \pm 0.08$ | $1.39 \pm 0.06$ | $3,000 - 16,600$ | **This Work** | - |
| XRF 100316D / 2010bh | Ic-BL | 25,000 | $0.12 \pm 0.01$ | $1.93 \pm 0.07$ | $1.20 \pm 0.05$ | $1,900 - 16,600$ | **This Work** | - |

[a] Physical parameters determined in: (1) this paper, or (2) taken from the literature.

(1) Richmond et al. (1996), (2) Galama et al. (1998), (3) McKenzie & Schaefer (1999), (4) Sollerman et al. (2002), (5) Patat et al. (2001), (6) Deng et al. (2005), (7) Campana et al. (2006), (8) Sollerman et al. (2006), (9) Cobb et al. (2006), (10) Ferrero et al. (2006), (11) Pignata et al. (2011), (12) Iwamoto et al. (1994), (13) Iwamoto et al. (1998), (14) Nakamura et al. (2001), (15) Mazzali et al. (2006a), (16) Richardson et al. (2006) and references therein, (17) Mazzali et al. (2002), (18) Benetti et al. (2011), (19) Valenti et al. (2008), (20) Taubenberger et al. (2006), (21) Sahu et al. (2011)

### 4.5.2 Bolometric Light Curves of other GRB-SNe and local stripped-envelope SNe

It is seen that, despite the similarities of the high-energy properties of XRF 100316D and XRF 060218 (Starling et al. 2011), the accompanying supernovae are somewhat different, with SN 2010bh being much more energetic. In this section we compare the bolometric properties of SN 2010bh with other GRB-SNe, as well as non-GRB-SNe: Ic SN 1994I and broad-lined Ic SN 2009bb. We investigate whether clear differences exist in the bolometric properties of stripped-envelope, core-collapse SNe that produce a GRB and those that do not.

To date, numerous authors have determined physical parameters of Ibc SNe by modeling photometry *and* spectroscopy. Many authors employ analytical models that are derived from the original prescription of Arnett (1982) to describe bolometric LCs, however very often these models only supplement the derivations of physical parameters that are determined by comparing synthetic spectra created by synthesis codes to observations. In this work we are attempting to obtain physical parameters of previously-studied events by modeling only the bolometric LCs (although including the peak photospheric velocity determined by Chornock et al. 2010), and compare our results with those obtained via more detailed analyses.

Figure 4.9 shows the optical & infrared bolometric light curves of XRF 100316D / SN 2010bh in relation to three spectroscopically-confirmed GRB-SNe: 1998bw, 2003dh & 2006aj, as well as broad-lined Ic SN 2009bb and Ic SN 1994I, neither of which were accompanied by an observed GRB-trigger. All of the bolometric light curves have been determined by integrating the flux in the optical and infrared $UBVR_cI_cJH$ filters, and have been corrected for foreground and host extinction. We have constructed the bolometric light curves of SN 1998bw (Galama et al. 1998a; McKenzie & Schaefer 1999; Sollerman et al. 2002; Patat et al. 2001), SN 2006aj (Campana et al. 2006; Sollerman et al. 2006; Cobb et al. 2006; Ferrero et al. 2006) & SN 2009bb (Pignatta et al. 2011) from the individual light curves presented in the literature. We have also incorporated the bolometric data presented in Deng et al. (2005) for SN 2003dh and the bolometric data presented in Richmond et al. (1996) for SN 1994I. When constructing the bolometric light curve of SN 1998bw, for which very little infrared data are available in the literature, we estimated the amount of flux emitted at IR wavelengths using the behaviour of SNe 2006aj, 2009bb and 2010bh as a proxy. Note also that the bolometric LCs of SN 1994I and SN 2003dh are in filters $BVR_cI_c$ and $UBVR_cI_c$ respectively. Finally, using the same method that we applied to SN 2010bh in Section 4.5.1, we have estimated the amount of flux emitted at UV wavelengths for SN 1998bw by using SN 2006aj as a template/proxy.

As for SN 2006aj, we fit our analytical model to the bolometric light curves of the comparison SNe. We have used values of the photospheric velocities given in the literature for each event. The results of our fits are presented in Table 4.6, where the quoted errors are statistical only.

When we compare the values of the nickel mass ($M_{Ni}$), ejected mass ($M_{ej}$) and explosion energy ($E_k$) determined from our model with those in the literature, it is seen that our model is able to recover the physical parameters of these events that are determined by more detailed spectral and photometric modeling. We note that on average, we tend to under-estimate the amount of ejected mass, which in turn under-estimates the explosion energy in these events, apart for SN 2006aj, where we over-estimate the explosion energy. There are several assumptions that go into our model, such as no asphericity and the central location of the nickel mass during the explosion (that does not mix with the expanding ejecta), as well as a constant value for the optical opacity. Even if the opacity is approximately constant during the photospheric phase, it may be a slightly different value to that we are using here.

Finally, the nickel mass determined from our fit of SN 2010bh (Section 5) reiterates the faint nature of SN 2010bh in relation to the other GRB-SNe. While the nickel mass produced by SN 2010bh is slightly greater than that of SN 1994I, it is less than that seen in other GRB-SNe. Indeed, for all of the GRB-SNe that have been investigated through spectral and photometric modeling, SN 2010bh appears to be the faintest (i.e. has created the least amount of nickel during the explosion).

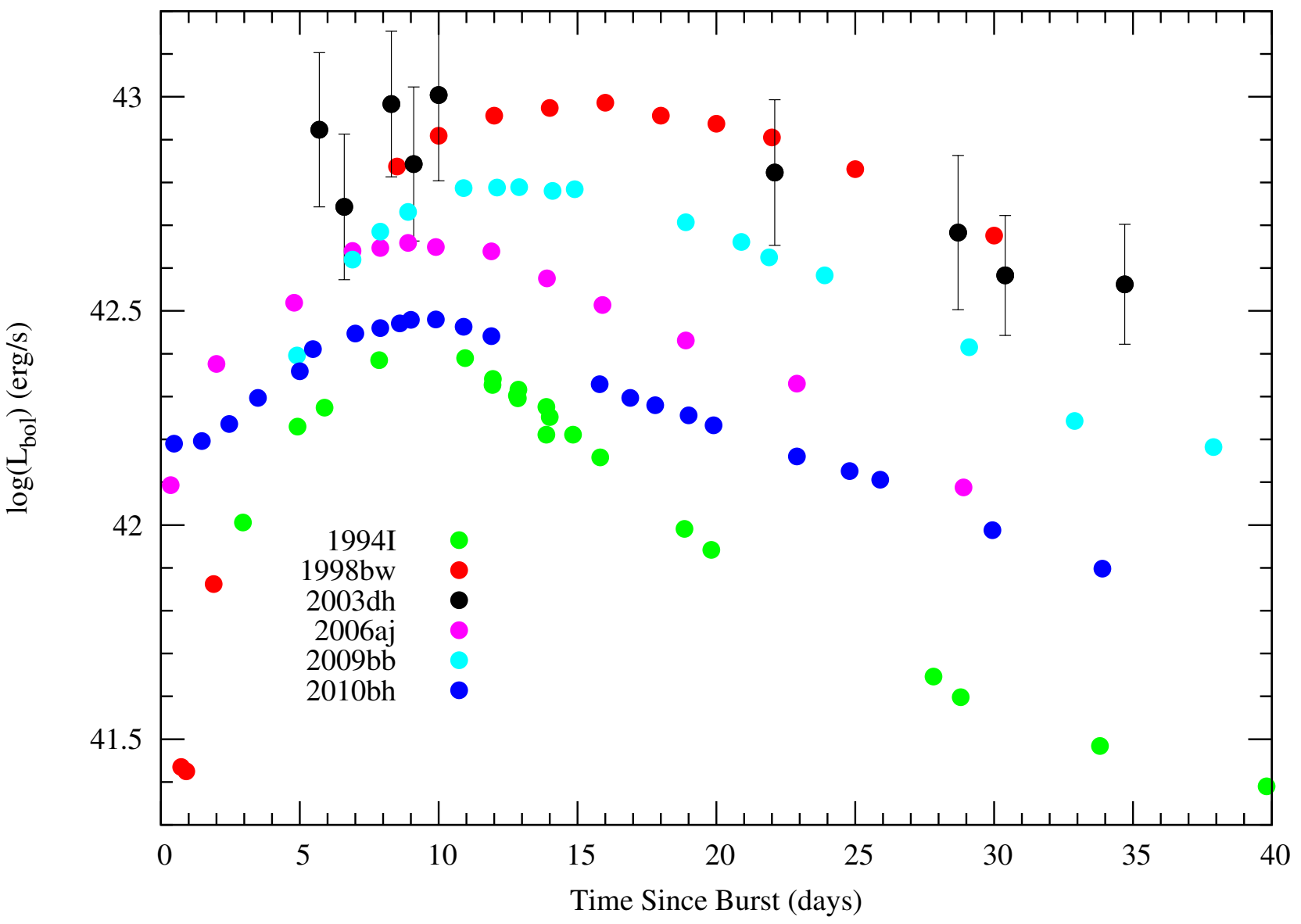

Figure 4.9 The quasi-bolometric $UBVR_ciJH$ light curves of XRF 100316D / SN 2010bh as well as three spectroscopically-confirmed GRB-SNe: 1998bw, 2003dh & 2006aj and two local type Ic SNe that were not accompanied by a GRB-trigger: Ic SN 1994I & broad-lined Ic SN 2009bb. All of the bolometric LCs have been corrected for foreground and host extinction. Note that the bolometric LC of SN 1994I and SN 2003dh are in filters $BVR_cI_c$ and $UBVR_cI_c$ respectively.

In conclusion, the energetics of GRB-SNe and Ic-BL SNe such as SN 2009bb and SN 2002ap are comparable, and are considerably higher than that of other Ibc SNe such as Ic SN 1994I and Ib SN 1999ex. Additionally, the ejected masses of the GRB-SNe and Ic-BL SNe are commensurate and somewhat higher than some Ibc SNe (e.g. SN 1994I and SN 1999ex), though some Ibc SNe do seem to eject comparable masses (e.g. SN 2009jf and SN 1999dn). Finally, the nickel masses of all these events span a range of $0.07 - 0.52 M_{\odot}$, with no clear distinction between GRB/XRF-SNe, Ic-BL SNe and Ibc SNe.

### 4.5.3 Stretch and Luminosity Factors of the GRB-SNe

In an attempt to put the luminosity and stretch factors of SN 2010bh (Table 4.5) into context, we assembled from the literature all of the photometric data available for most of the previously detected GRB-SNe. Note that we have not pursued every event but only included those where several measurements were made of the supernova "bump" that would allow us to constrain the properties of the associated SN.

For every GRB-SN event we assume that light is coming from three sources: (1) the afterglow, (2) the supernova and (3) the host galaxy. First we remove the constant source of flux from the host galaxy, either by using magnitudes determined by image subtraction by the authors (e.g. XRF 020903; Bersier et al. 2006), or by mathematically subtracting the flux from each measurement using observations made of the host galaxy long after the OT had faded below the brightness of the galaxy (e.g. GRB 090618; Cano et al. 2011).

Then, we use the early-time data for each event, and model the behaviour of the afterglow using single or broken power-laws:

$$flux\,(t) = f_o \times \left( \left( \frac{t}{T_{break}} \right)^{\alpha_1} + \left( \frac{t}{T_{break}} \right)^{\alpha_2} \right)^{-1} \tag{4.7}$$

where $f_o$ is the flux zeropoint, unique to each filter, $t$ is the time since the burst and $T_{break}$ is the time when the power-law changes from temporal index $\alpha_1$ to $\alpha_2$. The afterglow parameters $f_o$, $T_{break}$, $\alpha_1$ and $\alpha_2$ were allowed to vary in the fit for each event, and the results of our modeling are listed in Table 4.7.

In all cases the values we derived are similar to those seen in the literature. For example, for GRB 990712 we find that a single power-law fits both the $V$ and $R$-band data well with a decay constant of $\alpha = -0.95 \pm 0.02$. In comparison, Bjornsson et al. (2001) find that the $V$-band data can be fit with either a single or broken power-law, where the value in the former is $\alpha_V - 0.82 \pm 0.03$. They find that the $R$-band photometry can be fit with a single power-law with a slightly steeper decay index of $\alpha_R = -0.91 \pm 0.06$. Both of these values are fully consistent with our own fit of the optical data.

Additionally, for GRB 020405 Price et al. (2003) found from their $R$-band pho-tometry a decay constant of $\alpha = -1.41$ when they limited the range of the fit up to 10 days after the burst, while Masetti et al. (2003) find $\alpha = -1.54 \pm 0.06$ in all optical filters for the same time period. However, when Price et al. (2003) re-fit the data with a broken power-law, they find a better fit with parameters $\alpha_1 = -0.94 \pm 0.25$, $\alpha_2 = -1.93 \pm 0.25$, and the time the LC breaks from $\alpha_1$ to $\alpha_2$ of $T_{break} = 1.67 \pm 0.52$ days ($\chi^2$/d.o.f $= 35.9/28$). Similarily, Masetti et al. (2003) find that at times $> 20$ days, $\alpha = -1.85 \pm 0.15$. When we restrict our fit to $2 - 10$ days, we find that a single power-law provides a suitable fit to the data with $\alpha = -1.72 \pm 0.08$, $\chi^2$/d.o.f $= 6.16/4$. This value for the temporal index is consistent with the value of $\alpha_2$ found by Price et al. (2003) and the value of $\alpha$ found by Masetti et al. (2003) at late times.

Once the afterglow parameters were determined, we then subtract Equation (4.7) from the OT (optical transient, e.g. afterglow and supernova) flux light curve to obtain host- and afterglow-subtracted "supernova" light curves (e.g. Cano et al. 2011). The resultant light curves are displayed in Figure 4.10. Note that in Figure 4.10 for XRF 060218 / SN 2006aj, we have included an additional power-law component of $\alpha = -0.2$, which was fixed in the fit, and is similar to that determined by Ferrero et al. (2006) of $\alpha = -0.20 \pm 0.10$.

For each of the host- and afterglow-subtracted "supernova" light curves, we determined the stretch and luminosity factor for each GRB-SN in relation to SN 1998bw in various filters, repeating the same process we followed for SN 2010bh (Section 3.5). The results of our fit are listed in Table 4.8, with the *R*-band stretch and luminosity factors displayed in Figure 4.11, and the host- and afterglow-subtracted "supernova" light curves displayed in Figure 4.10.

A previous study by Zeh et al. (2004), which was extended by Ferrero et al. (2006) (F06 hereon), also determined the stretch and luminosity factors of GRB-SN in relation to SN 1998bw[5] by modeling the SN "bumps" seen in the GRB light curves. For many events where we have determined the stretch and luminosity factors from the afterglow- and host-subtracted "supernova" light curves, and F06 from the supernova "bumps", we find similar results: e.g. GRB 050525A / SN 2005nc: this paper: $k = 0.69 \pm 0.03$,

[5] only in the *R* band

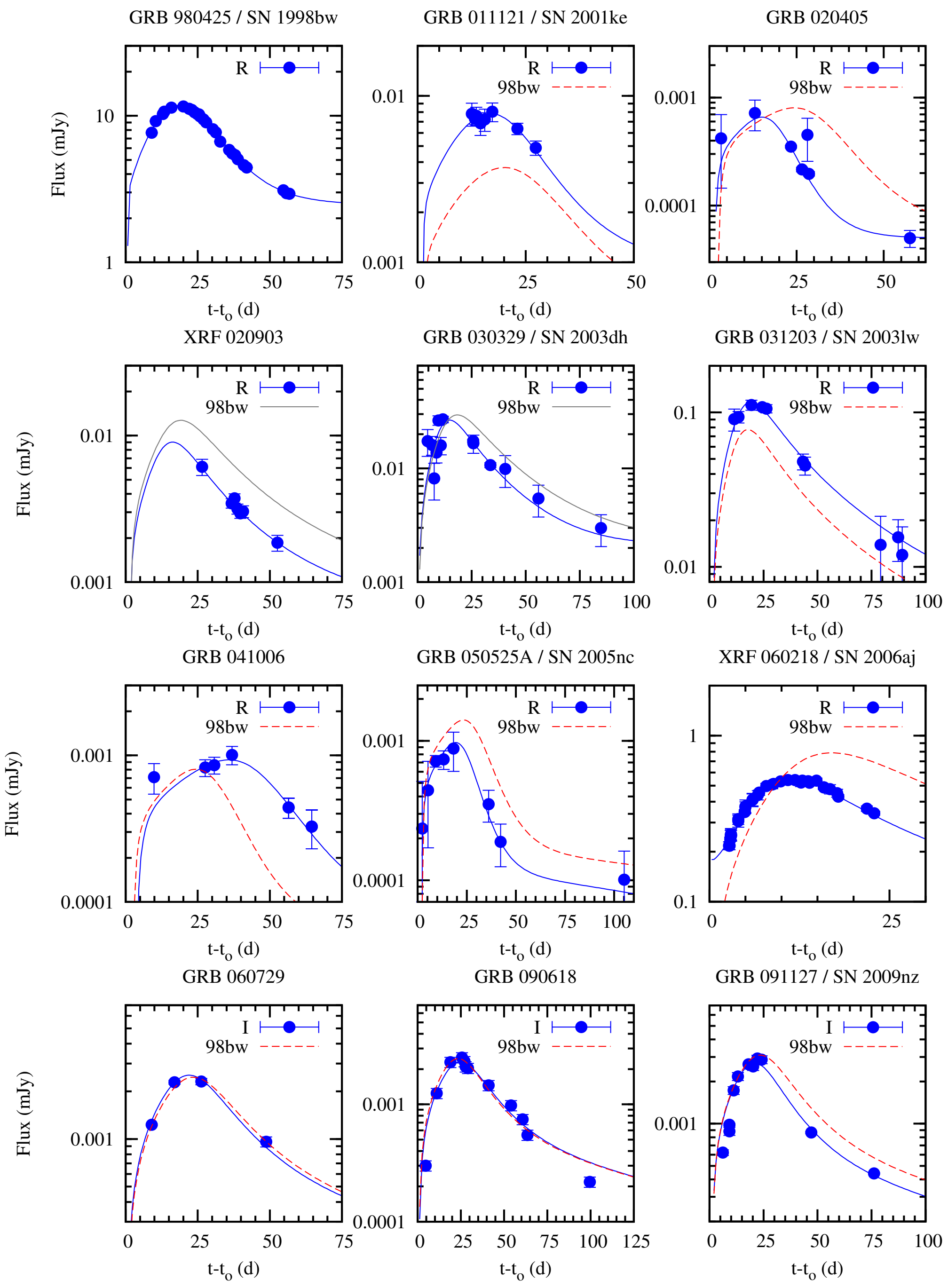

Figure 4.10 Mosaic of GRB-SNe. For each event the host has been subtracted, either by image subtraction by the original authors or by mathematically subtracting the flux due to the host. The OT (afterglow and supernova) light curves were then modeled, with the parameters of the afterglow determined and then subtracted out to create host- and afterglow-subtracted "supernova" light curves. Plotted for each event (dashed line) is how SN 1998bw would appear if it occurred at the redshift of the event. The solid-line is the light curve of SN 1998bw corrected by the luminosity and stretch factors listed in Table 4.8. For XRF 060218 / SN 2006aj, we have included an additional power-law component to the SN with index $\alpha = -0.2$, similar to that found by Ferrero et al. (2006).

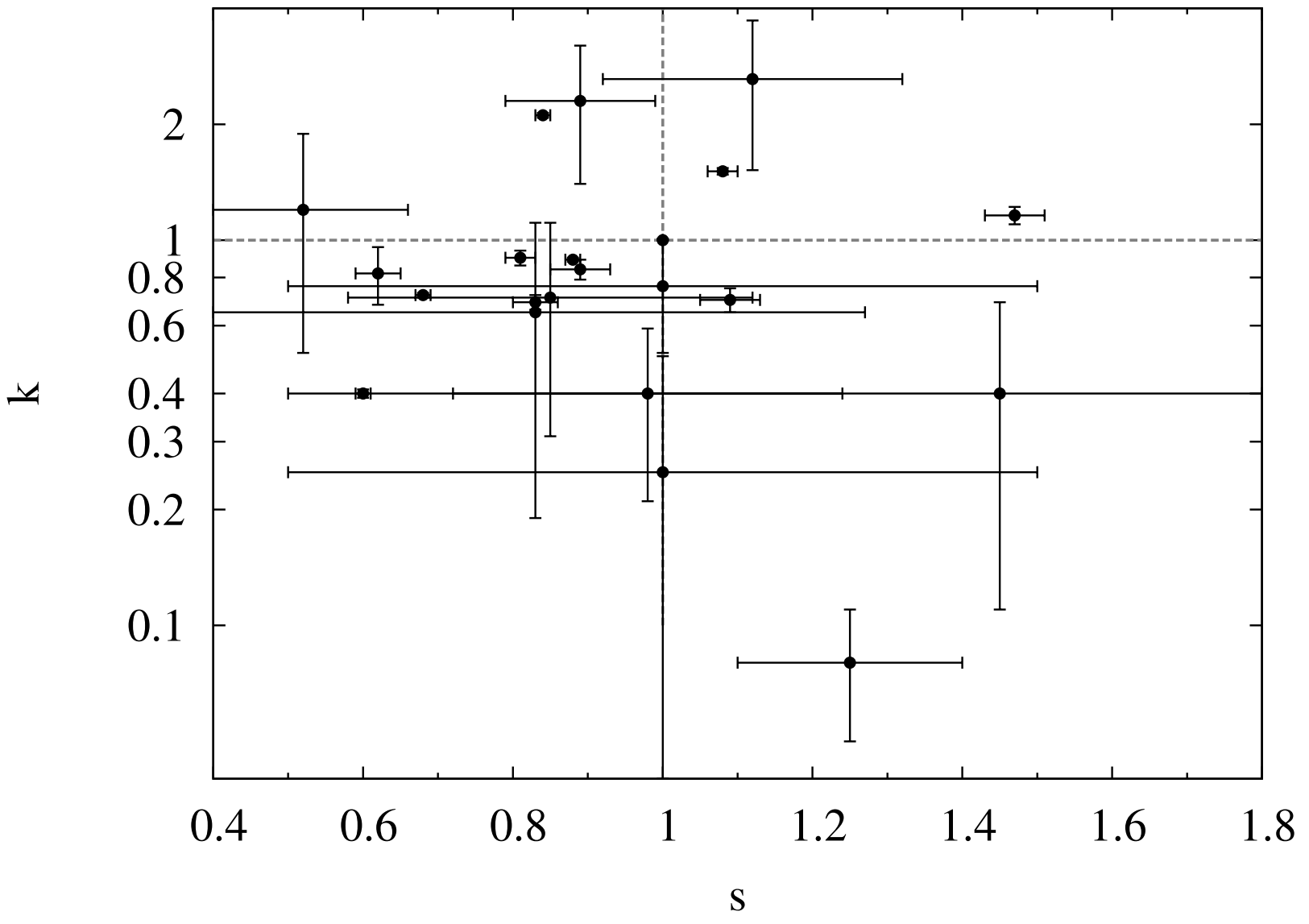

Figure 4.11 Luminosity factor ($k$) versus stretch factor ($s$) in the $R$ band for all of the GRB-SNe in Tables 4.5 and 4.8. We find a correlation/Pearson coefficient for the entire sample of $c = -0.05$, with a 95% confidence interval of $(-0.46, 0.38)$ for a significance of $p > 0.05$. As the critical value for the Pearson correlation coefficient for $N = 22$, d.o.f. $= 20$ and $p > 0.05$ is 0.492, we can conclude there is not a statistically significant correlation.

$s = 0.83 \pm 0.03$; F06[6] : $k = 0.66^{+0.10}_{-0.08}$, $s = 0.77 \pm 0.04$; and XRF 060218 / SN 2006aj ($V$ band): $k = 0.72 \pm 0.02$, $s = 0.65 \pm 0.02$; F06: $k = 0.724 \pm 0.007$, $s = 0.682 \pm 0.005$, with similar results for the other filters.

However, for some events we find different properties of the GRB-SN: for GRB 011121 / SN 2001ke, we find similar stretch factor to F06 (this paper: $s = 0.84 \pm 0.01$; F06: $s = 0.80 \pm 0.02$), however we find SN 2001ke to be much brighter once correcting for host-extinction: (this paper: $k = 2.11 \pm 0.03$; F06: $k = 0.88^{+0.08}_{-0.07}$). Additionally, for GRB 020405 we find a similar luminosity factor as F06: (this paper: $k = 0.82 \pm 0.14$; F06: $k = 0.90^{+0.15}_{-0.11}$), however we find that the SN evolves more quickly: (this paper: $s = 0.62 \pm 0.03$; F06: $s = 0.97 \pm 0.07$). However, GRB 020405 aside, for every other GRB-SN event we find the same value for the stretch factor as F06 within our respective errorbars.

We note that when using this method to determine the luminosity factor relative

[6] in the $R$ band and corrected for host-extinction

to SN 1998bw, the SN brightness depends on how accurately we have modeled the afterglow. For instance, if we over-estimate the relative contribution of the afterglow, then we will correspondingly under-estimate the SN brightness.

We find that the stretch factor of each GRB-SN event is approximately the same in each filter (e.g. GRB 990712, GRB 031203, XRF 060218, GRB 060729 & GRB 090618), though this trend applies less in the bluer filters. For example, for XRF 060218 / SN 2006aj, we find stretch factors in the $R$ and $I$ filters of $s = 0.68 \pm 0.02$, while in $V$ and $B$ we find slightly smaller values of $s = 0.65 \pm 0.02$ and $s = 0.60 \pm 0.02$ respectively.

We have investigated whether there might be a correlation between the $k$ and $s$ factors of the GRB-SNe, perhaps something akin to that established for type Ia SNe (i.e. the possibility of using GRB-SNe as "standard candles"). We have used all of the $R$-band data, of which there are 18 GRB-SNe. For GRB/XRFs 021211, 040924, 080319B and 091127, for which no $R$ band data exists, we have assumed that the stretch and luminosity factors in the $V$ band (GRB 021211) and the $I$ band (GRBs 040924, 080319B and 091127) are approximately the same as in the $R$ band. As already noted for events such as XRF 060218 and GRB 090618, the stretch and luminosity factors in filters $VR_cI_c$ are approximately the same and differ by only a few percent. The inclusion of these events brings the total sample size up to $N = 22$, and for the entire sample we find a correlation/Pearson coefficient of $c = -0.05$ and a 95% confidence interval of $(-0.46, 0.38)$ for a significance of $p > 0.05$. As the critical value for the Pearson correlation coefficient for $N = 22$, d.o.f. $= 20$ and $p > 0.05$ is 0.492, we can conclude that a statistically significant correlation does not exist.

Stanek et al (2005) noted that a relation may exist between light curve shape and luminosity, which the results of F06 could neither confirm nor refute (total sample of $N = 13$). F06 noted that there might be a trend of rising $k$ with rising $s$, however the addition of 9 more GRB-SNe to the total sample shows that it is highly unlikely that such a correlation exists. When this result is taken in tandem with the results of Drout et al. (2011), who found no statistically-significant correlation between peak luminosity and light-curve decay shapes for a sample of Ibc SNe, indicates that Ibc

Table 4.7 Afterglow parameters of the GRB-SNe

| GRB/XRF/SN | Filter | $\alpha_1$ | $\alpha_2$ | $T_{break}$ (days) | $\chi^2$/d.o.f |
|---|---|---|---|---|---|
| 011121/2001ke | $R_c$ | $-1.70 \pm 0.01$ | - | - | 3.56/4 |
| 020405 | $R_c$ | $-1.72 \pm 0.16$ | - | - | 6.16/4 |
| 020903 | $R_c$ | $-1.87 \pm 0.05$ | - | - | 1.86/2 |
| 041006 | $R_c$ | $-0.47 \pm 0.35$ | $-1.47 \pm 0.39$ | $0.30 \pm 0.78$ | 3.24/3 |
| 050525A/2005nc | $R_c$ | $-0.89 \pm 0.22$ | $-2.13 \pm 0.30$ | $0.43 \pm 0.35$ | 4.68/3 |
| 060729 | $U$ * | $-0.01 \pm 0.03$ | $-1.65 \pm 0.05$ | $0.75 \pm 0.08$ | 37.99/29 |
| 090618 | $R_c$, $i$ † | $-0.65 \pm 0.07$ | $-1.57 \pm 0.07$ | $0.50 \pm 0.11$ | 151.42/113 |

* Cano et al. (2011) determined the afterglow parameters using the $U$-band data due to the lack of early-time data in $R$ and $I$.
† Cano et al. (2011) fit all of the optical data simultaneously.

SNe (including GRB-SNe and Ic-BL SNe) cannot be used as standardizable candles.

SECTION 4.6

## Summary

We have presented optical and near infrared photometry of XRF 100316D / SN 2010bh obtained on the Faulkes Telescope South, Gemini South and *HST*, with the data spanning from $\mathrm{t} - \mathrm{t_o} = 0.5 - 47.3$ days. It was shown that the optical light curves of SN 2010bh evolve more quickly than the archetype GRB-SN 1998bw, and at a similar rate to SN 2006aj, which was associated with XRF 060218, and non-GRB associated type Ic SN 1994I. In terms of peak luminosity, SN 2010bh is the faintest SN yet associated (either spectroscopically or photometrically) with a long-duration GRB or XRF, and has a peak, $V$-band, absolute magnitude of $M_V = -18.62 \pm 0.08$.

SN 2010bh appears to be redder than GRB-SNe 1998bw and 2006aj, where the colour curves of SN 2010bh are seen to be redder early on, though at late times the $V - R$ and $R - i$ colour curves matched that of SN 1994I. The red nature of SN 2010bh is also demonstrated in Figure 4.8, where it is seen that more of the bolometric flux is emitted at infrared wavelengths (and less at optical wavelengths) than in the broad-lined Ic SN 2009bb.

Table 4.8 Luminosity ($k$) and Stretch ($s$) Factors of the GRB-SNe with respect to SN 1998bw

| GRB/XRF/SN | Filter | $s$ | $k$ | Ref ($s$ & $k$) | Ref (Photometry) |
|---|---|---|---|---|---|
| 970228 | $R$ | $1.45 \pm 0.95$ | $0.40 \pm 0.29^{*}$ | (1) | - |
| 980425 | $BVRi$ | 1.00 | 1.00 | - | (4) |
| 990712 | $V$ | $0.83 \pm 0.44$ | $0.65 \pm 0.46$ | This paper | (5), (6), (7), (8) |
| 990712 | $R$ | $0.90 \pm 0.07$ | $1.45 \pm 0.20$ | This paper | (5), (6), (7), (8) |
| 991208 | $R$ | $1.12 \pm 0.20$ | $2.62 \pm 1.10$ | (1) | - |
| 000911 | $R$ | $1.40 \pm 0.32$ | $0.85 \pm 0.44$ | (1) | - |
| 011121/2001ke | $R$ | $0.84 \pm 0.01$ | $2.11 \pm 0.03$ | This paper | (9), (10) |
| 020405 | $R$ | $0.62 \pm 0.03$ | $0.82 \pm 0.14$ | This paper | (11), (12) |
| 020903 | $R$ | $0.85 \pm 0.27$ | $0.71 \pm 0.40$ | This paper | (13) |
| 021211/2002lt | $V$ | $0.98 \pm 0.26$ | $0.40 \pm 0.19^{*}$ | (1) | - |
| 030329/2003dh | $R$ | $0.81 \pm 0.02$ | $0.90 \pm 0.04$ | This paper | (14), (15) |
| 031203/2003lw | $R$ | $1.08 \pm 0.02$ | $1.51 \pm 0.03$ | This paper | (16), (17), (18) |
| 031203/2003lw | $I$ | $1.07 \pm 0.04$ | $1.29 \pm 0.04$ | This paper | (16), (17), (18) |
| 040924 | $I,z$ | 1.00 (fixed) | $\approx 0.25$ | (2) | - |
| 041006 | $R$ | $1.47 \pm 0.04$ | $1.16 \pm 0.06$ | This paper | (19) |
| 050525A/2005nc | $R$ | $0.83 \pm 0.03$ | $0.69 \pm 0.03$ | This paper | (20) |
| 050824 | $R$ | $0.52 \pm 0.14$ | $1.20 \pm 0.69$ | (23), (24) | - |
| 060218/2006aj** | $B$ | $0.60 \pm 0.01$ | $0.67 \pm 0.02$ | This paper | (1) |
| 060218/2006aj** | $V$ | $0.65 \pm 0.01$ | $0.72 \pm 0.01$ | This paper | (1) |
| 060218/2006aj** | $R$ | $0.68 \pm 0.01$ | $0.72 \pm 0.01$ | This paper | (1) |
| 060218/2006aj** | $I$ | $0.68 \pm 0.01$ | $0.76 \pm 0.01$ | This paper | (1) |
| 060729 | $R$ | $0.89 \pm 0.04$ | $0.84 \pm 0.05$ | This paper | (21) |
| 060729 | $I$ | $0.94 \pm 0.02$ | $1.03 \pm 0.02$ | This paper | (21) |
| 080319B | $I$ | $0.89 \pm 0.10$ | $2.30 \pm 0.90$ | (3), (25) | - |
| 090618 | $R$ | $1.09 \pm 0.04$ | $0.70 \pm 0.05$ | This paper | (21) |
| 090618 | $I$ | $1.06 \pm 0.03$ | $0.94 \pm 0.03$ | This paper | (21) |
| 091127 | $I$ | $0.88 \pm 0.01$ | $0.89 \pm 0.01$ | This paper | (22) |
| 101225A | $R,i,z$ | $1.25 \pm 0.15$ | $0.08 \pm 0.03$ | (26) | - |

* Not corrected for host-extinction.
** Fit **excluding** additional decaying power-law component (see text).

(1) Ferrero et al. (2006), (2) Soderberg et al. (2006), (3) Tanvir et al. (2010), (4) Galama et al. (1998), (5) Sahu et al. (2000), (6) Christensen et al. (2004), (7) Björnsson et al. (2001), (8) Hjorth et al. (2000), (9) Garnavich et al. (2003), (10) Küpcü Yoldaş et al. (2007), (11) Bersier et al. (2003), (12) Price et al. (2003), (13) Bersier et al. (2006), (14) Matheson et al. (2003), (15) Deng et al. (2005), (16) Malesani et al. (2004), (17) Mazzali et al. (2006a), (18) Margutti et al. (2007), (19) Stanek et al. (2005), (20) Della Valle et al. (2006a), (21) Cano et al. (2011), (22) Cobb et al. (2010), (23) Sollerman et al. (2007), (24) Kann et al. (2010), (25) Bloom et al. (2009), (26) Thöne et al. (2011)

We also gave evidence of the detection of light coming from the shock breakout at $t - t_o = 0.598$ days. The brightness of the $B$-band light curve at this epoch, as well as the shape of the optical SED, which with a spectral power-law index of $\beta = +0.94 \pm 0.05$ (which is harder than is expected for synchrotron radiation), implies that the source of light at this epoch is not synchrotron in origin and is likely coming from the shock-heated, expanding stellar envelope.

We then applied a simple physical model to the bolometric light curve of SN 2010bh. When we include all photometry in the optical and infrared regime (in the $3,000$Å – $16,600$Å wavelength range) we find physical parameters of $M_{Ni} = 0.10 \pm 0.01\ M_{\odot}$, $M_{ej} = 2.24 \pm 0.08\ M_{\odot}$, $E_k = 1.39 \pm 0.06 \times 10^{52}$ erg. The faint nature of SN 2010bh becomes again apparent when the nickel mass synthesized during the explosion is compared with other GRB-SNe such as SN 1998bw, where it is believed that $M_{Ni} \approx 0.4 M_{\odot}$ was created in the explosion. Indeed SN 2010bh synthesized only a marginally larger amount of nickel than local type Ic SN 1994I ($M_{Ni} \approx 0.07\ M_{\odot}$). The investigation of SN 2010bh in relation to the general population of type Ic SNe has once again shown that type Ic SNe are a very heterogeneous class of supernovae, spanning a wide range of luminosities and ejected masses.

Finally, we assembled from the literature all of the available photometry of previously detected GRB-SNe. For each of these events we assumed that light is coming from three sources: from the host galaxy, the afterglow and the supernova. First we removed the constant flux due to the host galaxy, then we modeled the optical afterglows, and then subtracted the light due to the afterglow, which resulted in host- and afterglow-subtracted "supernova" light curves (Figure 4.10). We then created synthetic light curves of SN 1998bw as it would appear if it occurred at the redshift of each given GRB-SNe. We then compared the brightness and shape of the GRB-SNe with that of the shifted LCs of SN 1998bw, and derived stretch ($s$) and luminosity ($k$) factors for each GRB-SNe. The results of our method, when compared to previous studies performed by Ferrero et al. (2006) showed similar values for the stretch and luminosity factors, as well as including more events than the previous studies. When we checked for the possibility of a correlation between $k$ and $s$ in the $R$-band data, we find an in-significant correlation/Pearson coefficient of $c = -0.05$, suggesting that it is highly unlikely that such a relation exists.

# 5

# GRB-SNe vs non-GRB-SNe

SECTION 5.1

## Comparison of peak, $V$-band magnitudes

Many attempts to quantify the subtle differences between GRB-SNe and those of local stripped-envelope, core-collapse supernovae that were not accompanied by a GRB trigger (i.e. non-GRB-SNe) have been undertaken, with the possible differences in the progenitors of GRB-SNe and local Ibc SNe being attributed to binarity (e.g. Blinnikov et al. 1984; Paczyński 1986, 1991; Eichler et al. 1989; Narayan et al. 1991, Narayan, Paczyński & Piran 1992), rapid rotation (e.g. Yoon & Langer 2005; Woosley & Heger 2006), asphericity (e.g. Mesźaros & Rees 1992; Woosley & Baron 1992), magnetic fields (e.g., Narayan, Paczyński & Piran 1992) and very likely metallicity (e.g. Modjaz et al. 2008). As the GRB-SNe sample is still of a modest size ($N \approx 22$, though a few more have been detected but not included in this sample due to the sparseness of the dataset for each event; e.g. GRB 081007 & SN 2008hw; Della Valle et al. 2008), statistical analyses of the sample number only a few.

To put our detections made of the GRB-SNe in the preceding chapters into context, we have compiled from the literature (up to 2009) peak, restframe $V$-band absolute magnitudes for two samples of type Ibc SNe, incorporating the values of the

Table 5.1 Peak Rest-Frame $V$-band Absolute Magnitudes for GRB & XRF-producing SNe

| GRB | SN | $z$ | $A_{V,f}$ [a] | $A_{V,h}$ [b] | $M_V^{peak}$ (mag)[c,d] | Reference |
|---|---|---|---|---|---|---|
| GRB 970228 | - | 0.695 | 0.543 | 0.15 | $-18.56 \pm 0.30$ | (1), (2), (3) |
| GRB 980326 | - | $\approx 1$ | 0.26 | - | $\approx$ -19.5 | (4) |
| GRB 980425 | 1998bw | 0.0085 | 0.18 | 0.05 | $-19.42 \pm 0.30$ | (3), (5), (6), (7), (8), (31) |
| GRB 990712 | - | 0.434 | 0.09 | 1.67 | $-20.22 \pm 0.20$ | (3), (10), (11), (12), (31) |
| GRB 991208 | - | 0.706 | 0.05 | 0.76 | $-19.46 \pm 0.75$ | (9), (16) |
| GRB 000911 | - | 1.058 | 0.38 | 0.20 | $-18.31 \pm 0.15$ | (9), (16) |
| GRB 011121 | 2001ke | 0.36 | 1.33 | 0.39 | $-19.59 \pm 0.33$ | (3), (13), (14), (16) |
| GRB 020405 | - | 0.698 | 0.14 | 0.15 | $-19.46 \pm 0.25$ | (3), (15), (16), (31) |
| GRB 020410 | - | $\approx$ 0.5 | 0.40 | 0.0 | $\approx -17.6$ | (3), (17) |
| XRF 020903 | - | 0.251 | 0.09 | 0.0 | $-18.89 \pm 0.30$ | (3), (18), (31) |
| GRB 021211 | 2002lt | 1.006 | 0.08 | 0.0 | $-18.27 \pm 0.60$ | (9), (19), (16) |
| GRB 030329 | 2003dh | 0.169 | 0.07 | 0.39 | $-19.14 \pm 0.25$ | (3), (16), (20), (31), (32) |
| XRF 030723 | - | $\approx$ 0.4 | 0.089 | 0.23 | $\approx -17.9$ | (3), (9), (34) |
| GRB 031203 | 2003lw | 0.1055 | 2.77 | 0.85 | $-20.39 \pm 0.50$ | (3), (21), (22), (31) |
| GRB 040924 | - | 0.859 | 0.18 | 0.16 | $-17.47 \pm 0.48$ | (23) |
| GRB 041006 | - | 0.716 | 0.07 | 0.11 | $-19.57 \pm 0.30$ | (3), (16), (24) |
| GRB 050525A | 2005nc | 0.606 | 0.25 | 0.32 | $-18.76 \pm 0.28$ | (3), (25), (26), (33) |
| XRF 060218 | 2006aj | 0.033 | 0.39 | 0.13 | $-18.76 \pm 0.20$ | (3), (27), (28), (31) |
| GRB 060729 | - | 0.54 | 0.11 | 0.18 | $-19.43 \pm 0.06$ | This work |
| GRB 080319B | - | 0.931 | 0.03 | 0.05 | $-19.12 \pm 0.40$ | (29), (33) |
| GRB 090618 | - | 0.54 | 0.27 | 0.3 | $-19.75 \pm 0.14$ | This work |
| GRB 091127 | 2009nz | 0.49 | 0.12 | 0.0 | $-19.00 \pm 0.20$ | (30) |

[a] Foreground extinction calculated from the dust maps of Schlegel et al. (1998).
[b] Host extinction where available.
[c] Cosmological Parameters used: $H_o = 71$ km s$^{-1}$ Mpc$^{-1}$, $\Omega_M = 0.27$, $\Omega_\Lambda = 0.73$.
[d] Wherever errors are not quoted in the literature conservative errors of 0.4 mag are used.

(1) Galama et al. (2000), (2) Castander & Lamb (1999), (3) Richardson (2009), (4) Bloom et al. (1999), (5) Galama et al. (1998), (6) McKenzie & Schaefer (1999), (7) Sollerman et al. (2000), (8) Nakamura et al. (2001), (9) Zeh et al. (2004), (10) Sahu et al. (2000), (11) Fruchter et al. (2000), (12) Christensen et al. (2004), (13) Bloom et al. (2002b), (14) Garnavich et al. (2003), (15) Mazetti et al. (2003), (16) Kann et al. (2006), (17) Levan et al. (2005), (18) Bersier et al. (2006), (19) Della Valle et al. (2003), (20) Matheson et al. (2003), (21) Malesani et al. (2004), (22) Mazzali et al. (2006a), (23) Soderberg et al. (2006), (24) Stanek et al. (2005), (25) Della Valle et al. (2006a), (26) Blustin et al. (2006), (27) Sollerman et al. (2006), (28) Modjaz et al. (2006), (29) Tanvir et al. (2010), (30) Cobb et al. (2010), (31) Levesque et al. (2010), (32) Deng et al. (2005), (33) Kann et al. (2010), (34) Butler et al. (2005).

Table 5.2 Peak Rest-Frame $V$-band Absolute Magnitudes for Local type Ibc & Ic SNe

| Type | SN | $z$ | $A_{V,f}$ [a] | $A_{V,h}$ [b] | $M_V^{peak}$ (mag)[c,d] | Reference |
|---|---|---|---|---|---|---|
| Ib | 1954A | 0.000977 | 0.07 | - | $-18.75\pm0.40$ | (1) |
| Ic | 1962L | 0.00403 | 0.12 | - | $-18.83\pm0.83$ | (2), (3) |
| Ic | 1964L | 0.002702 | 0.07 | - | $-18.38\pm0.65$ | (2), (4) |
| Ib | 1966J | 0.002214 | 0.04 | - | $-19.00\pm0.4$ | (4) |
| Ib | 1972R | 0.002121 | 0.05 | - | $-17.44\pm0.4$ | (5) |
| Ic | 1983I | 0.002354 | 0.04 | - | $-18.73\pm0.45$ | (2), (6) |
| Ib | 1983N | 0.001723 | 0.20 | 0.3 | $-18.58\pm0.57$ | (7) |
| Ib | 1983V | 0.005462 | 0.06 | 1.18 | $-19.12\pm0.41$ | (2), (8) |
| Ib | 1984I | 0.0107 | 0.33 | - | $-17.50\pm0.40$ | (9) |
| Ib | 1984L | 0.005281 | 0.08 | 0.0 | $-18.84\pm0.40$ | (10) |
| Ib | 1985F | 0.00167 | 0.06 | 0.63 | $-20.19\pm0.50$ | (11) |
| Ic | 1987M | 0.004419 | 0.08 | 1.28 | $-18.33\pm0.71$ | (2), (12), (13) |
| Ic | 1990B | 0.007518 | 0.10 | 2.53 | $-19.49\pm1.02$ | (2), (14) |
| Ib | 1991D | 0.041752 | 0.19 | 0.0 | $-20.01\pm0.60$ | (15) |
| Ic | 1991N | 0.003319 | 0.07 | - | $-18.67\pm1.06$ | (2) |
| Ic | 1992ar | 0.1451 | 0.30 | 0.0 | $-18.84\pm0.42$ | (2), (16) |
| Ic | 1994I | 0.001544 | 0.11 | 1.39 | $-17.49\pm0.58$ | (2), (17), (18) |
| Ic BL | 1997ef | 0.011693 | 0.13 | 0.55 | $-17.80\pm0.21$ | (2), (19), (34) |
| Ic pec | 1999as | 0.127 | 0.09 | 0.0 | $-21.21\pm0.20$ | (20) |
| Ib/c | 1999cq | 0.026309 | 0.16 | - | $-19.75\pm0.72$ | (2), (21) |
| Ib | 1999dn | 0.00938 | 0.16 | - | $-17.17\pm0.40$ | (22) |
| Ib/c | 1999ex | 0.011401 | 0.06 | - | $-17.67\pm0.26$ | (23) |
| Ib | 2001B | 0.005227 | 0.39 | - | $-17.13\pm0.40$ | (24) |
| Ic BL | 2002ap | 0.002187 | 0.29 | 0.0 | $-17.73\pm0.21$ | (2), (25) |
| Ic | 2003L | 0.021591 | 0.06 | - | $-18.90\pm0.40$ | (27) |
| Ic BL | 2003jd | 0.018826 | 0.14 | 0.29 | $-19.50\pm0.30$ | (19), (26), (34) |
| Ic | 2004aw | 0.0175 | 1.15 | 0.0 | $-18.05\pm0.39$ | (28) |
| Ic | 2004ib | 0.056 | 0.07 | - | $-16.94\pm0.40$ | (32) |
| Ib pec | 2005bf | 0.018913 | 0.14 | - | $-18.23\pm0.40$ | (33) |
| Ic BL | 2005fk | 0.2643 | 0.19 | - | $-20.41\pm0.40$ | (29) |
| Ic BL | 2005kr | 0.13 | 0.31 | 0.27 | $-19.08\pm0.40$ | (19), (29), (34) |
| Ic BL | 2005ks | 0.10 | 0.17 | 0.79 | $-18.41\pm0.40$ | (19), (29), (34) |
| Ib/c | 2007gr | 0.001728 | 0.19 | - | $-16.74\pm0.40$ | (30) |
| Ic BL | 2007ru | 0.01546 | 0.89 | 0.0 | $-19.09\pm0.20$ | (31) |

[a] Foreground extinction calculated from the dust maps of Schlegel et al. (1998).
[b] Host extinction where available.
[c] Cosmological Parameters used: $H_o = 71$ km s$^{-1}$ Mpc$^{-1}$, $\Omega_M = 0.27$, $\Omega_\Lambda = 0.73$.
[d] Wherever errors are not quoted in the literature conservative errors of 0.4 mag are used.

(1) Wild (1960), (2) Richardson (2002), (3) Bertola (1964), (4) Miller & Branch (1990), (5) Barbon (1973), (6) Tsvetkov (1983), (7) Clocchiatti et al. (1996a), (8) Clocchiatti et al. (1997), (9) Binggeli et al. (1984), (10) Tsvetkov (1987), (11) Filippenko et al. (1986), (12) Filippenko et al. (1990), (13) Nomoto et al. (1990), (14) Clocchiatti et al. (2001), (15) Benetti et al. (2002), (16) Clocchiatti et al. (2000), (17) Yokoo et al. (1994), (18) Iwamoto et al. (1994), (19) Modjaz et al. (2008), (20) Hatano et al. (2001), (21) Matheson et al. (2000), (22) Qiu et al. (1999), (23) Martin et al. (1999), (24) BAOSS, (25) Mazzali et al. (2002), (26) Valenti et al. (2008), (27) Soderberg et al. (2003), (28) Taubenberger et al. (2006), (29) Barentine et al. (2005), (30) Foley et al. (2007), (31) Sahu et al. (2009), (32) Adelman et al. (2005), (33) Anupama et al. (2005), (34) Levesque et al. (2010).

host/restframe extinction (taken at the location of the GRB or SN when available; i.e., Kann et al. 2006; Modjaz et al. 2008; Levesque et al. 2010), and applying the SMC reddening law. The two samples are: (1) those associated with GRBs & XRFs, and (2) local type Ibc SNe (i.e. non-GRB-SNe). We note that we have limited this analysis to incorporate only GRB/XRF events where an optically-bright SN has been positively detected (i.e. not included upper limits of non-detections). These samples, along with their respective references, are listed in Tables 5.1 and 5.2.

When comparing these two samples we are attempting to determine whether the peak SN brightness is enough to distinguish GRB/XRF-SNe from Ibc SNe that do not have a GRB-trigger. We will test if the distribution of the peak magnitudes of the two samples of supernovae are different by performing a Kolmogorov-Smirnov (KS) test on the two samples. First we compared the entire GRB/XRF sample ($N = 22$) with *all* of the type Ibc SNe ($N = 34$), finding a modest probability that the two samples are drawn from the same parent population of $P = 0.12$. When we compare the GRB/XRF sample with only the type Ic SNe ($N = 19$) we find a similar probability of $P = 0.16$.

However, when we limited the samples to include only those SNe where an estimation of the host/restframe extinction has been made, we see an increased probability between the datasets: $P = 0.88$ between the GRB/XRF SNe ($N = 21$) and all of the type Ibc SNe ($N = 17$), and $P = 0.54$ between the GRB/XRF SNe and only the type Ic SNe ($N = 12$). For these samples a higher average peak magnitude is also seen among the local type Ibc SNe sample, with the local type Ibc SNe having $M_{V,\mathrm{Ibc}} = -18.93$. This is an increase in average peak brightness of $\sim 0.3$ mag, and is likely due to the inclusion of five additional bright type Ib events with known host-extinction (SN: 1983N, 1983V, 1984L, 1985F & 1991D) with the local Ic SNe sample (N=12). The average peak magnitude of these five SNe is $M_{V,\mathrm{Ib}} = -19.34$, which places them among the brightest of the type Ib events. For comparison, a study by Richardson et al. (2002) analyzed the absolute peak magnitudes of samples of all types of nearby SNe, and distinguished between "Normal" and "Bright" type Ibc events, with the latter having $M_{V,\mathrm{Ibc}} = -19.72 \pm 0.24$, comparable with the average peak magnitudes of these five SNe. Thus the inclusion of these 5 bright Ib events has increased the average peak brightness of the type Ibc SNe sample, as well as increased the probability of association between our samples of host extinction-corrected type Ibc SNe and the GRB/XRF-SNe. The results of the KS test are summarized in Table 5.3 and cumulative fraction plots are shown in Figure 5.1.

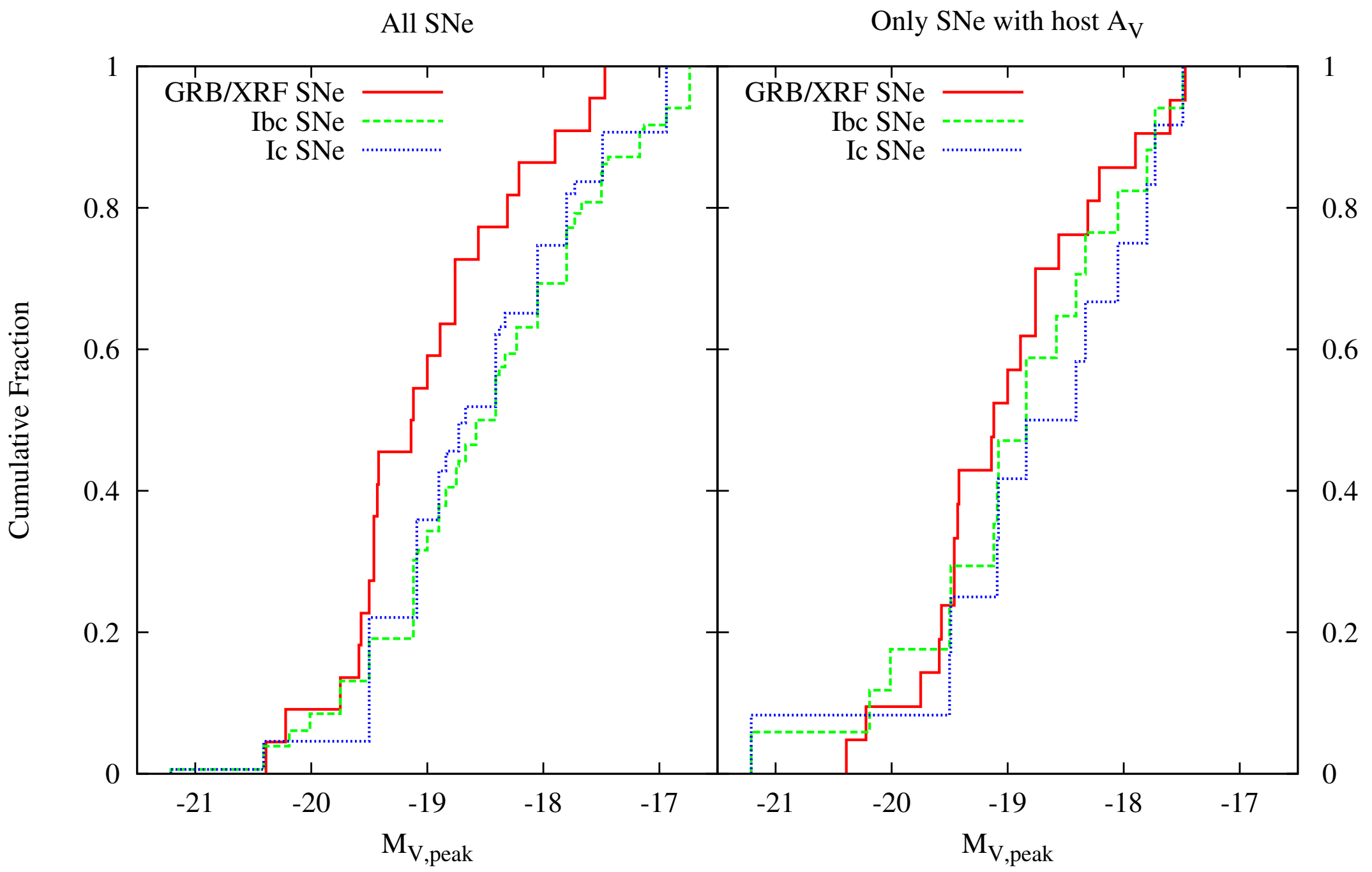

Figure 5.1 Cumulative fraction plots of the absolute, $V$-band, peak magnitudes of the GRB-SNe and non-GRB-SNe for the two comparison scenarios: (*right*) between *all* SNe and (*left*) considering only those events where an estimation of the host/restframe extinction has been made. In the latter case, the probabilities that the three sets of SNe are drawn from the same parent population are (1) GRB/XRF SNe & All Ibc SNe: $P = 0.88$, (2) GRB/XRF SNe & only Ic SNe: $P = 0.54$, which further supports the idea that GRB-SNe and type Ic SNe have similar progenitors.

Table 5.3 Kolmogorov-Smirnov test results of the peak $V$band comparison

| Dataset | $N$ | Mean | Standard Deviation | $P^a$ | $D^a$ | comments |
|---|---|---|---|---|---|---|
| GRB/XRF-associated SNe | 22 | -19.02 | 0.77 | - | - | all events |
| Local type Ic SNe | 19 | -18.73 | 1.00 | 0.16 | 0.33 | all events |
| Local type Ibc SNe | 34 | -18.59 | 1.04 | 0.12 | 0.31 | all events |
| GRB/XRF-associated SNe | 21 | -19.00 | 0.78 | - | - | only those with host $A_V$ |
| Local type Ic SNe | 12 | -18.75 | 1.03 | 0.54 | 0.27 | only those with host $A_V$ |
| Local type Ibc SNe | 17 | -18.93 | 0.97 | 0.88 | 0.18 | only those with host $A_V$ |

[a]Probability and maximum difference between the GRB/XRF SNe sample and the local SNe sample.

It is worth noting, however, that while the increase in probability may really be due to an increased association between the datasets, the caveat of smaller sample sizes is that as one goes to smaller samples it is harder to obtain a statistically significant discrepancy.

Our analysis is not unique, with previous studies having performed similar analyses, which our results generally support. Richardson 2009 (R09) found from a sample of 14 GRB/XRF-SNe with (mostly) known values of the host extinction an average $\langle M_{V,\rm peak}\rangle = -19.2 \pm 0.2$, with $\sigma = 0.7$, which our results are in agreement with. R09 also compared the GRB/XRF-SNe sample with a sample of stripped-envelope SNe, which included types Ib, Ic and IIb, finding the GRB/XRF-SNe sample brighter by $\sim 0.8$ mag. We find that our complete GRB/XRF-SNe sample is $\sim 0.4$ mag brighter than the complete local type Ibc SNe (but only $\sim 0.1$ mag when only events with known host extinction are considered), which is somewhat less than R09, however our sample does not include any type IIb SNe.

Ferrero et al. 2006 (F06) undertook a slightly different study, comparing the existing GRB/XRF-SNe sample (to date) with that of XRF 060218. They furthered an analysis originally performed by Zeh et al. (2004), and calculated for each GRB-SN event the luminosity ratio $k$ and stretch factor $s$ in comparison with SN1998bw.

They find for their host extinction-corrected sample of GRB/XRF-SNe clustering in the range $0.6 < k < 1.5$, which is in good agreement with our results. They also conclude that there was no evidence that the luminosity function evolved with redshift. Our analysis concurs with both of these results.

If we take our analysis a step further, we can use the peak $V$-band magnitudes as a proxy for the peak bolometric magnitude of these events (e.g. using Arnett (1982) and Equation 4.3 in Chapter 4). If we make the assumption that the distribution of peak $V$-band magnitudes will be the same as the distribution of the peak bolometric magnitudes, then our conclusion that all Ibc have similar peak magnitudes in turn implies that similar amounts of nickel are created in all Ibc SNe.

In conclusion, when we compare the host-corrected samples of the various Ibc SNe, it is seen that one cannot use only the peak, $V$-band magnitudes to distinguish between different sub-classes. Additionally, our results suggests that GRB/XRF-SNe do not, on average, eject more nickel during the explosion than typical Ibc SNe.

SECTION 5.2

## Comparison of $\varnothing m_{15}$ for GRB-SNe and non-GRB-SNe

The analysis performed in the preceding section has led us to the conclusion that one cannot use only the peak magnitudes of the different samples of SNe to distinguish between them. This result was echoed in a recent paper by Drout et al. (2010), who performed a systematic analysis of local type Ibc SNe and concluded that engine-driven SNe (e.g. GRB-SNe) cannot be distinguished from "normal" Ibc SNe by peak luminosity alone.

It has been noted in the literature, since the discovery of SN 1998bw, that the photospheric velocities of GRB-SNe that can exceed velocities of $30,000$ km s$^{-1}$ (e.g. SN 1998bw, Galama et al. 1998). These velocities are larger than those seen in other type Ic SNe without a GRB-trigger. Matheson et al. (2001) found for a sample of local Ibc SNe (excluding Ic-BL SNe) expansion velocities in the range $6,000$ km s$^{-1}$ to $12,000$ km s$^{-1}$.

However, there exists not just a simple dichotomy between typical type Ic SNe such as SN 1994I (Richmond et al. 1996; Iwamoto et al. 1994) that have modest peak magnitudes and ejecta velocities, and GRB-SNe that have comparable peak magnitudes but higher ejecta velocities. Indeed the type Ic SNe class is quite heterogeneous, especially when the discoveries of broad-lined SNe that are not accompanied by a GRB-trigger are also considered. Events such as SN 2009bb (Strizinger et al. 2009; Pignata et al. 2011) reveal that relativistic ejecta are present in some non-GRB associated SNe. Even more intriguingly, Soderberg et al. (2010) showed that SN 2009bb most likely had a prolonged central engine. The multi-frequency radio observations presented by Soderberg et al. (2010) were well described by a self-absorbed synchrotron spectrum which they suggest is produced as the shock break-out accelerated electrons in the local circumstellar medium.

Thus it seems that the special requisites for GRB-production are very atypical of the general Ic SNe population; indeed it is possible to have an energetic type Ic SN with a bright maximum and broad spectral features, *and* prolonged central engine activity, without the production of a GRB.

As mentioned in the introductory chapter, every supernova associated with a GRB to date has been type Ic. It is thought that the progenitor stars of type Ic SNe have undergone extensive stripping of their outer envelopes, even more so than the progenitors of type Ib SNe, which accounts for the lack of detected H or He in the spectra of type Ic SNe (and lack of H in type Ib SNe). Thus, a naïve expectation would be that, on average, type Ib SNe eject more mass than type Ic SNe *if* the progenitor stars of all type Ibc SNe have similar main sequence masses.

To estimate the relative amount of mass ejected in the various Ibc SNe sub-types, and in turn estimate the main-sequence mass of the progenitors, we have undertaken an investigation to determine the physical parameters that lead to the creation of a SN light curve. In terms of the bolometric light curve, the overall peak is determined by the amount of nickel that is created during the explosion. We have seen in the previous section that peak brightness of the SNe LCs is not enough to distinguish between GRB-SNe and type Ibc SNe. If we take the distribution of the peak $V$-band magnitude as a proxy for the peak bolometric magnitude, our conclusion in the previous section implies GRB-SNe and type Ibc SNe produce similar amounts of nickel during the explosions.

The width of the bolometric LC is dictated by the amount of mass ejected during the explosion, as well as the velocity of the ejecta and the opacity. Thus, for events that have similar ejecta velocities, and assuming the densities of the ejecta are approximately the same, light curves that are wider and fade slower imply larger ejected masses. A simple prediction is that, on average, type Ib SNe will have wider light curves than type Ic SNe due to the former ejecting more mass during the explosion.

To verify this prediction it would be ideal to assemble bolometric light curves of a large sample of type Ibc SNe, and then measure the widths. Then, if we know the velocity at which the ejecta is moving in each event, and we make an estimate of the opacity, it is possible to derive the amount of ejecta mass in each event. However, it is not currently possible to create a statistically large sample of bolometric light curves from the available data in the literature due to the lack of many events where there is enough multi-wavelength data to assemble a bolometric LC. Instead, what we can do is compare the widths of the light curves of type Ibc SNe in individual filters, of which there is much more available data, and use them as a proxy for the width of the bolometric light curves.

One way to measure the widths of any light curve is to determine the $\emptyset m_{15}$ parameter (i.e. the amount the LC fades from peak to fifteen days post-peak). A wider light curve will have a smaller $\emptyset m_{15}$ parameter as it will have faded less than that for a SN with a larger $\emptyset m_{15}$ value. Thus, if our naive prediction is true, than on average, type Ib SNe should have wider light curves and smaller $\emptyset m_{15}$ values than type Ic SNe.

It has been noted in the literature the varying risetimes of Ibc SNe LCs (e.g. Richardson et al. 2006; Hunter et al. 2009) between roughly $-10$ to $-15$ days (w.r.t $B_{\rm max}$ or $V_{\rm max}$). However in several cases, especially for some of the "historic" SNe in the sample (e.g. SN 1954A, Wild 1960; SN 1964L, Miller & Branch 1990) the amount of data is much more sparse before the peak than after. Thus measuring $\varnothing m_{15}$ (which occurs after the peak) allows a more accurate measurement of the width of a given SN LC due to more datapoints, and more importantly, allows us to use the same LC width measurement for all of the SNe in the sample.

We have assembled from the literature all of the photometric data available for a relatively well-observed sample of type Ibc SNe ($N = 32$). Note that we are using the classification for each SN as determined by the various authors of the published literature; we have not specified here a specific cutoff velocity between Ic and Ic-BL (see Section 5.2.2 for further discussion). For each of these events we fit simple polynomials to the data and measured in the $V$- and $R$-band filters: (1) the peak time and peak magnitude, (2) the magnitude fifteen days after the peak and thus (3) $\varnothing m_{15}$. The use of polynomials to fit the SN light curves was chosen because differentiation of the polynomials is trivial, as is finding the roots of the differentiated equation, thus providing an easy yet reliable way to find the peak time and peak magnitudes of the SN light curves. The $\varnothing m_{15}$ parameters of the Ibc SNe are found in Table 5.4 where the errors quoted are statistical and have been determined by using polynomials of different orders and taking the scatter that arises as a 1-$\sigma$ estimate of the uncertainty.

Comparing our values of $\varnothing m_{15}$ with those in the literature, we find that our method has produced results in most cases that are similar to those determined by other authors. For example, for SN 2007ru, we find $\varnothing m_{15,V} = 0.93 \pm 0.03$ mag, and $\varnothing m_{15,R} = 0.71 \pm 0.03$ mag, while Sahu et al. (2009) find $\varnothing m_{15,V} = 0.92$ mag, and $\varnothing m_{15,R} = 0.69$ mag. Additionally, for SN 2004aw, we find $\varnothing m_{15,V} = 0.55 \pm 0.02$ mag, and $\varnothing m_{15,R} = 0.45 \pm 0.02$ mag, and Taubenberger et al. (2006) find $\varnothing m_{15,V} = 0.62 \pm 0.03$ mag, and $\varnothing m_{15,R} = 0.41 \pm 0.03$ mag.

For SN 2009jf, we find $\varnothing m_{15,V} = 0.56 \pm 0.05$ mag, and $\varnothing m_{15,R} = 0.43 \pm 0.04$ mag, while Sahu et al. (2011) find $\varnothing m_{15,V} = 0.50$ mag, and $\varnothing m_{15,R} = 0.31$ mag. Here we have found a larger value for $\varnothing m_{15}$ in the $R$ band compared with that found by Sahu et al. (2011). In our analysis, we find that the peak in the $R$ band occurred at JD $2455123.61 \pm 0.51$ while Sahu et al. (2011) determined the peak time in the $R$ band to be at JD 2455121.44 $\pm$ 1.03 (i.e. $\approx 2.2$ days earlier). If we take their time of maximum we find $\varnothing m_{15} = 0.32$ mag, fully consistent with their result. We note that Sahu et al. (2011) determined the peak times and peak magnitudes from fitting cubic splines to the LC in each filter, while we have used polynomials of different orders to determine the peak times and peak magnitudes. As we have used polynomials for every Ibc SNe in our sample, to maintain consistency in our method we adopt our value of $\varnothing m_{15,R}$ in our analysis.

We have also determined the $\varnothing m_{15}$ parameter for 20 of the GRB-SNe in filters $V$ and $R$, using SN 1998bw and the stretch factors determined in Section 3.9.3 in the previous chapter. First, we take the light curve of SN 1998bw and then stretch it for each GRB-SN by the value listed in Table 4.8, then we followed the same procedure as before (i.e. fit simple polynomials to the stretched LC and determine the peak time, peak magnitude as well as the magnitude at fifteen days after the peak, thus calculating $\varnothing m_{15}$). This method was followed for all of the GRB-SNe apart from SN 2003dh, SN 2003lw & SN 2006aj, as the LCs in these events are very well sampled (i.e. there are enough data points to constrain the shape of the LC accurately and thus use polynomials to determine the peak times, magnitudes, etc.).

We note that we have used the stretched LCs of SN 1998bw to determine $\varnothing m_{15}$ for each of the GRB-SNe as opposed to using the actual data-sets for each of these events because the data-sets in most of the GRB-SN events are not very well sampled. By using the stretched LCs of SN 1998bw, which are very well sampled, it allows us to use polynomials much more accurately when determining the peak times and peak magnitudes.

Ideally we want to compare the $\varnothing m_{15}$ values of each GRB-SNe in $V$ and $R$, however for many GRB-SNe we only have observations in one filter (e.g. GRB 091127, Cobb et al. 2010). As mentioned at the end of Section 4.5.3 in Chapter 4, it appears that the stretch factor is approximately the same (within a few percent) in the redder filters ($R$, $i$ and $V$) for many GRB-SN events (e.g. SN 2006aj and GRB 090618). Thus for the GRB-SN events where we only have one measurement of the stretch factor, we have used this value for the other filters (e.g. for GRB 011121, we found $s = 0.84 \pm 0.01$ in $R$, and have used this value of the stretch factor when calculating $\varnothing m_{15}$ in $V$). All of the values of $\varnothing m_{15}$ found for the GRB-SNe are displayed in Table 5.5.

The limitations of using SN 1998bw as a template for the GRB-SNe needs to be considered. In Chapter 4, it is seen in Figure 4.1 that SN 2010bh fades differently than the stretched LC of SN 1998bw. For example, in the $R$ band, between 15 and 30 days the LC of SN 2010bh fades slower than the stretched LC of SN 1998bw. We found in Chapter 4 for SN 2010bh (in the $R$ band), $\varnothing m_{15} = 0.90 \pm 0.05$ mag. When we instead use the $R$-band stretch factor of SN 2010bh ($s = 0.62 \pm 0.02$) and SN 1998bw as a template, we calculate $\varnothing m_{15} = 1.03 \pm 0.06$ mag. So, when using the stretch factors listed in Table 4.8 in the preceding chapter, we are assuming that the SN fade in the same manner as SN 1998bw, which might not always be the case. Thus, the intrinsic error in applying the stretch factors in Table 4.8 to SN 1998bw and then calculating $\varnothing m_{15}$ is larger than the statistical error and is of order $0.1 - 0.2$ mag.

We have calculated the average values of the $\varnothing m_{15}$ parameter for the various Ibc SNe in both filters. The results are presented in Table 5.6. We find that the average values of $\varnothing m_{15}$ of the GRB-SNe sample are similar to those of the general Ibc SN population in both filters. However, we find that the light curves (in both filters) of the Ib SNe fade *slower* than those of the Ic SNe sample (not including broad-lined Ic events). If we make the approximation that the average ejecta velocities are same in these events (we will discuss this in more depth later on), this result implies that the Ic SNe eject *less* mass than the Ib SNe, thus confirming expectations that the progenitors of Ic SNe retain less of their outer envelope prior to explosion than those of Ib SNe.

We then compared the $\varnothing m_{15}$ parameters of the various samples of SNe by performing a KS test on pairs of SN samples to determine if the different types of SNe are drawn from the same parent population. The cumulative fraction plots of the $\varnothing m_{15}$ comparison are displayed in Figure 5.2. In each filter we compared the $\varnothing m_{15}$ values between the GRB-SNe and (1) all of the type Ibc, (2) the Ib, (3) the Ic & Ic-BL, and (4) between Ib and Ic. In both filters there was a high probability that the $\varnothing m_{15}$ values of GRB-SNe and type Ibc are drawn from the same parent population: $P = 0.87$, $D = 0.17$; $P = 0.92$, $D = 0.15$ in filters $V$ and $R$ respectively. Similarly large probabil-

Table 5.4 $\varnothing m_{15}$ of local Ibc SNe

| SN | SN Type | $\varnothing m_{15}$ ($V$) | $\varnothing m_{15}$ ($R$) | Ref ($\varnothing m_{15}$) | Ref (Photometry) |
|---|---|---|---|---|---|
| 1954A | Ib | $1.10 \pm 0.20$ * | - | This work | (2), (3) |
| 1983N | Ib | $0.73 \pm 0.30$ | - | This work | (3) |
| 1984I | Ib | $0.63 \pm 0.40$ | - | This work | (3) |
| 1991D | Ib | $1.56 \pm 0.22$ | - | This work | (4) |
| 1999di | Ib | - | $0.39 \pm 0.25$ ** | This work | (5) |
| 1999dn | Ib | - | $0.49 \pm 0.04$ ** | This work | (5) |
| 1999dt | Ib | - | $0.51 \pm 0.27$ ** | This work | (5) |
| 1999ex | Ib | $0.82 \pm 0.05$ | $0.75 \pm 0.04$ | This work | (6) |
| 2001B | Ib | - | $0.61 \pm 0.05$ | This work | (7) |
| 2004dk | Ib | $0.56 \pm 0.12$ | $0.45 \pm 0.13$ | (1) | - |
| 2004gq | Ib | $0.78 \pm 0.22$ | $0.60 \pm 0.22$ | (1) | - |
| 2005az | Ib | - | $0.42 \pm 0.22$ | (1) | - |
| 2005bf | Ib pec | $0.60 \pm 0.35$ | $0.45 \pm 0.34$ | This work | (8) |
| 2005hg | Ib | $0.79 \pm 0.15$ | $0.62 \pm 0.15$ | (1) | - |
| 2006F | Ib | - | $0.74 \pm 0.24$ | (1) | - |
| 2006dn | Ib | - | $0.44 \pm 0.26$ | (1) | - |
| 2007C | Ib | $0.94 \pm 0.14$ | $0.72 \pm 0.14$ | (1) | - |
| 2007Y | Ib | $1.00 \pm 0.03$ | $0.77 \pm 0.03$ | This work | (9) |
| 2008D | Ib | $0.55 \pm 0.08$ | $0.55 \pm 0.13$ | (1) | - |
| 2009jf | Ib | $0.56 \pm 0.05$ | $0.43 \pm 0.04$ | This work | (22) |
| 1962L | Ic | $1.24 \pm 0.25$ | - | This work | (3), (10) |
| 1990B | Ic | $1.09 \pm 0.15$ | - | This work | (11) |
| 1994I | Ic | $1.66 \pm 0.03$ | $1.45 \pm 0.03$ | This work | (12) |
| 1997ef | Ic BL | $0.44 \pm 0.05$ | - | This work | (13), (14), (15), (16) |
| 2002ap | Ic BL | $0.90 \pm 0.04$ | $0.69 \pm 0.06$ | (1) | - |
| 2003jd | Ic BL | $0.84 \pm 0.07$ | $0.69 \pm 0.06$ | This work | (17) |
| 2004aw | Ic | $0.55 \pm 0.02$ | $0.45 \pm 0.02$ | This work | (18) |
| 2004dn | Ic | $0.76 \pm 0.15$ | $0.61 \pm 0.14$ | (1) | - |
| 2004fe | Ic | $0.87 \pm 0.16$ | $0.80 \pm 0.15$ | (1) | - |
| 2004ff | Ic | - | $0.85 \pm 0.25$ | (1) | - |
| 2004ge | Ic | - | $0.55 \pm 0.27$ | (1) | - |
| 2005eo | Ic | - | $0.71 \pm 0.25$ | (1) | - |
| 2005kz | Ic BL | - | $0.41 \pm 0.29$ | (1) | - |
| 2005mf | Ic | $0.78 \pm 0.19$ | $0.66 \pm 0.19$ | (1) | - |
| 2006ab | Ic | - | $0.56 \pm 0.21$ | (1) | - |
| 2007D | Ic BL | - | $0.67 \pm 0.29$ | (1) | - |
| 2007gr | Ic | $0.93 \pm 0.03$ | $0.71 \pm 0.03$ | This work | (19) |
| 2007ru | Ic BL | $0.90 \pm 0.03$ | $0.71 \pm 0.03$ | This work | (20) |
| 2009bb | Ic BL | $1.10 \pm 0.03$ | $0.90 \pm 0.03$ | This work | (21) |
| 2010ah | Ic BL | - | $0.49 \pm 0.04$ | This work | (23) |

* Measured on photographic plates.
** Measured in **clear** filter, approximated to R-filter.

(1) Drout et al. (2010), (2) Wild (1960), (3) Cadonau & Leibundgut (1990) (and references therein), (4) Benetti et al. (2002), (5) Matheson et al. (2010), (6) Stritzinger et al. (2002), (7) Xu et al. (2001), (8) Anupama et al. (2005), (9) Stritzinger et al. (2009), (10) Bertola (1964), (11) Clocchiatti et al. (2001), (12) Richmond (1996), (13) Wei et al. (1997), (14) Hu et al. (1997), (15) Garnavich et al. (1997a), (16) Garnavich et al. (1997b), (17) Valenti et al. (2008), (18) Taubenberger et al. (2006), (19) Hunter et al. (2009), (20) Sahu et al. (2009), (21) Pignata et al. (2011), (22) Sahu et al. (2009), (23) Corsi et al. (2011).

Table 5.5 $\varnothing m_{15}$ of the GRB-SNe

| GRB/XRF/SN | $\varnothing m_{15}$ ($V$) | $\varnothing m_{15}$ ($R$) | $\varnothing m_{15}$ ($I$) |
|---|---|---|---|
| 970228 | $0.42\pm0.27$ | $0.28\pm0.18$ | $0.21\pm0.13$ |
| 980425/1998bw | $0.75\pm0.01$ | $0.54\pm0.01$ | $0.45\pm0.01$ |
| 990712 | $0.86\pm0.45$ | $0.64\pm0.04$ | $0.54\pm0.03$ |
| 991208 | $0.63\pm0.10$ | $0.45\pm0.07$ | $0.36\pm0.06$ |
| 000911 | $0.42\pm0.09$ | $0.30\pm0.06$ | $0.23\pm0.05$ |
| 011121/2001ke | $0.94\pm0.01$ | $0.71\pm0.01$ | $0.61\pm0.01$ |
| 020405 | $1.31\pm0.05$ | $1.03\pm0.04$ | $0.92\pm0.03$ |
| 020903 | $0.93\pm0.28$ | $0.69\pm0.21$ | $0.59\pm0.18$ |
| 021211/2002lt | $0.77\pm0.20$ | $0.56\pm0.14$ | $0.46\pm0.12$ |
| 030329/2003dh | $0.98\pm0.02$ | $0.74\pm0.01$ | $0.64\pm0.01$ |
| 031203/2003lw | $0.67\pm0.01$ | $0.48\pm0.02$ | $0.39\pm0.01$ |
| 040924 | $0.74\pm0.18$ | $0.54\pm0.13$ | $0.45\pm0.11$ |
| 041006 | $0.41\pm0.01$ | $0.27\pm0.01$ | $0.20\pm0.01$ |
| 050525A/2005nc | $0.95\pm0.02$ | $0.72\pm0.02$ | $0.62\pm0.02$ |
| 060218/2006aj | $1.35\pm0.02$ | $0.93\pm0.02$ | $0.83\pm0.01$ |
| 060729 | $0.83\pm0.03$ | $0.62\pm0.02$ | $0.52\pm0.01$ |
| 080319B | $0.74\pm0.18$ | $0.54\pm0.13$ | $0.45\pm0.11$ |
| 090618 | $0.67\pm0.02$ | $0.48\pm0.02$ | $0.39\pm0.01$ |
| 091127 | $0.88\pm0.01$ | $0.66\pm0.01$ | $0.56\pm0.01$ |
| 100316D/2010bh | $0.64\pm0.04$ | $0.62\pm0.02$ | $0.60\pm0.01$ |

Photometry of these events are the same as those listed in Table 5.1.

Table 5.6 Average $\varnothing m_{15}$ values of the various Ibc SNe types

| Filter | Dataset | $N$ | $\langle\varnothing m_{15}\rangle$ | $\sigma$ | Filter | Dataset | $N$ | $\langle\varnothing m_{15}\rangle$ | $\sigma$ |
|---|---|---|---|---|---|---|---|---|---|
| $V$ | GRB | 20 | 0.81 | 0.25 | $R$ | GRB | 20 | 0.60 | 0.20 |
| $V$ | Ibc | 26 | 0.87 | 0.29 | $R$ | Ibc | 33 | 0.63 | 0.20 |
| $V$ | Ib | 13 | 0.82 | 0.28 | $R$ | Ib | 16 | 0.56 | 0.13 |
| $V$ | Ic (incl. Ic BL) | 13 | 0.93 | 0.30 | $R$ | Ic (incl. Ic BL) | 17 | 0.70 | 0.23 |
| $V$ | Ic (excl. Ic BL) | 8 | 0.99 | 0.34 | $R$ | Ic (excl. Ic BL) | 10 | 0.74 | 0.28 |
| $V$ | Ic BL | 5 | 0.84 | 0.24 | $R$ | Ic BL | 7 | 0.65 | 0.16 |

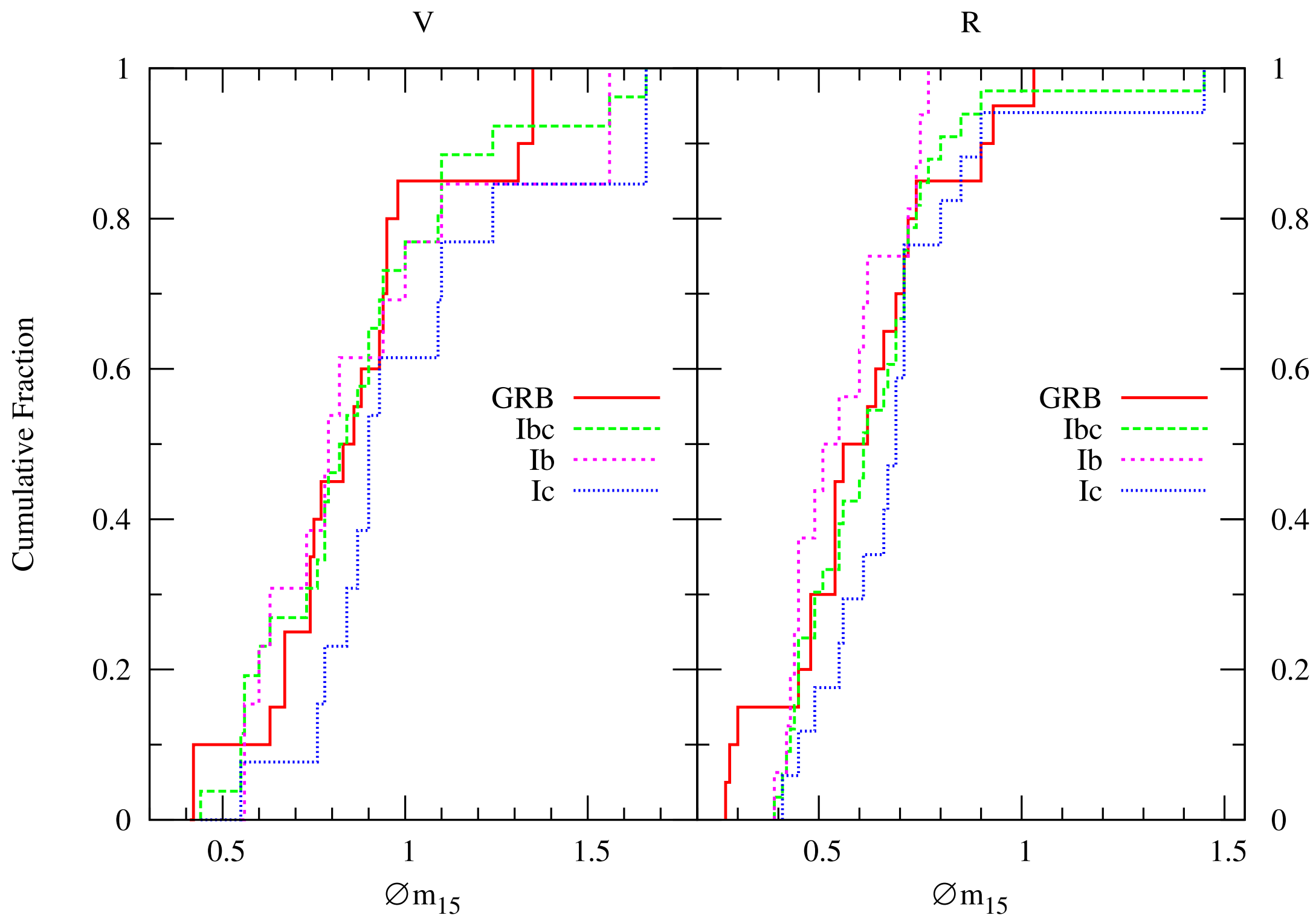

Figure 5.2 Cumulative fraction plots of the $\varnothing m_{15}$ comparison. It is seen that the change in magnitude from peak to fifteen days post-peak is similar between GRB-SNe and the complete non-GRB-SNe, with a probability that the two samples of SNe are drawn from the same parent population of $P = 0.87$ ($V$ band) and $P = 0.92$ ($R$ band). It is also seen that the probability that the type Ib SNe and Ic SNe are drawn from the same parent population is quite low ($P = 0.22$, $V$ band; $P = 0.11$, $R$ band). If we assume that the ejecta velocities are approximately the same in these two samples of SNe, the lower $\langle \varnothing m_{15} \rangle$ value of the type Ib SNe ($\langle \varnothing m_{15,R} \rangle = 0.56$ mag) compared to the type Ic SNe (not including the Ic-BL events: $\langle \varnothing m_{15,R} \rangle = 0.70$ mag) implies that that type Ib SNe eject more mass than type Ic SNe. This result echoes the expectations that the progenitors stars of type Ic SNe are more stripped before explosion than type Ib progenitors.

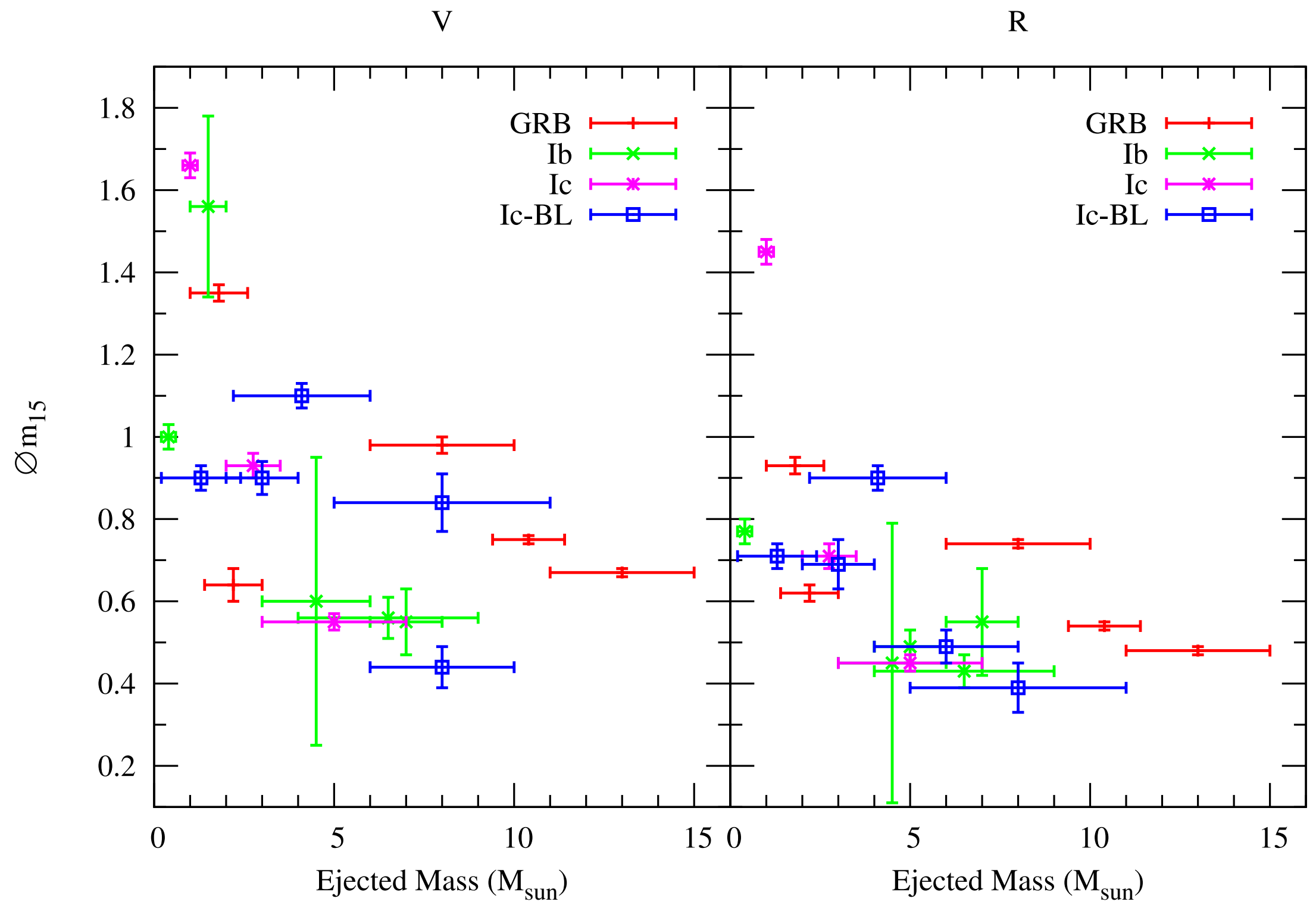

Figure 5.3 *Left*: $\varnothing m_{15,V}$ vs. ejected mass for Ibc SNe listed in Tables 5.4, 5.5 and 5.10 (see text). *Right*: $\varnothing m_{15,R}$ vs. ejected mass for the same sample of Ibc SNe. A general trend of decreasing $\varnothing m_{15}$ with increasing ejected mass is seen in both plots.

ities are seen when comparing the GRB-SNe with the Ib and Ic SNe.

More interestingly, when we compare the $\varnothing m_{15}$ parameters of the Ib and Ic SNe (including broad-lined Ic) we found smaller probabilities that the two types of SNe are drawn from the same parent population: $P = 0.22$, $D = 0.38$; $P = 0.11$, $D = 0.40$ in filters $V$ and $R$ respectively.

However, the ejecta velocities of broad-lined Ic SNe can be more than twice that of Ib SNe, so we compared the Ib SNe with all of Ic SNe that do not display broad spectral features. Again, we find smaller probabilities that the two samples are drawn from the same parent population: $P = 0.21$, $D = 0.40$ in the $R$ band. (Note that we limit the lower size of the samples to $N = 10$ when performing the KS tests, and as the number of "non broad-lined" Ic SNe is $N = 8$ in the $V$ band, we were only able to perform the KS test on the $R$-band sample).

Table 5.7 KS test results of $\Delta m_{15}$ comparison

| Filter | Dataset | $P$ | $D$ | Filter | Dataset | $P$ | $D$ |
|---|---|---|---|---|---|---|---|
| $V$ | GRB vs Ibc | 0.87 | 0.17 | $R$ | GRB vs Ibc | 0.92 | 0.15 |
| $V$ | GRB vs Ib | 0.90 | 0.19 | $R$ | GRB vs Ib | 0.82 | 0.20 |
| $V$ | GRB vs Ic | 0.42 | 0.29 | $R$ | GRB vs Ic | 0.43 | 0.27 |
| $V$ | Ib vs Ic (incl. Ic BL) | 0.22 | 0.38 | $R$ | Ib vs Ic (incl. Ic BL) | 0.11 | 0.40 |
| $V$ | Ib vs Ic (excl. Ic BL) | † | † | $R$ | Ib vs Ic (excl. Ic BL) | 0.21 | 0.40 |

† Too few points to perform KS test (minimum of 10 needed).

We note that comparison of the widths of the LCs via the $\Delta m_{15}$ parameter is perhaps less appropriate between events where ejecta velocities are different, as the overall LC width is (partially) determined by the ratio $\frac{\mathrm{M_{ej}}}{\mathrm{v_{ph}}}$ (see equation 4.5 from the previous chapter, and also below). So while the widths of the GRB-SNe and type Ibc SNe are quite similar, the reason for the similar values is down to two variables. It is only when we compare the $\Delta m_{15}$ values of the type Ib and Ic SNe, that have approximately the same average ejecta velocities (see next section), can we say something concrete about the physical parameters of these events. In the next section we try and break the degeneracy in the ratio $\frac{\mathrm{M_{ej}}}{\mathrm{v_{ph}}}$ by calculating the average ejecta velocities in each sample of SNe, and derive mass fractions of the different types of SNe.

Finally, we have plotted (in Figure 5.3) $\Delta m_{15}$ vs. ejected mass, in filters $V$ and $R$, for Ibc SNe listed in Tables 5.4, 5.5 and 5.10, where the ejected masses have been taken from the literature (see Table 5.10 for the respective references). It is seen that for each Ibc subtype (in both $V$ and $R$) that there is a general trend of *decreasing* $\Delta m_{15}$ with *increasing* ejected mass (i.e. a trend of increasing ejected mass with increasing LC width). There is some scatter in the plots however. For example, it is estimated that type Ic-BL SNe 1997ef and 2003jd both ejected $\sim 8M_{\odot}$ of material, however it is seen that $\Delta m_{15,V} = 0.44$ and 0.84 respectively. So, while the number of events considered here are still very low, and also considering the spread of the data in the plots, it appears that it is still possible to get a general understanding of the bolometric behaviour of a given Ibc SN when inspecting LCs taken in single filters.

Table 5.8 Peak Expansion Velocities of type Ibc & GRB-SNe

| SN | Type | $\langle v_{exp} \rangle$ (km s$^{-1}$) | Ref. | SN | Type | $\langle v_{exp} \rangle$ (km s$^{-1}$) | Ref. |
|---|---|---|---|---|---|---|---|
| 1998bw | GRB | 18,000 | (1) | 1988L | Ic | 6,400 | (6) |
| 2003dh | GRB | 20,000 | (2) | 1990B | Ic | 7,400 | (6) |
| 2003lw | GRB | 18,000 | (3) | 1990U | Ic | 6,700 | (6) |
| 2006aj | XRF | 20,000 | (4) | 1990aa | Ic | 8,300 | (6) |
| 2010bh | XRF | 25,000 | (5) | 1994I | Ic | 10,000 | (10) |
| 1984L | Ib | 7,200 | (6) | 1995F | Ic | 10,100 | (6) |
| 1991ar | Ib | 8,000 | (6) | 1997ei | Ic | 9,700 | (6) |
| 1997dc | Ib | 8,2000 | (6) | 2004aw | Ic | 11,800 | (11) |
| 1998I | Ib | 7,700 | (6) | 2007gr | Ic | 8,000 | (12) |
| 1998dt | Ib | 7,2000 | (6) | 1997dq | Ic-BL | 12,300 | (6) |
| 1999di | Ib | 6,500 | (6) | 1997ef | Ic-BL | 18,200 | (6) |
| 1999dn | Ib | 8,500 | (6) | 2002ap | Ic-BL | 18,000 | (13) |
| 2005bf | Ib | 8,000 | (7) | 2003jd | Ic-BL | 15,000 | (14) |
| 2007Y | Ib | 7,000 | (8) | 2007ru | Ic-BL | 20,000 | (15) |
| 2009jf | Ib | 12,000 | (9) | 2009bb | Ic-BL | 15,000 | (16) |

(1) Galama et al. (1998), (2) Matheson et al. (2003), (3) Mazzali et al. (2006a), (4) Pian et al. (2006), (5) Chornock et al. (2010), (6) Matheson et al. (2001), (7) Anupama et al. (2005), (8) Stritzinger et al. (2009), (9) Sahu et al. (2011), (10) Richmond (1996), (11) Taubenberger et al. (2006), (12) Hunter et al. (2009), (13) Mazzali et al. (2002), (14) Valenti et al. (2008), (15) Sahu et al. (2009), (16) Pignata et al. (2011).

### 5.2.1 Ejected mass estimates

Figure 5.4 shows the approximate ejecta velocities at peak (or near peak) for a sample of Ib, Ic, Ic-BL and GRB-SNe. The ejecta velocities have been taken from spectroscopic observations published in the literature. It is seen that the average velocities of the samples of SNe are: $\langle v_{GRB} \rangle = 20,200$ km s$^{-1}$ with $\sigma = 2,864$ km s$^{-1}$ and $N = 5$; $\langle v_{Ib} \rangle = 8,030$ km s$^{-1}$ with $\sigma = 651$ km s$^{-1}$ and $N = 10$; $\langle v_{Ic} \rangle = 8,711$ km s$^{-1}$ with $\sigma = 1,799$ km s$^{-1}$ and $N = 9$; $\langle v_{IcBL} \rangle = 16,417$ km s$^{-1}$ with $\sigma = 2,810$ km s$^{-1}$ and $N = 6$.

Immediately it is clear that the average ejecta velocities of type Ib and Ic SNe (excluding Ic-BL events) are quite similar, while the Ic-BL have faster moving ejecta than type Ibc and GRB-SNe having the fastest average ejecta velocities. If we now take these average velocities as representative of the ejecta velocities for each SNe type, we can estimate mass ratios between the different SNe types.

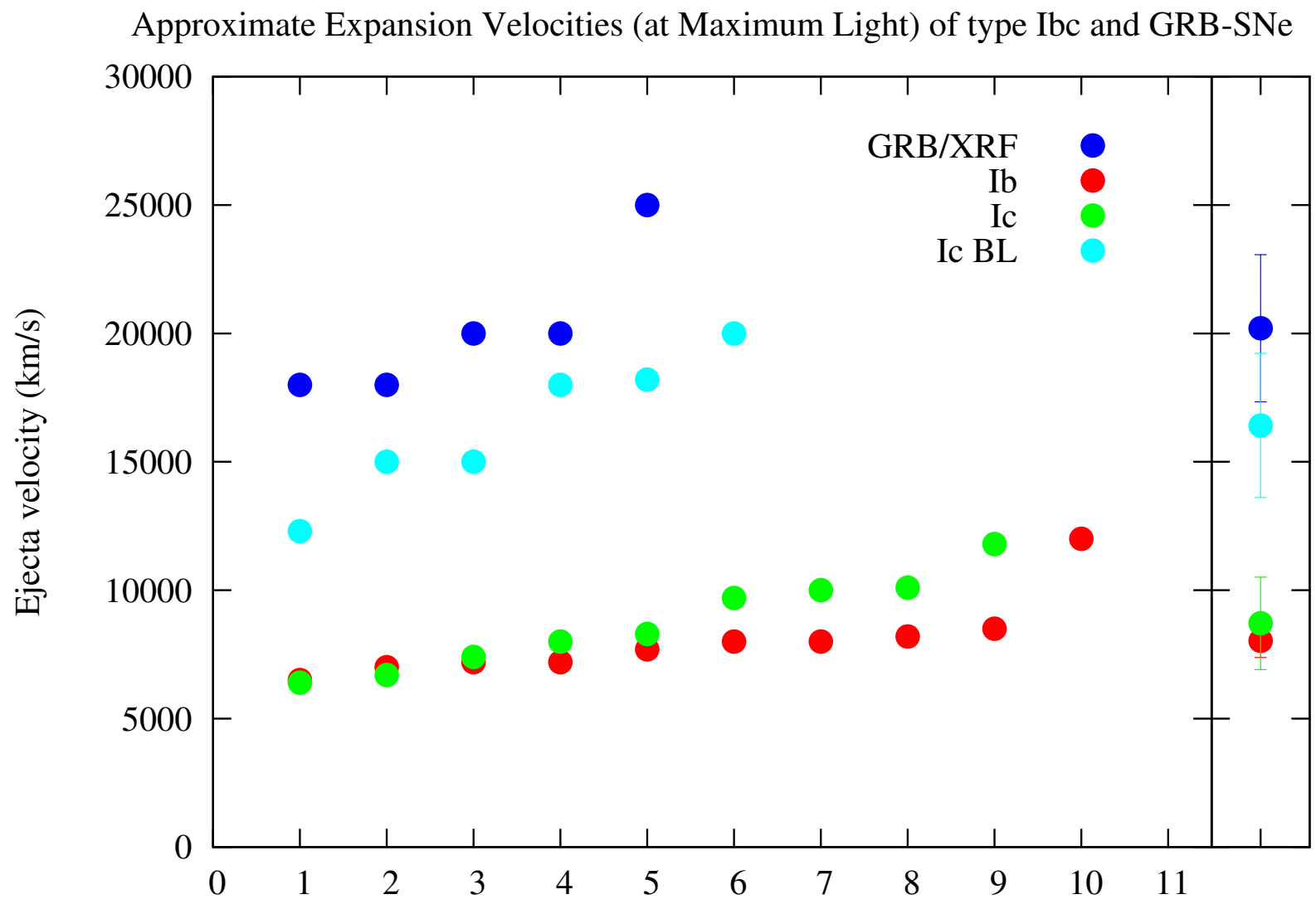

Figure 5.4 Approximate expansion velocities (at maximum photometric light) for a sample of GRB-SNe and local Ibc SNe, including several broad-lined Ic events. Shown in the right column are the average expansion velocities for each sub-sample along with the standard deviation as a 1-$\sigma$ error: $\langle v_{GRB} \rangle = 20,200$ km s$^{-1}$ with $\sigma = 2,864$ km s$^{-1}$ and $N = 5$; $\langle v_{Ib} \rangle = 8,030$ km s$^{-1}$ with $\sigma = 651$ km s$^{-1}$ and $N = 10$; $\langle v_{Ic} \rangle = 8,711$ km s$^{-1}$ with $\sigma = 1,799$ km s$^{-1}$ and $N = 9$; $\langle v_{IcBL} \rangle = 16,417$ km s$^{-1}$ with $\sigma = 2,810$ km s$^{-1}$ and $N = 6$. References for the spectroscopic measurements of the various SN in the samples are the same as those in Table 5.8.

Recall equation 4.5 from the previous chapter:

$$\tau_m \approx \left(\frac{\kappa}{\beta c}\right)^{1/2} \left(\frac{\mathrm{M_{ej}^3}}{\mathrm{E_k}}\right)^{1/4} \tag{5.1}$$

Substituting $\mathrm{E_k} = 1/2\mathrm{M_{ej}}\mathrm{v_{ph}^2}$, where $\mathrm{v_{ph}}$ is the peak photospheric velocity, we find that the width of the bolometric light curve is related to:

$$\tau_\mathrm{m} \approx \left(\frac{\kappa}{\beta c}\right)^{1/2} \left(\frac{\mathrm{M_{ej}}}{\mathrm{v_{ph}}}\right)^{1/2} \tag{5.2}$$

If we assume that the opacity, $\kappa$, is approximately the same in all of our SNe events, and using the average peak photospheric velocities determined above (Figure 5.4), we can derive mass estimates between the different SN types, e.g.:

$$\frac{M_{ej,Ib}}{M_{ej,Ic}} = \left(\frac{\tau_{m,Ib}}{\tau_{m,Ic}}\right)^2 \times \frac{v_{ph,Ib}}{v_{ph,Ic}} \tag{5.3}$$

Table 5.9 Mass Ratios

| | *V* band‡ | *R* band‡ |
|---|---|---|
| $\frac{M_{ej,GRB}}{M_{ej,Ib}}$ | 2.5 | 2.0 |
| $\frac{M_{ej,GRB}}{M_{ej,Ic}}$ | 3.4 | 3.4 |
| $\frac{M_{ej,GRB}}{M_{ej,Ic-BL}}$ | 1.3 | 1.5 |
| $\frac{M_{ej,Ib}}{M_{ej,Ic}}$ | 1.4 | 1.7 |
| $\frac{M_{ej,Ib}}{M_{ej,Ic-BL}}$ | 0.5 | 0.7 |
| $\frac{M_{ej,Ic}}{M_{ej,Ic-BL}}$ | 0.4 | 0.4 |

‡ As derived from the *V*-band and *R*-band data.

We have estimated the LC width ratios (e.g. $\frac{\tau_{m,Ib}}{\tau_{m,Ic}}$) from the $\varnothing m_{15}$ values in Table 5.7. For example, when comparing the Ib SNe with the Ic SNe (excl. the Ic-BL events) in the $V$ band, it is seen that $\langle \varnothing m_{15,Ib} \rangle = 0.82$ and $\langle \varnothing m_{15,Ic} \rangle = 0.99$. Thus the Ib SNe fade slower and have wider LCs by a factor of $\frac{\tau_{m,Ib}}{\tau_{m,Ic}} \approx \frac{0.92}{0.82} \approx 1.21$. By using these ratios and the average ejecta velocities, we have determined mass ratios between the different SN types, and our results are displayed in Table 5.9.

Our results both confirm existing expectations and reveal intriguing new ideas. As expected, our results show that type Ib SNe eject more mass than type Ic SNe (excluding Ic-BL events), where we find: $M_{ej,Ib} = 1.4 - 1.7\ M_{ej,Ic}$. This reinforces the idea that the progenitor stars of type Ib SNe are less stripped then their Ic counterparts.

More intriguingly, and perhaps unexpectedly, we find that of all of the SN events, GRB-SNe actually eject the *most* mass (on average), more than Ib SNe: $M_{ej,GRB} =$

$2.0 - 2.5$ $M_{ej,Ib}$, more than type Ic SNe (not including Ic-BL events): $M_{ej,GRB} \approx 3.4$ $M_{ej,Ic}$), and more than type Ic-BL SNe: $M_{ej,GRB} = 1.3 - 1.5$ $M_{ej,Ic-BL}$.

It is seen that the Ic-BL events also eject more mass than types Ib ($M_{ej,Ic-BL} = 1.4 - 2.0$ $M_{ej,Ib}$) and Ic ($M_{ej,Ic-BL} \approx 2.5$ $M_{ej,Ic}$). Thus it seems that the progenitors of type Ic-BL SNe and GRB-SNe eject more mass during the explosion compared to the progenitors of "normal" Ibc SNe. While a linear relationship between ejecta mass and the original main sequence mass does not exist, it is generally accepted that SNe that ejected more mass implies they arise from more massive main sequence stars. Thus, in conclusion, we find that GRB-SNe and Ic-BL SNe likely arise from more massive progenitor stars than "normal" type Ibc SNe as they have retained more mass before exploding in comparison with "normal" Ibc SNe. This result will be discussed in more depth in the following section.

## 5.2.2 Discussion

### Different progenitors?

The progenitors of type Ibc supernova are thought to arise from either massive Wolf-Rayet (WR) stars (Gaskell et al. 1986; Woosley, Langer & Weaver 1993) or from close interacting binaries of less massive stars (Shigeyama et al. 1990; Podsiadlowski et al. 1992, Nomoto et al. 1994; Iwamoto 1994). WR stars are massive, evolved stars that have shed most or all of their hydrogen and helium envelopes though strong, line-driven stellar winds (e.g. Meynet et al. 1994; Heger et al. 2003), and are a popular, possible progenitor of GRBs. Alternatively, it is believed that Ibc SNe can also arise from the interaction of stars with lower initial masses in close binaries that have had their envelopes stripped through interactions, possibly through Roche lobe overflow or common envelope evolution.

Massive, single star progenitor models of stripped-envelope, core-collapse SNe

Table 5.10 Ejected Masses and ZAMS Masses of type Ibc & GRB-SNe

| SN | Type | $M_{ej}$ ($M_\odot$) | error ($M_{ej}$) | $M_{ZAMS}$ ($M_\odot$) | error ($M_{ZAMS}$) | Ref. |
|---|---|---|---|---|---|---|
| 1998bw | GRB-SN | 10.4 | 1.0 | 40 | 5 | (1), (2), (3), (4) |
| 2003dh | GRB-SN | 8.0 | 2.0 | 35 | 5 | (5) |
| 2003lw | GRB-SN | 13.0 | 2.0 | 45 | 5 | (6) |
| 2006aj | XRF-SN | 1.8 | 0.8 | 20 | 2 | (7), (8) |
| 2010bh | XRF-SN | 2.2 | 0.8 | - | - | (9) |
| 1990I | Ib | 3.7 | 1.0 | $15-20$ | - | (10) |
| 1991D | Ib | $1-2$ | - | - | - | (11) |
| 1999dn | Ib | $4-6$ | - | $23-25$ | - | (12) |
| 2005bf | Ib | $3-7$ | 1 | $23-30$ | - | (13), (14), (15) |
| 2007Y | Ib | 0.4 | - | $10-13$ | - | (16) |
| 2008D | Ib | 7 | 1 | 25 | 5 | (17), (18) |
| 2009jf | Ib | $4-9$ | - | $20-25$ | - | (19) |
| 1987M | Ic | $\sim 1.0$ | - | $12-15$ | - | (20), (21) |
| 1994I | Ic | 1.0 | 0.2 | 17 | 1 | (22), (23), (24) |
| 2004aw | Ic | 5.0 | 2.0 | 28 | 3 | (25) |
| 2007gr | Ic | $2.0-3.5$ | - | - | - | (26) |
| 1997ef | Ic-BL | 8.0 | 2.0 | 35 | 2 | (27), (28) |
| 2002ap | Ic-BL | 3.0 | 1.0 | 24 | 2 | (29), (30), (31), (32) |
| 2003jd | Ic-BL | 8.0 | 3.0 | 27 | 3 | (33) |
| 2007ru | Ic-BL | 1.3 | 1.1 | - | - | (34) |
| 2009bb | Ic-BL | 4.1 | 1.9 | - | - | (35) |
| 2010ah | Ic-BL | 6.0 | 2.0 | $\sim 35$ | - | (36) |

(1) Galama et al. (1998), (2) Iwamoto et al. (1998), (3) Nakamura et al. (2001), (4) Maeda et al. (2006), (5) Deng et al. (2005), (6) Mazzali et al. (2006a), (7) Mazzali et al. (2006b), (8) Modjaz et al. (2006), (9) Cano et al. (2011), (10) Elmhamdi et al. (2004), (11) Benetti et al. (2002), (12) Benetti et al. (2011), (13) Tominaga et al. (2005), (14) Anupama et al. (2005) (15) Folatelli et al. (2006), (16) Stritzinger et al. 2009, (17) Mazzali et al. (2008), (18) Soderberg et al. (2008), (19) Sahu et al. (2011), (20) Nomoto et al. (1990), (21) Filippenko et al. (1990), (22) Richmond (1996), (23) Nomoto et al. (1994), (24) Sauer et al. (2006), (25) Taubenberger et al. (2006), (26) Hunter et al. (2009), (27) Mazalli et al. (2000), (28) Iwamoto et al. (2000), (29) Mazalli et al. (2007), (30) Foley et al. (2003), (31) Tomita et al. (2006), (32) Yoshii et al. (2003), (33) Valenti et al. (2008), (34) Sahu et al. (2009), (35) Pignata et al. (2011), (36) Corsi et al. (2011).

(i.e. "collapsars") can be roughly divided into two groups: (1) those with no rotation and strong stellar winds that eject the entire hydrogen envelope (e.g. Fryer et al. 1999; Heger et al. 2003), and (2) those with rotation that extensively mix the stellar interior (e.g. Yoon & Langer 2005; Woosley & Heger 2006). Massive stars of the latter type undergo so much mixing that most of the hydrogen is burned into helium, implying that they do not lose their hydrogen envelopes to stellar winds, but rather retain this mass before exploding. The former models predict that "normal" Ibc SNe do not arise from single stars unless the metallicity is above solar, and the authors conclude that Ibc SNe are produced in binary systems. These non-rotating models also predict single stars are only a small subset of GRB progenitors, with the bulk of the population formed in binary systems.

In massive, single star models that include rotation, it is seen that it is possible to impart enough rotation to the core to power a GRB after the star explodes. In their model, Woosley & Heger (2006) find that the maximum angular momentum and slowest rotational periods are always at the surface of the star regardless of the initial parameters (i.e. progenitor mass, initial rotational velocities, etc.). The authors claim that this result makes it difficult to envision how any binary interaction could impart faster rotation to the inner core than that is calculated in their models.

Furthermore, a recent paper by Yoon et al. (2010) investigates the properties of the progenitors of Ibc SNe in binary systems and include rotational effects into their model. They conclude that the angular momentum retained is large enough to produce millisecond pulsars, but too small to produce magnetars and GRBs.

From an observational stand point, in a recent review paper discussing the progenitors of core-collapse supernova, Smartt (2009) suggests that most type Ibc SNe arise from moderate mass interacting binaries, while Ic-BL SNe are likely produced by massive WR stars. Smartt (2009) investigates the probability that WR stars are the sole progenitors of all type Ibc SNe, concluding that the hypothesis is false at the 90% confidence level, implying that at least some of the progenitors of type Ibc SNe arise from binary stars. Moreover, Smartt (2009) states that the relative frequencies of discovered Ibc SNe are strongly suggestive that at least a fraction come from interacting binaries. This suggestion was derived from earlier work done in the 1990s, where interacting binaries were suggested as a common channel for Ibc SNe production (e.g. Nomoto et al. 1995, Podsiadlowski et al. 1992).

Additionally, it has been observed that at least 60% (and can be as much as 100%) of all massive stars are in binaries that will undergo mass transfer (Kobulnicky & Fryer 2007). While the actual percentages are sensitive to many factors, Kobulnicky & Fryer (2007) conclude from their Monte Carlo simulations that massive stars preferentially have massive companions.

The rates at which Ibc SNe occur provide additional clues to their likely progenitors, where it is observed that Ic-BL SNe comprise only 5% – 10% of all Ic SNe, with GRB-SNe only 20% of the Ic-BL fraction, and thus only 1% of all Ic SNe (e.g. Fryer et al. 2007). So while it is highly likely that some, if not most Ibc SNe arise from binary systems, it is well known from the rates of GRB-SNe that very special conditions are

Table 5.11 Average Ejected Masses, Average ZAMS Masses & Mass Fractions of type Ibc & GRB-SNe

| SN type | N | $\langle M_{ej} \rangle$ ($M_{\odot}$) | $\sigma$ |
|---|---|---|---|
| GRB/XRF-SNe | 5 | 7.1 | 5.0 |
| GRB-SNe | 3 | 10.5 | 2.5 |
| XRF-SNe | 2 | 2.0 | 0.3 |
| Ib | 7 | 4.3 | 2.6 |
| Ic | 4 | 2.4 | 1.9 |
| Ic-BL | 6 | 5.1 | 2.7 |
| SN type | N | $\langle M_{ZAMS} \rangle$ ($M_{\odot}$) | $\sigma$ |
| GRB/XRF-SNe | 4 | 35 | 11 |
| Ib | 6 | 19.7 | 6.0 |
| Ic | 3 | 19.3 | 7.8 |
| Ic-BL | 4 | 30.3 | 5.6 |

$\frac{M_{ej,GRB}}{M_{ej,Ib}} = 1.7\ (2.0-2.5)^{\dagger}$ $\quad$ $\frac{M_{ej,Ib}}{M_{ej,Ic}} = 1.8\ (1.4-1.7)^{\dagger}$

$\frac{M_{ej,GRB}}{M_{ej,Ic}} = 3.0\ (3.4)^{\dagger}$ $\quad$ $\frac{M_{ej,Ib}}{M_{ej,Ic-BL}} = 0.8\ (0.5-0.7)^{\dagger}$

$\frac{M_{ej,GRB}}{M_{ej,Ic-BL}} = 1.4\ (1.3-1.5)^{\dagger}$ $\quad$ $\frac{M_{ej,Ic}}{M_{ej,Ic-BL}} = 0.5\ (0.4)^{\dagger}$

† Mass fractions in parentheses are from Table 5.9 (i.e. derived from the light curve widths in the $V-$ and $R-$ bands).

needed for GRB production.

To compare the results of our derived ejected mass ratios between the different types of Ibc SNe that were determined from the widths of the $V$- and $R$-band LCs, we have compiled from the literature results from modelling photometric and spectroscopic data for many Ibc SNe and displayed them in Table 5.10 along with their respective references. We have compared the ejected masses and zero-age main-sequence (ZAMS) masses for different SN types and calculated (1) average ejected masses, and (2) average ZAMS masses of the SN-progenitors, and the results are shown in Table 5.11. Using these average ejected masses and average ejecta velocities in Table 5.8, we have also re-calculated the mass fractions between the various SN types, and compared them with those obtained via comparing the widths of the $V$- and $R$-band LCs. The mass fractions are also displayed in Table 5.11.

The first thing that becomes apparent is that the mass ratios obtained via different methods are remarkably similar. Both analyses show that GRB/XRF-SNe, on average, ejected more mass than the other type Ibc SNe. Interestingly, if the GRB/XRF-SNe sample is divided into GRB-SNe and XRF-SNe, it is seen that the former (of which there are three: SNe 1998bw, 2003dh and 2003lw) eject, on average, **five times** more mass ($\langle \mathrm{M}_{\mathrm{ej,GRB}} \rangle = 10.5\ M_{\odot}$) than the two SNe associated with XRFs 060218 and 100316D ($\langle \mathrm{M}_{\mathrm{ej,XRF}} \rangle = 2.0\ M_{\odot}$). It is also seen that the Ic-BL SNe eject more mass than the "normal" Ibc SNe, and as expected, Ib SNe eject more mass than Ic SNe.

Additionally, comparison of the average ZAMS masses of the various SN types reveal what we concluded at the end of the previous section, namely that GRB-SNe and Ic-BL SNe arise from more massive progenitors than "normal" type Ibc SNe. The average main sequence mass of GRB-SNe progenitors is $\langle \mathrm{M}_{\mathrm{ZAMS,GRB}} \rangle = 35\ M_{\odot}$. If we further divide them into GRB-SNe and XRF-SNe, it is seen that the former arise from more massive progenitors: $\langle \mathrm{M}_{\mathrm{ZAMS,GRB}} \rangle = 40\ M_{\odot}$, $\langle \mathrm{M}_{\mathrm{ZAMS,XRF}} \rangle = 20\ \mathrm{M}_{\odot}$. However the only XRF-SNe to have been modeled and to have had its ZAMS mass estimated is XRF060218 / SN 2006aj, so this comparison is obviously very limited at this time.

The average ZAMS mass of the Ic-BL SNe progenitors is found to be $\langle \mathrm{M}_{\mathrm{ZAMS,Ic-BL}} \rangle = 30\ M_{\odot}$. It has been observed that the ZAMS mass range of WR stars in the Milky Way is $35 - 40\ M_{\odot}$ (Humphreys et al. 1985; Massey 1995, 2001; Massey et al. 2003; Crowther et al. 2006, 2007). The progenitor masses calculated here for GRB-SNe and Ic-BL SNe are approximately those needed to form a WR star.

Finally, the average ZAMS masses of the type Ib SNe ($\langle \mathrm{M}_{\mathrm{ZAMS,Ib}} \rangle = 20\ M_{\odot}$) and type Ic SNe ($\langle \mathrm{M}_{\mathrm{ZAMS,Ic}} \rangle = 19\ M_{\odot}$) are considerably less than those calculated for the GRB-SNe and Ic-BL SNe, but similar to the ZAMS mass found for XRF 060218 / SN 2006aj.

Here, we have found via different methods that GRB-SNe and Ic-BL SNe eject more mass than type Ibc SNe. This result is perhaps unexpected if it is assumed that all Ibc SNe arise from non-rotating, massive stars, where mass is lost via strong stellar winds. Logic then dictates that type Ib SNe should eject the most mass as they have undergone the least amount of mass-loss before exploding.

However, our result that Ic-BL SNe and GRB-SNe arise from more massive main sequence progenitor stars than type Ibc SNe, as well as ejecting more mass during the supernova explosion, supports the idea that there are likely at least two different progenitor channels for stripped-envelope, core-collapse supernova. While observations support the notion that massive stars preferentially occur in binary systems, it is not unreasonable to suggest that the bulk of Ibc SNe arise from massive stars (though not as massive as the progenitors of Ic-BL and GRB-SNe) in binary systems. Moreover, Yoon et al. (2010) has shown that the angular momentum retained by the exploding star in the binary system is not enough to produce a magnetar nor a GRB, suggesting that GRB progenitors arise via a different process.

One clue is how much more mass GRB progenitors retain prior to explosion. The models of Yoon & Langer (2005) and Woosley & Heger (2006) have shown that massive stars that undergo rapid rotation can retain enough angular momentum to then power a GRB after exploding. Furthermore, these models show that the extensive mixing that these stars undergo as a consequence of the rapid rotation means that the hydrogen in the outer layers is almost entirely burned into helium. The key point is that instead of ejecting the hydrogen envelope via stellar winds (though some mass-loss still occurs, and is strongly dependent on the metal content of the progenitor star), the bulk of the mass is retained, with more of it being ejected during the supernova in comparison with the progenitors of Ibc SNe that have had their hydrogen envelopes stripped before exploding.

It is worth remembering here that mass-loss occurs in *both* models, and is strongly dependent on the metal content of the progenitor star. Massive stars with higher metallicities will lose more mass via line-driven winds than massive stars with lower metallicities (e.g. Puls et al. 1996; Kudritzki & Puls 2000; Mokiem et al. 2007). In the models of Yoon & Langer (2005) and Woosley & Heger (2006), the rapidly-rotating massive stars do still undergo some mass-loss, but it is considerably less than the mass-loss considered in the non-rotating models. So while the non-rotating stars might lose more mass in comparison to stars in the rapidly-rotating models, metallicity plays a key role in the amount of mass that is retained or ejected. Thus a rapidly-rotating star with high metal content will still lose mass from its outer layers even though the star is thoroughly mixed.

From our results it can be argued that Ic-BL SNe might also arise from rapidly-rotating, massive stars due to the higher ejected mass ratios in these events compared with "normal" Ibc SNe. Moreover, if GRB/XRF-SNe do eject more mass than Ic-BL SNe this could be due to the progenitors of GRB/XRF-SNe having undergone less mass-loss prior to explosion (which is crucial, as it is seen that mass-loss via stellar winds removes vital angular momentum that is needed for producing an accretion disk around the newly-formed BH and powering a GRB), due to the progenitors of GRB/XRF-SNe having lower metal abundances than the progenitors of type Ic-BL SNe. The suggestion that the progenitors of GRBs are low in metallicity is amply supported in the literature, both from theoretical expectations (e.g. Heger et al. 2003; Woosley & Heger 2006) and through observations (e.g. Prochaska et al. 2004; Modjaz et al. 2008; 2011). Work presented by Modjaz et al. (2008) has shown via spectroscopic observations at the sites of 12 Ic-BL SNe (after the SN had faded away) that the progenitors of local Ic-BL SNe have higher metallicity abundances than those of GRB progenitors. This result is robust and is independent of the metallicity diagnostic used.

Therefore, we suggest a general picture where most Ibc SNe arise from the interaction of binary stars, where the primary that explodes has a mass lower than that usually associated with WR stars. However, the progenitors of GRB-SNe, and possibly Ic-BL SNe, arise from the collapse of a rapidly-rotating, massive WR stars. The difference between GRB-SNe and Ic-BL SNe could be due to mass loss (where mass loss also removes angular momentum), which is governed by the metallicity of the progenitor. As GRB-SNe arise from more metal-poor progenitors than Ic-BL SNe, if we assume that rapid rotation is present in these events, the key difference is that the higher metallicity of Ic-BL SNe leads to more mass loss and crucially, more loss of angular momentum than those of GRB-SNe. It is because GRB-SNe lose less mass due to their lower metal content, and hence retain more angular momentum before exploding, that they are able to power a GRB at the time of death, while Ic-BL SNe do not.

## Caveats of our method

While our results are potentially quite interesting, we must also retain a sense of caution due to the many assumptions we have made in this analysis. We have already discussed in Section 5.2 the limitations of using the stretch factors in Table 4.8 of Chapter 4 and applying them to SN 1998bw to determine the $\varnothing m_{15}$ parameter of the GRB-SNe. We have also discussed in Section 4.4 of Chapter 4 the limitations of using the model developed by Arnett (1982). Briefly to recap, the model assumes, among other things, homologous expansion of the ejected material and *spherical symmetry*. For Ibc SNe the the assumption of spherical (or near-spherical) expansion is perhaps valid, however for GRB-SNe, it is expected that the ejecta is quite aspherical, which makes the use of the model of Arnett (1982) less valid.

For example, if the ejecta in GRB-SNe is collimated into a jet, then the amount of ejecta measured along this line of sight is not valid for all sight-lines, thus the Arnett (1982) model will over-predict the total ejecta mass. Similarly for the explosion energy, if the fastest-moving ejecta is only travelling along (or near to) the jet-axis, then the model over-predicts the explosion energy as it assumes that all of the ejecta moves at the same photospheric velocity. In highly-aspherical explosions it is quite obvious that this assumption is not correct. Further still, the assumption of a homologous density structure that all travels at the photospheric velocity is an over-simplification of a more complex situation where "shells" of material (e.g. carbon, oxygen, etc.) travel at different speeds and have different densities. Thus this assumption will also over-predict the explosion energy of the SN.

In addition to these, there are several additional caveats that need to also be considered when interpreting the mass ratios determined in the preceding section. First, Equation 4.5 applies to *bolometric* light curves, however here we have used the $V$- and $R$-band light curves as proxies for the bolometric light curve in each event. Though we find similar mass ratios using data taken in each filter, the validity of using data taken in a single filter and assuming that the behaviour is the same as the bolometric be-haviour is debatable. For example, we have determined that the bolometric light curve of SN 2010bh (Figure 4.5) peaks around $\mathrm{t-t_o} \approx 8.6$ days, while the peak times in the *V* and *R* filters were found to be, respectively, $\mathrm{t-t_o} \approx 8.9$ and $\mathrm{t-t_o} \approx 11.2$ days. So while the peak time in the *V* filter coincides with the peak bolometric light, the same is not so for the *R*-band LC.

Further still, we have assumed that the fraction of light emitted at optical and infrared wavelengths is the same in all Ibc SNe. If we assume that the emitted radiation during the first few weeks is roughly that of a blackbody (which is more valid early on, and certainly less valid at late times during the "nebular phase", where the observed flux is due mostly to emission lines), we are thus assuming that the photospheric temperature is approximately the same in each Ibc SN event. In terms of blackbody radiation (i.e. a Planck spectrum), if each SN has roughly the same photospheric temperature then they will emit radiation that peaks at approximately the same frequency. However, if the photospheric temperatures are drastically different between events, then the fraction of light emitted at optical and infrared wavelengths will also differ. This will in turn have an effect on the evolution of the LCs in individual filters. For example, if a given sub-class of SNe has, on average, cooler photospheric temperatures than another sub-class, the former will emit more light at longer wavelengths than the latter. This would then imply that the LCs of the cooler SNe sub-class will fade faster in the bluer filters relative to the other sub-class, thus the faster fading has nothing to do with different ejected masses.

Next, we have approximated from a range of ejecta velocities, an average ejecta velocity and applied this average velocity to all events in a particular SN sample. For the Ib events, the standard deviation of the sample was only 651 $\mathrm{kms^{-1}}$, however for the other samples the scatter was much more considerable. For Ic-BL events, the ejecta velocities ranged from $12,300$ km $\mathrm{s^{-1}}$ to $20,000$ km $\mathrm{s^{-1}}$, with a standard deviation of $2,810$ km $\mathrm{s^{-1}}$. So using an average value for the ejecta velocities over-simplifies the rich diversity in these events.

Similarly, there is a spread of ejected masses for a given Ibc subtype. For example, we have shown in Table 5.11 that on average, type Ib SNe eject $\sim 4M_{\odot}$ of material.

However it is seen in Table 5.10 that SN 2007Y ejected only $\sim 0.4 M_{\odot}$ of material, which is much lower than the average ejected mass for Ib SNe. Thus using an average value for the ejected masses over-simplifies the diversity in these events, though it is still worth noting that it was seen in Figure 5.3 that a trend of increasing ejecta mass with increasing LC width in filters $V$ and $R$ was observed, and SN 2007Y does not deviate from this general trend.

Additionally, in the literature, the peak ejecta velocities have been determined using mostly He and Fe lines (e.g. Matheson et al. 2001; Pian et al. 2006, etc.). It has not been possible to create a homogeneous sample of SNe where the ejecta velocity has been determined using the same spectroscopic line, thus adding another degree of uncertainty into the average peak velocity estimate. Moreover, we are using the SN classification determined by the various authors in the literature. It is seen in Figure 5.4 that the peak ejecta velocities of Ic (including Ic-BL) range, in an apparent continuum, from $\sim 6,000-20,000$ km s$^{-1}$, with no clear division between the two sub-classes. To date no clear dividing line has been established, either observationally or theoretically, between the ejecta velocities of Ic and Ic-BL events, and while it might be safe to assume that an event with an ejecta velocity of say $\sim 8,000$ km s$^{-1}$ is a Ic SNe and another event with an ejecta velocity of $\sim 20,000$ km s$^{-1}$ is clearly a Ic-BL, events with ejecta velocities $\sim 10,000-12,000$ km s$^{-1}$ are less easily classified, and some confusion may exist in the literature in the exact classification of a handful of events.

It also needs to be remembered that the modelling results, and measurements of peak photospheric velocities from the literature are still relatively few. Throughout this analysis, the computed statistics are derived from quite small sample sizes, and thus we are dealing with small number statistics. Clearly the addition of many more events is needed to either verify or refute the conclusions drawn in this work.

Next, we have assumed that the opacity in these events is roughly the same, which of course is an over-simplification. While it may be more appropriate to assume a constant opacity within a certain class of SNe (i.e. all Ib, or all Ic), the assumption of constant opacity across all classes is more questionable. Consider Ib SNe, where the ejecta has a much higher fraction of helium in it than those of Ic, Ic-BL and GRB-SNe, which are rich in carbon and oxygen but deficient in helium. However, it is expected that the overall opacity of He-rich Ib SNe and C/O-rich Ic SNe will be similar as the opacity is primarily dominated by the large number of Doppler-broadened iron lines, and it is expected that similar abundances of iron group elements exist in the ejecta of Ib and Ic SNe. The electron scattering opacity does also contribute to the bulk opacity of the outflow however, and because it is somewhat harder to ionize helium than carbon and oxygen, the electron scattering opacity in He-rich events will be a bit lower.

So while there are several caveats that need to be considered in this somewhat speculative analysis, it is none the less encouraging that we have derived similar mass ratios via different methods despite the assumptions made.

SECTION 5.3

## Summary

In this Chapter we have drawn several conclusions including:

1. It is not possible to distinguish between GRB-SNe and Ibc SNe using only their peak $V$-band magnitudes. If use the peak, $V$-band magnitudes as a proxy for the peak bolometric luminosities, this implies that similar amounts of nickel are created in all Ibc SNe.

2. GRB-SNe eject more mass than Ib, Ic and Ic-BL SNe. This has been determined via two methods: (1) by comparing the widths of the $V$- and $R$-band LCs and using them as proxies for the widths of the bolometric LCs, and (2) using the values for the ejected masses that are determined via spectroscopic and photometric modeling and published in the literature.

3. As expected, Ib SNe eject more mass than Ic SNe due to more stripping of the outer stellar envelope of the progenitor star prior to explosion.

4. We propose a speculative picture where the bulk of “normal” Ibc SNe have different progenitors than Ic-BL and GRB-SNe, where the former arise from binary stars, and the latter arise from the explosion of rapidly-rotating, chemically-homogeneous, massive WR stars,.

# 6

# Summary & Future Work

SECTION 6.1

## Summary

6.1.1

### GRB 060729

We presented optical photometry of GRB 060729 obtained with *HST*, as well as a few epochs on *Gemini*-S and CTIO. We performed image subtraction on our *HST* observations using our final epoch of observations taken on WFPC2 as template images for the subtraction. However, as our earlier epochs were taken on ACS in slightly different filters, we had to be very careful with our image-subtraction technique. We checked for the presence of a colour-term in the residuals of the objects in the subtracted images. For very red objects ($F622W - F850LP > 1.5$), a small colour-dependent effect was seen, however when we restricted the colour range to include stars with $0 \leq F622W - F850LP \leq 1.0$ we found no statistically-significant colour term in the residuals. The colour of the OT was never greater than $F622W - F850LP \sim 0.8$ mag, thus by using stars in this colour range to match the ACS images to the WFPC2 template in our image-subtraction procedure we are confident with our resultant magnitudes.

For this event (and for any long, soft GRB) we attribute light coming from three sources: (1) the afterglow, (2) the supernova, and (3) the host galaxy. For the *HST* images, we performed image subtraction to remove the constant source of flux from the hot galaxy. However, for the few *Gemini*-S and single CTIO observations, we were not able to perform image subtraction due to the lack of late-time images to use as templates. However, we measured the brightness of the host galaxy in our WFPC2 images, and mathematically subtracted the host flux from the fluxes of the afterglow, SN and host determined from the *Gemini*-S and CTIO images, thus obtaining fluxes of just the OT.

Next, we determined the behaviour of the optical afterglow using the $U$-band data, incorporating our *HST* observations with the *Swift* $U$-band data published by Grupe et al. (2007). We combined the two datasets using IRAF task **calphot** to convert the two sets of magnitudes into Johnson-$U$. A broken power-law was then fit to the $U$-band LC (Figure 2.1) to determine the rate at which the LC faded. We then made the assumption that the rate at which the LC of GRB 060729 fades is approximately the same in all optical passbands. This assumption is perhaps quite valid in this event (it may not always be valid, see Melandri et al. 2008) when we consider Figure 8 in Grupe et al. (2007). Here the authors found that the rate at which the UV, optical and X-ray light curves faded at late times (i.e. $\geq 2$ days) was approximately the same in all energy-bands. Thus we were confident in assuming that the *R*- and *I*-band LCs will fade at a similar rate to the $U$-band LC.

After determining the decay rate, we subtracted the afterglow model to obtain "SN" LCs (Figure 2.2). Both the shape and brightness of the LCs closely resembled the LCs of SN 1998bw up to $\sim 30$ days (after they had been corrected for foreground and host extinction, and placed at the same redshift as GRB 060729). We also plotted the SED of the OT (i.e. afterglow and SN) of GRB 060729 (Figure 2.3), which resembled the SEDs of SN 1998bw and SN 1994I at similar epochs, albeit with less curvature.

For the associated SN, we found peak apparent magnitudes of $R_c = 23.80 \pm 0.08$ and $I_c = 23.20 \pm 0.06$. When restframe extinction was considered using the analysis performed by Schady et al. (2010) (who found $E(B-V)_{restframe} \leq 0.06$ mag), the peak, restframe absolute $V$-band magnitude was shown to be $M_V = -19.43 \pm 0.06$, which is approximately the same peak brightness as SN 1998bw (in the $V$-band).

Finally, we used our *HST* observations at $\mathrm{t} - \mathrm{t_o} \sim 425$ days in filters $F222W$, $F850LP$ and $F160W$ to constrain the SED of the host galaxy of GRB 060729. The paucity of optical points didn't allow us to constrain the SED very well, however the best-fitting template is for a galaxy with a young stellar population, low global metallicity, a modest stellar mass ($\sim 10^9 M_\odot$) and a star formation rate (SFR $\sim 0.13 M_\odot$ yr$^{-1}$) that is typical of many other GRB host galaxies.

### 6.1.2 GRB 090618

We presented extensive optical, X-ray and radio observations of GRB 090618. Our vast optical data-set allowed us to present well-sampled early-time LCs in filters $BVR_ci$ (Figure 3.1), as well as late-time LC in filters $R_ci$ that clearly show supernova "bumps" (Figure 3.3). As for GRB 060729, we attribute light coming from three sources: (1) the afterglow, (2) the supernova, and (3) the host galaxy. This time, we have not been able to carry out image-subtraction on our optical images, however we have measured the host brightness at $\sim 355$ days post burst in filters $R_c$ and $i$ using the WHT. We converted the host magnitudes into fluxes, and subtracted them from the earlier $R_c$ and $i$ observations, producing LCs of just the OT (Figure 3.3). We then modeled the early-time optical data by fitting broken power-laws to the optical data in all filters simultaneously to determine the rate at which the afterglow faded. Finally, we subtracted this model from the OT LC to obtain "afterglow-subtracted", supernova light curves (Figure 3.4).

For the SN associated with GRB 090618, we found peak apparent magnitudes of the SN to be $R_c = 23.45 \pm 0.08$ and $i = 23.00 \pm 0.06$. Then, using the value of the restframe extinction determined from the optical-to-X-ray SEDs (Figure 3.6), we determined the peak, restframe, absolute $V$-band magnitude of the associated SN to be $M_V = -19.75 \pm 0.13$, which is $\sim 0.3$ mag brighter than SN 1998bw in the $V$-band.

In addition to constraining the restframe extinction using our optical-to-X-ray SEDs, we also concluded that: (1) each epoch where we modeled our contemporaneous optical and X-ray data was well described by a broken power-law, with a cooling break between the optical and X-ray energy bands, and (2) the cooling frequency clearly decreased in time, indicating an ISM environment (rather than a wind environment). In our SED at $\mathrm{t} - \mathrm{t_o} = 1.68$ days, we included a contemporaneous radio observation at 4.86 GHz and determined the typical synchrotron frequency at this epoch. When then used this result to model our radio light curves, and when we considered both a spherical and jet-like evolution of the fireball, we found that the observations were consistent with a jet-like evolution of the fireball. This result, when taken in tandem with the conclusions made from the SED modeling, support the notion that GRB 090618 is typical of other cosmological GRBs.

### 6.1.3 XRF 100316D / SN 2010bh

We have presented optical and near infrared photometry of XRF 100316D / SN 2010bh obtained on the Faulkes Telescope South, Gemini South and *HST*, with the data spanning from $\mathrm{t} - \mathrm{t_o} = 0.5 - 47.3$ days. It was shown that the optical light curves of SN 2010bh evolve more quickly than the archetype GRB-SN 1998bw, and at a similar rate to SN 2006aj, which was associated with XRF 060218, and non-GRB associated type Ic SN 1994I. In terms of peak luminosity, SN 2010bh is the faintest SN yet associated (either spectroscopically or photometrically) with a long-duration GRB or XRF, and has a peak, $V$-band, absolute magnitude of $M_V = -18.62 \pm 0.08$.

SN 2010bh appears to be redder than GRB-SNe 1998bw and 2006aj, where the colour curves of SN 2010bh are seen to be redder early on, though at late times the $V - R$ and $R - i$ colour curves matched that of SN 1994I. The red nature of SN 2010bh is also demonstrated in Figure 4.8, where it is seen that more of the bolometric flux is emitted at infrared wavelengths (and less at optical wavelengths) than in the broad-lined Ic SN 2009bb.

We also gave evidence of the detection of light coming from the shock break-out at $\mathrm{t} - \mathrm{t_o} = 0.598$ days. The brightness of the $B$-band light curve at this epoch, as well as the shape of the optical SED, with a spectral power-law index of $\beta = +0.94 \pm 0.05$ (which is harder than is expected for synchrotron radiation), implies that the source of light at this epoch is not synchrotron in origin and is likely coming from the shock-heated, expanding stellar envelope.

We then applied a simple physical model to the bolometric light curve of SN 2010bh. When we include all photometry in the optical and infrared regime (in the $3,000$Å – $16,600$Å wavelength range) we find physical parameters of $\mathrm{M_{Ni}} = 0.10 \pm 0.01\ M_{\odot}$, $\mathrm{M_{ej}} = 2.24 \pm 0.08\ M_{\odot}$, $\mathrm{E_k} = 1.39 \pm 0.06 \times 10^{52}$ erg. The faint nature of SN 2010bh becomes again apparent when the nickel mass synthesized during the explosion is compared with other GRB-SNe such as SN 1998bw, where it is believed that $\mathrm{M_{Ni}} \approx 0.4\mathrm{M_{\odot}}$ was created in the explosion. Indeed SN 2010bh synthesized only a marginally larger amount of nickel than local type Ic SN 1994I ($\mathrm{M_{Ni}} \approx 0.07\ M_{\odot}$). The investigation of SN 2010bh in relation to the general population of type Ic SNe has once again shown that type Ic SNe are a very heterogeneous class of supernovae, spanning a wide range of luminosities and ejected masses.

Finally, we assembled from the literature all of the available photometry of previously detected GRB-SNe. For each of these events we assumed that light is coming from three sources: from the host galaxy, the afterglow and the supernova. First we removed the constant flux due to the host galaxy, then we modeled the optical afterglows, and then subtracted the light due to the afterglow, which resulted in host- and afterglow-subtracted "supernova" light curves (Figure 4.10). We then created synthetic light curves of SN 1998bw as it would appear if it occurred at the redshift of each given GRB-SNe. We then compared the brightness and shape of the GRB-SNe with that of the shifted LCs of SN 1998bw, and derived stretch ($s$) and luminosity ($k$) factors for each GRB-SNe. The results of our method, when compared to previous studies performed by Ferrero et al. (2006) showed similar values for the stretch and luminosity factors, as well as including more events than the previous studies. When we checked for the possibility of a correlation between $k$ and $s$ in the $R$-band data, we find an in-significant correlation/Pearson coefficient of $c = -0.05$, suggesting that it is highly unlikely that such a relation exists.

### 6.1.4 GRB-SNe vs non-GRB-SNe

In this section we attempted to determine various global properties of GRB-SNe in relation to other Ibc SNe that were not accompanied by a GRB-trigger. In the first section we compiled from the literature peak, restframe $V$-band absolute magnitudes for most of the known GRB-SNe (where there were enough observations to constrain the peak brightness) and for a sample of local Ibc SNe. We incorporated estimates of the restframe extinction wherever it had been determined in the literature.

We then compared the peak, $V$-band absolute magnitudes of the various Ibc SNe subtypes, and performed a KS test to determine if all of the SNe are drawn from the same parent population by testing if the distribution of the peak magnitudes are different. We found that when we compared samples of SN that had been corrected for foreground and host extinction, that the peak magnitudes are not enough to differentiate between GRB-SNe and non-GRB-SNe (i.e. there was a large probability that all samples of host-corrected SNe are drawn from the same parent population).

We then took this result a step further and suggested that if we take the distribution of the peak $V$-band magnitudes as a proxy for the distribution of the peak *bolometric* magnitudes, then the similar distributions imply that similar amounts of nickel are created in all Ibc SNe.

In the following Section, we derived ejected-mass ratios between the various Ibc subtypes. We used the values of the stretch and luminosity factors (in relation to GRB-SN 1998bw) determined for the GRB-SNe in Chapter 4 (Table 4.8). By using SN 1998bw as a template, we applied the stretch and luminosity factors to the flux LC of SN 1998bw, and then fit the modified LC with polynomials of differing orders to determine (1) the peak time, (2) the peak magnitude, (3) the magnitude at 15 days after the peak, and thus (4) $\varnothing m_{15}$. We performed this procedure for all of the GRB-SNe in filters $V$ and $R$, deriving $\varnothing m_{15}$ values that are displayed in Table 5.5.

We also collected $V$- and $R$-band photometric data published in the literature for a sample of local Ibc SNe. Again we fit the LCs with polynomials of differing orders to determine (1) the peak time, (2) the peak magnitude, (3) the magnitude at 15 days after the peak, and thus (4) $\varnothing m_{15}$. We displayed our results in Table 5.4.

We then compared the distribution of the $\varnothing m_{15}$ values between the different Ibc subtypes. We performed a KS test and found that the distribution of $\varnothing m_{15}$ between the GRB-SNe and the Ibc SNe sample have a very high probability that they are drawn from the same parent population. However, when we compared the Ib SNe with the Ic SNe we found a a much lower probability that the two subtypes are drawn from the same parent population. We interpret this result being due to the progenitors of Ib SNe being *less-stripped* before explosion than those of Ic SNe.

We noted that the $\varnothing m_{15}$ parameter is a measurement of the width of the LC. In terms of the bolometric LC, the overall width is related to the ratio $\frac{M_{ej}}{v_{ph}}$ (i.e. the amount of ejected mass and the photospheric velocity). Thus when we compared the $\varnothing m_{15}$ parameters of GRB-SNe with those of the Ibc SNe, as the average ejecta velocity in these events can differ by a factor of 2 or so, we are not really comparing the same quantity.

So, we gathered from the literature peak ejecta velocities for a sample of GRB-SNe as well as a sample of local Ibc SNe. We then determined the average peak ejecta velocity in each of the Ibc subtypes. We also made the assumption that the behaviour of the LC in the $V$ and $R$ band is the same as the behaviour of the bolometric LC. Thus, by using the widths of the LCs in $V$ and $R$, as well as the average peak ejecta velocities, we were able to break the degeneracy in the ratio $\frac{M_{ej}}{v_{ph}}$ and determine ratios of $M_{ej}$ between the Ibc SN subtypes.

Our results were very interesting. First, they confirmed what was already expected in that Ib SNe eject more mass on average than Ic SNe (as the Ib arise from progenitors that undergo less stripping of their outer envelopes before exploding). We also found a potentially very interesting result in that GRB-SNe eject *more* mass than all of the other Ibc subtypes. We also found that Ic-BL SNe eject more mass than Ib and Ic SNe, and only a little less than GRB-SNe.

We also collected from the literature results from modeling spectroscopic and photometric observations of GRB-SNe and local Ibc SNe. Again we derived mass ratios between the Ibc subtypes, finding nearly the exact same ejected mass ratios as those found via the $V$- and $R$-band LC widths. Thus we found via two different methods that: (1) Ib SNe eject more mass than Ic SNe, (2) GRB-SNe eject the most mass of all the Ibc subtypes, and (3) Ic-BL SNe eject more mass than Ib and Ic SNe, and only a little less than GRB-SNe.

We then put our results into context with the existing progenitor models for GRB-SNe and Ibc SNe. Currently three models are presented as possible evolutionary paths leading to the formation of GRB-SNe: (1) arising from non-rotating, single massive stars, (2) arising from rapidly-rotating, single massive stars, and (3) arising from binary massive stars. The last evolutionary path is proposed by many to be the main progenitor channel for producing Ibc SNe, while it is less certain for GRB-SNe.

Using our conclusion that GRB-SNe eject more mass than all of the other Ibc subtypes, we tentatively suggested a general picture where most Ibc SNe arise from the interaction of binary stars, where the primary that explodes has a mass lower than is usually associated with WR stars. However, GRB-SNe, and possibly Ic-BL SNe, arise from the collapse of a rapidly-rotating, massive WR stars. The difference between GRB-SNe and Ic-BL SNe could be due to mass loss (where mass loss also removes angular momentum), which is governed by the metallicity of the progenitor. As GRB-SNe arise from more metal-poor progenitors than Ic-BL SNe, if we assume that rapid-rotation is present in these events, the key difference is that the higher metallicity of Ic-BL SNe leads to more mass loss and crucially, more loss of angular momentum than those of GRB-SNe. It is because GRB-SNe lose less mass due to their lower metal content, and hence retain more angular momentum before exploding, that they are able to power a GRB at the time of death, while Ic-BL SNe do not.

SECTION 6.2

## Conclusions

The work presented in this thesis supports the existing notion that the progenitors of long, soft GRBs are massive stars. While spectroscopy of nearby GRB-SNe is the ideal way to unambiguously show the presence of a SN with a GRB/XRF, it is also possible to detect the presence of a SN photometrically via "bumps" in the LCs of GRB afterglows, as we have done for GRB 060729 and very successfully for GRB 090618.

GRBs 060729 and 090618 fall in the category of "cosmological" GRBs as they obey the Amati relation (Amati 2002) and have an isotropic energy release of roughly $\sim 10^{52} - 10^{53}$ erg. XRF 100316D however is very reminiscent of other nearby GRB/XRF-SNe that are under-luminous by a factor of $100 - 1000$ in relation to cosmological GRBs. The differences in the high-energy properties of the GRBs in these events displays much more variation than the properties of the associated SNe, which display broadly-similar LC shapes and a much smaller range in the peak magnitudes.

If we take the conclusions in Chapter 5 at face value, it would appear that if GRB-SNe do arise from single, star progenitors than the progenitor must have been rotating very rapidly before exploding. The consequence of the rapid-rotation is that the progenitor undergoes extensive mixing, and instead of ejecting an outer envelope of hydrogen before exploding, the hydrogen is retained and burned into helium. Thus GRB-SNe eject more mass at the time of exploding than Ibc SNe that have likely had their outer layers of hydrogen stripped either by strong stellar winds or by an orbiting companion.

SECTION 6.3

## Future Work

While considerable progress has been made in recent decades regarding the nature of the progenitors of long, soft GRBs, many questions still remain, including:

1. Why hasn't a GRB or XRF been associated with a Ib SNe?
2. What is the mass range and evolutionary stage of the progenitor stars?
3. How many different types of progenitors are there for GRB-SNe? Between collapsars and millisecond magnetars, which are more common? How many arise from binary systems?
4. How similar are the evolutionary paths of GRB-SNe and Ic-BL SNe?
5. Do GRB-SNe and XRF-SNe arise from the same progenitors? Why do the former have larger ejecta masses and arise from more massive progenitors?
6. Why do some long, soft GRBs have no accompanying optically-bright SNe?
7. Is there a difference between low and high-redshift GRB-SNe? Is there a distinct subclass of low-energy GRBs in the local universe (i.e. GRB 980425)?

All of these questions can be answered in time, given that more GRB/XRF-SNe and non-GRB Ibc SNe are observed. So far, for each GRB-SNe that has been spectroscopically linked, the amount of photometric and spectroscopic data has allowed astronomers to determine physical parameters of the SN (explosion energy, density structure, abundances of different elements in the ejecta) and progenitor (mainly the ZAMS and pre-explosion masses). The same applies to nearby Ibc SNe without a GRB-trigger, where for SN events such as Ic-BL SN 2002ap (Mazzali et al. 2002) it has been possible to determine physical properties of the SN and progenitor.

However, with more observations have also come more questions. For example, observations of recent GRB101225A and its associated SN have led Thöne et al.

(2011) to surmise that the progenitor system in this event was not a single massive star (like those of previously detected GRB-SNe and Ibc SNe), but likely a helium-neutron star binary which underwent a common envelope phase, expelling its hydrogen envelope prior to explosion. The final merging process created a GRB-like event.

If the progenitor scenario of Thöne et al. (2011) is correct, then GRB101225A has shown that GRBs (or GRB-like events) can arise from a multitude of progenitor configurations. Already it has been proposed that at least some, if not most, Ibc SNe arise from binary systems (e.g. Smartt 2009), and now there is observational evidence that suggests some GRB-like SNe may also arise from binary systems. Additionally, as XRT 080109 / SN 2008D is also likely not to be a true GRB-event (e.g. Van der Horst 2011), this opens the possibility that all SNe may be related to detectable high-energy events (that are not just the shock-breakout) and that GRBs are only one manifestation of a larger continuum of high-energy properties of SN events.

While even more revelations are likely to be made over the coming years, this is only going to be possible for the occurrence of nearby events where we can collect as much spectroscopic and photometric data as possible. This will allow astronomers to determine properties of the individual progenitors, and over time, as more events occur and are observed, it will be possible to start making statistical conclusions of the general properties of the progenitors.

So, while intense scrutiny of individual events is one avenue for continued research regarding the progenitors of GRB-SNe and those of local Ibc SNe without a GRB-trigger, indirect measurements of the environments of GRBs can also provide insight. Existing studies have already shown that a metallicity difference exists between the environments of the progenitors of GRB-SNe and those of Ic-BL SNe, with the former arising from more metal-poor environments (e.g. Modjaz et al. 2008). Thus, as astronomers work towards the possible progenitor scenarios of GRB-SNe and local Ibc SNe, metallicity measurements can teach us a lot about stellar evolution.

# 7

# Appendix: Photometric Methods

All photometric data presented in this thesis has been obtained by using tasks within the software package IRAF (Image Reduction and Analysis Facility).[1] Here we present an example of using IRAF to perform standard star photometry in order to calibrate instrumental magnitudes of XRF 100316D / SN 2010bh. To to this we will take the following general steps:

1. Perform standard star photometry to obtain a set of transformations between the instrumental and catalog magnitudes for a series of Landolt (1992) standard photometric fields.
2. Create a catalog of calibrated, secondary standards in the field of XRF 100316D / SN 2010bh.
3. Perform image subtraction to remove light from the host galaxy.
4. Perform aperture photometry on the host-subtracted images and calibrate them to the secondary standards in the field of the XRF.

In Section 7.1 we will describe the method we used to calibrate a set of stars in the field of XRF 100316D. In Section 7.2 we will describe the image subtraction method

[1] IRAF is distributed by the National Optical Astronomy Observatory, which is operated by the Association of Universities for Research in Astronomy, Inc., under cooperative agreement with the National Science Foundation

we used to remove the light from the host galaxy.

SECTION 7.1

## Using IRAF to perform standard star photometry

The purpose of this task is to create a catalog of standard stars in the field of the XRF. These will then be used to calibrate individual epochs of XRF 100316D / SN 2010bh. In order to create our set of secondary standards in the field of the XRF we have obtained (on a single night):

1. Images of photometric standards PG-1047 and PG-1323 (which are photometric standard stars observed by Landolt 1992).
2. Images of the field of the XRF.

It is vital to obtain observations of the photometric standards and the XRF in the same filters and taken close together in time. Here we present observations made on 02-May-2010 with the Faulkes Telescope South (FTS) of photometric standard fields PG-1047, PG-1323 as well as the XRF field, in filters $BVR_ci$.

We are quite fortunate that the images downloaded from the FTS image archive have already been dark and bias frame corrected, as well being flat-fielded, and are ready to be used straight away for photometry.

Once we obtained our images, our first task was to align all of the images taken in a single field. This was done using a pre-existing shell-script ("imregister.pl") that uses IRAF routines to create a list of objects in each image and calculates the difference in the coordinates of the objects in each list to those of a pre-determined reference/template image. The script then performs the transformation and output images that are aligned to the specified template/reference image. This script is performed outside of IRAF and is run via the command line in a terminal window, and can be used to align a single image to the template:

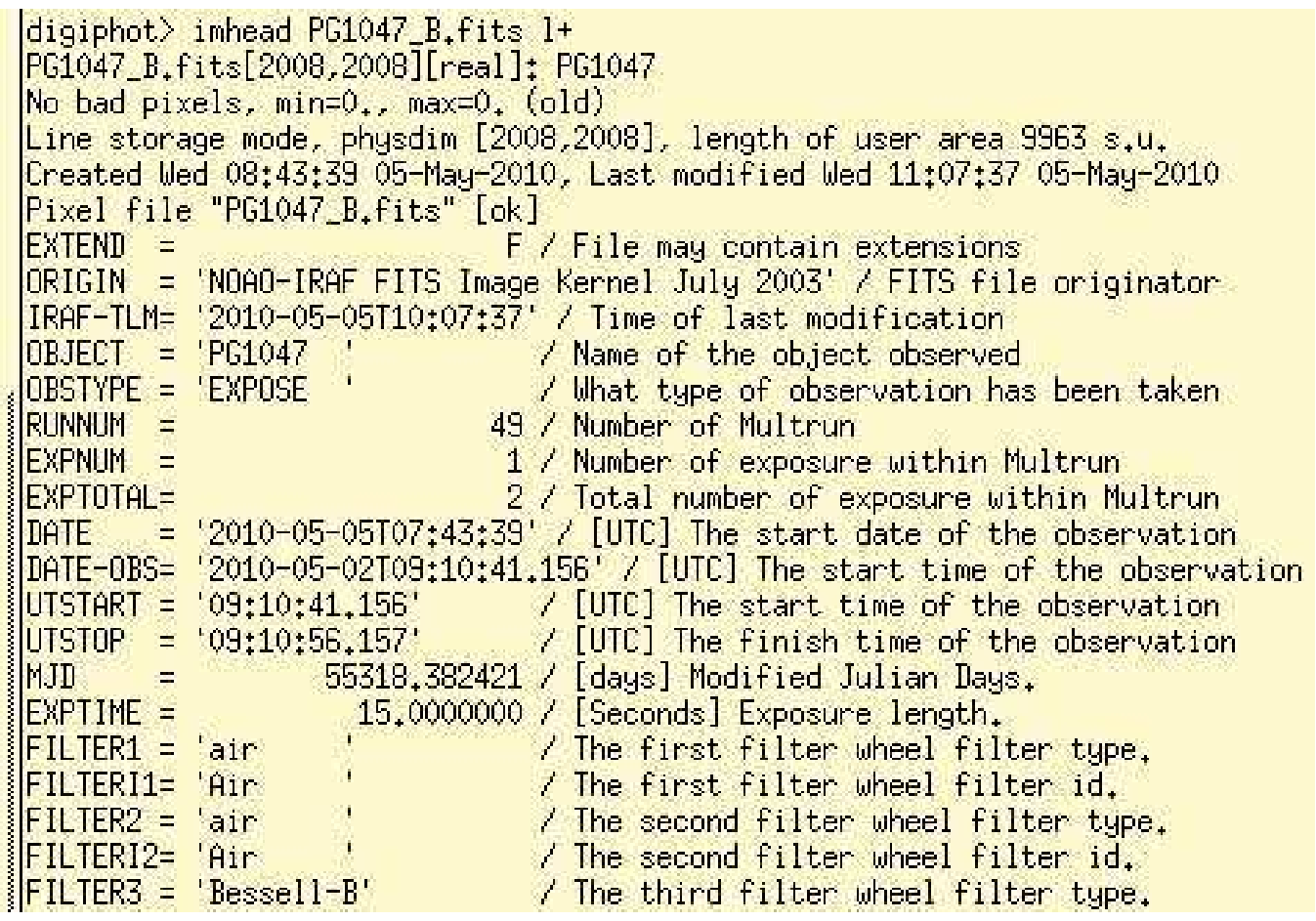

```
digiphot> imhead PG1047_B.fits l+
PG1047_B.fits[2008,2008][real]: PG1047
No bad pixels, min=0., max=0. (old)
Line storage mode, physdim [2008,2008], length of user area 9963 s.u.
Created Wed 08:43:39 05-May-2010, Last modified Wed 11:07:37 05-May-2010
Pixel file "PG1047_B.fits" [ok]
EXTEND  =                    F / File may contain extensions
ORIGIN  = 'NOAO-IRAF FITS Image Kernel July 2003' / FITS file originator
IRAF-TLM= '2010-05-05T10:07:37' / Time of last modification
OBJECT  = 'PG1047  '           / Name of the object observed
OBSTYPE = 'EXPOSE  '           / What type of observation has been taken
RUNNUM  =                   49 / Number of Multrun
EXPNUM  =                    1 / Number of exposure within Multrun
EXPTOTAL=                    2 / Total number of exposure within Multrun
DATE    = '2010-05-05T07:43:39' / [UTC] The start date of the observation
DATE-OBS= '2010-05-02T09:10:41.156' / [UTC] The start time of the observation
UTSTART = '09:10:41.156'       / [UTC] The start time of the observation
UTSTOP  = '09:10:56.157'       / [UTC] The finish time of the observation
MJD     =         55318.382421 / [days] Modified Julian Days.
EXPTIME =           15.0000000 / [Seconds] Exposure length.
FILTER1 = 'air     '           / The first filter wheel filter type.
FILTERI1= 'Air     '           / The first filter wheel filter id.
FILTER2 = 'air     '           / The second filter wheel filter type.
FILTERI2= 'Air     '           / The second filter wheel filter id.
FILTER3 = 'Bessell-B'          / The third filter wheel filter type.
```

Figure 7.1 Part of the fits image header for PG1047_B.fits. The keywords DATE-OBS, UTSTART, EXPTIME and FILTER3 are noted and used in the IRAF task **datapars**.

>> imregister.pl 'image.fits' 'template.fits'

or a list of images to a template:

>> imregister.pl @list_of_images.txt 'template.fits'

Once each set of images are aligned, it's time to open IRAF and DS9. For this example we will be using the following packages: **DIGIPHOT**, **DAOPHOT** and **PHOTCAL**. The general procedure is similar to the method described in "A User's Guide to Stellar CCD Photometry with IRAF" by Philip Massey and Lindsey E. Davis[2] .

The next step is to check the image headers and obtain the image header keywords for the airmass, filter, start time of observation, date of observation, gain, read noise and exposure time. For image 'PG1047_B.fits', these correspond respectively to **AIRMASS**, **FILTER3**, **UTSTART**, **DATE-OBS**, **GAIN**, **RDNOISE** and **EXPTIME** (Figure 7.1).

To check if the keywords are present in the fits header, the task **imhead** is used:

>> imhead image.fits l+

Where "l+" tells IRAF to print the entire fits header. If any of the keywords are missing, they can be updated and/or added to the header keywords via:

[2] http://iraf.net/irafdocs/

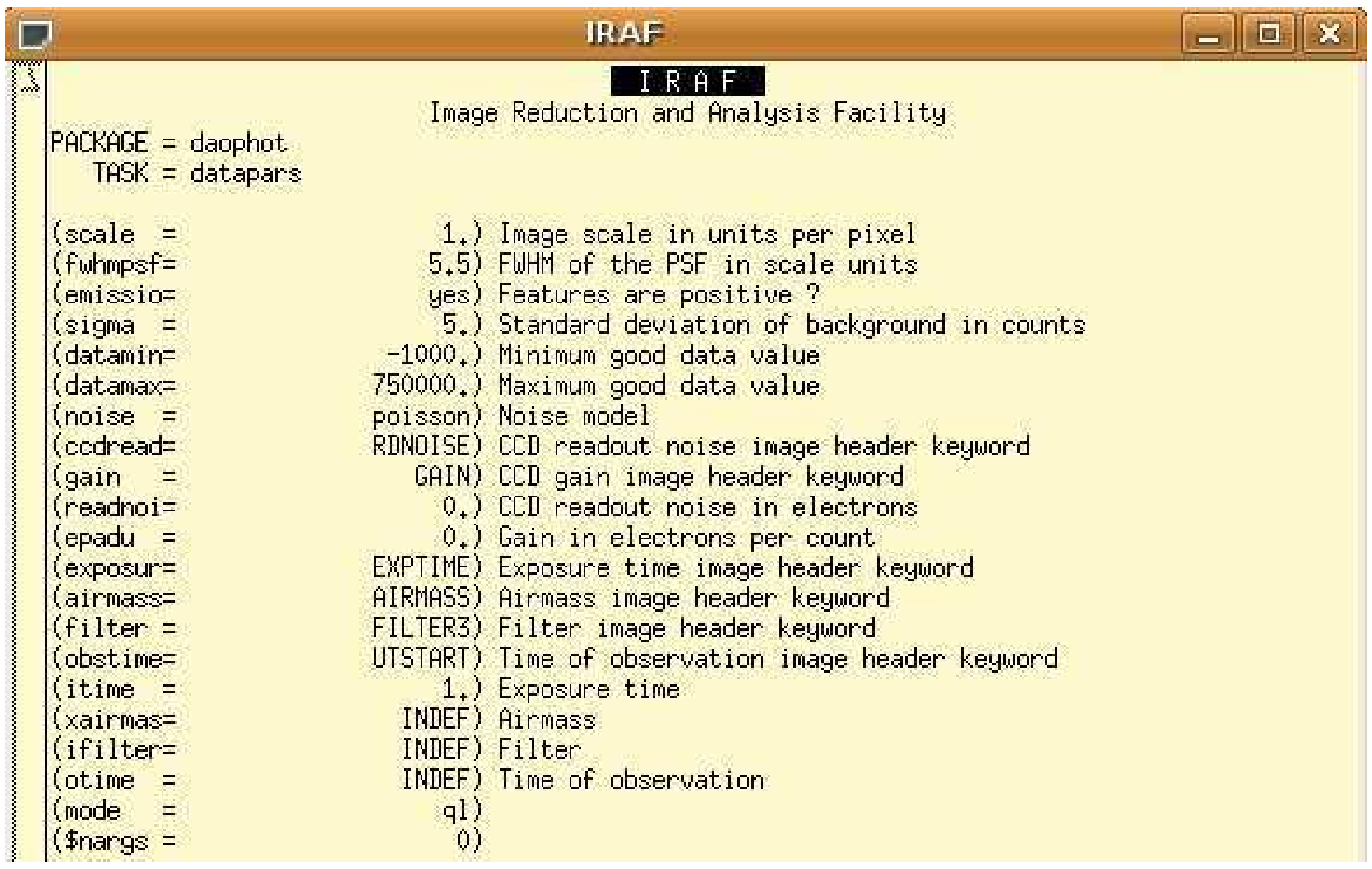

Figure 7.2 Example input parameters for the IRAF task **datapars**.

>> hedit image.fits "image header word" add+

(NB. only use "add+" to add the keyword to the header, otherwise omit it if you only wish to update an existing keyword.)

Once the keywords have been updated/added, its time to let IRAF know what these keywords are. In IRAF task **DIGIPHOT/DIGIPHOT/datapars**, the corresponding keywords have been inputed (Figure 7.2).

It's now time to examine each image individually to measure the minimum and maximum pixel values as well as the standard deviation of the background pixels. We also want to determine the peak pixel value for objects in our image that are not saturated, as well as those that *are* saturated, as to determine the saturation level for each image. To determine these values we will use the IRAF task **imexamine**:

>> imexamine image.fits

There are several keystrokes that can be used to measure pixel brightness, etc., and the reader is encouraged to use the "help" task whenever uncertain.

We also want to measure the FWHM of the point spread function (psf; see Figure 7.3) of several bright stars (not extended objects such as galaxies) in the image. Once we have an idea of the average FWHM of the bright (but not saturated) stars in a given

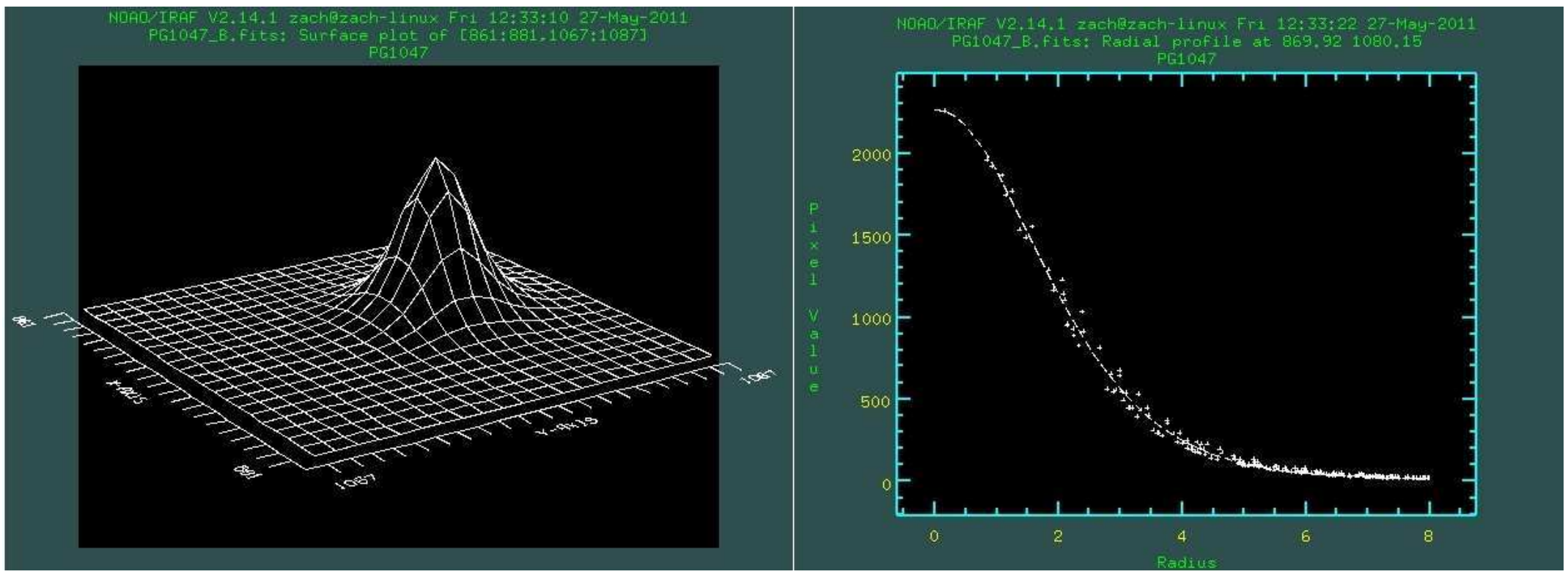

Figure 7.3 Stellar profile of a bright star in the field of PG-1047. *Left:* Surface profile of the star. *Right:* Radial profile of the same star. The FWHM is measured with the task **imexamine** to be 4.01 pixels.

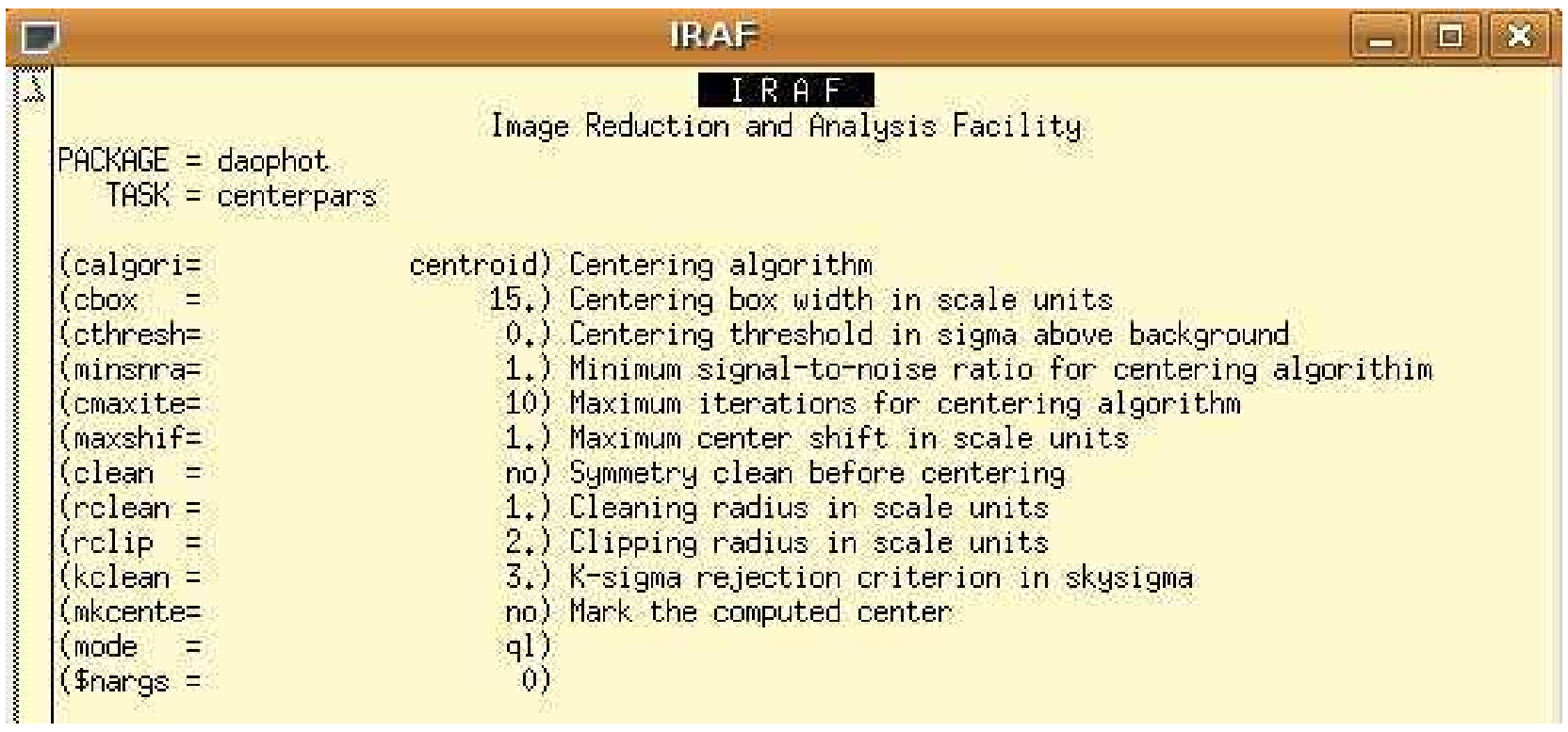

```
IRAF
                         I R A F
              Image Reduction and Analysis Facility
PACKAGE = daophot
   TASK = centerpars

(calgori=           centroid) Centering algorithm
(cbox   =                15.) Centering box width in scale units
(cthresh=                 0.) Centering threshold in sigma above background
(minsnra=                 1.) Minimum signal-to-noise ratio for centering algorithim
(cmaxite=                 10) Maximum iterations for centering algorithm
(maxshif=                 1.) Maximum center shift in scale units
(clean  =                 no) Symmetry clean before centering
(rclean =                 1.) Cleaning radius in scale units
(rclip  =                 2.) Clipping radius in scale units
(kclean =                 3.) K-sigma rejection criterion in skysigma
(mkcente=                 no) Mark the computed center
(mode   =                 ql)
($nargs =                  0)
```

Figure 7.4 Example input parameters for the IRAF task **centerpars**.

image, we can then choose an appropriate aperture size to use for photometry later on.

Once we have measured these values for a set of images (i.e. for images of PG-1047 in filters $BVR_ci$), we then need to update the keywords in IRAF task **DIGIPHOT/DIGIPHOT/datapa** (Figure 7.2). We also need to specify other parameters in tasks **centerpars, fitskypars and photpars** (all are within **DIGIPHOT/DIGIPHOT/**), that will be used by the **phot** task when we perform photometry later on. In **centerpars** (Figure 7.4) we need to specify the centering algorithm to be used (if any), as well as the size of the centering box (in pixels). Generally, the "centroid" centering-algorithm is used with a centering box of $10-20$ pixels. In **fitskypars** (Figure 7.5) we need to specify the radius and width of the sky annulus that is used to measure the background level. Generally speaking, the radius of the sky annulus should be $\approx 4$ times the radius of the chosen aperture. The width of the sky annulus is generally between $10-20$ pixels.

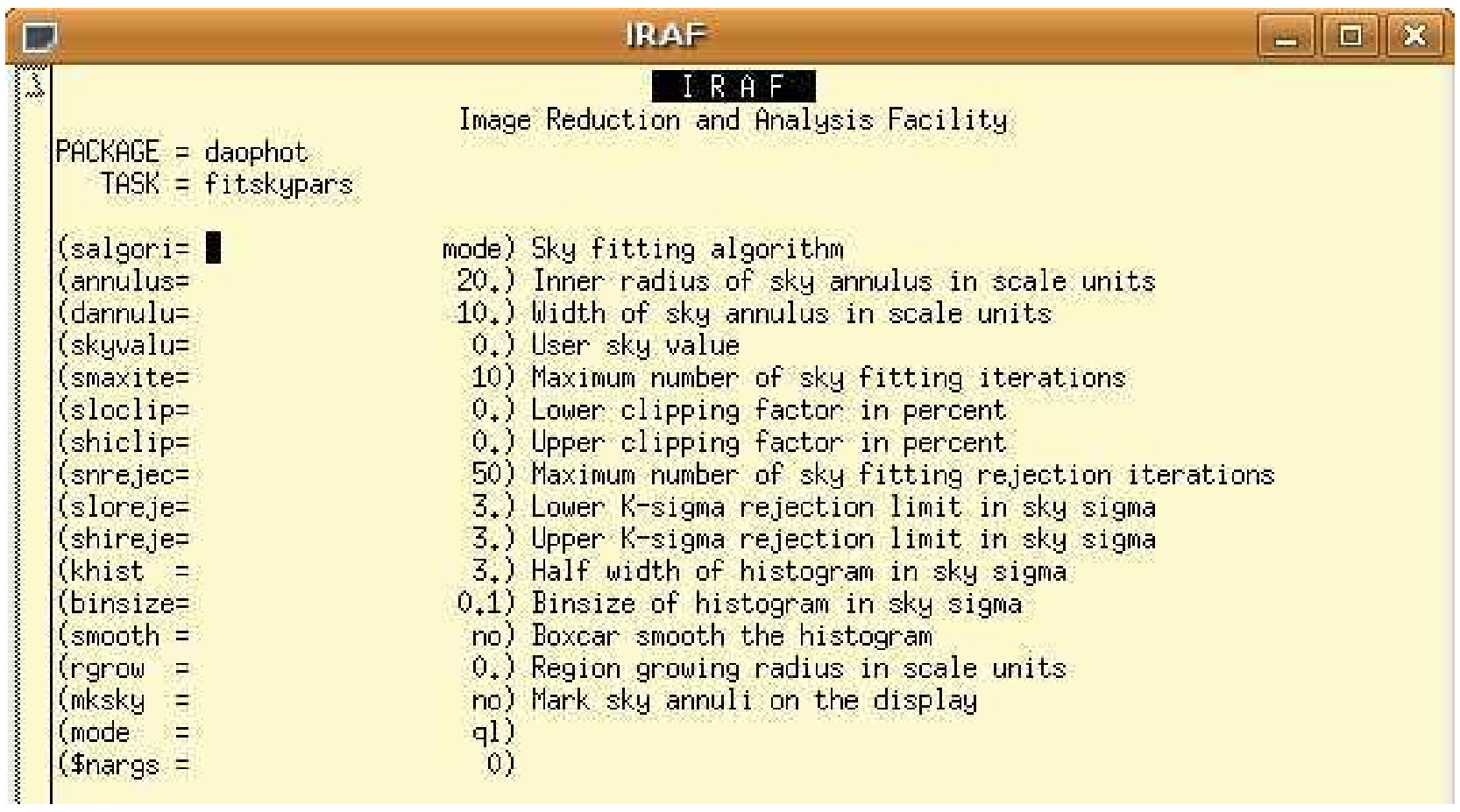

Figure 7.5 Example input parameters for the IRAF task **fitskypars**.

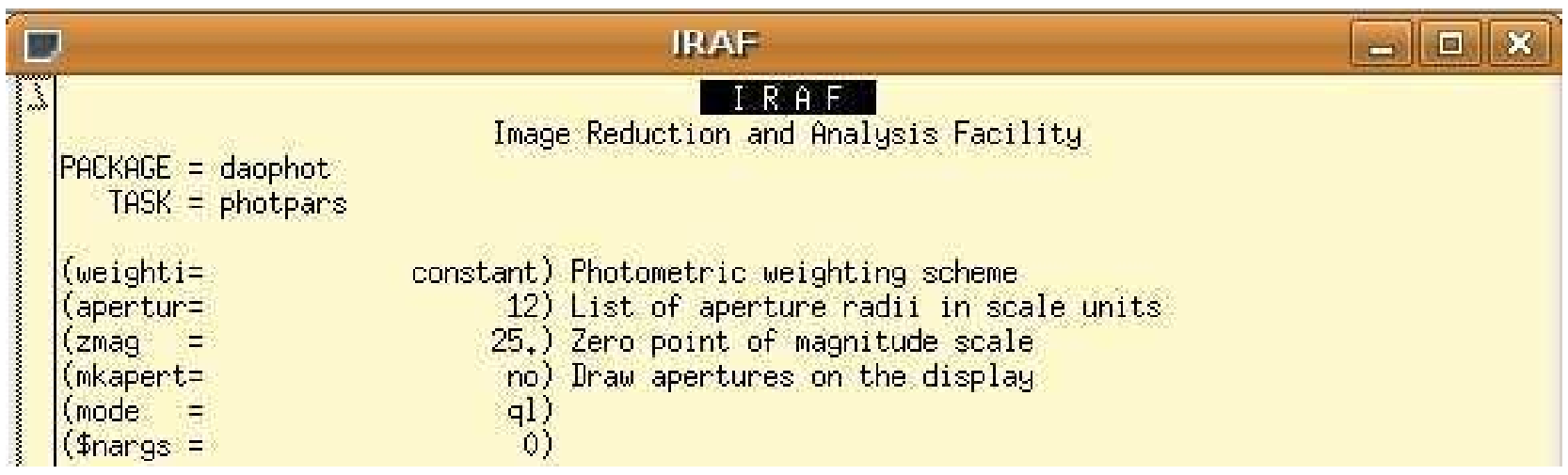

Figure 7.6 Example input parameters for the IRAF task **photpars**.

Next, we want to use **photpars** (Figure 7.6) to specify the radius of the aperture to be used for photometry. There is no ideal way to chose the optimum aperture radius, however a good general rule is to choose an aperture that is 1-2 times larger than the average image psf FWHM. However, as we want to perform photometry on a set of images, we need to choose an aperture that will be suitable for all images. For example, our $BVR_ci$ images of PG-1047 have average psf FWHMs of 5.6,5.5,6.2,5.9 pixels respectively. If we approximate the average psf FWHM of these four images to be $\sim 6$ pixels, then we might choose an aperture of 8 pixels. This aperture will not contain all of the light for a given point source as the psf (see Figure 7.3) has "stellar wings" that contain light up to a radius of $20-30$ pixels. To account for this we will compute and apply an aperture correction (see below).

So, for our set of $BVR_ci$ images of PG-1047, we will use a single aperture of radius 8 pixels for our photometry, and we specify this value in **photpars** (Figure 7.6). Now we are almost ready to perform photometry on our images, but first we must make a list of all of the images we will perform photometry on (rather than performing photometry on single images separately which will take longer). This is done via the following command:

>> files *.fits > list_name

This take will put all files ending in the suffix ".fits" into a text file of a specified name. Next, we need to specify which stars we want to measure the magnitudes of. One way is to allow IRAF to choose the stars by running the task **daofind**. However, we can have more control over which stars are chosen by picking them ourselves.

So, first load an image into DS9 by:

>> display PG1047_B.fits 1

Where the "1" specifies which DS9 window the image is displayed into. Once the image has been loaded, we then use IRAF task **tvmark** to choose the objects by hand in the image and put them into a coordinate list:

>> tvmark coordinate_list.coo.1 interac+

The command "interac+" tells IRAF that we will manually choose the stars in the interactive mode. Once the command has been written in the IRAF terminal, the cursor will appear in the DS9 window, and a star can be added to the coordinate list by pressing "a". Once the coordinates of all the stars have been added to the list, it is time to perform photometry on our images. This is done by using the task **DIGIPHOT/DAOPHOT/phot**:

>> phot @list_name 'coordinate_list.coo.1'

The task **phot** will perform aperture photometry on the images listed in "list_name" and on the objects listed in the coordinate file. **phot** will then output instrumental magnitudes of the specified objects in each image into magnitude files (e.g. for PG1047_B.fits, a file with the instrumental magnitudes will be output into a text file called PG1047_B.fits.mag.1; and similarly for the other images).

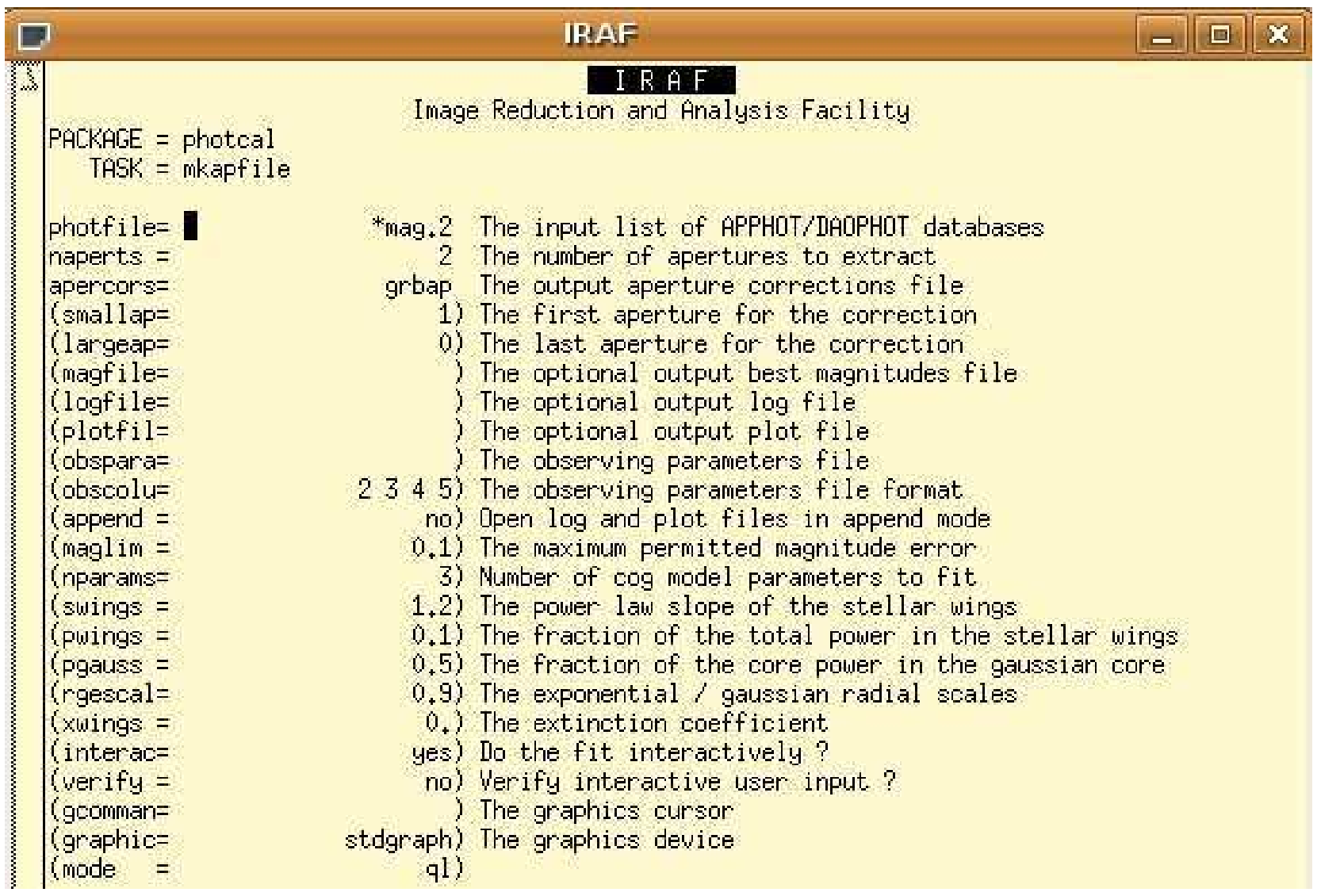

Figure 7.7 Example of the IRAF task **mkapfile**.

Finally, we need to compute an aperture correction. This correction is included to account for the finite size of the aperture, and the fact that not all of the light emitted by the object is contained within it. This correction will be applied to all point sources (i.e. not extended objects) in the images. The aperture correction needs to be computed in each filter and it needs to be done using a dozen or so bright (but not saturated!) stars in a given field. Once the bright stars are chosen, we perform photometry on the set of images, but this time using two aperture sizes: a small aperture of 8 pixels (as chosen earlier) and a larger aperture of 30 pixels (of which we are confident that all of the light emitted by the point sources will be contained within). In **photpar** a sequence of aperture radii can be specified (e.g. in our case we input 8,30).

Once the aperture sizes have been chosen and inputted into **photpar**, we run **phot** again. This time **phot** will output two instrumental magnitudes, one for light contained within a radius of 8 pixels, and another for 30 pixels. As this is our second pass at doing photometry on these images, the outputted magnitude files will have a suffix ending in ".mag.2".

Now, using the task **DIGIPHOT/PHOTCAL/mkapfile**, we will interactively compute the aperture correction:

>> epar mkapfile

In this example (Figure 7.7), we are extracting the magnitudes from the ".mag.2" files, and putting the computed aperture corrections into a file called "grbap". We are also doing this interactively, so that we can remove any significant outliers from the computed aperture correction (e.g. in case we mistakenly used a galaxy instead of a star). The aperture correction file (which in this example has been called "grbap"; see Figure 7.7) will then be used later on during the calibration.

OK, for the $BVR_ci$ images of PG-1047, we have computed instrumental magnitudes of selected objects in the images, as well as computing an aperture correction in each filter. The next thing is to now repeat the procedure for the set of images taken of PG-1323. Once this has been completed we are now ready to calibrate these two photometric standards with the Landolt Catalog (Landolt 1992).

IRAF already has the catalog magnitudes for these fields saved in a catalog file called "nlandolt.txt", so now we need to make the standard star observations files.

The first file we want to make is an "image set" file (Figure 7.8). This file gives the standard star names and the corresponding image names. Next we need to create a configuration file, that specifies the format of the data listed in the catalog (i.e. tells IRAF which column contains the object name, catalog magnitude, magnitude error, etc.). We also need to specify (within the configuration file) the way in which we would like to calibrate our data: do we want to only compute an offset/zero-point between the catalog and instrumental magnitudes, or do we also want to consider the varying airmasses as well as including a colour correction. When performing the calibration between images taken in the same filter, it is not uncommon to compute only a zero-point, however when calibrating a given filter to a slightly different one (e.g. calibrating the SDSS $r$ filter to the Johnson-Counsins $R_c$ filter) it is necessary to include a colour term. The IRAF task to create a configuration file is **DIGIPHOT/PHTOCAL/mkconfig**, and an example of a standard configuration file is displayed in Figure 7.9.

So, to briefly re-cap, we have now created:

1. Magnitude files for an aperture of radius 8 pixels.

```
grbset
PG1047 : PG1047_B.fits PG1047_V.fits PG1047_R.fits PG1047_i.fits
PG1323 : PG1323_B.fits PG1323_V.fits PG1323_R.fits PG1323_i.fits
```

Figure 7.8 An example of an image set file. For each set of photometric standards the images in $BVR_ci$ are listed along with the name of the standard star.

```
Ti           24          # time of observation in filter i
Xi           25          # airmass in filter i
xi           26          # x coordinate in filter i
yi           27          # y coordinate in filter i
mi           28          # instrumental magnitude in filter i
error(mi)    29          # magnitude error in filter i

# Sample transformation section for the new Landolt UBVRI system

transformation

fit   v1=0.0, v2=0.08, v3=0.000
const v4=0.0
VFIT : mV = (I+VI) + v1 + v2 * VI +v3*XV

fit   b1=0.0, b2=0.08, b3=0.000
const b4=0.0
BFIT : mB = (I+BI) + b1 + b2 * BI +b3*XB

fit   r1=0.0, r2=0.08, r3=0.000
const r4=0.0
RFIT : mR = (I+RI) + r1 + r2 * RI +r3*XR

fit   i1=0.0, i2=0.03, i3=0.000
const i4=0.0
IFIT : mi = I + i1 + i2 * RI +i3*Xi
```

Figure 7.9 Example of a configuration file that specifies the format of the data in the catalog file and the form of the transformation equations.

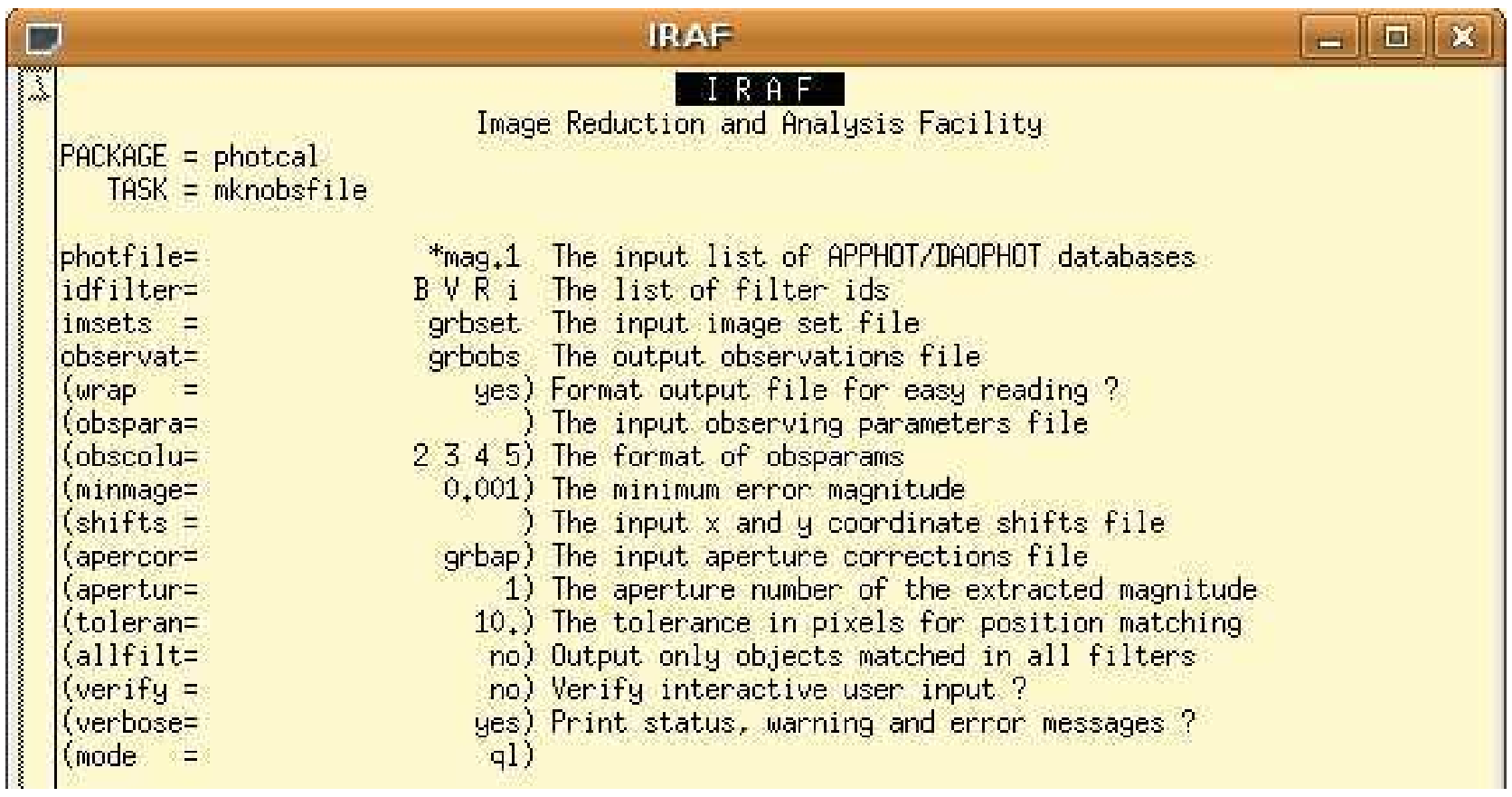

Figure 7.10 Example of using the IRAF task **mknobsfile** to create an observations file.

2. An aperture correction file (for all filters).
3. An image set file.
4. A configuration file that specifies the format of the catalog as well as the transformation equations.

We are now ready to use **DIGIPHOT/PHOTCAL/mknobsfile** to make our observation file. The example in Figure 7.10 shows the usage of **mknobsfile**. Here we are extracting the instrumental magnitudes from the “.mag.1” magnitude files, which are taken in filters $BVR_ci$. The image set file is called “grbset”, the aperture correction file is called “grbap”, and the output of this task will be an observations file called “grbobs” (Figure 7.11).

As we have aligned the images it makes tidying up the observations file trivial as all of the objects in all of the images all have the same x and y coordinates. Its now just a matter of correctly identifying each of the photometric standards by their x and y coordinates and labeling them *as they appear in the catalog file*. (NB. if they are labeled incorrectly then IRAF will not be able to match them and will instead return error messages.)

Now that the observations file has been created, its time to compute the transformation between the instrumental and catalog values. This is done using the task **DIGIPHOT/PHOTCAL/fitparams** (Figure 7.12). This task takes the observations file “grbobs”, and compares it against the catalog file “nlandolt”, computing the values of the variables in the transformation that are specified in the configuration file ’grb.cfg’ (Figure 7.9). This task is usually run interactively, as to remove any significant outliers from the fit. The computed transformations are then output into a parameters file, which here is called “grb.ans” and is displayed in Figure 7.13.

| # FIELD | FILTER | OTIME | AIRMASS | XCENTER | YCENTER | MAG | MERR |
|---|---|---|---|---|---|---|---|
| PG1047-3 | B | 9:10:41.2 | 1.214 | 1405.760 | 1065.526 | 16.477 | 0.003 |
| * | V | 9:12:03.4 | 1.212 | 1405.738 | 1065.694 | 16.299 | 0.003 |
| * | R | 9:13:15.3 | 1.211 | 1405.806 | 1065.672 | 15.909 | 0.003 |
| * | i | 9:14:29.8 | 1.209 | 1405.757 | 1065.674 | 16.309 | 0.003 |
| PG1047-3A | B | 9:10:41.2 | 1.214 | 1266.109 | 952.937 | 17.299 | 0.004 |
| * | V | 9:12:03.4 | 1.212 | 1266.179 | 952.962 | 16.378 | 0.003 |
| * | R | 9:13:15.3 | 1.211 | 1266.152 | 952.932 | 15.421 | 0.002 |
| * | i | 9:14:29.8 | 1.209 | 1266.185 | 952.902 | 15.325 | 0.002 |
| PG1047-3B | B | 9:10:41.2 | 1.214 | 1155.410 | 778.014 | 18.449 | 0.007 |
| * | V | 9:12:03.4 | 1.212 | 1155.384 | 778.081 | 17.598 | 0.005 |
| * | R | 9:13:15.3 | 1.211 | 1155.409 | 778.083 | 16.672 | 0.004 |
| * | i | 9:14:29.8 | 1.209 | 1155.433 | 778.081 | 16.626 | 0.004 |
| PG1047-3C | B | 9:10:41.2 | 1.214 | 869.940 | 1080.109 | 16.097 | 0.002 |
| * | V | 9:12:03.4 | 1.212 | 869.902 | 1080.240 | 15.290 | 0.002 |
| * | R | 9:13:15.3 | 1.211 | 869.914 | 1080.184 | 14.400 | 0.001 |
| * | i | 9:14:29.8 | 1.209 | 869.936 | 1080.183 | 14.359 | 0.001 |
| PG1047-S2 | B | 9:10:41.2 | 1.214 | 1626.045 | 1661.523 | 20.986 | 0.031 |
| * | V | 9:12:03.4 | 1.212 | 1625.961 | 1661.431 | 20.180 | 0.020 |
| * | R | 9:13:15.3 | 1.211 | 1626.031 | 1661.404 | 19.253 | 0.014 |
| * | i | 9:14:29.8 | 1.209 | 1625.843 | 1661.472 | 19.223 | 0.014 |

Figure 7.11 Example of an observations file produced by the IRAF task **mknobsfile**.

At this point the transformation has now been computed. However it is wise to check how well the transformation has been calculated by “invert-fitting” the computed transformations onto the original instrumental magnitudes (which will transform the instrumental magnitudes into catalog magnitudes). We will then manually compare the “invertfitted” magnitudes against the actual calibrated magnitudes in the Landolt catalog.

The IRAF task **DIGIPHOT/PHOTCAL/invertfit** will apply the transformations to the instrumental magnitudes (Figure 7.12), and output the calibrated magnitudes into a file, which for this example we have called “grb.cal” (Figure 7.14). When comparing our transformed magnitudes with the Landolt catalog, it is seen that they are all within $\approx 0.03$ mag of the catalog values (not shown here), which is a decent calibration.

OK, so we have computed our transformation. We will now apply this transformation to a set of secondary standards in the field of XRF 100316D (i.e. create a catalog

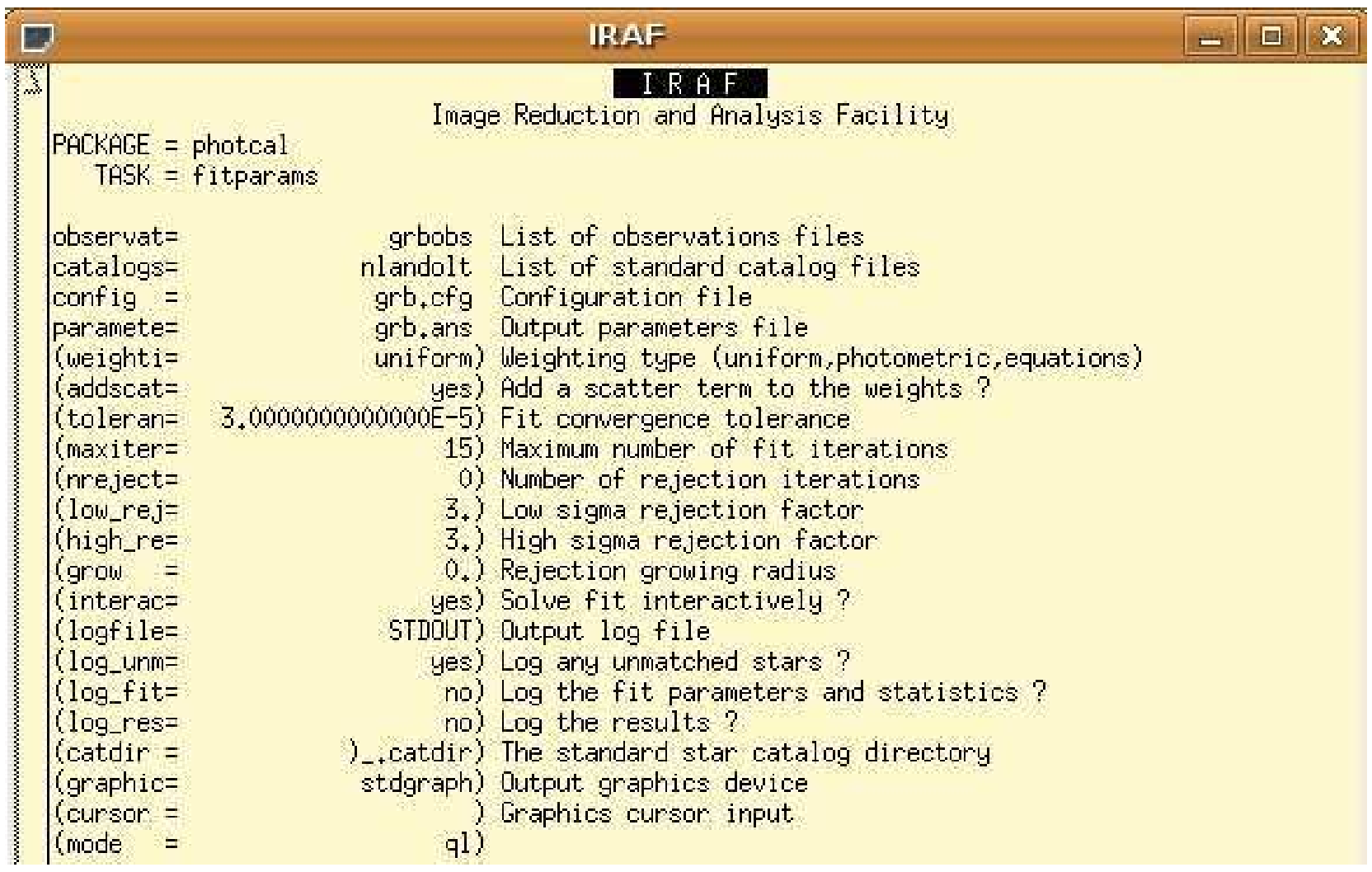

Figure 7.12 Example of the usage of the IRAF task **fitparams**.

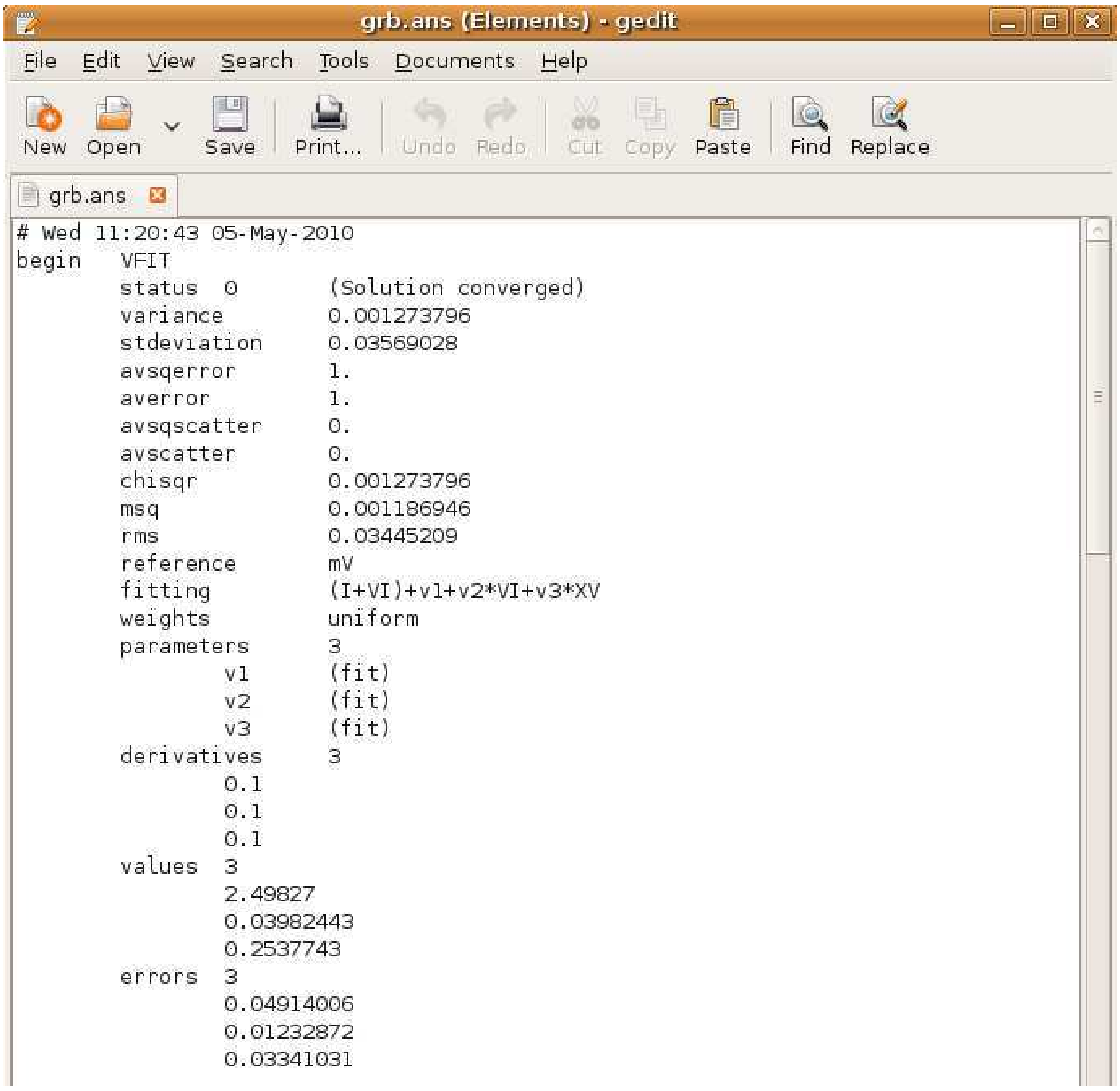

Figure 7.13 Example of the output file from **fitparams**. This file, called "grb.ans" displays the transformation specified in the configuration file as well as the values of the coefficients determined from the fit. Here we can see that the rms of the fit is $\sim 0.03$ mag, which is a decent calibration.

```
|
# Wed 11:21:31 05-May-2010
# List of observations files:
#               grbobs
# Config:       grb.cfg
# Parameters:   grb.ans
#
# Computed indices for program and standard objects
#
# Columns:
#       1       object id
#       2       I
#       3       error(I)
#       4       RI
#       5       error(RI)
#       6       BI
#       7       error(BI)
#       8       VI
#       9       error(VI)


PG1047-3    13.763  0.004   -0.155  0.005   -0.527  0.005   -0.260  0.005
PG1047-3A   12.678  0.002   0.425   0.003   1.610   0.005   0.860   0.004
PG1047-3B   13.989  0.005   0.366   0.007   1.429   0.010   0.772   0.007
PG1047-3C   11.724  0.001   0.360   0.002   1.333   0.003   0.732   0.002
PG1047-S2   16.590  0.017   0.347   0.023   1.358   0.040   0.754   0.025
PG1047-S6   15.645  0.011   1.141   0.018   3.703   0.081   2.086   0.027
PG1047-S7   16.928  0.022   0.416   0.029   1.374   0.051   0.802   0.032
PG1047-S12  15.983  0.013   0.706   0.019   2.816   0.058   1.563   0.026
PG1047-S13  15.755  0.011   0.419   0.015   1.637   0.026   0.883   0.017
```

Figure 7.14 The output file from the IRAF task **invertfit**. This file is the result of taking the instrumental magnitudes in the file "grbobs" and applying the transformation computed in the file "grb.ans". These are our calibrated magnitudes.

of standard stars that we can use to calibrate images of XRF 100316D taken on *any* telescope against). We will apply the transformation to the $BVR_ci$ images taken on FTS, and to do this, we need to:

1. Align the images.
2. Hand pick the stars to be the secondary standards.
3. Measure the psf FWHM and pick a suitable aperture.
4. Perform aperture photometry on the set of images.
5. Determine the aperture correction.
6. Make an image set file.
7. Make an observations file, using the same configuration file as before.

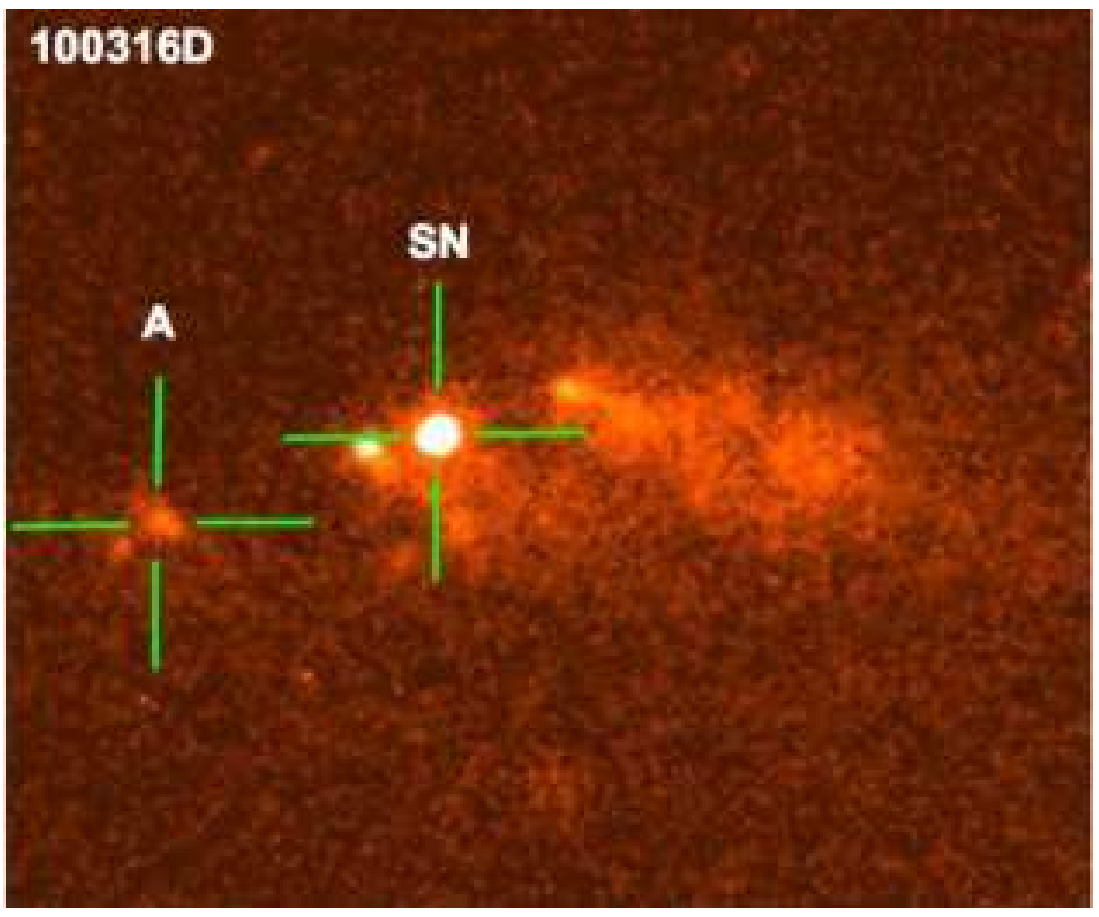

Figure 7.15 The host galaxy system of XRF 100316D / SN 2010bh. The *HST* image is taken from Starling et al. (2011).

8. Invertfit the transformations determined from PG-1047 and PG-1323 onto the instrumental magnitudes of the secondary standards in the field of XRF 100316D to obtain calibrated magnitudes.

After the last task has been completed, we now have a catalog of standard stars in the field of our XRF. We can now use these magnitudes to calibrate all of the epochs of observations of XRF 100316D / SN 2010bh against, regardless of what telescope is used to take the image.

However, photometry of the XRF cannot be done yet due to its position inside a bright galaxy (see Figure 7.15). To isolate the flux coming only from the OT, we need to first perform image subtraction on the FTS images.

SECTION 7.2

## Image Subtraction

Image subtraction was carried out on the FTS and Gemini-S images using the method prescribed in Alard (2000), in which images are matched against each other by using a space-varying convolution kernel.

Late-time images were obtained on both telescopes, in the same filters as those of the earlier epochs, and taken long after light from the SN had faded away. These were then used as templates that are subtracted from each of the earlier epochs for which light from the OT can still be detected.

The general method for performing the image subtracting is as follows:

1. Obtain template images in the same filters as the earlier epochs.
2. Subtract the template image from the earlier epochs to remove light from the host galaxy.

While the method is conceptually straight-forward, there are numerous parameters that need to be determined so to fully and effectively optimize the image subtraction procedure. Basically, the program looks for a pre-specified number of bright objects and defines small windows called *stamps* around each of them. The user has the option to either let the program decide which stamps are used (and therefore the user need to specify how may stamps are to be used), or the user can specify a pre-chosen list of stamps (which are the coordinates of the stars to be used) via a text file. Additionally, the program has other parameters that need to be specified including the: template image, epoch image, saturation level of the template and epoch image, the psf FWHM of the template and epoch images as well as the "mesh-size", which is the size of the space-varying kernel.

Here we have hand chosen our list of stars instead of letting the program choose them. We have examined the individual template images (in all filters) in IRAF using **imexamine** to check each bright object if they have a stellar psf (i.e. they are a point source and not a galaxy). Once the list was chosen, using as many stars as possible, the program was run in a terminal window:

```
>> runalard.pl template_image epoch_image −saturation_template −saturation_image
−FWHM_template −FWHM_image −mesh_size −list_of_stamps
```

In our example, the image subtraction procedure was run many times for each individual epoch, where the kernel mesh size was varied so to obtain the optimum subtracted image. The subtracted image must be free of systematic pattern (such as a gradient across the image), and with a noise level that did not vary greatly. Thus, for each subtracted image, we measured the magnitude of the OT, as well as the standard deviation of the background pixels. Images that had a clear gradient, and a large standard deviation of the background pixels were immediately rejected.

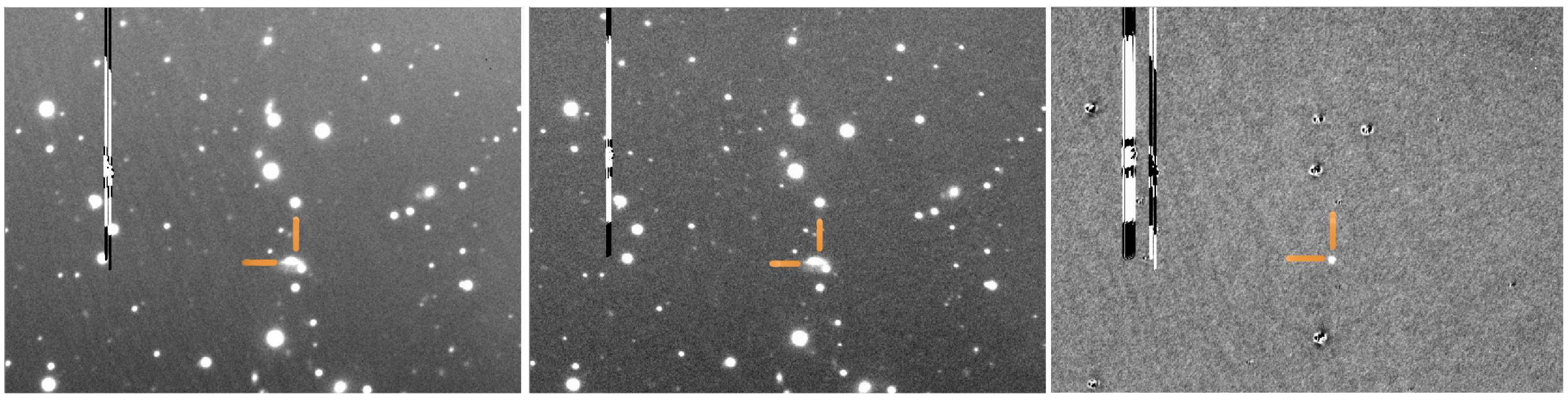

Figure 7.16 Example of image subtraction using the *i*-band images of XRF 100316D / SN 2010bh taken on FTS. The host galaxy system can be seen just below the center of the template and single epoch images, while in the subtracted image light coming from the OT is clearly visible. *Left:* Template image, taken long after the SN had faded away. *Center:* Single epoch for which light from the OT is contributing to the total flux. *Right:* Subtracted image. Light from the host galaxy has been removed and all that remains is light from the OT.

Additionally, the appearance of the residuals of the subtracted stars was noted. If we examine Figure 7.16, we can see that in the subtracted image (far right) there are a few residuals of some bright stars. Ideally, if all the stars in an image are not saturated and have similar psf FWHMs, then the image subtraction procedure should completely remove them. However we can clearly see some residuals, which is primarily due to these stars being saturated in the template and single epoch images.

We can also see in Figure 7.16 that the image subtraction procedure has correctly subtracted the amount of light due to the host galaxy, leaving light only from the OT behind in the subtracted image.

Once all of the images have undergone the image-subtraction process, the resultant, image-subtracted images were calibrated. The general procedure for calibrating the subtracted images is as follows:

1. Using the single epoch (un-subtracted) images, the image psf FWHM was mea-

sured, and a suitable aperture was determined.

2. An aperture correction was determined for each of the single epoch (un-subtracted) images.

3. Instrumental magnitudes of the OT in the subtracted image were obtained using the aperture determined in step (1) and IRAF task **phot**.

4. Use the catalog of secondary standard stars in the field of the XRF, a zero-point was determined.

5. The aperture correction and zero-point are applied to the instrumental magnitudes to produce calibrated magnitudes of the XRF.

# Bibliography

Achterberg, A., Gallant, Y. A., Kirk, J. G., & Guthmann, A. W. 2001, MNRAS, 328, 393

Adelman-McCarthy et al 2005 IAU Circ., 8481

Alard, C. 2000, A& AS, 144, 363

Amati, L., et al. 2002, A&A, 390, 81

Amati, L. 2006, MNRAS, 372, 233

Anderson, J. P., & James, P. A. 2009, MNRAS, 399, 559

Anderson, J. P., Covarrubias, R. A., James, P. A., Hamuy, M., & Habergham, S. M. 2010, MNRAS, 407, 2660

Anupama, G. C., Sahu, D. K., Deng, J., Nomoto, K., Tominaga, N., Tanaka, M., Mazzali, P. A., & Prabhu, T. P. 2005, ApJL, 631, L125

Anupama, G. C., Gurugubelli, U. K., & Sahu, D. K. 2009, GRB Coordinates Network, 9576, 1

Arnett, W. D. 1969, Ap&SS, 5, 180

Arnett, W. D. 1982, ApJ, 253, 785

Barbon, R., Ciatti, F., & Rosino, L. 1973, Memorie della Societa Astronomica Italiana, 44, 65

Barbon, R., Ciatti, F., & Rosino, L. 1979, A&A, 72, 287

Barentine, J., et al. 2005, Central Bureau Electronic Telegrams, 304, 1

Barlow, M. J., Smith, L. J., & Willis, A. J. 1981, MNRAS, 196, 101

Beardmore, A. P., & Schady, P. 2009, GRB Coordinates Network, 9528, 1

Benetti, S., Branch, D., Turatto, M., Cappellaro, E., Baron, E., Zampieri, L., Della Valle, M., & Pastorello, A. 2002, MNRAS, 336, 91

Benetti, S., et al. 2011, MNRAS, 411, 2726

Berger, E., Fox, D. B., Kulkarni, S. R., Frail, D. A., & Djorgovski, S. G. 2007, ApJ, 660, 504

Bersier, D., et al. 2003, ApJL, 583, L63

Bersier, D., et al. 2004, GRB Coordinates Network, 2544, 1

Bersier, D., et al. 2006, ApJ, 643, 284

Bertola, F. 1964, Annales d'Astrophysique, 27, 319

Bethe, H. A. 1990, Reviews of Modern Physics, 62, 801

Beuermann, K., et al. 1999, A&A, 352, L26

Björnsson, G., Hjorth, J., Jakobsson, P., Christensen, L., & Holland, S. 2001, ApJL, 552, L121

Blandford, R. D., & Znajek, R. L. 1977, MNRAS, 179, 433

Blandford, R. D., & Payne, D. G. 1982, MNRAS, 199, 883

Binggeli, B., Leibundgut, B., & Tammann, G. A. 1984, IAU Circ., 3944, 1

Blinnikov, S. I., Novikov, I. D., Perevodchikova, T. V., & Polnarev, A. G. 1984, Soviet Astronomy Letters, 10, 177

Blondin, S., & Calkins, M. 2008, Central Bureau Electronic Telegrams, 1191, 2

Bloom, J. S., et al. 1999, Nat, 401, 453

Bloom, J. S., Frail, D. A., & Sari, R. 2001, AJ, 121, 2879

Bloom, J. S., Kulkarni, S. R., & Djorgovski, S. G. 2002a, AJ, 123, 1111

Bloom, J. S., et al. 2002b, ApJL, 572, L45

Bloom, J. S., et al. 2009, ApJ, 691, 723

Blustin, A. J., et al. 2006, ApJ, 637, 901

Boissier, S., & Prantzos, N. 2009, A&A, 503, 137

Branch, D., Livio, M., Yungelson, L. R., Boffi, F. R., & Baron, E. 1995, PASP, 107, 1019

Brown, P. J., et al. 2009, AJ, 137, 4517

Bruenn, S. W., Mezzacappa, A., Hix, W. R., Blondin, J. M., Marronetti, P., Messer, O. E. B., Dirk, C. J., & Yoshida, S. 2009, Journal of Physics Conference Series, 180, 012018

Bruzual, G., & Charlot, S. 2003, MNRAS, 344, 1000

Bufano, F., Benetti, S., Sollerman, J., Pian, E., & Cupani, G. 2011, Astronomische Nachrichten, 332, 262

Burrows, D. N., Hill, J. E., Nousek, J. A. et al., 2005, SSRv, 120, 165

Burrows, A., Livne, E., Dessart, L., Ott, C. D., & Murphy, J. 2006, ApJ, 640, 878

Butler, N. R., et al. 2005, ApJ, 621, 884

Cadonau, R., & Leibundgut, B. 1990, A&AS, 82, 145

Campana, S., et al. 2006, Nat, 442, 1008

Cano, Z., Guidorzi, C., Bersier, D., Melandri, A., Steele, I. A., Smith, R. J., & Mundell, C. 2009, GRB Coordinates Network, 9531, 1

Cano, Z., et al. 2011, MNRAS, 413, 669

Cano, Z., et al. 2011, arXiv:1104.5141

Castander, F. J., & Lamb, D. Q. 1999, ApJ, 523, 602

Castro-Tirado, A. J., et al. 2005, A&A, 439, L15

Castro-Tirado, A. J., et al. 2007, A&A, 475, 101

Cenko, S. B., Perley, D. A., Junkkarinen, V., Burbidge, M., Diego, U. S., & Miller, K. 2009, GRB Coordinates Network, 9518, 1

Chandra, P., & Frail, D. A. 2009, GRB Coordinates Network, 9533, 1

Chevalier, R. A. 1976, ApJ, 207, 872

Chornock, R., et al. 2010, arXiv:1004.2262

Christensen, L., Hjorth, J., & Gorosabel, J. 2004, A&A, 425, 913

Clocchiatti, A., Wheeler, J. C., Benetti, S., & Frueh, M. 1996A, ApJ, 459, 547

Clocchiatti, A., et al. 1996B, AJ, 111, 1286

Clocchiatti, A., et al. 1997, ApJ, 483, 675

Clocchiatti, A., & Wheeler, J. C. 1997, ApJ, 491, 375

Clocchiatti, A., et al. 2000, ApJ, 529, 661

Clocchiatti, A., et al. 2001, ApJ, 553, 886

Cobb, B. E., Bailyn, C. D., van Dokkum, P. G., & Natarajan, P. 2006, ApJL, 645, L113

Cobb, B. E., Bloom, J. S., Perley, D. A., Morgan, A. N., Cenko, S. B., & Filippenko, A. V. 2010, ApJ, 718, L150

Colgate, S. A., & White, R. H. 1966, ApJ, 143, 626

Colgate, S. A. 1968, Canadian Journal of Physics, 46, 476

Colgate, S. A. 1971, ApJ, 163, 221

Conselice, C. J., et al. 2005, ApJ, 633, 29

Corsi, A., et al. 2011, arXiv:1101.4208

Costa, E., et al. 1997, Nature, 387, 783

Covino, S., et al. 2006, A&A, 447, L5

## Bibliography

Christensen, L., Hjorth, J., & Gorosabel, J. 2004, A& A, 425, 913

Christensen, L., Hjorth, J., Gorosabel, J., Vreeswijk, P., Fruchter, A., Sahu, K., & Petro, L. 2004, A& A, 413, 121

Crockett, R. M., et al. 2008, MNRAS, 391, L5

Crowther, P. A., Hillier, D. J., Evans, C. J., Fullerton, A. W., De Marco, O., & Willis, A. J. 2002, ApJ, 579, 774

Crowther, P. A., Hadfield, L. J., Clark, J. S., Negueruela, I., & Vacca, W. D. 2006, MNRAS, 372, 1407

Crowther, P. A. 2007, ARA&A, 45, 177

Curran, P. A., Evans, P. A., de Pasquale, M., Page, M. J., & van der Horst, A. J. 2010, ApJL, 716, L135

Dado, S., & Dar, A. 2010, ApJL, 708, L112

Della Valle, M., Malesani, D., Benetti, S., Testa, V., & Stella, L. 2003, IAU Circ., 8197, 2

Della Valle, M., et al. 2006a, ApJL, 642, L103

Della Valle, M., et al. 2006b, Nat, 444, 1050

Della Valle, M., et al. 2008, Central Bureau Electronic Telegrams, 1602, 1

Deng, J., Tominaga, N., Mazzali, P. A., Maeda, K., & Nomoto, K. 2005, ApJ, 624, 898

Djorgovski, S. G., Frail, D. A., Kulkarni, S. R., Bloom, J. S., Odewahn, S. C., & Diercks, A. 2001, ApJ, 562, 654

Drenkhahn, G., & Spruit, H. C. 2002, A&A, 391, 1141

Drout, M. R., et al. 2010, arXiv:1011.4959

Eichler, D., Livio, M., Piran, T., & Schramm, D. N. 1989, Nature, 340, 126

Elias, J. H., Matthews, K., Neugebauer, G., & Persson, S. E. 1985, ApJ, 296, 379

Elmhamdi, A., Danziger, I. J., Cappellaro, E., Della Valle, M., Gouiffes, C., Phillips, M. M., & Turatto, M. 2004, A&A, 426, 963

Evans, P. A., Goad, M. R., Osborne, J. P., & Beardmore, A. P. 2009, GRB Coordinates Network, 9521, 1

Evans, P. A., et al. 2009, MNRAS, 397, 1177

Fatkhullin, T., Moskvitin, A., Castro-Tirado, A. J., & de Ugarte Postigo, A. 2009, GRB Coordinates Network, 9542, 1

Fernandez-Soto, A., Peris, V., & Alonso-Lorite, J. 2009, GRB Coordinates Network, 9536

Ferrero, P., et al. 2006, A&A, 457, 857

Filippenko, A. V., Porter, A. C., Sargent, W. L. W., & Schneider, D. P. 1986, AJ, 92, 1341

Filippenko, A. V., Porter, A. C., & Sargent, W. L. W. 1990, AJ, 100, 1575

Filippenko, A. V., Matheson, T., & Ho, L. C. 1993, ApJL, 415, L103

Filippenko, A. V. 1997, ARA&A, 35, 309

Folatelli, G., et al. 2006, ApJ, 641, 1039

Foley, R. J., et al. 2003, PASP, 115, 1220

Foley, R. J., Smith, N., Ganeshalingam, M., Li, W., Chornock, R., & Filippenko, A. V. 2007, ApJL, 657, L105

Foley, R. J., et al. 2009, AJ, 138, 376

Foley, R. J., Brown, P. J., Rest, A., Challis, P. J., Kirshner, R. P., & Wood-Vasey, W. M. 2010, ApJL, 708, L61

Fong, W., Berger, E., & Fox, D. B. 2010, ApJ, 708, 9

Fox, D. B., et al. 2005, Nature, 437, 845

Frail, D. A., Kulkarni, S. R., Nicastro, L., Feroci, M., & Taylor, G. B. 1997, Nature, 389, 261

Frail, D. A., et al. 2001, ApJL, 562, L55

Fruchter, A., Vreeswijk, P., Hook, R., & Pian, E. 2000, GRB Coordinates Network, 752, 1

Fruchter, A.S., et al. 2006, Nat, 441, 463

Fryer, C. L., & Warren, M. S. 2004, ApJ, 601, 391

Fryer, C. L., et al. 2007, PASP, 119, 1211

Fryer, C. L., et al. 2009, ApJ, 707, 193

Fukugita, M., Shimasaku, K., & Ichikawa, T. 1995, PASP, 107, 945

Fynbo, J. P. U., et al. 2003, A&A, 406, L63

Fynbo, J. P. U., et al. 2006, Nat, 444, 1047

Fynbo, J. P. U., et al. 2009, ApJS, 185, 526

Gal-Yam, A., et al. 2005, ApJL, 630, L29

Gal-Yam, A., et al. 2006, Nat, 444, 1053

Gal-Yam, A., et al. 2007, ApJ, 656, 372

Galama, T. J., et al. 1998, Nat, 395, 670

Galama, T. J., et al. 2000, ApJ, 536, 185

Garnavich, P., Jha, S., Kirshner, R., Challis, P., Balam, D., Brown, W., & Briceno, C. 1997a, IAU Circ., 6786, 1

Garnavich, P., Jha, S., Kirshner, R., Challis, P., Balam, D., Berlind, P., Thorstensen, J., & Macri, L. 1997b, IAU Circ, 6798, 2

Garnavich, P. M., et al. 2003, ApJ, 582, 924

Gaskell, C. M., Cappellaro, E., Dinerstein, H. L., Garnett, D. R., Harkness, R. P., & Wheeler, J. C. 1986, ApJL, 306, L77

Gehrels, N., et al. 2005, Nature, 437, 851

Gehrels, N., et al. 2006, Nature, 444, 1044

Gezari, S., et al. 2008, ApJL, 683, L131

Gezari, S., et al. 2010, ApJL, 720, L77

Ghirlanda, G., Ghisellini, G., & Lazzati, D. 2004, ApJ, 616, 331

Ghirlanda, G., Nava, L., & Ghisellini, G. 2010, A& A, 511, A43

Gilmozzi, R., & Panagia, N. 1999, MEMSAI, 70, 583

Goodman, J. 1986, ApJL 308, L47

Gorosabel, J., et al. 2003, A&A, 400, 127

Gorosabel, J., et al. 2003, A& A, 400, 127

Gorosabel, J., et al. 2005, A&A, 444, 711

Graham, J. F., Fruchter, A. S., Kewley, L. J., Levesque, E. M., Levan, A. J., Tanvir, N. R., Reichart, D. E., & Nysewander, M. 2009, American Institute of Physics Conference Series, 1133, 269

Granot, J., Piran, T., & Sari, R. 1999, ApJ, 527, 236

Greiner, J., Peimbert, M., Estaban, C., Kaufer, A., Jaunsen, A., Smoke, J., Klose, S., & Reimer, O. 2003, GRB Coordinates Network, 2020, 1

Groot, P. J., et al. 1998, ApJL, 493, L27

Grupe, D., et al. 2006, GRB Coordinates Network, 5365, 1

Grupe, D., et al. 2007, ApJ, 662, 443

Grupe, D., et al. 2010, ApJ, 711, 1008

Hakobyan, A. A. 2008, Astrophysics, 51, 69

Hamuy, M., Suntzeff, N. B., Gonzalez, R., & Martin, G. 1988, AJ, 95, 63

Hatano, K., Branch, D., Nomoto, K., Deng, J. S., Maeda, K., Nugent, P., & Aldering, G. 2001, Bulletin of the American Astronomical Society, 33, 838

Heger, A., Fryer, C. L., Woosley, S. E., Langer, N., & Hartmann, D. H. 2003, ApJ, 591, 288

Hendry, M. A., et al. 2006, MNRAS, 369, 1303

Hjorth, J., Holland, S., Courbin, F., Dar, A., Olsen, L. F., & Scodeggio, M. 2000, ApJL, 534, L147

Hjorth, J., et al. 2003, Nature, 423, 847

Holland, S. T., et al. 2010, ApJ, 717, 223

Holtzman, J. A., Burrows, C. J., Casertano, S., Hester, J. J., Trauger, J. T., Watson, A. M., & Worthey, G. 1995, PASP, 107, 1065

Hoyle, F., & Fowler, W. A. 1960, ApJ, 132, 565

Hu, J. Y., et al. 1997, IAU Circ., 6783, 1

Humphreys, R. M., Nichols, M., & Massey, P. 1985, AJ, 90, 101

Hunter, D. J., et al. 2009, A&A, 508, 371

Iben, I., Jr., & Tutukov, A. V. 1984, ApJS, 54, 335

Im, M., Park, W. K., & Urata, Y. 2009a, GRB Coordinates Network, 9522, 1

Im, M., Jeon, Y.-B., & Urata, Y. 2009b, GRB Coordinates Network, 9541, 1

Im, M., Ko, J., Cho, Y., Choi, C., Jeon, Y., Lee, I., & Ibrahimov, M. 2010, Journal of Korean Astronomical Society, 43,

Iwamoto, K., Nomoto, K., Hoflich, P., Yamaoka, H., Kumagai, S., & Shigeyama, T. 1994, ApJL, 437, L115

Iwamoto, K., et al. 1998, Nature, 395, 672

Iwamoto, K., et al. 2000, ApJ, 534, 660

Jakobsson, P., Hjorth, J., Fynbo, J. P. U., Watson, D., Pedersen, K., Björnsson, G., & Gorosabel, J. 2004, ApJL, 617, L21

Janka, H.-T., Langanke, K., Marek, A., Martínez-Pinedo, G., Muumlller, B. 2007, Phys Rev P, 442, 38

Joseph, R. D., Meikle, W. P. S., Robertson, N. A., & Wright, G. S. 1984, MNRAS, 209, 111 the

Jordi, K., Grebel, E. K., & Ammon, K. 2006, A& A, 460, 339

Kalberla, P. M. W., Burton, W. B., Hartmann, D., Arnal, E. M., Bajaja, E., Morras, R., P&oumlppel, W. G. L. 2005, A& A, 440, 775

Kamble, A., van der Horst, A. J., & Wijiers, R. 2009, GRB Coordinates Network, 9538, 1

Kann, D. A., Klose, S., & Zeh, A. 2006, ApJ, 641, 993

Kann, D. A., et al 2010, ApJ, 720, 1513

Kelly, P. L., Kirshner, R. P., & Pahre, M. 2008, ApJ, 687, 1201

Kennicutt, R. C., Jr., & Keel, W. C. 1984, ApJL, 279, L5

Kirk, J. G., Guthmann, A. W., Gallant, Y. A., & Achterberg, A. 2000, ApJ, 542, 235

Klebesadel, R. W., Strong, I. B., & Olson, R. A. 1973, ApJL, 182, L85

Klose, S., et al. 2004, AJ, 128, 1942

Klunko, E., Volnova, A., & Pozanenko, A. 2009, GRB Coordinates Network, 9613, 1

Koekemoer, A. M., Fruchter, A. S., Hook, R. N., & Hack, W. 2002, The 2002 HST Calibration Workshop : Hubble after the Installation of the ACS and the NICMOS Cooling System, 337

Kotak, R., Meikle, W. P. S., Adamson, A., & Leggett, S. K. 2004, MNRAS, 354, L13

Kouveliotou, C., Meegan, C. A., Fishman, G. J., Bhat, N. P., Briggs, M. S., Koshut, T. M., Paciesas, W. S., & Pendleton, G. N. 1993, ApJL, 413, L101

Kudritzki, R.-P., & Puls, J. 2000, ARA&A, 38, 613

Kumar, P., & Panaitescu, A. 2000, ApJL, 541, L9

Küpcü Yoldaş, A., Salvato, M., Greiner, J., Pierini, D., Pian, E., & Rau, A. 2007, A&A, 463, 893

Landolt, A. U. 1992, AJ, 104, 340

Lee, I., Im, M., & Urata, Y. 2010, Journal of Korean Astronomical Society, 43, 95

Leloudas, G., Sollerman, J., Levan, A. J., Fynbo, J. P. U., Malesani, D., & Maund, J. R. 2010, A&A, 518, A29

Leloudas, G., Gallazzi, A., Sollerman, J., et al. 2011, A&A, 530, A95

Levan, A., et al. 2005, ApJ, 624, 880

Levan, A., et al. 2006, ApJ, 647, 471

Levesque, E. M., Berger, E., Kewley, L. J., & Bagley, M. M. 2010, AJ, 139, 694

Lewis, J. R., et al. 1994, MNRAS, 266, L27

Li, W., Van Dyk, S. D., Filippenko, A. V., Cuillandre, J.-C., Jha, S., Bloom, J. S., Riess, A. G., & Livio, M. 2006, ApJ, 641, 1060

Li, W., Wang, X., Van Dyk, S. D., Cuillandre, J.-C., Foley, R. J., & Filippenko, A. V. 2007, ApJ, 661, 1013

Li, W., et al. 2011, MNRAS, 412, 1441

Litvinova, I. I., & Nadezhin, D. K. 1983, Ap&SS, 89, 89

Lyutikov, M., & Blackman, E. G. 2001, MNRAS, 321, 177

Lyutikov, M., & Blandford, R. 2003, arXiv:astro-ph/0312347

MacFadyen, A. I., & Woosley, S. E. 1999, ApJ, 524, 262

MacFadyen, A. I., Woosley, S. E., & Heger, A. 2001, ApJ, 550, 410

MacFadyen, A. I. 2003, From Twilight to Highlight: The Physics of Supernovae, 97

Maeda, K., Mazzali, P. A., & Nomoto, K. 2006, ApJ, 645, 1331

Malesani, D., et al. 2004, ApJL, 609, L5

Malesani, D., et al. 2008, GRB Coordinates Network, 7169, 1

Margutti, R., et al. 2007, A&A, 474, 815

Margutti, R., Genet, F., Granot, J. et al., 2010, MNRAS, 402, 46

Martin, R., Williams, A., Woodings, S., Biggs, J., & Verveer, A. 1999, IAU Circ., 7310, 1

Massey, P., Johnson, K. E., & Degioia-Eastwood, K. 1995, ApJ, 454, 151

Massey, P. 2003, ARA&A, 41, 15

Masetti, N., et al. 2003, A& A, 404, 465

Matheson, T., Filippenko, A. V., Li, W., Leonard, D. C., & Shields, J. C. 2001, AJ, 121, 1648

Matheson, T., Filippenko, A. V., Chornock, R., Leonard, D. C., & Li, W. 2000, AJ, 119, 2303

Matheson, T., et al. 2003, ApJ, 599, 394

Maza, J., & van den Bergh, S. 1976, ApJ, 204, 519

Mazzali, P. A. Iwamoto, K., & Nomoto, K. 2000, ApJ, 545, 407

Mazzali, P. A., et al. 2002, ApJL, 572, L61

Mazzali, P. A., et al. 2006a, ApJ, 645, 1323

Mazzali, P. A., et al. 2006b, Nature, 442, 1018

Mazzali, P. A., et al. 2007, ApJ, 670, 592

Mazzali, P. A., et al. 2008, Science, 321, 1185

Maund, J. R., Smartt, S. J., & Danziger, I. J. 2005, MNRAS, 364, L33

Maury, A., Hook, I., Gorki, R., Selman, F., & Dennefeld, M. 2000, IAU Circ., 7528, 2

McKenzie, E. H., & Schaefer, B. E. 1999, PASP, 111, 964

Meegan, C. A., Fishman, G. J., Wilson, R. B., Horack, J. M., Brock, M. N., Paciesas, W. S., Pendleton, G. N., & Kouveliotou, C. 1992, Nature, 355, 143

Melandri, A., et al. 2008, ApJ, 686, 1209

Melandri, A., Guidorzi, C., Bersier, D., Cano, Z., Steele, I. A., Mundell, C. G., O'Brien, P., & Tanvir, N. 2009, GRB Coordinates Network, 9520, 1

Meszaros, P., & Rees, M. J. 1997, ApJL, 482, L29

Metzger, M. R., Djorgovski, S. G., Kulkarni, S. R., Steidel, C. C., Adelberger, K. L., Frail, D. A., Costa, E., & Frontera, F. 1997, Nature, 387, 878

Meynet, G., Maeder, A., Schaller, G., Schaerer, D., & Charbonnel, C. 1994, A&AS, 103, 97

Miller, D. L., & Branch, D. 1990, AJ, 100, 530

Minkowski, R. 1941, PASP, 53, 224

Mochkovitch, R., Hernanz, M., Isern, J., & Martin, X. 1993, Nature, 361, 236

Modjaz, M., et al. 2006, ApJL, 645, L21

Modjaz, M., et al. 2008, AJ, 135, 1136

Modjaz, M., Kewley, L., Bloom, J. S., et al. 2011, ApJL, 731, L4

Moretti et al., 2005, Proceeding of SPIE, Vol. 5898, pp. 360-368

Mokiem, M. R., et al. 2007, A&A, 465, 1003

Moriya, T., Tominaga, N., Tanaka, M., Nomoto, K., Sauer, D. N., Mazzali, P. A., Maeda, K., & Suzuki, T. 2010, ApJ, 719, 1445

Nakagawa, Y. E., et al. 2006, PASJ, 58, L35

Nakamura, T., Mazzali, P. A., Nomoto, K., & Iwamoto, K. 2001, ApJ, 550, 991

Nakano, S., Kadota, K., Itagaki, K., & Corelli, P. 2008, IAU Circ., 8908, 2

Nakar, E. 2007, Phys Rev P., 442, 166

Narayan, R., Piran, T., & Shemi, A. 1991, ApJL, 379, L17

Narayan, R., Paczynski, B., & Piran, T. 1992, ApJL, 395, L83

Nomoto, K., Sugimoto, D., & Neo, S. 1976, Ap&SS, 39, L37

Nomoto, K. 1982, ApJ, 253, 798

Nomoto, K., Filippenko, A. V., & Shigeyama, T. 1990, A& A, 240.

Nomoto, K., Yamaoka, H., Pols, O. R., van den Heuvel, E. P. J., Iwamoto, K., Kumagai, S., & Shigeyama, T. 1994, Nature, 371, 227

Nomoto, K. I., Iwamoto, K., & Suzuki, T. 1995, PHYSREP, 256, 173

Norris, J. P., Gehrels, N., & Scargle, J. D. 2010, ApJ, 717, 411

Nousek, J. A., et al. 2006, ApJ, 642, 389

Oates, S. R., et al. 2006, MNRAS, 372, 327

Ofek, E. O., et al. 2007, ApJ, 662, 1129

Ofek, E. O., et al. 2010, ApJ, 724, 1396

Paragi, Z., et al. 2010, Nat, 463, 516

Paczynski, B. 1986, ApJL, 308, L43

Paczynski, B. 1991, ACTAA, 41, 257

Paczynski, B., & Xu, G. 1994, ApJ, 427, 708

Paczynski, B. 1998, ApJL, 494, L45

Patat, F., et al. 2001,ApJ, 555, 900

Pei, Y. C. 1992, ApJ, 395, 130

Pellizza, L. J., et al. 2006, A&A, 459, L5

Perley, D. A., et al. 2009, AJ, 138, 1690

Perlmutter, S., et al. 1997, ApJ, 483, 565

Perlmutter, S., et al. 1999, ApJ, 517, 565

Phillips, M. M. 1993, ApJL, 413, L105

Pian, E., et al. 2006, Nature, 442, 1011

Piran, T. 2004, Reviews of Modern Physics, 76, 1143

Piro, L., et al. 2002, ApJ, 577, 680

Pignata, G., et al. 2011, ApJ, 728, 14

Podsiadlowski, P., Joss, P. C., & Hsu, J. J. L. 1992, ApJ, 391, 246

Pollock, A. M. T. 1987, ApJ, 320, 283

Poole, T. S., et al. 2008, MNRAS, 383, 627

Pooley, G. 2009, GRB Coordinates Network, 9532, 1

Pooley, G. 2009, GRB Coordinates Network, 9592, 1

Popham, R., Woosley, S. E., & Fryer, C. 1999, ApJ, 518, 356

Prantzos, N., & Boissier, S. 2003, A&A, 406, 259

Price, P. A., et al. 2003, ApJ, 589, 838

Prieto, J. L., Stanek, K. Z., & Beacom, J. F. 2008, ApJ, 673, 999

Prochaska, J. X., et al. 2004, ApJ, 611, 200

Proga, D., MacFadyen, A. I., Armitage, P. J., & Begelman, M. C. 2003, ApJL, 599, L5

Puls, J., et al. 1996, A&A, 305, 171

Qiu, Y. L., Hatano, K., Branch, D., & Baron, E. 1999, Bulletin of the American Astronomical Society, 31, 1425

Rees, M. J., & Meszaros, P. 1992, MNRAS, 258, 41P

Rees, M. J., & Meszaros, P. 1994, ApJL, 430, L93

Reichart, D. E., & Price, P. A. 2002, ApJ, 565, 174

Rhoads, J. E. 1997, ApJL, 487, L1

Rhoads, J. E. 1999, ApJ, 525, 737

Richardson, D., Branch, D., Casebeer, D., Millard, J., Thomas, R. C., & Baron, E. 2002, AJ, 123, 745

Richardson, D., Branch, D., & Baron, E. 2006, AJ, 131, 2233

Richardson, D. 2009, AJ, 137, 347

Richmond, M. W., et al. 1996, AJ, 111, 327

Riess, A. G., Press, W. H., & Kirshner, R. P. 1996, ApJ, 473, 88

Riess, A. G., et al. 1998, AJ, 116, 1009

Romano, P., et al. 2006, A&A, 456, 917

Rumyantsev, V., & Pozanenko, A. 2009, GRB Coordinates Network, 9539, 1

Sahu, K. C., et al. 2000, ApJ, 540, 74

Sahu, D. K., Anupama, G. C., Srividya, S., & Muneer, S. 2006, MNRAS, 372, 1315

Sahu, D. K., Tanaka, M., Anupama, G. C., Gurugubelli, U. K., & Nomoto, K. 2009, ApJ, 697, 676

Sahu, D. K., Gurugubelli, U. K., Anupama, G. C., & Nomoto, K. 2011, MNRAS, 383

Sakamoto, T., Ukwatta, T. N., & Barthelmy, S. D. 2009, GRB Coordinates Network, 9534, 1

Sari, R., Piran, T., & Narayan, R. 1998, ApJL, 497, L17

Sari, R., Piran, T., & Halpern, J. P. 1999, ApJL, 519, L17

Sari, R., & Piran, T. 1999, ApJ, 520, 641

Sauer, D. N., Mazzali, P. A., Deng, J., Valenti, S., Nomoto, K., & Filippenko, A. V. 2006, MNRAS, 369, 1939

Savaglio, S., Glazebrook, K., & Le Borgne, D. 2009, ApJ, 691, 182

Schady, P., et al. 2007, MNRAS, 377, 273

Schady, P., et al. 2009, GRB Coordinates Network, 9512, 1

Schlegel, E. M. 1990, MNRAS, 244, 269

Schlegel, D. J., Finkbeiner, D. P., & Davis, M. 1998, ApJ, 500, 525

Schmidt, B. P., et al. 1993, Nature, 364, 600

Schawinski, K., et al. 2008, Science, 321, 223

Shen, R., Kumar, P., & Robinson, E. L. 2006, MNRAS, 371, 1441

Shigeyama, T., Nomoto, K., Tsujimoto, T., & Hashimoto, M.-A. 1990, ApJL, 361, L23

Sirianni, M., et al. 2005, PASP, 117, 1049

Smartt, S. J., Maund, J. R., Hendry, M. A., Tout, C. A., Gilmore, G. F., Mattila, S., & Benn, C. R. 2004, Science, 303, 499

Smartt, S. J. 2009, ARA&A, 47, 63

Smith, N., Li, W., Foley, R. J., et al. 2007, ApJ, 666, 1116

Soderberg, A. M. 2003, arXiv:astro-ph/0310305

Soderberg, A. M., et al. 2006, ApJ, 650, 261

Soderberg, A. M., et al. 2006, ApJ, 636, 391

Soderberg, A. M., et al. 2008, Nature, 453, 469

Soderberg, A. M., Brunthaler, A., Nakar, E., Chevalier, R. A., & Bietenholz, M. F. 2010, ApJ, 725, 922

Soderberg, A. M., et al. 2010, Nat, 463, 513

Sollerman, J., Kozma, C., Fransson, C., Leibundgut, B., Lundqvist, P., Ryde, F., & Woudt, P. 2000, ApJL, 537, L127

Sollerman, J., et al. 2002, A&A, 386, 944

Sollerman, J., Östlin, G., Fynbo, J. P. U., Hjorth, J., Fruchter, A., & Pedersen, K. 2005, New Astronomy, 11, 103

Sollerman, J., et al. 2006, A& A, 454, 503

Sollerman, J., et al. 2007, A&A, 466, 839

Stanek, K. Z., et al. 2003, ApJL, 591, L17

Stanek, K. Z., et al. 2005, ApJL, 626, L5

Stanek, K. Z., et al. 2006, Acta Astronomica, 56, 333

Starling, R. L. C., Wijers, R. A. M. J., Wiersema, K., Rol, E., Curran, P. A., Kouveliotou, C., van der Horst, A. J., & Heemskerk, M. H. M. 2007, ApJ, 661, 787

Starling, R. L. C., van der Horst, A. J., Rol, E., Wijers, R. A. M. J., Kouveliotou, C., Wiersema, K., Curran, P. A., & Weltevrede, P. 2008, ApJ, 672, 433

Starling, R. L. C., et al. 2010, arXiv:1004.2919

Stritzinger, M., et al. 2002, AJ, 124, 2100

Stritzinger, M., et al. 2009, ApJ, 696, 713

Strong, I. B.,lebesadel, R. W., & Olson, R. A. 1974, ApJL, 188, L1

Svensson, K. M., Levan, A. J., Tanvir, N. R., Fruchter, A. S., & Strolger, L.-G. 2010, MNRAS, 405, 57

Tanvir, N. R., et al. 2008, MNRAS, 388, 1743

Tanvir, N. R., et al. 2010, ApJ, 725, 625

Taubenberger, S., et al. 2006, MNRAS, 371, 1459

Thatte, D. and Dahlen, T. et al. 2009, NICMOS Data Handbook, version 8.0, (Baltimore, STScI).

Thöne, C. C., et al. 2006, GRB Coordinates Network, 5373, 1

Thöne, C. C., et al. 2011, arXiv:1105.3015

Thompson, C. 1994, MNRAS, 270, 480

Tinney, C., Stathakis, R., Cannon, R., & Galama, T. 1998, IAU Circ., 6896, 3

Tominaga, N., et al. 2005, ApJL, 633, L97

Tominaga, N., Maeda, K., Umeda, H., Nomoto, K., Tanaka, M., Iwamoto, N., Suzuki, T., & Mazzali, P. A. 2007, ApJL, 657, L77

Tomita, H., et al. 2006, ApJ, 644, 400

Tsvetkov, D. Y. 1983, Peremennye Zvezdy, 22, 39

Tsvetkov, D. Y. 1987, Soviet Astronomy Letters, 13, 376

Tsvetkov, D. Y., Pavlyuk, N. N., & Bartunov, O. S. 2004, Astronomy Letters, 30, 729

Turatto, M., Benetti, S., & Cappellaro, E. 2003, From Twilight to Highlight: The Physics of Supernovae, 200

Turatto, M., Benetti, S., & Pastorello, A. 2007, Supernova 1987A: 20 Years After: Supernovae and Gamma-Ray Bursters, 937, 187

Uomoto, A., & Kirshner, R. P. 1985, A&A, 149, L7

Urata, Y., et al. 2007, PASJ, 59, L29

Usov, V. V. 1992, Nat, 357, 472

Valenti, S., et al. 2008, MNRAS, 383, 1485

Valenti, S., et al. 2009, Nat, 459, 674

van den Bergh, S. 1997, AJ, 113, 197

van den Bergh, S., Li, W., & Filippenko, A. V. 2005, PASP, 117, 773

van der Horst, A. J., Kouveliotou, C., Gehrels, N., Rol, E., Wijers, R. A. M. J., Cannizzo, J. K., Racusin, J., & Burrows, D. N. 2009, ApJ, 699, 1087

van der Horst, A. J., et al. 2011, ApJ, 726, 99

van Dyk, S. D. 1992, AJ, 103, 1788

Vaughan, S., Goad, M. R., Beardmore, A. P., et al., 2006, ApJ, 638, 920

Vinkó, J., Blake, R. M., Sárneczky, K., et al. 2004, A&A, 427, 453

Vreeswijk, P. M., et al. 2008, GRB Coordinates Network, 7444, 1

Wainwright, C., Berger, E., & Penprase, B. E. 2007, ApJ, 657, 367

Webbink, R. F. 1984, ApJ, 277, 355

Wei, J. Y., Hu, J. Y., Qiu, Y. L., Qiao, Q. Y., Deng, J. S., & You, J. H. 1997, IAU Circ., 6797, 2

Wheeler, J. C., & Levreault, R. 1985, ApJL, 294, L17

Wheeler, J. C., et al. 1993, ApJL, 417, L71

Wheeler, J. C., Yi, I., Höflich, P., & Wang, L. 2000, ApJ, 537, 810

Whelan, J., & Iben, I., Jr. 1973, ApJ, 186, 1007

Wild, P. 1960, PASP, 72, 97

Woosley, S. E., & Weaver, T. A. 1986, ARA&A, 24, 205

Woosley, S. E., & Baron, E. 1992, ApJ, 391, 228

Woosley, S. E. 1993, ApJ, 405, 273

Woosley, S. E., & Weaver, T. A. 1995, ApJS, 101, 181

Woosley, S. E., Heger, A., & Weaver, T. A. 2002, Reviews of Modern Physics, 74, 1015

Woosley, S. E., & Bloom, J. S. 2006, ARA& A, 44, 507

Woosley, S. E., & Heger, A. 2006, ApJ, 637, 914

Xu, M., Huang, Y.-F., & Lu, T. 2009, Research in Astronomy and Astrophysics, 9, 1317

Xu, D. W., & Qiu, Y. L. 2001, IAU Circ., 7555, 2

Yokoo, T., Arimoto, J., Matsumoto, K., Takahashi, A., & Sadakane, K. 1994, PASJ, 46, L191

Yoon, S.-C., & Langer, N. 2005, A&A, 443, 643

Yoshii, Y., et al. 2003, ApJ, 592, 467

Young, T. R., & Branch, D. 1989, ApJL, 342, L79

Zampieri, L., Pastorello, A., Turatto, M., Cappellaro, E., Benetti, S., Altavilla, G., Mazzali, P., & Hamuy, M. 2003, MNRAS, 338, 711

Zeh, A., Klose, S., & Hartmann, D. H. 2004, ApJ, 609, 952

Zeh, A., Klose, S., & Kann, D. A. 2006, ApJ, 637, 889

Zhang, B., & Mészáros, P. 2001, ApJL, 552, L35

Zhang, B., & Mészáros, P. 2004, International Journal of Modern Physics A, 19, 2385

Zhang, B., Fan, Y. Z., Dyks, J., Kobayashi, S., Mészáros, P., Burrows, D. N., Nousek, J. A., & Gehrels, N. 2006, ApJ, 642, 354